\documentclass[useAMS,usenatbib]{mn2e}
\usepackage{lscape,graphicx,amssymb,aas_macros,bm}
\usepackage{amsmath} 
\usepackage{indentfirst} 
\usepackage{pdflscape} 
\usepackage{placeins} 
\usepackage{nicefrac} 
\usepackage{afterpage} 
\usepackage{booktabs} 
\usepackage{morefloats} 
\usepackage{color} 
\usepackage{float,subfloat}
\usepackage[bottom]{footmisc}

\newcommand{\apg}{\:^{>}_{\sim}\:}

\newcommand{\cmjj}{\mbox{${\rm cm^{-2}}$}}
\newcommand{\cmjjj}{\mbox{${\rm cm^{-3}}$}}
\newcommand{\etal}{et al.}

\newcommand{\kms}{\mbox{km\ s${^{-1}}$}}

\newcommand{\MgI}{{\mbox{Mg\,{\scriptsize I}}}}

\title[CGM in Massive Quiescent Halos II]{Characterizing circumgalactic gas around massive ellipticals at $\bm{z}\sim 0.4$ - II. Physical properties and elemental abundances\thanks{Based on data gathered with the 6.5m Magellan Telescopes located at Las Campanas Observatory, the W.~M.~Keck Observatory, and the NASA/ESA Hubble Space Telescope operated by the Space Telescope Science Institute and the Association of Universities for Research in Astronomy, Inc., under NASA contract NAS 5-26555.}}

\author[Zahedy et al.]{Fakhri S.\ Zahedy$^{1}$\thanks{E-mail: fsz@uchicago.edu}, 
Hsiao-Wen Chen$^{1}$, Sean D.\ Johnson$^{2,3}$\thanks{Hubble \& Carnegie-Princeton Fellow}, Rebecca M.\ Pierce$^{1,4}$, \newauthor  Michael Rauch$^{3}$, Yun-Hsin Huang$^{5}$, Benjamin J.\ Weiner$^{5}$, and Jean-Ren\'{e} Gauthier$^{6}$ \\ \\
$^{1}$Department of Astronomy \& Astrophysics, The University of Chicago, Chicago, IL 60637, USA \\
$^{2}$Department of Astrophysical Sciences, Princeton University, Princeton, NJ 08544, USA \\
$^{3}$The Observatories of the Carnegie Institution for Science, 813 Santa Barbara Street, Pasadena, CA 91101, USA \\
$^{4}$Department of Aerospace Engineering, University of Maryland, MD 20742, USA \\
$^{5}$Steward Observatory, University of Arizona, Tucson, AZ 85721, USA \\
$^{6}$Oracle Corporation, Culver City, CA 90230, USA \\
}

\begin{document}

\pagerange{\pageref{firstpage}--\pageref{lastpage}} \pubyear{2018}

\maketitle

\label{firstpage}

\begin{abstract}

We present a systematic investigation of the circumgalactic medium
(CGM) within projected distances $d<160$ kpc of luminous red galaxies
(LRGs).  The sample comprises 16 intermediate-redshift ($z=0.21-0.55$)
LRGs of stellar mass $M_\mathrm{star}>10^{11}\,\mathrm{M}_\odot$. 
Combining far-ultraviolet Cosmic Origin
Spectrograph spectra from the {\it Hubble Space Telescope} and optical
echelle spectra from the ground enables a detailed ionization analysis
based on resolved component structures of a suite of absorption
transitions, including the full \ion{H}{I} Lyman series and various ionic 
metal transitions.  By 
comparing the relative abundances of different ions in individually-matched components, 
we show that cool gas ($T\sim10^4$ K) density and metallicity can vary by more than a
factor of ten in an LRG halo.  Specifically, metal-poor
absorbing components with $<1/10$ solar metallicity are seen in 50\%
of the LRG halos, while gas with solar and super-solar metallicity is
also common. 
These results indicate a
complex multiphase structure and poor chemical mixing in these
quiescent halos.  
We calculate the total surface mass density of cool
gas, $\Sigma_\mathrm{cool}$, by applying the estimated ionization
fraction corrections to the observed \ion{H}{I} column densities. 
The radial profile of $\Sigma_\mathrm{cool}$ is best-described by a 
projected Einasto profile of slope $\alpha=1$ and scale radius $r_s=48$ kpc. 
We find that typical LRGs at $z\sim0.4$ contain cool gas mass of
$\mathrm{\mathit{M}_{cool} = (1-2)\times10^{10}\,\mathrm{M_\odot}}$
at $d<160$ kpc (or as much as $\mathrm{\mathit{M}_{cool} \approx 4\times10^{10}\,\mathrm{M_\odot}}$ at $d<500$ kpc), comparable to the cool
CGM mass of star-forming galaxies.  
Furthermore, we show that high-ionization \ion{O}{VI} and
low-ionization absorption species exhibit distinct velocity profiles, highlighting their different physical 
origins. We discuss the implications of our findings for the origin and fate of
cool gas in LRG halos.

\end{abstract}


\section{Introduction}
\label{section:introduction}

Substantial efforts have been made in the last two decades to identify and characterize the physical processes which are at play in the gaseous halo surrounding galaxies, known as the circumgalactic medium (CGM; see recent reviews by Chen 2017 and Tumlinson \etal\ 2017 and references therein). The CGM is situated between the intergalactic medium (IGM), where most baryons in the Universe reside, and galaxies, where star formation occurs and heavy metals are synthesized. This unique characteristic makes the CGM a prime location to investigate the intricate interplay between gas accretion from the IGM and feedback processes originating in galaxies, in order to understand the baryon cycles that regulate galaxy evolution over cosmic time. 

Some of the major unanswered questions in the study of galaxy evolution concern the origin and nature of cool ($T\sim10^{4-5}\,$K) gas in and around massive quiescent galaxies. Among the most massive galaxies in the Universe, they consist of predominantly old ($\gtrsim1\,$Gyr) stars and do not show any recent star formation (e.g., Eisenstein \etal\ 2003; Roseboom \etal\ 2006; Gauthier \& Chen 2011). While it is tempting to attribute the ``red and dead" nature of quiescent galaxies as due to the absence of cool gas needed to fuel star formation, successive QSO absorption-line studies probing the CGM of luminous red galaxies (LRGs) have established that a significant fraction of these $z\sim0.5$ massive elliptical galaxies host chemically enriched cool gas (e.g., Gauthier \etal\ 2009, 2010; Lundgren \etal\ 2009; Bowen \& Chelouche 2011; Gauthier \& Chen 2011; Thom \etal\ 2012; Zhu \etal\ 2014; Huang \etal\ 2016; Chen \etal\ 2018). These findings at intermediate redshifts are consistent with observations in the local Universe, where \ion{H}{I} and CO surveys found that at least a third of nearby ellipticals contain abundant atomic or even molecular gas (e.g., Serra \etal\ 2012; Young \etal\ 2014; 2018). The high incidence of cool gas in massive quiescent halos is puzzling, and it presents a challenge to our current understanding of galaxy formation. 

\emph{First, how does cool gas survive in massive halos?} The strong clustering of LRGs indicates that these galaxies reside inside massive dark matter halos with $M_\mathrm{h}\gtrsim10^{13}\,\mathrm{M}_\odot$, where gas accreted from the IGM is expected to be shock-heated to the virial temperature of the halo, $T\sim10^{6.5-7}\,$K (see Faucher-Gigu\`ere 2017 for a recent review). Recent cosmological simulations predict that massive galaxies at high redshifts can still acquire cool gas via dense and narrow filaments that penetrate deep into the halo (e.g., Dekel \etal\ 2009; Kere{\v s} \etal\ 2009; van de Voort \etal\ 2012; Nelson \etal\ 2013; Shen \etal\ 2013), but they also show that this mechanism may not be effective for massive dark matter halos hosting LRGs at $z<1$ (e.g., Kere{\v s} \etal\ 2009). Alternatively, thermal instabilities may cause cool clumps to condense from the hot halo and fall toward the galaxy (e.g., Mo \& Miralda Escude 1996; Maller \& Bullock 2004; Sharma \etal\ 2012; Voit \etal\ 2015). Although some observational results suggest this mechanism as a promising explanation (Huang \etal\ 2016), infalling cool clumps of gas are subject to disruption from ram pressure drag and thermal conduction with the hot medium. For that reason, it is still unclear whether cool clumps in the gaseous halo of LRGs will survive their journey to the center of the halo. The detection of high-column density cool gas within projected distances $d<10$ kpc from $z\sim0.5$ massive quiescent galaxies (Zahedy \etal\ 2016; 2017a) indicates that some cool gas may survive, but \emph{to address this question quantitatively requires knowledge of the density and size distributions of cool clumps in LRG halos}. 

\emph{Secondly, what are the dominant feedback mechanisms in massive quiescent halos?} The quiescent nature of both local and intermediate-redshift massive ellipticals indicates that some form of energetic feedback is effective at preventing the cooling of the hot halo over cosmic time, which would otherwise trigger continuing star formation. At the same time, the absence of young stellar populations and strong active galactic nuclei (AGNs) in typical LRGs (e.g., Roseboom \etal\ 2006; Sadler \etal\ 2007; Hodge \etal\ 2009; Gauthier \& Chen 2011; Huang \etal\ 2016) makes it difficult to invoke starburst-driven outflows or AGN feedback to explain the high incidence of chemically enriched cool gas in and around massive quiescent galaxies.  On the other hand, recent observational and theoretical studies have emphasized the importance of the old stellar population themselves in providing the necessary heating, through energy injection from Type Ia supernovae (SNe Ia) and/or winds from asymptotic giant branch (AGB) stars (e.g., Conroy et al. 2015; Zahedy \etal\ 2016, 2017b; Li \etal\ 2018; and references therein). Further insights into the dominant feedback mechanisms in LRGs can be obtained by directly comparing observations in the CGM with theoretical predictions for different feedback prescriptions.\emph { Doing so requires knowledge of the ionization states and chemical abundances in the CGM of LRGs}.

A systematic study is necessary to characterize the physical properties and chemical abundances in the CGM of LRGs. This is a primary motivation behind our COS-LRG survey, a comprehensive survey of the gaseous halos of 16 LRGs at $z\sim0.4$ using a combination of far-ultraviolet (FUV) spectra from the Cosmic Origins Spectrograph (COS) on board the {\it Hubble Space Telescope} ({\it HST}) and ground-based optical echelle spectra.  \emph{The COS-LRG sample was selected without any prior knowledge of the absorption properties of the LRGs}. In Chen \etal\ (2018a, hereafter Paper I), we presented the initial results of our study, which we summarize here. First, high \ion{H}{I} column density gas is common in the CGM of LRGs, with a median of $\langle\,\log\,N\mathrm{(\ion{H}{I})/\cmjj}\rangle = 16.6$ at $d<160$ kpc. Secondly, we measured a high covering fraction of optically thick gas ($\log\,N\mathrm{(\ion{H}{I})/\cmjj}\apg 17.2$) of $\langle\kappa\rangle_{\rm LLS}=0.44^{+0.12}_{-0.11}$ at $d<160$ kpc, which increases to $\langle\kappa\rangle_{\rm LLS}=0.71^{+0.11}_{-0.20}$ at $d< 100$ kpc. Moreover, the CGM of LRGs contains widespread chemically enriched gas traced by low-, intermediate-, and high-ionization metals. The most prominent metal transitions in LRG halos are those of intermediate-ionization species such as \ion{C}{III} and \ion{Si}{III}, with a high covering fraction of $\langle\kappa\mathrm{(\ion{C}{III})}\rangle=0.75^{+0.08}_{-0.13}$ within $d<160$ kpc, comparable to what have been observed in the CGM of star-forming galaxies (e.g., Werk \etal\ 2013). In this paper, we expand our investigation with absorption-line and ionization analyses of both metal and \ion{H}{I} absorption in LRG halos, in order to characterize the physical properties and elemental abundances in the CGM of LRGs. 

The paper is organized as follows. In Section 2 we discuss the COS-LRG sample and the spectroscopic observations and data reduction of the background QSOs. We describe the absorption-line and ionization analyses in Section 3. In Section 4, we characterize the physical properties and elemental abundances in the gaseous halos of LRGs. Finally, we discuss the implications of our findings in Section 5 and present a summary of our results/conclusions in Section 6. In addition, we discuss the results of the analysis for each individual LRG halo in Appendix A. A standard $\Lambda$ cosmology is adopted throughout the paper, with $\Omega_M$ = 0.3, $\Omega_\Lambda$=0.7, and a Hubble constant of $H_{\rm 0} = 70\rm \,km.\,s^{-1}\,Mpc^{-1}$.
 
\section{Sample and Data}

In this section we summarize the COS-LRG sample and the observations of background QSOs. We refer the readers to Paper I for a more detailed discussion on the program design, sample selection, and data reduction of the far-ultraviolet COS spectra and optical echelle spectra of background QSOs in our sample. 

The COS-LRG sample was established by cross-correlating spectroscopically identified LRGs in the Sloan Digital Sky Survey (SDSS; York \etal\
 2000) archive and the literature with all known UV-bright QSOs with FUV $\lesssim 18.5$ mag.\footnote{Because the FUV bandpass of GALEX has a minimum wavelength of $\approx1350$\,\AA, our FUV-bright selection for the background QSOs does not bias against optically thick Lyman-limit systems at $z\lesssim0.5$, which is coincident with the redshift range of COS-LRG galaxies.}  
No prior knowledge of the absorption properties of the LRGs was used in selecting all the LRG-QSO pairs that make up our sample. The UV magnitude cut was chosen to ensure that high-quality and high-resolution spectra of the background QSOs could be obtained with the Cosmic Origins Spectrograph (COS; Green \etal\ 2012) onboard the {\it HST}.  Furthermore, we imposed a lower limit on the LRG stellar 
mass of log\,$M_\mathrm{star}/\mathrm{M}_\odot>11$, and a maximum projected distance of $d=160$ kpc from the QSO. Both choices
were informed by the well-known finding of a significant incidence ($>10\%$) of \ion{Mg}{II} absorbers at $d<120$ kpc from 
massive LRGs (e.g., Gauthier \etal\ 2010; Huang \etal\ 2016). These selection criteria resulted in a mass-limited sample of 16  
quiescent galaxies at $0.21<z<0.56$, each probed by a background QSO at $d<160$ kpc (which corresponds to roughly 1/3 
of the virial radius, $R_\mathrm{h}$, of a $10^{13}\,M_\odot$ dark-matter halo).

All 16 QSOs in the COS-LRG sample were observed with COS, either during our own observing program (PID: 14145) or 
previously available  from the {\it HST} data archive. {\it HST}/COS with the G130M and G160M gratings provides high-resolution 
(${\rm FWHM}\approx17\, \kms $) FUV spectra of the QSOs over a nearly contiguous wavelength coverage between $\lambda\approx1150$ \AA\ 
and $\lambda \approx 1780$ \AA, allowing us to probe halo gas using observations of the full \ion{H}{I} Lyman series and corresponding
low-, intermediate-, and high-ionization metal absorption features at the LRG redshift, including \ion{C}{III} $\lambda977$, the \ion{O}{VI} 
$\lambda\lambda1031,1037$ doublet, \ion{Si}{III} $\lambda1206$, and \ion{Si}{II} $\lambda1260$. The COS data were downloaded from the {\it HST}  
archive and processed using our custom software. The data reduction steps were previously described in detail in Paper I.  To summarize, an important aspect of our custom data reduction software is a recalibration of the COS wavelength solution, which was done in two steps. First, relative wavelength offsets between different exposures of the same QSO were corrected using a low-order polynomial that best describes the offsets of common narrow absorption features found in different exposures. Next, different exposures were co-added and an absolute wavelength correction was performed on the combined spectrum by registering non-saturated, low-ionization Galactic absorption lines to their known vacuum wavelengths. The final wavelength solution for our FUV COS spectra is accurate to within $\pm3$\,\kms, based on a comparison with low-ionization absorption features seen in the ground-based optical echelle spectra.

Optical echelle spectra of COS-LRG QSOs are available for 11 out the 16 QSOs in the sample. 
The echelle observations were obtained using two high-resolution spectrographs, MIKE (Bernstein \etal\ 2003) on the Magellan Clay telescope 
and HIRES (Vogt \etal\ 1994) on the Keck I telescope. The MIKE observations were obtained during our own observing program, whereas the HIRES data
were retrieved from the Keck Observatory Archive (KOA). The instrumental configuration chosen for our MIKE observations provides 
a spectral resolution of ${\rm FWHM}\approx 10$ \kms\ at wavelength $\lambda<5100$ \AA. The archival HIRES observations are characterized by 
a spectral resolution of ${\rm FWHM}\approx 6.5$ \kms\ at $\lambda<5900$ \AA. By extending the spectral coverage of the COS-LRG QSOs to optical wavelengths (from $\lambda \sim 3100$ \AA\ to well over $\lambda \sim 5000$ \AA), the echelle spectra of the QSOs allow access to additional prominent absorption features arising in low-ionization gas in LRG halos, especially the \ion{Mg}{II} $\lambda\lambda\,2796,2803$ doublet, the \MgI\,$\lambda\,2852$ transition, and a series of \ion{Fe}{II} transitions including \ion{Fe}{II}\,$\lambda2586$ and \ion{Fe}{II}\,$\lambda2600$. A detailed description of the data reduction for the MIKE and HIRES spectra can be found in Paper I. 

A summary of FUV and optical echelle spectroscopic observations is presented in Table 1, where we list for each background QSO the instrument used for the observations, the spectral coverage of the data, and the mean $S/N$ per resolution element in final reduced spectrum.

\begin{table}
\footnotesize
\centering
\caption{Summary of spectroscopy of background QSOs}
\label{table:cosobs}
\hspace{-2.5em}
\resizebox{3.5in}{!}{
\begin{tabular}{llccl}
\hline 
\multicolumn{1}{c}{QSO} & \multicolumn{1}{c}{Instrument} & \multicolumn{1}{c}{Spectral Window} & \multicolumn{1}{c}{$S/N$} & \multicolumn{1}{c}{Notes} \\
					&					  	 & \multicolumn{1}{c}{($\mathrm{\AA}$)} &  & \\

\hline \hline
SDSS\,J0246$-$0059	& COS 	& $1140-1790$	& 10	& PID: 14145  \\
					& MIKE 	& $3350-9400$	& 41	& \\
SDSS\,J0803$+$4332 	& COS	& $1160-1800$ & 9  	& PID: 11598 \\
					& HIRES 	& $3150-5870$	& 24	&  \\
SDSS\,J0910$+$1014 	& COS 	& $1140-1790$	& 7	& PID: 11598 \\
					& HIRES 	& $3150-5870$	& 15	&  \\
SDSS\,J0925$+$4004	& COS 	& $1160-1800$ & 6	& PID: 11598 \\
					& HIRES 	& $3240-5870$	& 16 &  \\
SDSS\,J0946$+$5123	& COS 	& $1140-1780$	& 7 	& PID: 14145\\
SDSS\,J0950$+$4831	& COS 	& $1070-1800$	& 10 & PID: 11598 \& 13033\\
					& HIRES 	& $3100-5870$	& 30 &  \\
SDSS\,J1111$+$5547	& COS 	& $1140-1800$	&15 	& PID: 12025 \\
SDSS\,J1127$+$1154	& COS 	& $1140-1780$	& 8 	& PID: 14145 \\
					& MIKE 	& $3350-9400$	& 17 & \\
SDSS\,J1243$+$3539	& COS 	& $1140-1780$	& 14	& PID: 14145 \\
SDSS\,J1244$+$1721	& COS 	& $1420-1780$	& 7 	& PID: 12466 \\
					& MIKE 	& $3350-9400$	& 33 & \\
SDSS\,J1259$+$4130	& COS 	& $1120-1790$	& 13 & PID: 13833 \\
SDSS\,J1357$+$0435 	& COS 	& $1130-1800$	& 13 & PID: 12264 \\
					& MIKE 	& $3350-9400$	& 25 &\\
SDSS\,J1406$+$2509 	& COS 	& $1140-1780$	& 6 	& PID: 14145 \\
					& MIKE 	& $3350-9400$	& 10 &\\
SDSS\,J1413$+$0920	& COS 	& $1130-1750$	& 17 & PID: 13833 \\
SDSS\,J1550$+$4001	& COS 	& $1140-1790$	& 8 	& PID: 11598 \\
					& HIRES 	& $3100-5870$	& 31 	& \\
SDSS\,J1553$+$3548	& COS 	& $1140-1790$ & 8 	& PID: 11598 \\
					& HIRES 	& $3100-5870$	& 36	&  \\
\hline
\end{tabular}}
\end{table}

\section{Analysis}

To promote a deeper understanding of the circumgalactic environment of massive halos, 
we assembled a mass-limited sample of 16 LRGs with log\,$M_\mathrm{star}/\mathrm{M}_\odot>11$. 
The LRGs were selected without prior knowledge of the presence or absence of CGM absorption features. This uniform sample
of galaxies allows an unbiased and accurate characterization of the gaseous halo of intermediate-redshift, massive elliptical galaxies. The two 
main objectives of the COS-LRG program are:
(1) to probe the bulk of cool gas in LRG halos by obtaining accurate measurements of $N\mathrm{(\ion{H}{I})}$; and 
(2) to constrain the physical properties and elemental abundances in massive quiescent halos by observing different ionic metal transitions
that probe a wide range of ionization states.

In Paper I, we presented $N\mathrm{(\ion{H}{I})}$ measurements for the sample and reported significant incidences ($>40\%$)
of low-, intermediate-, and high-ionization metal absorptions at $d<160$ kpc in massive quiescent halos. To investigate the physical properties of the CGM of LRGs and constrain the chemical abundance of the gas requires 
(1) accurate column density measurements for the observed metal absorption features, and 
(2) a detailed ionization modeling of the gas under different physical conditions (e.g., density, metallicity) to explain the observations.
Here we describe the analysis to first measure the ionic column densities and subsequently constrain the physical properties and metallicities of the gas. 

\subsection{Voigt Profile Analysis}

The available high-resolution FUV and optical echelle spectra of the QSOs enable us to resolve the component structures of different absorption transitions and measure the column densities of metal ions accurately. Utilizing a custom software previously 
developed by and described in Zahedy \etal\ (2016), we performed a forward modeling of Voigt profiles to constrain the ionic 
column densities of individual absorbing components in each LRG halo. The software was designed to analyze both well-sampled and under-sampled absorption spectra with known line-spread function (LSF), and to properly assess the confidence intervals of derived model parameters via a Markov Chain Monte Carlo (MCMC) analysis.

In summary, the Voigt profile of each absorption component is uniquely defined by three parameters: the column density $N_c$, the Doppler parameter $b_c$, and the velocity centroid $dv_c$ relative to the redshift of the strongest \ion{H}{I} component in the absorption system. To perform the fit, the program first generated a theoretical spectrum using the minimum number of components necessary to explain the observed absorption profile.  Then, this model spectrum was  
convolved by the appropriate instrumental LSF of the spectrograph used to collect the data, and binned to match the spectral pixel width of the data. 
Finally, the simulated absorption profile was compared to the observed absorption profile, and the best-fit model was found by minimizing the $\chi^2$ value. 

To assess uncertainties in the model parameters, we performed a MCMC analysis using the \textsc{emcee} package (Foreman-Mackey \etal\ 2013). The MCMC analysis allows us to construct the marginalized posterior probability distribution for each model parameter. Each MCMC run consisted of 500 steps performed by an ensemble of 250 walkers. To speed up convergence, the walkers were seeded in a tiny region within the parameter space which is centered at the minimum $\chi^2$ solution. 

The absorption transitions which were analyzed in a given absorption system include all observed transitions from the following list, ordered by increasing rest wavelength:
\ion{O}{I} $\lambda971$, \ion{C}{III} $\lambda977$, \ion{O}{I} $\lambda988$, \ion{N}{III} $\lambda989$, the \ion{O}{VI} $\lambda\lambda1031,1037$ doublet, \ion{C}{II} $\lambda1036$, \ion{N}{II} $\lambda1083$, \ion{Fe}{III} $\lambda1122$, \ion{Fe}{II} $\lambda1144$, \ion{Si}{II} $\lambda1190$, \ion{Si}{II} $\lambda1193$, \ion{Si}{III} $\lambda1206$, \ion{Si}{II} $\lambda1260$, \ion{O}{I} $\lambda1302$, \ion{C}{II} $\lambda1334$, \ion{Si}{IV}\,$\lambda\lambda1393,1402$, \ion{Fe}{II}\,$\lambda2382$, \ion{Fe}{II}\,$\lambda2586$, \ion{Fe}{II}\,$\lambda2600$, the \ion{Mg}{II}\,$\lambda\lambda2796, 2803$ doublet, and \ion{Mg}{I}\,$\lambda2852$. In our analysis, we required different transitions of the same species (e.g., \ion{Si}{II} $\lambda1190$ and \ion{Si}{II} $\lambda1193$) to have the same Voigt profile parameters. Furthermore, we imposed the same kinematic structure (i.e., number of components and velocity structure) among \ion{H}{I}, low-ionization, and intermediate-ionization species. This choice was justified by the excellent kinematic agreement among the observed absorption profiles of various low- and intermediate-ionization species, including \ion{Mg}{II}, \ion{Si}{II}, \ion{Si}{III}, and \ion{C}{III} (Paper I). Excepted from this requirement was high-ionization \ion{O}{VI} absorption, which is known to often exhibit distinct velocity profiles compared to lower-ionization gas (e.g., Savage \etal\ 2010; Werk \etal\ 2016). For that reason, we performed the Voigt profile analysis for the \ion{O}{VI} doublet independently from the analysis for \ion{H}{I} and lower-ionization metals. 

The results of our Voigt profile analysis are presented in Appendix A for each LRG. In Figures A1a to A16a, we present the continuum-normalized absorption profiles of different transitions, the best-fit Voigt profiles for individual components, and the integrated Voigt profile summed over all components. These figures show the excellent agreement in velocity centroids among individual components of different ionic species, including the \ion{H}{I} Lyman series, \ion{C}{II}, \ion{Mg}{II}, \ion{Si}{II}, and \ion{C}{III}, which demonstrates the high accuracy of our wavelength calibration. 
In Tables A1a to A16a, we report the best-fit Voigt profile parameters and the associated 68\% confidence intervals for each component identified in the Voigt profile analysis. For saturated components, we report the 95\% lower limits on the column density $N_c$ and the corresponding 95\% upper limits on Doppler parameter $b_c$. For non-detections, we report the 95\% upper limits on $N_c$ based on the error spectrum, calculated over a spectral window that is twice the FWHM of a line with $b_c=10\, \kms$ for low- and intermediate-ionization species and $b_c=30\, \kms$ for highly ionized \ion{O}{VI}. Finally, we also report in Table A1a to Table A16a the total $N\mathrm{(\ion{H}{I})}$ and ionic metal column densities summed over all components in each system.



\subsection{Ionization Analysis}

\begin{figure}
\hspace{-0.4em}
\includegraphics[scale=0.69]{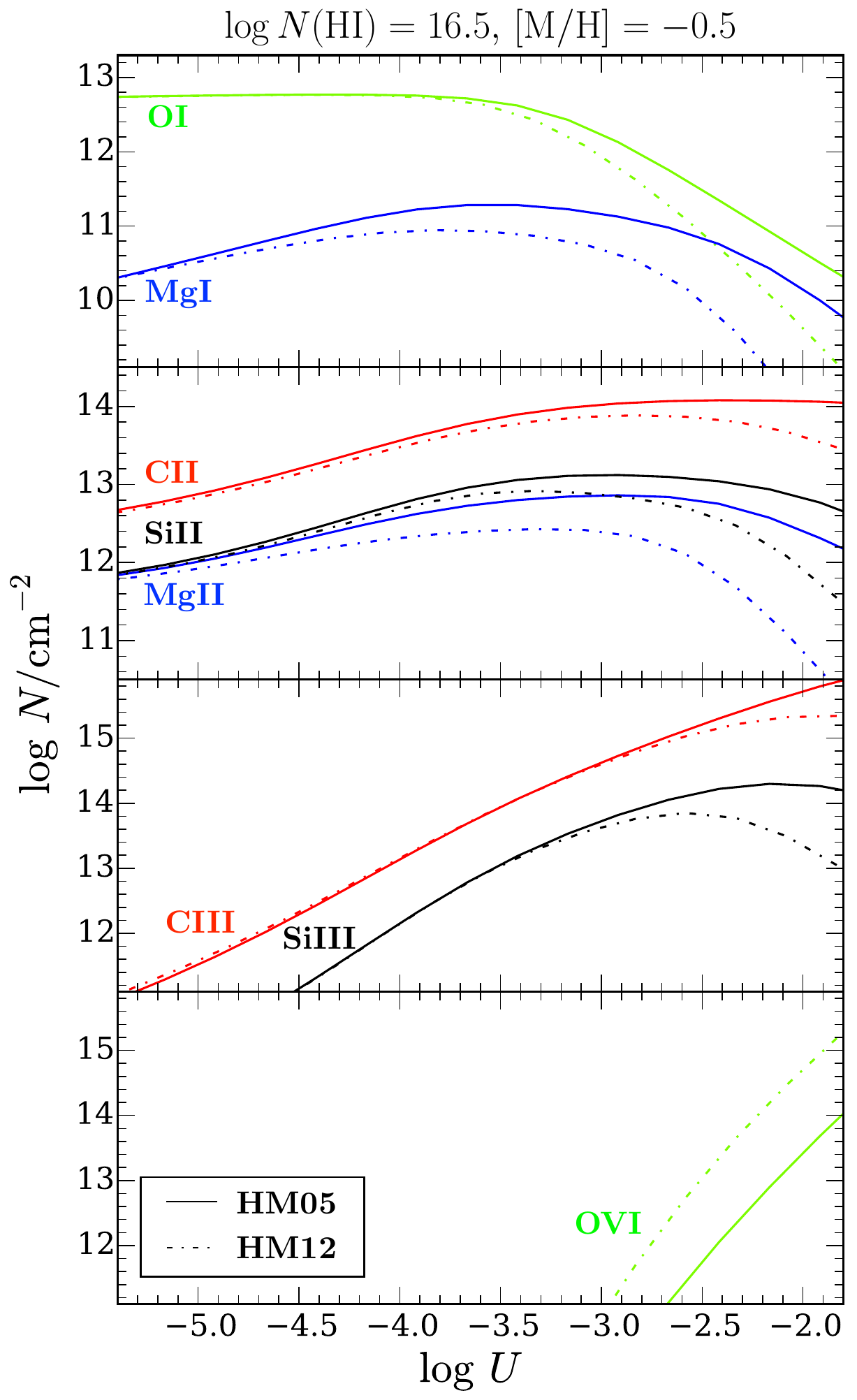}
\vspace{-1.5em}
\caption{Example predictions of ionic column densities as a function of ionization parameter $U$ from \textsc{cloudy} photoionization calculations. From top to bottom, the panels are ordered by increasing ionization state, shown here for common ionic species \ion{O}{I}, \ion{Mg}{I}, \ion{C}{II}, \ion{Mg}{II}, \ion{Si}{II}, \ion{C}{III}, \ion{Si}{III}, and \ion{O}{VI}. The prediction curves are shown for a gas with $\log N\,\mathrm{(\ion{H}{I})}=16.5$ and a 
metallicity of $\mathrm{[M/H]=-0.5}$, which are typical values in our sample. In solid lines, we show the predicted 
column densities for a gas irradiated by an updated Haardt \& Madau (2001) ionizing background radiation field (HM05 in \textsc{Cloudy})
at $z = 0.4$. In dashed-dotted lines, we show the corresponding predictions under the Haardt \& Madau (2012) background radiation field
(HM12 in \textsc{Cloudy}) at the same redshift. A solar abundance pattern is assumed for the model calculations shown here.}
\label{figure:ions}
\end{figure}

To constrain the metallicity and other physical quantities of the CGM, it is necessary to determine the ionization state of the gas. 
The inferred cool temperature of the gas ($T\lesssim$ a few $\times 10^4$ K, see \S 4.2 and Appendix A) is consistent with a photoionized
gas. The ionization state of the gas can be determined by comparing the observed column densities of 
different ionic species to predictions from photoionization calculations (e.g., Chen \etal\ 2017). An important physical quantity in 
photoionized gaseous environment is the ionization parameter $U$, defined as the number of incident ionizing photons per hydrogen 
atom. For a fixed radiation field characterized by a total flux of hydrogen-ionizing ($\geq1$ Ryd) photons $\Phi$, the 
$U$ parameter is inversely proportional to the hydrogen number density $n_\mathrm{H}$, according to $U\equiv \Phi/c\,n_{\rm H}$. 
Higher gas density results in lower $U$, which leads to a more neutral gas, and vice versa. Another physical quantity 
which affects the observed ionic column densities is the metallicity of the gas $\mathrm{[M/H]}$. High-metallicity gas cools
more efficiently than low-metallicity gas, shifting the photoionization equilibrium toward lower ionization (i.e., more neutral) states. 

We performed a series of photoionization calculations using \textsc{Cloudy} v.13.03 (Ferland \etal\ 2013) package. We considered a plane-parallel 
column of gas with uniform volume density $n_\mathrm{H}$, which was irradiated by an ultraviolet background (UVB) radiation field. 
To investigate how uncertainty on the UVB affects the derived gas density and metallicity (see Chen \etal\ 2017 for an extensive discussion),
we performed two sets of calculations using two different UVBs: (1) the updated Haardt \& Madau (2001) UVB, known as HM05 in \textsc{Cloudy}; 
and (2) the Haardt \& Madau (2012) UVB, known as HM12 in \textsc{Cloudy}. The two radiation fields differ in both their spectral slopes and overall intensities between $1-10$ Ryd. While the HM05 spectrum is softer than HM12 within this energy regime, the HM05 UVB has more $1-3$ Ryd photons which have large photoionization cross-sections for neutral hydrogen atoms as well as low- to intermediate-ionization metals. Furthermore, the HM05 UVB has about 2.5 times (0.4 dex) the total number of hydrogen-ionizing photons of the HM12 UVB. 
In our ionization calculations, both UVBs were adopted at $z=0.4$, which is roughly the median redshift of the COS-LRG galaxies. 
At this fiducial redshift, the relationship between $U$ and $n_\mathrm{H}$ is log\,$U =-5.42  - \mathrm{log}\, n_\mathrm{H}$ for HM05, and 
log\,$U =-5.83 -\mathrm{log}\, n_\mathrm{H}$ for HM12. For example, a typical CGM gas density of  
$n_\mathrm{H}=0.01\, \cmjjj$ corresponds to log\,$U\approx-3.4$ and log\,$U\approx-3.8$ for the HM05 and HM12 UVBs, respectively. 

For each UVB, we constructed a grid of \textsc{Cloudy} models spanning a wide range of 
\ion{H}{I} column densities ($14\leq \mathrm{log}\,N\mathrm{(\ion{H}{I})/cm^{-2}}\leq 20$ in 0.25 dex steps), 
gas densities ($-5\leq \mathrm{log}\,n\mathrm{_H/cm^{-3}}\leq 1$ in 0.25 dex steps), 
and metallicities ($-3\leq\mathrm{[M/H]}\leq1$ in 0.25 dex steps). For each point in the grid, \textsc{Cloudy} calculated the expected
column densities and ionization fractions of different ionic species assuming photoionization equilibrium. 
We assumed a solar abundance pattern for the gas, although when the predictions were compared 
to observations, we relaxed this assumption whenever necessary and allowed by the data (see Appendix A).  

An example of \textsc{cloudy} calculations is presented in Figure 1, where the predicted column densities of different ions  
are plotted as a function of ionization parameter $U$. The column density curves are shown for a gas with log\,$N\mathrm{(\ion{H}{I})/\cmjj}=16.5$ 
and $\mathrm{[M/H]}=-0.5$, which are typical values for individual components in our sample. 
The model predictions for a gas irradiated by the HM05 UVB are shown in solid lines, whereas predictions
for a gas irradiated by the HM12 UVB are shown in dash-dotted lines. Comparing the model expectations under the two different UVBs, 
it is clear that at fixed ionization parameter, the predicted ionic abundances for neutral and singly ionized species (e.g., 
\ion{Mg}{I}, \ion{Si}{II} and \ion{Mg}{II}) are systematically lower under HM12 UVB than HM05 UVB. In addition, the decrements in HM12-predicted column densities relative to HM05 grow larger for higher $U$ parameter (or equivalently, lower $n_\mathrm{H}$). Similar, albeit more modest, trends 
are also predicted for doubly ionized species such as \ion{C}{III}, and \ion{Si}{III}. These trends result from of the harder HM12 UVB spectrum, which has a higher fraction of $>3$ Ryd photons that are needed to produce highly ionized (triply ionized or more) metal species compared to the HM05 UVB. As $U$ increases, both low- and intermediate-ionization species are preferentially lost to higher ionization states under HM12 UVB than HM05 UVB. As a consequence of these intrinsic differences between HM05 and HM12 UVBs, HM05 models require a higher gas metallicity than HM12 models to reproduce the observed ionic abundances (see also Wotta \etal\ 2016). 

To estimate the metallicity and density of the gas, we compared the resulting \textsc{Cloudy} grid of predictions to the data and performed 
a statistical analysis which took into account measurements as well as upper limits (non-detections) and lower limits 
(saturation) in the data. Given a suite of observed ionic transitions $\{y_i\}$ for a kinematically matched absorbing component with $n$ number of measurements, $m$ upper limits, and $l$ lower limits, the probability that the gas has a given density and metallicity is defined to be
{\small
\begin{align}
{\cal P}(n_\mathrm{H},\mathrm{[M/H]}\mid\{y_i\}) \propto  \left( \prod_{i=1}^{n} \exp \left\{ -\frac{1}{2} \left[ \frac{y_i -
\bar{y_i}(n_\mathrm{H},\mathrm{[M/H]})}{\sigma_i} \right]^2 \right\} \right)  \nonumber 
\\  
\times\left(\prod_{i=1}^{m} \int_{-\infty}^{y_i} dy' \exp \left\{ -\frac{1}{2} \left[ 
\frac{y' - \bar{y_i}(n_\mathrm{H},\mathrm{[M/H]})}{\sigma_i} \right]^2 \right\} \right)  \nonumber
\\
\times\left(\prod_{i=1}^{l} \int_{y_i}^{+\infty} dy' \exp \left\{ -\frac{1}{2} \left[ 
\frac{y' - \bar{y_i}(n_\mathrm{H},\mathrm{[M/H]})}{\sigma_i} \right]^2 \right\} \right),
\end{align}}
where $y_i=\log\,N_i$ is the observed column density of the  $i$-th ionic species, and $\sigma_i$ is the measurement uncertainty of $y_i$,
and $\bar{y_i}=\log\,\bar{N_i}$ is the corresponding model prediction. Note that in equation 1, the first product is equivalent to 
calculating $e^{-\frac{1}{2}\chi^{2}}$ for the $n$ ionic column density measurements, whereas the second and third products extend the calculation 
over the $m$ upper limits and $l$ lower limits, respectively (see also Chen \etal\ 2010; Crighton \etal\ 2015; Stern \etal\ 2016).

The statistical analysis described above was performed for each absorbing component identified in our Voigt profile analysis (\S3.1). 
For each component, all available column density measurements, upper limits, and lower limits for low- and intermediate-ionization species 
were compared to an interpolated grid of \textsc{Cloudy} models evaluated at the observed $N\mathrm{(\ion{H}{I})}$ of the data. We note that 
\ion{O}{VI} measurements were excluded from this analysis, not only because of the the well-known uncertainty in the ionization mechanism of \ion{O}{VI} absorbers, but also because of the observed kinematic misalignments between the absorption profiles of \ion{O}{VI}  and lower-ionization gas (\ion{H}{I} and metal ions; see \S 4.2 and Appendix A). As discussed in \S 5.3, our observations indicate that contributions from higher ionization gas phase to the observed column densities of lower ionization species are negligible, so the exclusion of higher ionization gas from our ionization analysis should not bias the inferred ionization parameter of cool and lower-ionization gas phase considered here. We discuss the possible origins of the high-ionization gas traced by \ion{O}{VI} absorbers in LRG halos in \S 5.3. 

The results of the ionization analysis are presented in Tables A1b to A16b in Appendix A, where for each individual component we report the number of detected metal species which are used to constrain the model, $N_\mathrm{metal}$, the most probable gas metallicity $\mathrm{[M/H]}$ and density $n_\mathrm{H}$ under both the HM05 and HM12 UVBs, as well as the estimated 68\% confidence intervals for $\mathrm{[M/H]}$ and $n_\mathrm{H}$. For components with $N_\mathrm{metal}<2$, we find that the inferred $\mathrm{[M/H]}$ and log\,$n_\mathrm{H}$ are subject to large uncertainties of $>0.5$ dex, and in a number of cases there is no clear point of maximum probability within the parameter space of the models. For these components, we report in Tables A1b to A16b the estimated 95\% upper or lower limits on the parameter values. In addition, we report in Tables A1b to A16b the inferred gas metallicity and density considering each absorber as a single clump. For the single-clump (SC) model, the aforementioned ionization analysis was performed using the integrated $N\mathrm{(\ion{H}{I})}$ and ionic column densities summed over all components in each system, to facilitate comparisons with existing CGM/IGM ionization studies in the literature (e.g., Werk \etal\ 2014; Prochaska \etal\ 2017; Muzahid \etal\ 2018). Finally, we present in Figures A1b to A16b the two-dimensional joint probability distribution of $\mathrm{[M/H]}$ and $n_\mathrm{H}$ for components with $N_\mathrm{metal}\geq 2$, under both the HM05 (black contours) and HM12 (blue contours) UVBs. The contours indicate the estimated 68\% and 95\% confidence levels for the model parameters.

\section{Physical Properties and Metallicities in LRG halos}

Our analysis of CGM absorption in the COS-LRG sample reveals a diversity of gas properties in massive quiescent halos $z\sim0.4$. A detailed discussion on the absorption and gas properties in individual LRG halos is presented in Appendix A, which we summarize as follows. First, a combined Voigt profile analysis on \ion{H}{I} and metal absorption lines shows that absorbers in LRG halos exhibit a multi-component structure that is distributed over up to $\pm$a few $\times100$ \kms\ in line-of-sight velocity relative to the LRGs. Furthermore, the excellent kinematic alignments between \ion{H}{I}, low ions (e.g., \ion{Mg}{II}), and intermediate ions (e.g., \ion{C}{III}) indicate a physical connection between these different species. In this section, we characterize the physical properties and chemical abundances in the gaseous halos of LRGs. 

\subsection {Column density profiles of \ion{H}{\uppercase {I}} and heavy ions}

In Figure 2, we present the spatial distribution of absorption column densities for various ions observed in the gaseous halos of COS-LRG galaxies.
From top to bottom panels, we show the integrated column densities versus projected distance $d$ for neutral \ion{H}{I}, low-ionization \ion{Mg}{II}, intermediate-ionization \ion{C}{III} and \ion{Si}{III}, and high ionization \ion{O}{VI} species. To facilitate comparisons with other surveys, we include a second horizontal axis showing the halo radius-normalized projected distance $d/R_\mathrm{h}$. Recall from Paper I that the COS-LRG sample of massive quiescent galaxies has a median stellar mass and dispersion of log\,$\langle M_\mathrm{star}/\mathrm{M}_\odot\rangle=11.2\pm0.2$, which corresponds to a typical halo mass of $M_\mathrm{h}\approx10^{13}\,\mathrm{M}_\odot$ according to the Kravtsov \etal\ (2018) stellar-to-halo-mass relation, and a halo radius of $R_\mathrm{h}\approx500$ kpc at $z=0.4$.\footnote{We approximate $R_\mathrm{h}$ as the region with average density of 200 times above the mean matter density of the Universe at a given epoch.} Given the narrow range in $M_\mathrm{star}$, we adopt this $R_h$ for all COS-LRG galaxies plotted in Figure 2.

For comparison, Figure 2 shows absorption measurements for the COS-Halos red galaxy subsample (Werk \etal\ 2013; Prochaska \etal\ 2017) for \ion{H}{I}, \ion{Mg}{II}, \ion{C}{III}, and \ion{Si}{III}. For \ion{O}{VI}, we also show measurements from the passive galaxy subsample of Johnson \etal\ (2015), which includes all COS-Halos red galaxies. Note that column density measurements from these studies are plotted versus the normalized projected distance of the galaxies, $d/R_\mathrm{h}$. Compared to COS-LRG, the COS-Halos subsample of red galaxies comprises predominantly lower mass galaxies, with a mass range of from log\,$M_\mathrm{star}/\mathrm{M}_\odot=10.3$ to log\,$M_\mathrm{star}/\mathrm{M}_\odot=11.3$, and a median of  log\,$\langle M_\mathrm{star}/\mathrm{M}_\odot\rangle=10.8$. 

Despite considerable scatter in the observed $N\mathrm{(\ion{H}{I})}$ radial profile, there is a general trend of declining $N\mathrm{(\ion{H}{I})}$ with increasing $d$ in the COS-LRG sample (top panel of Figure 2). This trend is consistent with what is seen in COS-Halos red galaxies as well as previous CGM surveys of the general galaxy populations (e.g., Chen \etal\ 1998; Johnson \etal\ 2015).  To further examine the decline of $N\mathrm{(\ion{H}{I})}$ with increasing distance, we divide our sample into two bins at $d=100$ kpc, which is approximately the median projected distance. At $d<100$ kpc from LRGs, the majority of \ion{H}{I} absorbers are optically thick (LLSs with log\,$N\mathrm{(\ion{H}{I})/\cmjj}>17.2$). In contrast, there is a significantly higher fraction of optically thin absorbers as well as sightlines with non-detections at $d>100$ kpc. The mean covering fraction of optically thick \ion{H}{I} gas is $\langle\kappa\rangle_{\rm LLS}=0.71^{+0.19}_{-0.26}$ at $d<100$ kpc, which declines to $\langle\kappa\rangle_{\rm LLS}=0.22^{+0.22}_{-0.14}$ at $d=100-160$ kpc. 

A trend of declining column density with increasing $d$ is also seen in low-ionization metal species such as \ion{Mg}{II}. At $d<100$ kpc, strong $N\mathrm{(\ion{Mg}{II})}$ absorbers with log\,$N\mathrm{(\ion{Mg}{II})/\cmjj}>13$ are common in COS-LRG. In contrast, absorbers at $d\gtrsim100$ kpc exhibit significantly lower $N\mathrm{(\ion{Mg}{II})}$, where log\,$N\mathrm{(\ion{Mg}{II})/\cmjj}<13$ is seen in all cases. For strong \ion{Mg}{II} absorbers in COS-LRG, we estimate a mean covering fraction of $\langle\kappa\mathrm{(\ion{Mg}{II})}\rangle_{13.0}=0.60^{+0.25}_{-0.30}$ at $d<100$ kpc. A caveat of this calculation is that two sightlines at $d<100$ kpc do not have any \ion{Mg}{II} constraints and consequently do not contribute to the covering fraction estimation. Including these two sightlines would lead to a mean \ion{Mg}{II} covering fraction of $\langle\kappa\mathrm{(\ion{Mg}{II})}\rangle_{13.0}\approx0.4-0.7$ at $d<100$ kpc, depending on whether these two absorbers satisfy the strong \ion{Mg}{II} absorption criterion or not. In contrast, the lack of strong \ion{Mg}{II} absorption At $d=100-160$ kpc from LRGs in our sample constrains the mean covering fraction to $\langle\kappa\mathrm{(\ion{Mg}{II})}\rangle_{13.0}\approx0.0-0.2$, for log\,$N\mathrm{(\ion{Mg}{II})/\cmjj}>13$. 

A surprising finding from Paper I is the high incidence of absorption from intermediate-ionization species \ion{C}{III} and \ion{Si}{III}, comparable to what have been observed around star-forming galaxies. While the high oscillator strength of the \ion{C}{III} $\lambda977$ transition makes \ion{C}{III} absorption easily detectable, it also means that C\,III absorption profiles are often saturated (see Appendix A). For that reason, it is difficult to draw a strong conclusion on possible radial trends in intermediate ionic column densities using \ion{C}{III} absorption. For the comparatively weaker \ion{Si}{III} absorption, it is clear that strong \ion{Si}{III} absorption with log\,$N\mathrm{(\ion{Si}{III})/\cmjj}>13.0$ are more prevalent at smaller $d$. At $d<100$ kpc from COS-LRG galaxies, absorbers with log\,$N\mathrm{(\ion{Si}{III})/\cmjj}>13.0$ are present in 5 out of 7 cases, which constraints the mean \ion{Si}{III} covering fraction to $\langle\kappa\mathrm{(\ion{Si}{III})}\rangle_{13.0}=0.71^{+0.19}_{-0.26}$. In contrast, \ion{Si}{III} absorption are generally weaker at $d=100-160$ kpc, with a high fraction (50 percent) of non-detections. The estimated mean \ion{Si}{III} covering fraction absorption at $d=100-160$ kpc is $\langle\kappa\mathrm{(\ion{Si}{III})}\rangle_{13.0}=0.37^{+0.24}_{-0.19}$, for log\,$N\mathrm{(\ion{Si}{III})/\cmjj}>13.0$.

\begin{figure}
\hspace{-1em}
\includegraphics[scale=1.04]{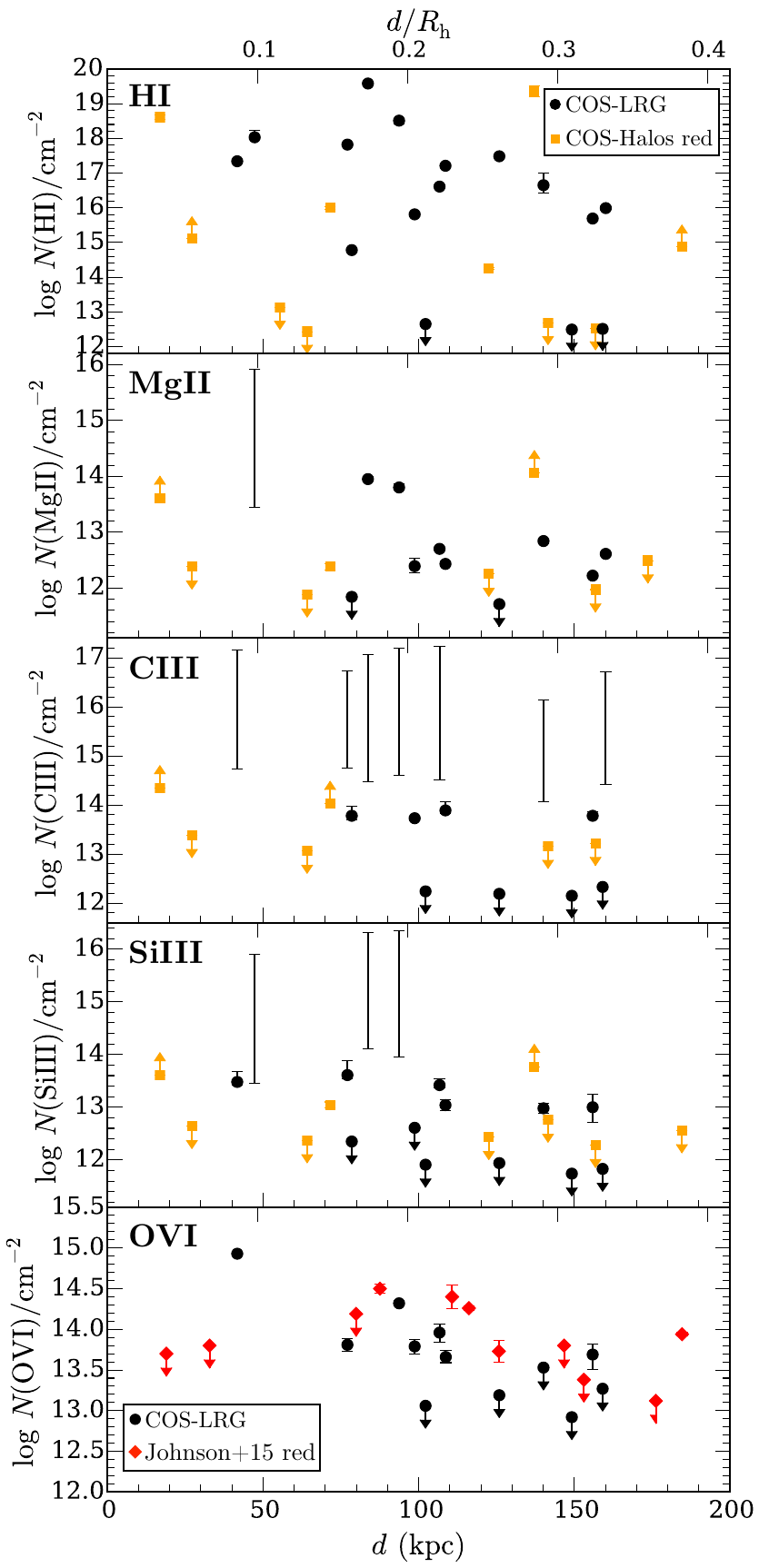}
\vspace{-1.5em}
\caption{Radial profile of integrated absorption column densities in the CGM of quiescent galaxies, shown for \ion{H}{I}, \ion{Mg}{II}, \ion{C}{III}, \ion{Si}{III}, and \ion{O}{VI}. COS-LRG measurements (black circles) are plotted versus projected distance $d$. For comparison, absorption measurements from passive galaxies in COS-Halos  (Werk \etal\ 2013; orange squares) and Johnson \etal\ (2015; red diamonds) are plotted versus normalized projected distance $d/R_\mathrm{h}$. We have excluded five COS-Halos red galaxies which overlap with our LRG sample. Non-detections are shown as downward arrows which represent the 2$\sigma$ upper limits on ionic column density. Meanwhile, the allowed column density range for saturated absorbers in COS-LRG are shown in empty vertical error bars. For COS-Halos red galaxies, saturated absorbers are represented by upward arrows, which show the lower limits on the absorption column density.}
\label{figure:ions}
\end{figure}

\begin{figure*}
\includegraphics[scale=0.412]{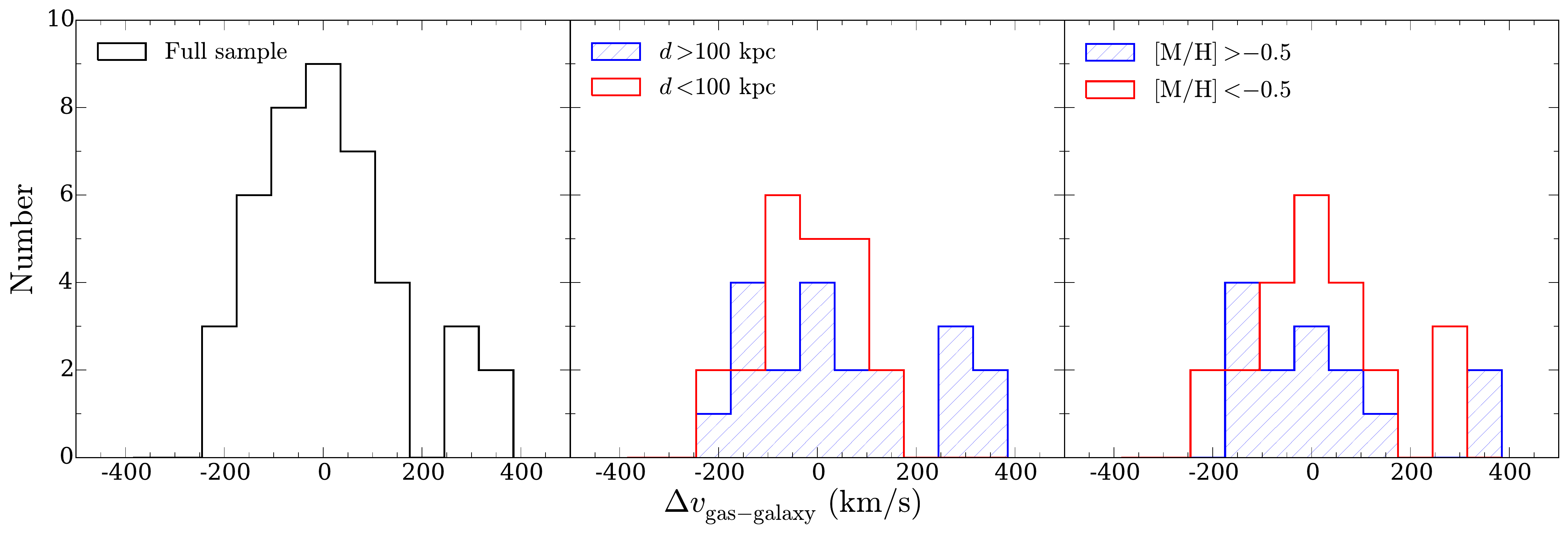}
\vspace{-0.5em}
\caption{Line-of-sight velocity distributions of individual absorption components relative to the LRG systemic redshifts. For the full COS-LRG sample (left panel), we find a mean and dispersion of $\langle \Delta v_\mathrm{gas-galaxy} \rangle =17$ \kms and $\sigma_{\Delta v_\mathrm{gas-galaxy}}=147$ \kms. In the middle panel, the sample is bisected by projected distance $d$, whereas in the right panel the sample is divided by metallicity. We find no statistically significant distinction between the subsamples in either case.}
\label{figure:ions}
\end{figure*}

\begin{figure*}
\includegraphics[scale=0.77]{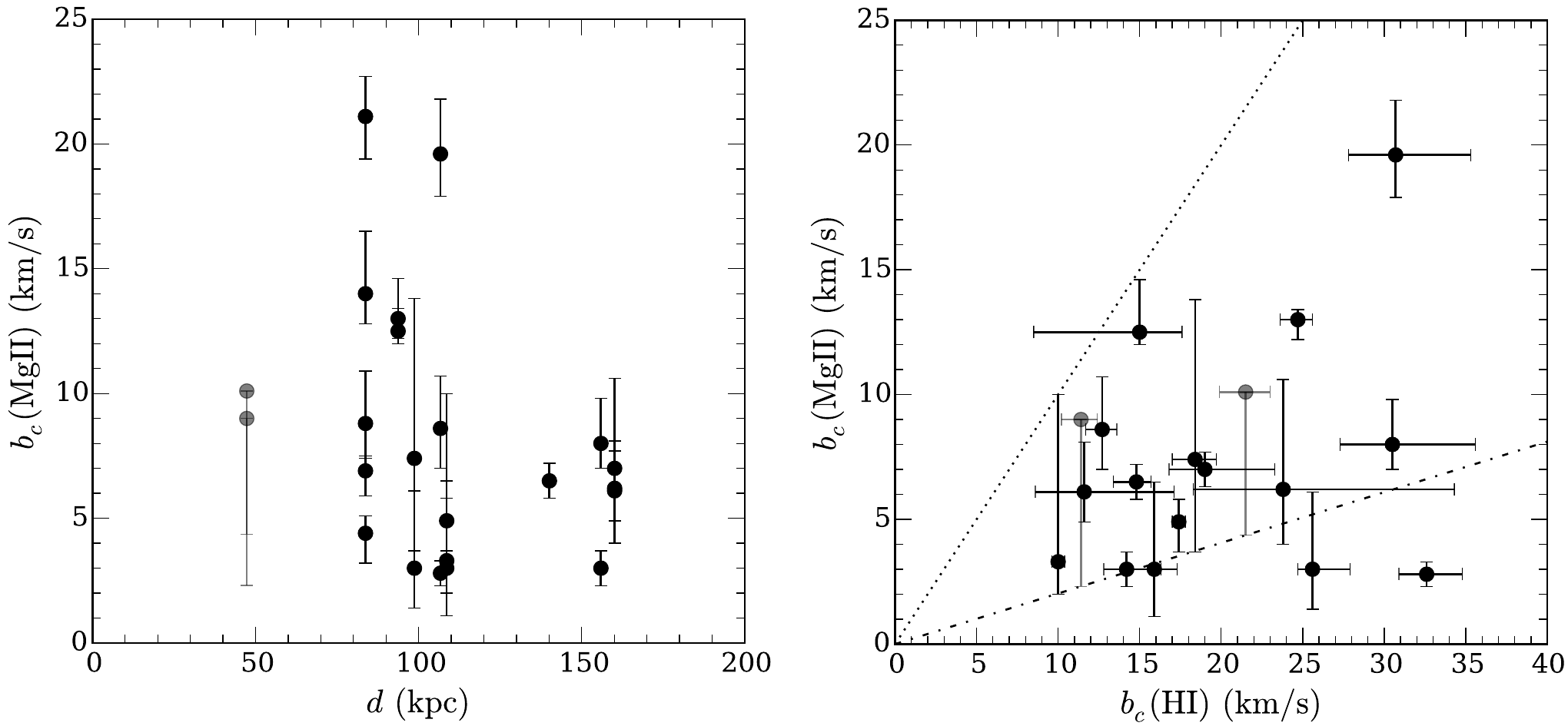}
\vspace{-0.5em}
\caption{{\it Left}: Doppler parameter $b_c$ plotted versus $d$ for individual \ion{Mg}{II} absorption components. Saturated components are represented by grayed out data points. {\it Right}: Distribution of Doppler linewidths for matched \ion{H}{I} and \ion{Mg}{II} components. The dashed-dotted line shows the expectation for a pure thermal broadening case, $b_c\mathrm{(\ion{Mg}{II})}\approx0.2\,b_c\mathrm{(\ion{H}{I})}$, whereas the dotted line shows the expected relationship when \ion{Mg}{II} and \ion{H}{I} linewidths are dominated by non-thermal broadening, $b_c\mathrm{(\ion{Mg}{II})}=b_c\mathrm{(\ion{H}{I})}$. We find that cool CGM gas around LRGs has a mean temperature and dispersion of $\langle T \rangle =2.0\times10^4\,$K and $\sigma_T =1.4\times10^4\,$K, with a modest non-thermal broadening of $\langle b_\mathrm{nt} \rangle =7\pm5\,$\kms.}
\label{figure:ions}
\end{figure*}

For high-ionization gas, measurements of \ion{O}{VI} column density in LRG halos are available for 12 out of 16 COS-LRG galaxies. We detect \ion{O}{VI} absorption in 7 sightlines at a detection threshold of log\,$N\mathrm{(\ion{O}{VI})/\cmjj}>13.5$, which translates to an estimated mean covering fraction of $\langle\kappa\mathrm{(\ion{O}{VI})}\rangle_{13.5}=0.58^{+0.17}_{-0.18}$ at $d<160$ kpc ($\sim0.3\,R_\mathrm{h}$). The mean \ion{O}{VI} covering fraction for LRGs is comparable to what Johnson \etal\ (2015) found at $d\lesssim0.3\,R_\mathrm{h}$ for their passive galaxy subsample, $\langle\kappa\mathrm{(\ion{O}{VI})}\rangle_{\mathrm{J15}}=0.62^{+0.13}_{-0.17}$ for log\,$N\mathrm{(\ion{O}{VI})/\cmjj}>13.5$. Note that the red galaxies in Johnson \etal\ (2015) are predominantly less massive than COS-LRG galaxies, with a median stellar mass and dispersion of  log\,$\langle M_\mathrm{star}/\mathrm{M}_\odot\rangle=10.7\pm0.5$.  The comparable \ion{O}{VI} covering fractions in massive quiescent halos spanning over an order of magnitude in halo mass (from $M_\mathrm{h}\sim10^{12}\,\mathrm{M}_\odot$ to $M_\mathrm{h}\gtrsim10^{13}\,\mathrm{M}_\odot$) suggest that \ion{O}{VI} absorbers in all quiescent halos may share a similar physical origin. A two-sample Kolmogorov-Smirnov (K-S) test demonstrates that we cannot rule out at high statistical significance ($>99$ percent) that the lower-mass (Johnson \etal\ 2015) and massive (COS-LRG) quiescent halo samples of \ion{O}{VI} absorbers are drawn from the same parent population. 

\subsection {Kinematic and thermal properties}

The line-of-sight kinematics of absorbing gas relay crucial information about the underlying motion of cool clumps within LRG halos.  Our discussion of individual LRG halos in Appendix A highlights the fact that cool gas absorption profiles in the CGM of LRGs consist of multiple components that are distributed within $\pm$ a few hundred \kms\ in line-of-sight velocity relative to the systemic redshift of the galaxy. The distribution of line-of-sight velocities of individual \ion{H}{I} components relative to the LRGs is shown in the left panel of Figure 3. The velocity distribution can be characterized by a mean and dispersion of $\langle \Delta v_\mathrm{gas-galaxy} \rangle =17$\,\kms and $\sigma_{\Delta v_\mathrm{gas-galaxy}}=147$\,\kms. The observed velocity dispersion is consistent with what have been reported for \ion{Mg}{II} absorbers around large samples of \ion{Mg}{II} absorbers around LRGs using low-resolution data (e.g., Zhu \etal\ 2014; Huang \etal\ 2016; Lan \etal\ 2018). To provide a physical context, the inferred mean mass of LRG halos is $M_\mathrm{h}\approx10^{13.4}\,\mathrm{M}_\odot$ (e.g., Mandelbaum \etal\ 2008; Gauthier \etal\ 2009), and the expected line-of-sight velocity dispersion for virialized motion in LRG halos is $\sigma_\mathrm{h}\approx260$ \kms. The observed line-of-sight velocity dispersion of the gas, $\sigma_{\Delta v_\mathrm{gas-galaxy}}=147$ \kms, is merely $\sim 60$ percent of the expectation from virial motion. The narrow distribution of line-of-sight velocities indicates that an effective dissipative mechanism is at play to slow down the motion of cool gas in the halo (e.g., Huang \etal\ 2016). 

To evaluate whether the observed velocity dispersion varies with projected distance, we divide the absorbing components into two subsamples on $d$, one for components at $d<100$ kpc and another for those at $d>100$ kpc. The resulting velocity distributions of the two subsamples are shown in the middle panel of Figure 3. While the velocity histograms are understandably noisy due to the smaller size of the two subsamples, no significant trend is detected between the line-of-sight velocity distributions at small and large $d$. Using a two-sided Kolmogorov-Smirnov (K-S) test, we cannot rule out that the two $d$ subsamples come from the same parent distribution in $\Delta v_\mathrm{gas-galaxy}$ at $>50$ percent confidence.  We also bisect the sample of individual components into two groups based on their metallicities, a low-metallicity subsample with $\mathrm{[M/H]}<-0.5$ and a high-metallicity subsample with $\mathrm{[M/H]}>-0.5$. Again, no statistically significant distinction can be made between the low- and high-metallicity subsamples, with components from each subsample occupying the full range of velocities with respect to the LRGs (Figure 3, right panel). A K-S test cannot rule out that the two $\mathrm{[M/H]}$ subsamples come from the same parent distribution in $\Delta v_\mathrm{gas-galaxy}$ at $>68$ percent confidence.

\begin{figure}
\hspace{0.5em}
\includegraphics[scale=0.56]{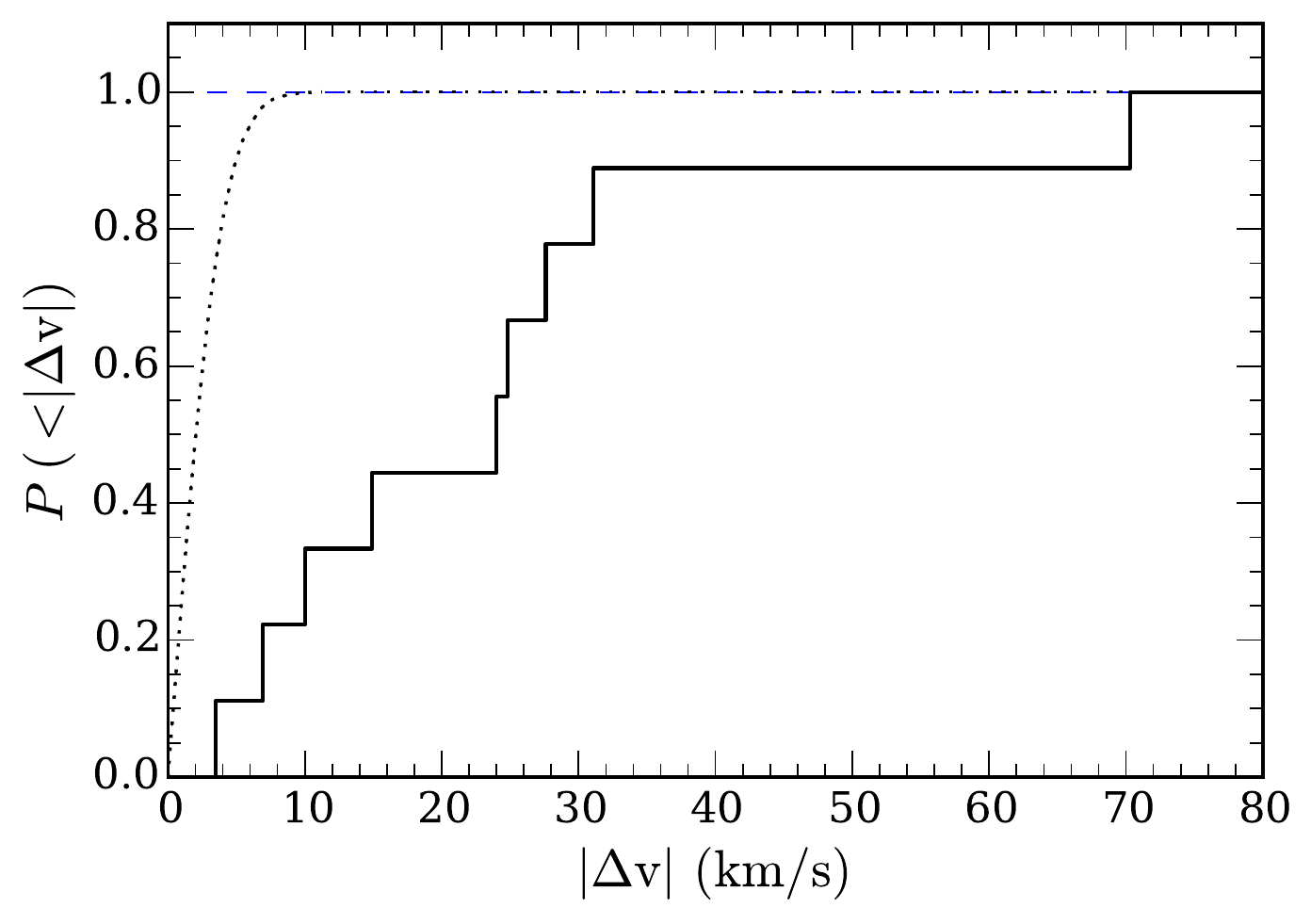}
\vspace{-0.7em}
\caption{
Cumulative fraction, $P$, of \ion{O}{VI} absorption components with absolute centroid velocity difference less than $|\Delta v|$ from the nearest low-ionization metal and \ion{H}{I} components (solid histogram). The mean/median value is $\langle |\Delta v| \rangle=24$ \kms, with a full range of from $|\Delta v|=4$ to  $|\Delta v|=71$ \kms. Note that the final wavelength solution for our FUV COS spectra is accurate to within $\pm3$\,\kms, and the expected $P\mathrm{(|\Delta v|)}$ for a normal distribution with a width of $\sigma=3$ \kms\ is shown in dotted curve for comparison. The mismatched kinematics between high-ionization and low-ionization gas in COS-LRG suggest different physical origins between the high- and low-ionization gas.}
\label{figure:ions}
\end{figure}

Our Voigt profile analysis also allows us to examine the thermal properties of cool clumps in LRG halos. In the left panel of Figure 4, we plot the Doppler linewidths of \ion{Mg}{II} components as a function of $d$. With the exception of a few broad components, most \ion{Mg}{II} components in LRG halos are narrow with $b_c\mathrm{(\ion{Mg}{II})}<10$ \kms. The narrow linewidths imply that the gas is both cool and kinematically quiescent. Furthermore, no trend in $b_c\mathrm{(\ion{Mg}{II})}$ in seen versus $d$, indicating that cool gas at small and large projected distances from LRG have similar thermal properties. 

Next, we show the distribution of Doppler linewidths for matched \ion{H}{I} and \ion{Mg}{II} absorption components in the right panel of Figure 4. Two straight lines are drawn to indicate two limiting cases. First, the dashed-dotted line in the bottom represents the expectation for a pure thermal-broadening case where the \ion{Mg}{II} and \ion{H}{I} linewidths are related by the square root of their mass ratio alone, giving $b_c\mathrm{(\ion{Mg}{II})}\approx0.2\,b_c\mathrm{(\ion{H}{I})}$. Secondly, the dotted line on top of the panel shows the expected relation when the \ion{Mg}{II} and \ion{H}{I} linewidths are dominated by non-thermal broadening, $b_c\mathrm{(\ion{Mg}{II})}\approx b_c\mathrm{(\ion{H}{I})}$.\footnote{While the parameter space outside the region bounded by the two limiting cases is unphysical, two components are found below the thermal-broadening line. One has a $b_c\mathrm{(\ion{Mg}{II})}$ that is consistent within 1-$\sigma$ with thermal broadening. The other component has a broad $b_c\mathrm{(\ion{H}{I})}=33$ \kms, but its \ion{Mg}{II} linewidth is only 3 \kms, which is narrower than expected from thermal broadening. The unphysical relationship between $b_c\mathrm{(\ion{H}{I})}$ and $b_c\mathrm{(\ion{Mg}{II})}$ for this component implies the presence of unresolved \ion{H}{I} components that are not \ion{Mg}{II}-bearing.} It is clear from the right panel of Figure 4 that a large majority data points are situated closer to the thermal-broadening line than to the non-thermal broadening line. This is consistent with a quiescent gas that is subject to little non-thermal broadening. The ratios of Doppler linewidths for matched \ion{H}{I} and \ion{Mg}{II} components in the COS-LRG sample show that the gas has a mean temperature and dispersion of $\langle T \rangle =2.0\times10^4\,$K and $\sigma_T =1.4\times10^4\,$K, with a modest mean non-thermal line broadening of $\langle b_\mathrm{nt} \rangle =7\pm5\,$\kms.

Finally, we find that \ion{O}{VI} absorption profiles in COS-LRG show distinct kinematic structures from the absorption profiles of lower-ionization metal and \ion{H}{I} (see Appendix A for a detailed description of individual absorbing systems). The mean/median absolute difference in centroid velocity between \ion{O}{VI} absorption components and the nearest \ion{H}{I} and low-ionization metal component is $\langle |\Delta v| \rangle=24$ \kms, with a full range of from $|\Delta v|=4$ to  $|\Delta v|=71$ \kms\ (Figure 5). Recall that the final wavelength solution for our FUV COS spectra is accurate to within $\pm3$\,\kms. The kinematic misalignments between high-ionization and low-ionization gas in the COS-LRG sample suggest that different phases of the CGM gas of LRGs have different physical origins (a more in-depth discussion is presented in \S 5.2 and \S 5.3).

\subsection {Metallicities and densities}

Our ionization analysis on matched absorption components reveals significant variations in gas metallicities and densities in the cool CGM of LRGs, both within individual halos and among different halos in the COS-LRG sample. We now discuss and investigate for trends in gas metallicities and densities in the COS-LRG ensemble of galaxies. We begin with a discussion on systematic errors in the ionization analysis. 

\subsubsection {Systematic errors arising from the uncertain UVB}

\begin{figure}
\hspace{-0.9em}
\includegraphics[scale=0.54]{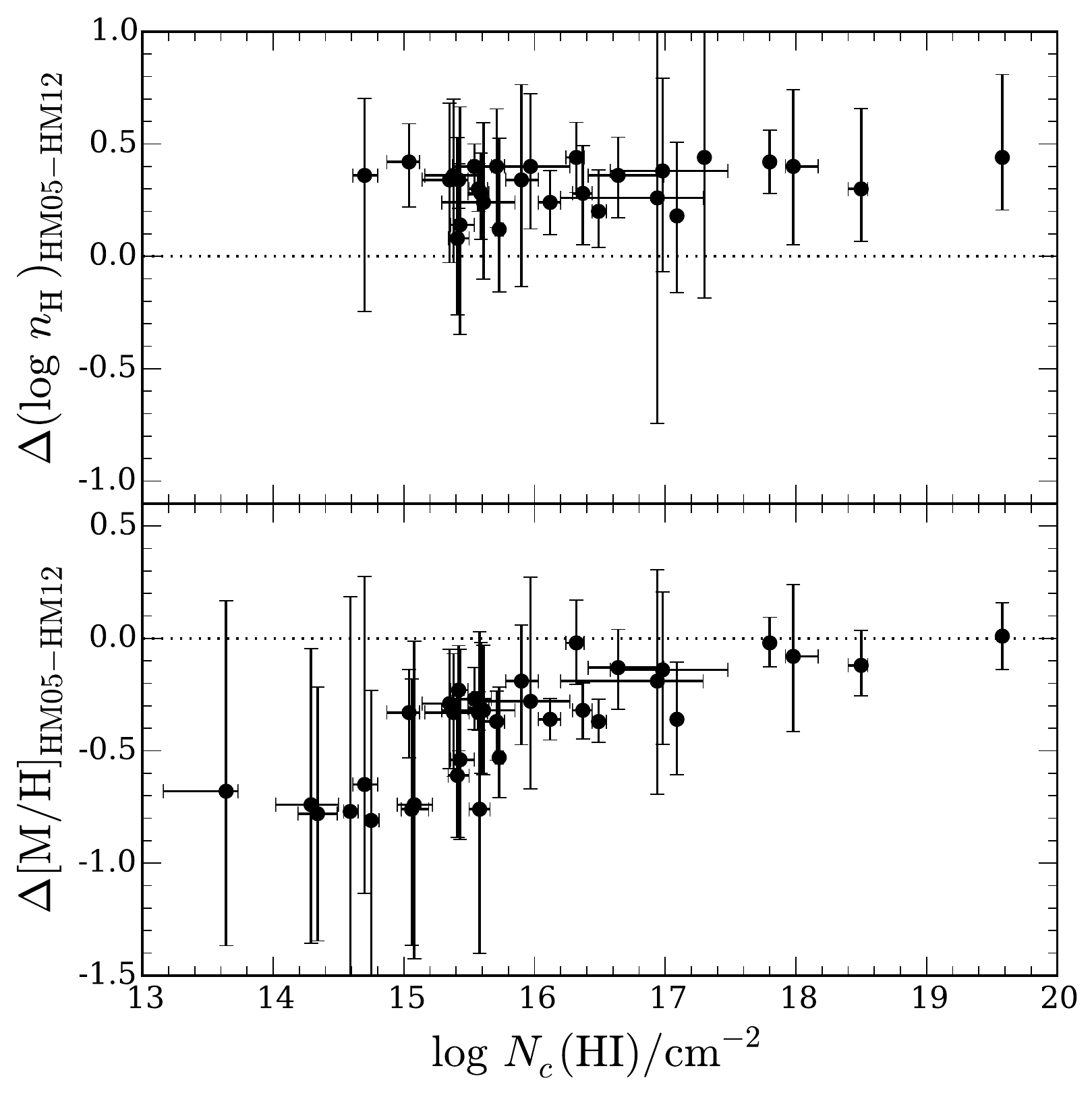}
\vspace{-1.9em}
\caption{
{\it Top}: Difference in gas densities derived under HM05 and HM12 UVBs, plotted versus $N_c\mathrm{(\ion{H}{I})}$. Compared to the HM12 UVB, the HM05 UVB leads to higher inferred $n_\mathrm{H}$ values, with a median difference and dispersion of $\mathrm{\langle\Delta log\,\mathit{n}_H \rangle}=0.34$ dex and $\mathrm{\sigma_{\Delta log\,\mathit{n}_H}}=0.10$ dex.
{\it Bottom}: Difference in metallicities derived under HM05 and HM12 UVBs, plotted versus individual component \ion{H}{I} column density, $N_c\mathrm{(\ion{H}{I})}$.  The HM05 UVB leads to lower inferred $\mathrm{[M/H]}$ than the HM12 UVB, with a metallicity difference that range from $\mathrm{\langle\Delta [M/H] \rangle}=-0.1$ for LLSs to $\mathrm{\langle\Delta [M/H] \rangle}=-0.7$ for optically thin gas. 
}
\label{figure:ions}
\end{figure}

Metallicity and density estimates in CGM studies are based on comparing the absorption column densities of ionic metals and neutral hydrogen. Because the gas is highly ionized in all but the highest column density absorbers, substantial ionization fraction corrections are necessary to convert the observed ionic column density ratios to the desired elemental abundances. A complicating factor in the ionization analysis of CGM gas is the well-known uncertainties in the shape, intensity, and redshift evolution of the extragalactic UVB (e.g., Faucher-Gigu\`ere \etal\ 2008; Haardt \& Madau 2012; Kollmeier \etal\ 2014; Shull \etal 2015), which affect the expected ionization fraction corrections in the gas. Adopting different UVBs for the ionization analysis can propagate to order-of-magnitude discrepancies in the inferred gas metallicity (see Chen 2017 for an extensive discussion on the subject). 

To explore how the uncertain UVB spectrum affects the derived gas densities and metallicities, we performed our ionization analysis using two different photoionizing background radiation fields, the HM05 and HM12 UVBs (see the discussion in \S 3.2). In the top panel of Figure 6, the difference in gas densities derived under HM05 and HM12 UVBs is plotted versus component \ion{H}{I} column density $N_c\mathrm{(\ion{H}{I})}$. Over almost 5 decades in $N_c\mathrm{(\ion{H}{I})}$, the gas densities inferred using HM05 UVB are systematically higher than gas densities inferred using the HM12 UVB, with a median difference and dispersion of $\mathrm{\langle\Delta log\,\mathit{n}_H \rangle}=0.34$ dex and $\mathrm{\sigma_{\Delta log\,\mathit{n}_H}}=0.10$ dex.  The higher inferred gas density under HM05 can be understood as due to the higher intensity of HM05 UVB compared to the HM12 UVB. Recall from our discussion in \S 3.2 that the total flux of hydrogen-ionizing photons in HM05 UVB is $\sim0.4$ dex higher than that of the HM12 UVB. Consequently, the higher-intensity HM05 UVB 
requires a higher underlying gas density than the HM12 UVB for fixed ionization parameter $U$, which describes the ionization state of the gas.

\begin{figure}
\includegraphics[scale=0.62]{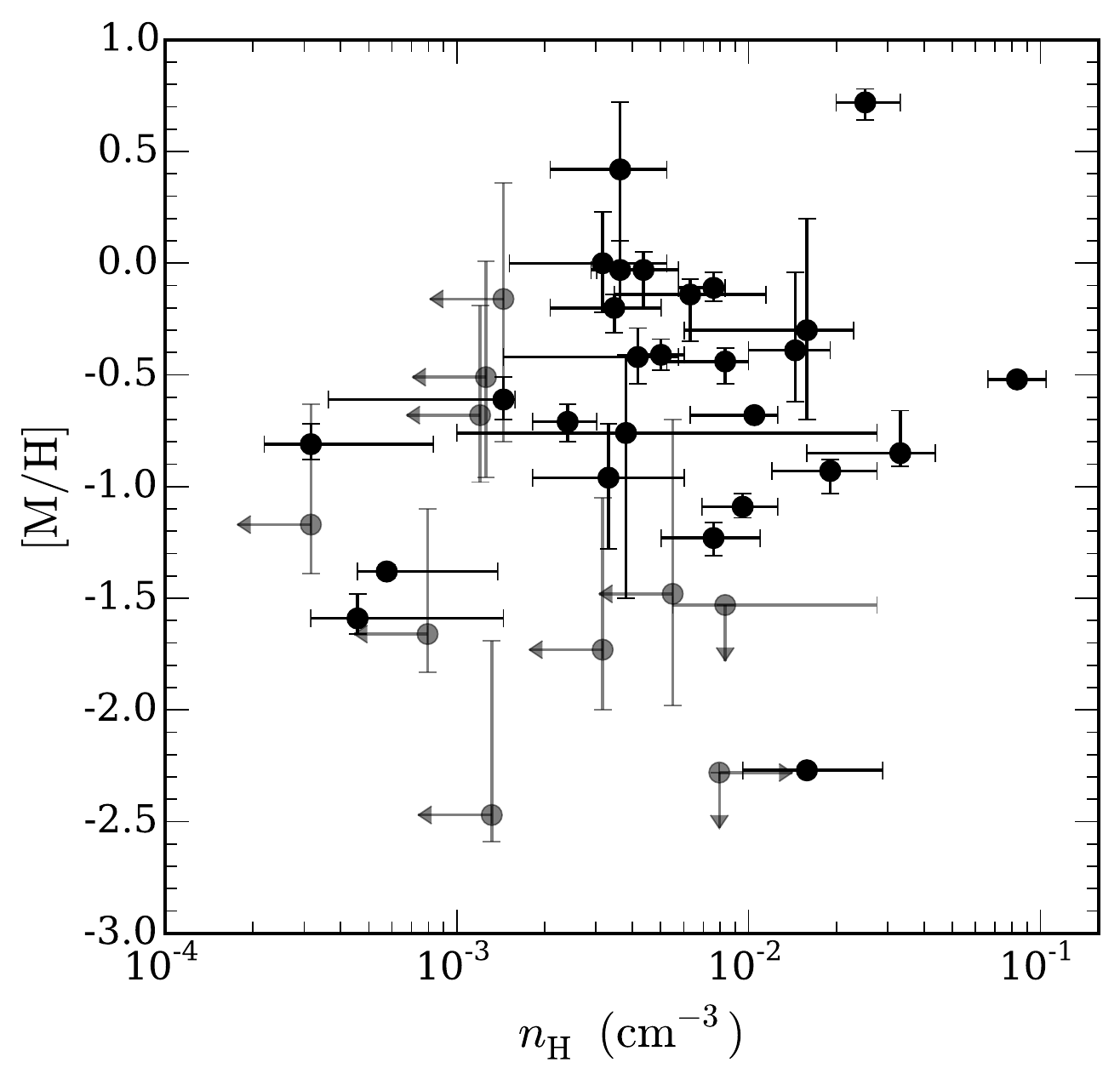}
\vspace{-0.5em}
\caption{Inferred gas metallicity $\mathrm{[M/H]}$ versus hydrogen density $n_\mathrm{H}$ for individual absorbing components in the COS-LRG sample. The vertical and horizontal error bars associated with each data point show the 68 percent confidence intervals for $\mathrm{[M/H]}$ and $n_\mathrm{H}$, respectively. Grayed out data points show components for which only upper/lower limits on $\mathrm{[M/H]}$ and/or $n_\mathrm{H}$ are available, with arrows indicating the 95 percent upper/lower limits. We find no statistically significant correlation between $\mathrm{[M/H]}$ and $n_\mathrm{H}$.}
\label{figure:ions}
\end{figure}

\begin{figure*}
\includegraphics[scale=0.79]{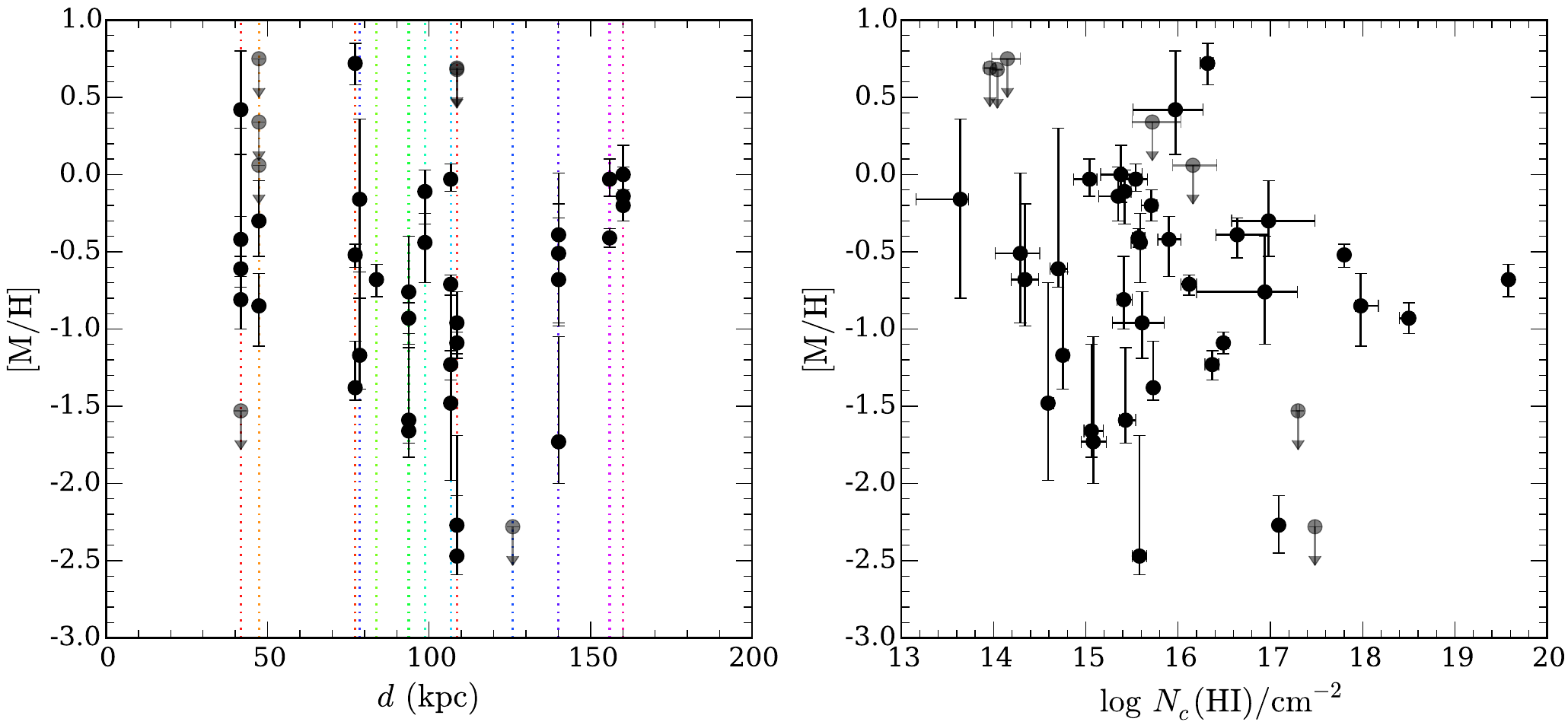}
\vspace{-0.5em}
\caption{
{\it Left}: Gas metallicity $\mathrm{[M/H]}$ versus $d$ in the COS-LRG sample. Each vertical colored line connects different absorption components detected within the same LRG halo. The vertical error bar associated with each data point shows the 68 percent confidence interval for $\mathrm{[M/H]}$. Grayed out data points are absorbing components with no metal ions detected, with downward arrows indicating the 95 percent upper limits on $\mathrm{[M/H]}$.  
Large ($\gtrsim 1$ dex) variations in gas metallicities within the CGM are seen in a majority of LRGs that exhibit multi-component absorption profiles.
{\it Right}: $\mathrm{[M/H]}$ versus component \ion{H}{I} column density $N_c\mathrm{(\ion{H}{I})}$. There is no evidence for any metallicity trend with $N_c\mathrm{(\ion{H}{I})}$. The median metallicity of individual components is $\mathrm{\langle [M/H] \rangle}=-0.7\pm0.2$, with an estimated 16-84 percentile range of $\mathrm{[M/H]}=(-1.6,-0.1)$ for the whole sample.}
\label{figure:ions}
\end{figure*}

In the bottom panel of Figure 6, we plot the difference in metallicities derived under HM05 and HM12 UVBs versus $N_c\mathrm{(\ion{H}{I})}$. Over more than 5 decades in \ion{H}{I} column density, not only is the $\mathrm{[M/H]}$ inferred under HM05 UVB systematically lower than inferred under HM12 UVB, but also the difference in metallicities depends on $N_c\mathrm{(\ion{H}{I})}$. For optically thick gas with log\,$N_c\mathrm{(\ion{H}{I})/\cmjj}\gtrsim17$, the typical metallicity difference between HM05 and HM12 is modest, $\mathrm{\langle\Delta [M/H] \rangle}=-0.1\pm0.1$ dex. The median metallicity difference is larger for lower $N_c\mathrm{(\ion{H}{I})}$ gas, ranging from $\mathrm{\langle\Delta [M/H] \rangle}\sim -0.3$ dex for gas with log\,$N_c\mathrm{(\ion{H}{I})/\cmjj}\sim16$ to $\mathrm{\langle\Delta [M/H] \rangle}\sim -0.7$ dex for optically thin gas with log\,$N_c\mathrm{(\ion{H}{I})/\cmjj}<15$. To understand the origin of this trend, recall from \S 3.2 that not only does the HM12 UVB have a harder spectrum than the HM05 UVB, but also it has a higher fraction of $>3$ Ryd photons which are required to produce high-ionization (triply ionized or more) metal species.  As $N_c\mathrm{(\ion{H}{I})}$ decreases and the gas becomes more highly ionized, more low- and intermediate-ionization metals are preferentially lost to higher ionization states under HM12 UVB than under HM05 UVB. Because metallicity estimates of cool CGM gas often rely on suite of low-ionization and intermediate-ionization metal species, the difference in metallicities inferred under HM05 and HM12 UVBs naturally increases with decreasing $N\mathrm{(\ion{H}{I})}$.

Finally, we note that in all ionization calculations performed in this work, both HM05 and HM12 UVBs were adopted at redshift $z=0.4$, which is roughly the median redshift of COS-LRG galaxies. Changing the adopted UVB redshift to $z=0.2$ or $z=0.6$ would change the intensity of each UVB by no more than  $\pm0.2$ dex. As a result, the inferred gas density  $n_\mathrm{H}$ would change by less than $\pm0.2$ dex by changing the adopted UVB redshift, which is smaller than the median difference in $n_\mathrm{H}$ derived under HM05 and HM12 UVBs. 

Using two different UVBs that are frequently utilized in CGM/IGM studies, we have quantified the systematic errors resulting from the uncertain shape and intensity of the extragalactic UVB radiation field. It must also be noted that a known issue with the HM12 UVB is that it over-predicts the amplitude of \ion{H}{I} column density distribution function in low-redshift ($z<1$) Ly$\alpha$ forest by a factor of $2-4$ (e.g., Kollmeier \etal\ 2014; Shull \etal\ 2015; Viel \etal\ 2017). The HM12 UVB does not match low-redshift IGM observations because of its low hydrogen photoionization rate ($\Gamma_\mathrm{H}$), which is a result of the adopted negligible escape fraction of Ly-continuum photons from low-redshift galaxies (e.g., Shull \etal\ 2015). In contrast, the HM05 UVB assumes a higher escape fraction of ionizing photons from galaxies, which has been shown to provide better agreement with observations (e.g., Kollmeier \etal\ 2014; Khaire \& Srianand 2015; Viel \etal\ 2017). For that reason, we adopt the gas metallicities and densities inferred using the HM05 UVB for subsequent analyses and discussions in this work.

\subsubsection {Trends in gas metallicities and densities}

 \begin{figure*}
\includegraphics[scale=0.79]{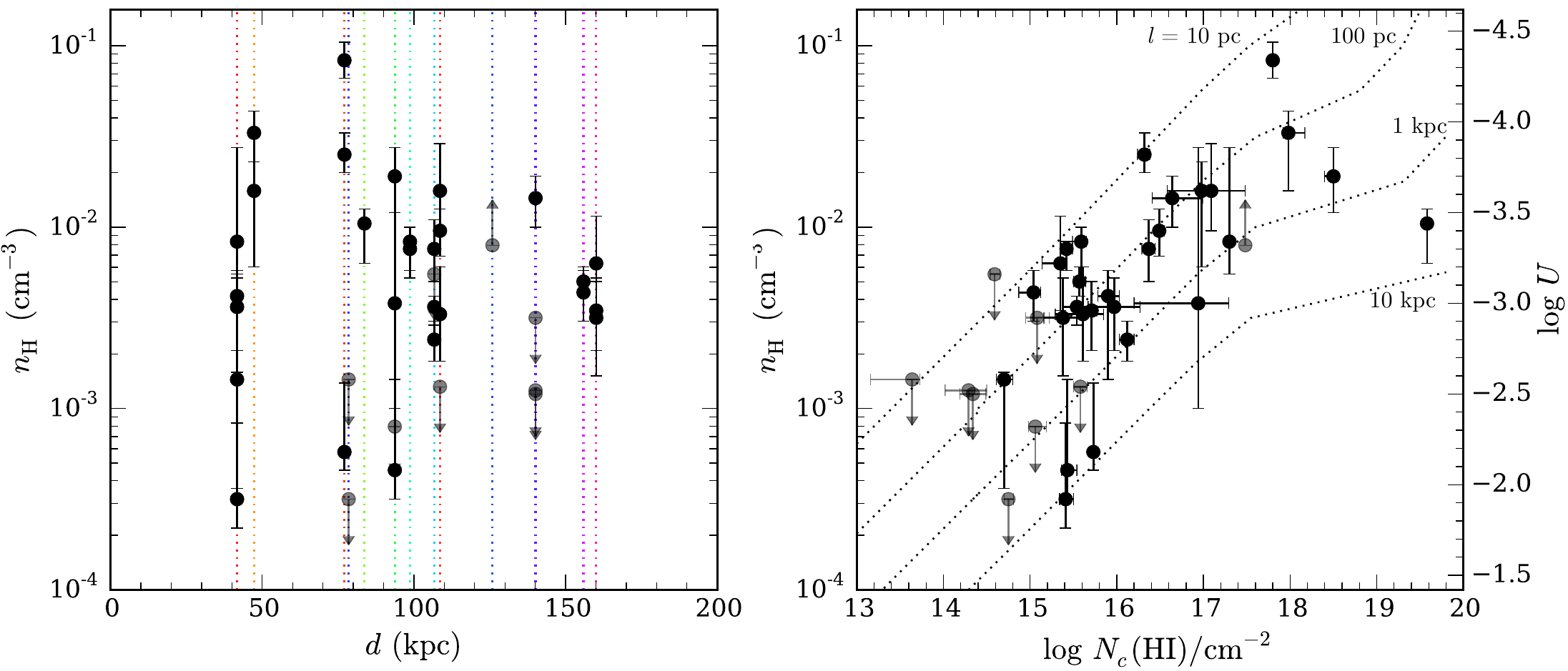}
\vspace{-0.5em}
\caption{
{\it Left}: Gas density $n_\mathrm{H}$ versus $d$ in the COS-LRG sample. Each vertical colored line connects different absorption components detected within the same LRG halo. The vertical error bar associated with each data point shows the 68 percent confidence interval for $n_\mathrm{H}$. Grayed out data points show components for which only upper/lower limits on $n_\mathrm{H}$ are available, with upward/downward arrows indicating the 95 percent upper/lower limits on the underlying gas density. Large ($\gtrsim 1$ dex) variations in $n_\mathrm{H}$ within the CGM are seen in half of COS-LRG galaxies that exhibit multi-component absorption profiles. The median gas density of individual components is log\,$\langle n_\mathrm{H} \rangle/ \cmjjj=-2.4\pm0.1$.
{\it Right}: $n_\mathrm{H}$ versus component \ion{H}{I} column density $N_c\mathrm{(\ion{H}{I})}$. The corresponding ionization parameter $U$ for a given $n_\mathrm{H}$ is indicated on the right y-axis. The trend of rising gas density with increasing \ion{H}{I} column density indicates that high $N_c\mathrm{(\ion{H}{I})}$ gas has lower ionization parameter $U$ and is therefore less ionized than low  $N_c\mathrm{(\ion{H}{I})}$ gas. The median $U$ in the COS-LRG sample is $\mathrm{log\,\langle U \rangle}\approx-3.0$. Finally, each dotted curve shows the expected $n_\mathrm{H}$-$N_c\mathrm{(\ion{H}{I})}$ relation for a cool cloud of a given line-of-sight thickness, from $l=10$ pc to $l=10$ kpc. The distribution of cool clump sizes shows a clear mode at $\sim100$ pc, with an estimated median of  $\langle l \rangle =120^{+80}_{-40}$ pc.
}
\label{figure:ions}
\end{figure*}

To illustrate the diversity of inferred gas metallicities and densities in the cool CGM of LRGs, we plot component $\mathrm{[M/H]}$ versus $n_\mathrm{H}$ in Figure 7.  While no evidence is seen for any correlation between $\mathrm{[M/H]}$ and $n_\mathrm{H}$, Figure 7 shows that cool gas in LRG halos occupy a wide range of metallicities (from less than 0.01 solar to solar and super-solar metallicities) and densities (from $n_\mathrm{H}\lesssim0.001\, \cmjjj$ to $n_\mathrm{H}\sim0.1\, \cmjjj$). 

We present the spatial distribution of component metallicity as a function of $d$ in the left panel of Figure 8. Two interesting features are revealed by this plot. First, $\mathrm{[M/H]}$ exhibits large variations among different components detected in the gaseous halo of an LRG, at small and large $d$ alike. A majority of LRG halos ($\sim60$ percent) that exhibit multi-component absorption profiles show over a factor of 10 difference in $\mathrm{[M/H]}$ between the most metal-rich and metal-poor components. Such large variations in $\mathrm{[M/H]}$ within the gaseous halo indicate poor chemical mixing in the CGM of LRGs and underscore the importance of resolving the component structures of CGM absorbers, which is afforded by our high-resolution absorption spectra (see also e.g., Churchill \etal\ 2012; Rosenwasser \etal\ 2018). In contrast, any information on intra-halo variations is lost if one utilizes only the integrated \ion{H}{I} and metal column densities along individual sightlines in the ionization analysis. 

Furthermore, while high-metallicity ($\mathrm{[M/H]}\gtrsim-1.0$) components are observed in most LRG halos, metal-poor ($\mathrm{[M/H]}\lesssim-1.0$) components are found in half of LRG halos, with a majority these low-metallicity components occuring at $d\gtrsim100$ kpc. Over the full sample, the median metallicity of individual components is $\mathrm{\langle [M/H] \rangle}=-0.7\pm0.2$, where the uncertainty is calculated using a combined bootstrap and Monte-Carlo resampling. In addition, we estimate the 16-84 percentile range in $\mathrm{[M/H]}$ to be  $\mathrm{[M/H]}=(-1.6,-0.1)$ for the whole sample. Note that components with poor constraints on $\mathrm{[M/H]}$ (those with metallicity upper limits which are higher than solar metallicity) are excluded from these estimates. 

In the right panel of Figure 8, we present a plot of $\mathrm{[M/H]}$ versus component \ion{H}{I} column density. We find no significant trend in $\mathrm{[M/H]}$ versus $N_c\mathrm{(\ion{H}{I})}$. This lack of correlation in our data stands in contrast to the anti-correlation between metallicity and \ion{H}{I} column density that were reported in a number of recent studies (e.g., Prochaska \etal\ 2017; Muzahid \etal\ 2018). Considering the known trend of declining $N\mathrm{(\ion{H}{I})}$ with $d$ in the CGM (e.g., Chen \etal\ 1998; Johnson \etal\ 2015; see also \S 4.1), the reported anti-correlation implies that metallicity increases with distance from galaxies, which is difficult to explain. This discrepancy can be attributed as due to two systematic effects. First, the HM12 UVB, which was the adopted UVB in Prochaska \etal\ (2017), predicts progressively higher metallicities (up to 0.7 dex) with decreasing $N\mathrm{(\ion{H}{I})}$ compared to the HM05 UVB (see \S 4.3.1). Secondly, the ionization analyses in these studies utilized integrated \ion{H}{I} and metal column densities summed over all components in each absorption system. As we approach lower column density regime, the required data quality ($S/N$) is higher to detect the gas.  Given a fixed $S/N$ and a system with multiple components, weaker metal components is more challenging to uncover. Consequently, relatively more metal-poor gas goes undetected more easily. By treating resolved components separately, we find that several low-$N\mathrm{(\ion{H}{I})}$ components only have non-constraining metallicity upper limits.  The combination of these two systematic effects explain the reported anti-correlation between $\mathrm{[M/H]}$ and $N\mathrm{(\ion{H}{I})}$. 


A surprising finding from our analysis is the significant incidence of low-metallicity LLSs in the COS-LRG. The right panel of Figure 8 shows that three optically thick components (out of seven overall) with log\,$N_c\mathrm{(\ion{H}{I})/\cmjj}\gtrsim17$ have very low metallicities, $\mathrm{[M/H]}\lesssim-1.5$ or less than 0.03 solar metallicity. Two of these components (component 2 along SDSS J0946$+$5123  and component 4 along SDSS J0246$-$0059, see \S A1 and \S A10, respectively) contain anomalously little ionic metals despite hosting the bulk of the total \ion{H}{I} column density in their respective absorbers. The other component, a remarkable metal-free LLS along  SDSS\, J1357$+$0435, has the lowest metallicity in the COS-LRG sample, with an estimated metallicity upper limit of $\mathrm{[M/H]}<-2.3$ or lower than $0.5$ percent of solar metallicity. Such low metallicities in low-redshift LLSs are consistent with recently accreted gas from the IGM (e.g., Hafen \etal\ 2017). We estimate the rate of very-low-metallicity LLS (with $\mathrm{[M/H]}\lesssim-1.5$) to be $0.43^{+0.25}_{-0.22}$ assuming binomial statistics (Gehrels 1986), which suggests that chemically pristine gas accreted from the IGM contributes to a substantial fraction of LLS population in LRG halos at $z\lesssim0.5$. A more in-depth discussion on the possible origins of low-metallicity gas in LRG halos is presented in a companion paper on the galaxy environment of the chemically pristine LLS observed along SDSS\, J1357$+$0435 (Chen \etal\ 2018b). 

Next, we present a plot of gas density $n_\mathrm{H}$ versus $d$ in the left panel of Figure 9. Similar to what is seen with gas metallicities, the inferred $n_\mathrm{H}$ shows substantial variations among different components detected within the gaseous halo of a given LRG. In half of LRG halos ($6/12$) that exhibit multi-component absorption profiles, we find over a factor of 10 difference in $n_\mathrm{H}$ between the highest and lowest density components. These large intra-halo variations in $n_\mathrm{H}$ are observed at both $d<100$ kpc and $d>100$ kpc. The median gas density of individual components is log\,$\langle n_\mathrm{H} \rangle/ \cmjjj=-2.4\pm0.1$, where the uncertainty is calculated using a combined bootstrap and Monte-Carlo resampling. In addition, the estimated 16-84 percentile range in gas density is  log\,$n_\mathrm{H}/ \cmjjj=(-3.0,-1.8)\, \cmjjj$.

\begin{figure}
\includegraphics[scale=0.80]{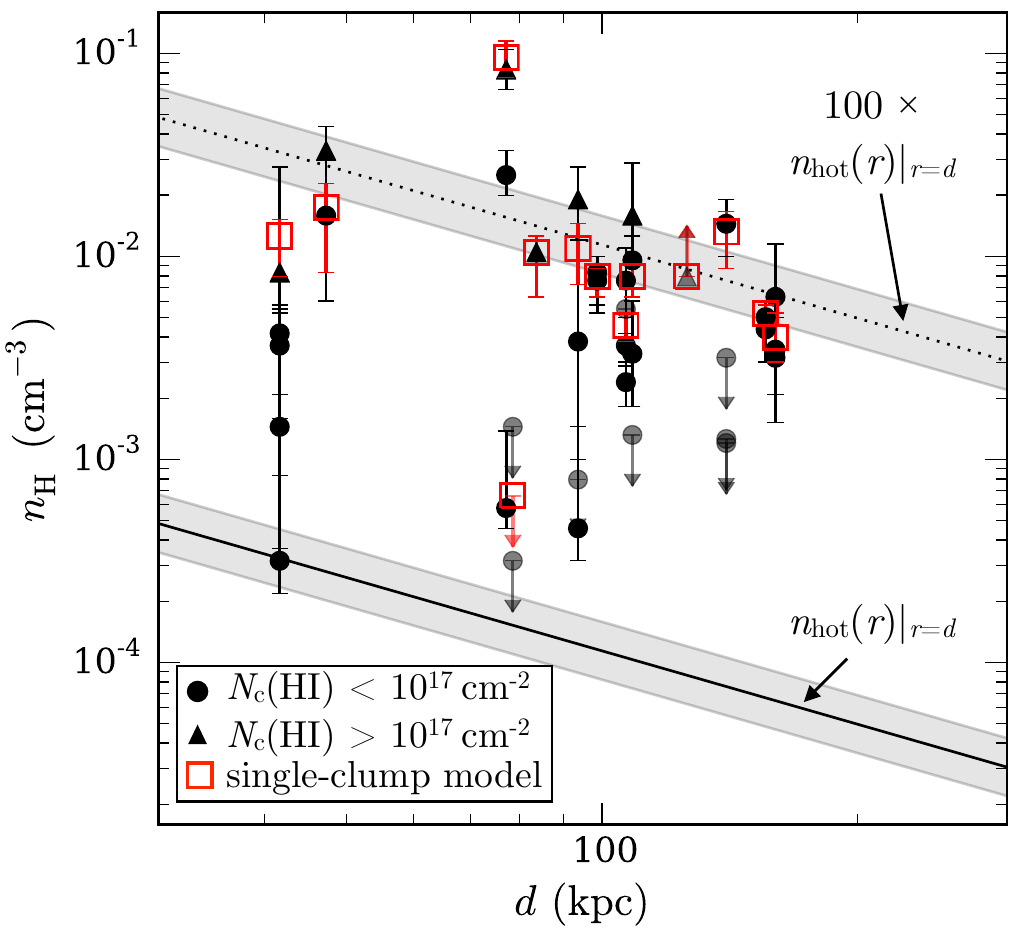}
\vspace{-1.em}
\caption{Spatial distribution of gas densities in the cool ($T\sim10^4\,$K) and hot ($T\sim10^6\,$K) CGM of LRGs. 
First, $n_\mathrm{H}$ is plotted versus $d$ and shown in circles/triangles for optically thin/thick components. In hollow red squares we show $n_\mathrm{H}$ derived for a single-clump model, where a single density is assumed for different components  within an individual halo. For comparison, the solid line shows ${n}_\mathrm{hot}(r)|_{r=d}$,  the radial profile of mean hot gas density in massive halos ($M_\mathrm{h}\sim10^{13}\,\mathrm{M}_\odot$; Singh \etal\ 2018) evaluated at $r=d$. The shaded gray area signifies the 68\% confidence region of this power-law density profile. Finally, in dashed line we show the same hot CGM density profile which has been scaled up by a factor of 100.}
\label{figure:ions}
\end{figure}

To investigate whether gas density varies with \ion{H}{I} column density, we plot $n_\mathrm{H}$ versus $N_c\mathrm{(\ion{H}{I})}$ in the right panel of Figure 9.  The corresponding ionization parameter $U$ for a given $n_\mathrm{H}$ is indicated on the right y-axis. The data points exhibit a clear trend of rising gas density with increasing \ion{H}{I} column density. Because the ionizing background radiation is fixed, the observed correlation is consistent with what is expected from a photoionized gas: more optically thick gas has lower ionization parameter $U$ and is therefore less ionized than optically thin gas. The inferred median $U$ for our sample is $\mathrm{log}\,\langle U \rangle=-3.0\pm0.1$. For stronger absorption components with log\,$N_c\mathrm{(\ion{H}{I})/\cmjj}\gtrsim16$, the median ionization parameter is lower,  $\mathrm{log} \langle U \rangle\approx-3.5$, which is comparable to what have been found in previous surveys of $z<1$ pLLSs/LLSs (e.g., Lehner \etal\ 2013).

The strong correlation between $n_\mathrm{H}$ and $N_c\mathrm{(\ion{H}{I})}$ also suggests that cool clumps in the CGM of LRGs follow a well-defined distribution of clump sizes. In the right panel of Figure 9, we plot the expected relationship between $n_\mathrm{H}$ and $N_c\mathrm{(\ion{H}{I})}$ for cool clumps of different thicknesses, from $l=10$ pc to $l=10$ kpc. It is clear that a large majority of clumps are between $\sim10$ pc and $\sim1$ kpc thick, with a mode of $\sim100$ pc. Furthermore, this characteristic clump thickness of $l\sim100$ pc is shared by both optically thin and thick clumps, covering a range of nearly three orders of magnitude in $N\mathrm{(\ion{H}{I})}$. The median clump size estimated for the COS-LRG sample is $\langle l \rangle =120^{+80}_{-40}$ pc, where the uncertainty is calculated using a combined bootstrap and Monte-Carlo resampling. In addition, we estimate that the range of $l$ containing 68 percent of individual components is $l=(20,800)$ pc. The range of inferred clump sizes in LRG halos is in excellent agreement with transverse clump sizes estimated directly from intervening low-ionization absorbers in the spectra of multiply lensed, high-redshift QSOs (e.g., Rauch \etal\ 1999; 2002).

To put the inferred gas densities of cool CGM clumps in a broader context, we compare the inferred $n_\mathrm{H}$ in the cool CGM with the expected gas densities in the hot CGM ($T\sim10^6\,$K) of LRGs in Figure 10. First, $n_\mathrm{H}$ is plotted versus $d$ and shown in circles/triangles for optically thin/thick cool gas. We also show $n_\mathrm{H}$ derived for a single-clump model in hollow red squares, where different components  within a given LRG halo are imposed to have the same density and metallicity.\footnote{Note that the single-clump model results in a positive bias on the inferred distribution of $n_\mathrm{H}$, because the inferred density in the single-clump model is driven predominantly by the densest cool absorption component in each halo.} For comparison, the solid line in Figure 10 represents ${n}_\mathrm{hot}(r)|_{r=d}$, the mean radial profile of mean hot gas density in LRG halos, evaluated at $r=d$. For ${n}_\mathrm{hot}(r)$, we chose a power-law model that describes the hot CGM of massive halos ($M_\mathrm{h}\sim10^{13}\,\mathrm{M}_\odot$) from Singh \etal\ (2018), which is based on a combined analysis of X-ray and Sunyaev-Zel'dovich (SZ) signals from a stack of $\sim10^5$ massive galaxies at $z\sim0.1$. The dotted line in Figure 10 represents a boosted Singh \etal\ (2018) hot gas density profile which has been scaled up by a factor of 100, for visual comparison. 

It is clear that the projected radial density profile of optically thick cool gas sits about 100 times higher than ${n}_\mathrm{hot}(r)|_{r=d}$. Considering the two orders of magnitude of temperature difference between cool CGM gas ($T\sim10^4$ K, see \S 4.2) and X-ray emitting hot gas ($T\sim 10^6$ K), the inferred density contrast of $\sim100$ indicates that optically thick cool CGM gas occurs at $r\sim d$ and is close to being in pressure equilibrium with the hot halo (e.g., Mo \& Miralda-Escud\'e 1996; Maller \& Bullock 2004; see also a more in-depth discussion in \S 5.2). In contrast, Figure 10 shows only $\sim40$ percent of optically thin components have densities consistent with being in thermal pressure equilibrium with the hot halo at $r\sim d$, which implies that a majority of optically thin absorbers likely occur at larger radii in the halo, $r>d$.

\section{Discussion}

The COS-LRG survey consists of a mass-limited sample of 16 LRGs with log\,$M_\mathrm{star}/\mathrm{M}_\odot>11$ and $d<160$ kpc from a background QSO, chosen without any prior knowledge of the presence or absence of absorption features in the LRG halos. This mass-limited and absorption-blind sample enables an unbiased and accurate characterization of the physical properties and metallicities in the CGM of these intermediate-redshift massive ellipticals. Our survey demonstrates that despite their quiescent nature, LRGs are surrounded by widespread and chemically enriched cool gas. By carrying out a detailed ionization analysis on the absorbers, we discover large variations in gas metallicities and number densities in the cool gas, both within individual LRG halos and across the entire sample. When compared with the expected gas densities in the hot halo, the inferred densities of the cool gas imply that cool clumps in the CGM of LRGs are likely supported by thermal pressure. In addition, we find kinematic mismatches between high-ionization \ion{O}{VI} gas and lower-ionization gas traced by \ion{H}{I} and associated metal ions, which suggest different physical origins of the gas. We now discuss the implications of our study. 

\subsection {Total mass in the cool CGM of LRGs}

The relative amounts of gas that reside in different phases of the CGM are governed by the interplay of accretion and feedback, as well as the detailed gas physics. Empirical constraints on the total mass of the gaseous halo around galaxies are therefore critical to test the validity of current theoretical models of galaxy formation. However, previous estimates of the total mass in the cool CGM of quiescent galaxies suffer from large uncertainties of up to two orders of magnitude, due to the unknown ionization state of the gas (e.g., Thom \etal\ 2012; Zhu \etal\ 2014). 

Here, we leverage the results of our ionization analysis of the COS-LRG dataset in order to infer the surface mass density profile and estimate the total gas mass in the cool CGM of LRGs. For each absorption system, we first calculate the total hydrogen column density, $N_\mathrm{H}$, according to the following equation,
\begin{equation}
\mathrm{log\,}N_\mathrm{H} 		=  \mathrm{log\,} \sum_{i} \frac{N_c\mathrm{(\ion{H}{I})}_i}{f_{\mathrm{H^0}_{i}}},
\end{equation}
where $f_{\mathrm{H^0}_{i}}$ is the hydrogen neutral fraction for component $i$ determined by our ionization analysis, and the sum is evaluated over all components in the absorption system. For components with poorly constrained ionization state, the range of allowed ionization fraction correction is computed by imposing that the corresponding clump size is not larger than 1 kpc. Once $N_\mathrm{H}$ is calculated, the corresponding cool gas surface mass density can be computed using the relation $\Sigma_\mathrm{cool} = 1.4\,m_\mathrm{H}\,N_\mathrm{H}$, where $m_\mathrm{H}$ is the mass of the hydrogen atom and a factor of 1.4 is introduced to account for the contribution of helium to the total gas mass. 

\begin{figure}
\includegraphics[scale=0.75]{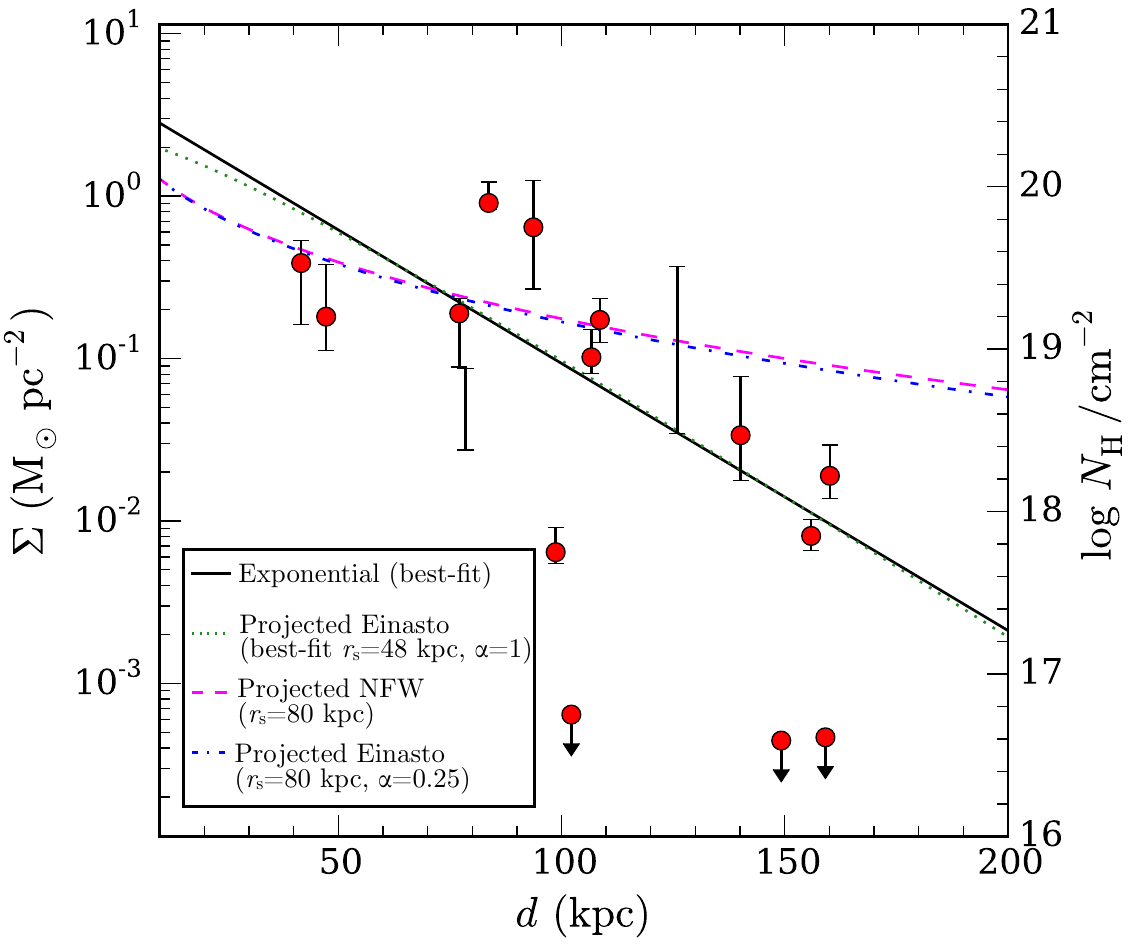}
\vspace{-1.4em}
\caption{Surface mass density profile of the cool CGM of LRGs. The surface mass density, $\Sigma_\mathrm{cool}$,  is estimated by calculating the total hydrogen column density along each sightline, $N_\mathrm{H}$, with the estimated ionization fraction corrections applied to the observed \ion{H}{I} column densities according to equation (2). Downward arrows represent estimated upper limits on $\Sigma_\mathrm{cool}$ for LRG halos with non-detected \ion{H}{I}. Empty vertical error bars show the range of allowed surface mass density for systems with weakly constrained $n_\mathrm{H}$, calculated by imposing that the corresponding clump size is $l\lesssim 1$ kpc. $\Sigma_\mathrm{cool}$ exhibits a steep decline with increasing $d$, which cannot be reproduced by a projected NFW profile or Einasto profile expected for $M_\mathrm{h}\approx10^{13}\,\mathrm{M}_\odot$ dark-matter halos (dashed and dash-dotted curves). On the other hand, the radial profile of $\Sigma_\mathrm{cool}$ is best described by an exponential profile in either 2D (solid line) or 3D (a projected Einasto profile with $\alpha\approx1$; dotted curve).}
\label{figure:ions}
\end{figure}

The spatial profile of cool gas surface mass density in the CGM is shown in Figure 11. It is clear that
$\Sigma_\mathrm{cool}$ exhibits a declining trend with $d$. At $d<100$ kpc, the mean $N_\mathrm{H}$ is
log\,$\langle N_\mathrm{H}\rangle/\cmjj=19.5\pm0.2$, which is equivalent to a mean cool gas surface density of 
$\langle\Sigma_\mathrm{cool}\rangle\approx 0.4^{+0.2}_{-0.1}\,\mathrm{M_\odot\,pc^{-2}}$. 
The mean $\Sigma_\mathrm{cool}$ at $d<100$ kpc is comparable to inferred surface mass densities in the predominantly neutral ISM of an LRG lensing galaxy at $z=0.4$ (Zahedy \etal\ 2017a). In contrast, the mean $N_\mathrm{H}$ and $\Sigma_\mathrm{cool}$ at $d=100-160$ kpc are significantly lower, log\,$\langle N_\mathrm{H}\rangle/\cmjj=18.7^{+0.2}_{-0.3}$ and $\langle\Sigma_\mathrm{cool}\rangle\approx (0.06\pm0.03)\,\mathrm{M_\odot\,pc^{-2}}$. 

To gain insights into the observed cool gas surface mass density profile in LRG halos, we compare the data with different analytic functions to obtain a best-fit model that characterizes the relationship between $\Sigma_\mathrm{cool}$ and $d$. We first consider a simple power-law in $d$, which has been used to describe the spatial distributions of \ion{H}{I} and metal equivalent widths in the cool CGM (e.g., Chen \etal\ 2001, 2010), and find that it cannot reproduce the rapid decline of $\Sigma_\mathrm{cool}$ with increasing $d$. In contrast, we find that the steepness of the $\Sigma_\mathrm{cool}$ profile is well-fitted by a exponential model in 2D, $\Sigma_\mathrm{cool} = \Sigma_0\,e^{-d/d_s}$, with best-fit parameters of $d_s=(27\pm4)$\,kpc and $\Sigma_0=(4.1\pm1.4)\,\mathrm{M_\odot\,pc^{-2}}$ determined from a likelihood analysis (Figure 11, solid line). 

\begin{figure}
\includegraphics[scale=0.74]{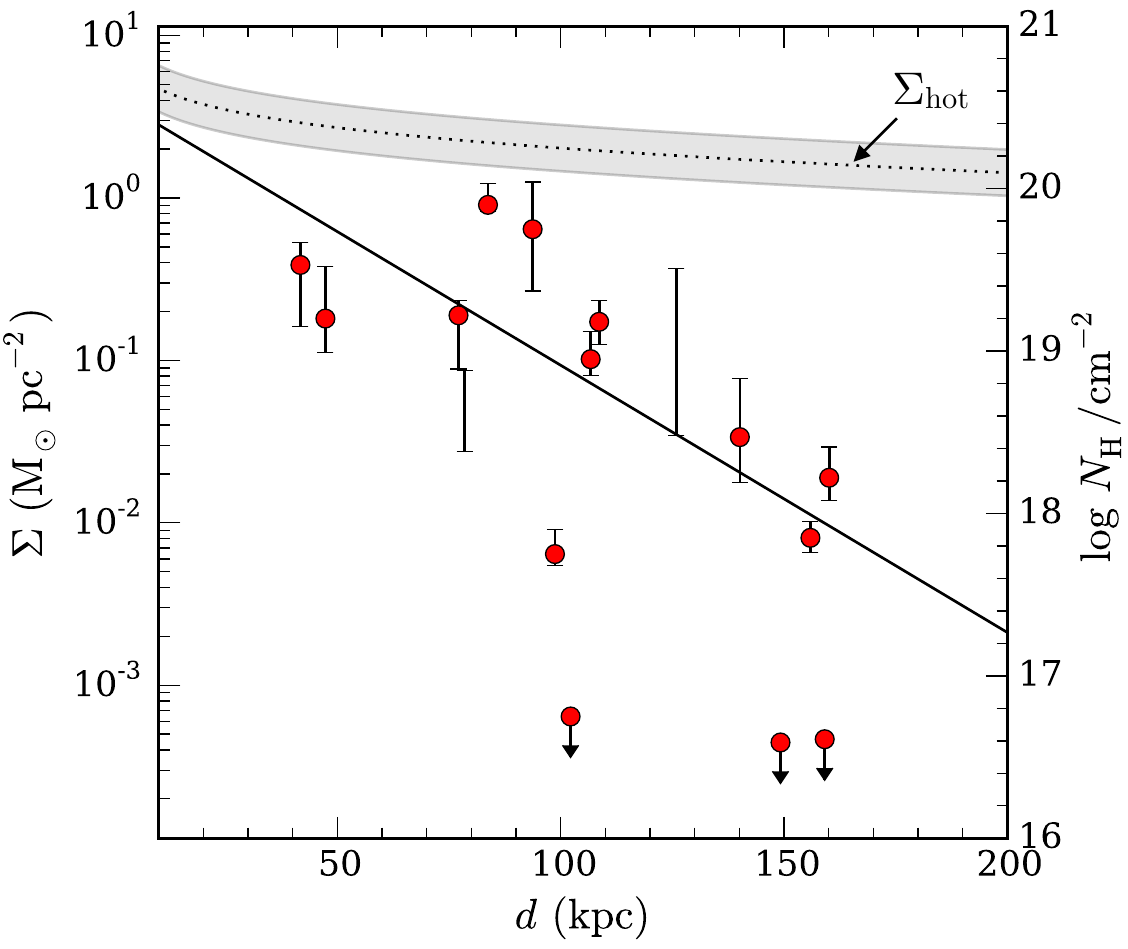}
\vspace{-1.4em}
\caption{Comparison between the surface mass density profiles of the cool and hot CGM of LRGs. Symbols are the same as those in Figure 11, with the solid black line showing the best-fit exponential model describing the relationship of $\Sigma_\mathrm{cool}$ with $d$. In contrast, the dotted curve shows the inferred surface mass density profile in the hot CGM of LRG-sized halos ($M_\mathrm{h}\approx10^{13}\,\mathrm{M}_\odot$), based on a combined X-ray and SZ analysis (Singh \etal\ 2018), with the 68\% confidence region shaded in gray.  Within $d<160$ kpc from LRGs, we estimate a total cool gas mass of $\mathrm{\mathit{M}_{cool} =  1.5^{+0.7}_{-0.3}\times10^{10}\,\mathrm{M_\odot}}$, which is $\sim6-13$ percent of the expected total mass in the hot CGM.}
\label{figure:ions}
\end{figure}

Next, to investigate whether cool baryons follows the large-scale dark matter mass distribution in the halo, we compare the $\Sigma_\mathrm{cool}$ profile in Figure 11 to the projected surface density of spherically symmetric functions commonly used to describe dark-matter mass distributions, including the Einasto and NFW profiles. The Einasto profile (Einasto 1965) is defined by a power-law logarithmic slope, $\mathrm{d\,ln}\,\rho/\mathrm{d\,ln}\,r \equiv-2\,(r/r_s)^\alpha$, in which the scale radius $r_s$ and shape parameter $\alpha$ are free parameters governing its shape.  Relatively shallow Einasto profiles with $\alpha\lesssim0.3$ have been found to produce good fits to the mass distribution of simulated dark-matter halos (e.g., Hayashi \& White 2008; Gao \etal\ 2008; Dutton \& Macci\`o 2014). By fitting a projected Einasto profile to our data, we find that $\Sigma_\mathrm{cool}$ requires a steep Einasto profile with $\alpha=1.0^{+0.6}_{-0.2}$ and $r_s=48^{+19}_{-8}\,$kpc (Figure 11, dotted curve). Note that $\alpha$ and $r_s$ is degenerate in a way that models with larger $r_s$ would require still higher values of $\alpha$ in order to fit the observations.

Because an Einasto profile with $\alpha=1.0$ is equivalent to an exponential profile in 3D, this exercise demonstrates that reproducing the observed $\Sigma_\mathrm{cool}$ requires an underlying density profile that is exponentially declining with radius. In contrast, neither a projected NFW profile (Navarro \etal\ 1997) with $r_s=80$\, kpc (expected for $M_\mathrm{h}\approx10^{13}\,\mathrm{M}_\odot$ halos at $z=0.5$, e.g., Dutton \& Macci\`o 2014) nor a shallow Einasto profile with $\alpha<0.3$ produces a good fit to the the data. As illustrated in Figure 11 (dashed and dashed-dotted curves), these dark-matter-like profiles can be ruled out because they cannot reproduce the sharp decline of $\Sigma_\mathrm{cool}$ with $d$. Therefore, it appears that the mass distribution of cool gas in the CGM of LRGs is different from the expected mass distribution of the underlying dark matter halo.  

The estimated total mass in cool CGM gas within $d=160$ kpc ($\sim 0.3\, R_\mathrm{h}$) of LRGs is
\begin{equation}
\mathrm{\mathit{M}_{cool}(<160\,kpc)} =  1.5^{+0.7}_{-0.3}\times10^{10}\,\mathrm{M_\odot},
\end{equation}
which is obtained by first multiplying the estimated $\langle\Sigma_\mathrm{cool}\rangle$ at $d<100$ kpc and $d=100-160$ kpc by their respective surface areas modulo the covering fraction inferred from Figure 11 (unity at $d<100$ kpc and $\sim0.7$ at $d=100-160$ kpc), and then summing them. We obtain a similar estimate of $\mathit{M}_\mathrm{cool} = (1-2)\times 10^{10}\,\mathrm{M_\odot}$ by integrating the best-fit exponential model for $\Sigma_\mathrm{cool}$ from $d=0$ to 160 kpc. Note that we choose to limit our mass estimate out to only $d=160$ kpc in the CGM because it is the largest projected distance probed in the COS-LRG data set. Our estimate above should therefore be considered as a lower limit on the total mass of cool, photoionized gas in massive quiescent halos. However, note that if we naively adopted the estimated $\langle\Sigma_\mathrm{cool}\rangle$ and gas covering fraction at $d=100-160$ kpc and extrapolated these values out to $d=500$ kpc, which is the typical virial radius of $z\sim0.4$ LRGs, the mass estimate in equation (3) would increase by a factor of three, to $\mathrm{\mathit{M}_{cool}\approx 4\times10^{10}\,\mathrm{M_\odot}}$. 

Our mass estimate demonstrates that despite their quiescent nature, LRGs at $z\sim0.4$ still host a significant reservoir of cool gas in their circumgalactic space. Furthermore, the estimated cool CGM mass of $\sim10^{10}\,\mathrm{M_\odot}$ is comparable to the inferred total mass in the cool CGM of lower-mass and predominantly star-forming $L^*$ galaxies (e.g., Chen \etal\ 2010; Prochaska \etal\ 2011; Werk \etal\ 2014; Stern \etal\ 2016). At the same time, the total baryon mass budget of the typical LRG in our sample is $\approx1.6\times10^{12}\,\mathrm{M_\odot}$ within the virial radius, which is estimated for the median halo mass in COS-LRG, $M_\mathrm{h}\approx10^{13}\,\mathrm{M}_\odot$ and adopting a baryon-to-dark-matter mass ratio of $\Omega_b/\Omega_\mathrm{DM}=0.16$. Thus, the inferred $M_\mathrm{cool}$ in the CGM is at most $\approx3$ percent of the total baryon budget for typical LRGs. 

It is also interesting to compare the total mass contained in cool clumps to the expected total mass of the hot CGM. The dotted curve in Figure 12 represents  $\Sigma_\mathrm{hot}$, the inferred hot gas surface mass density profile in massive $M_\mathrm{h}\sim10^{13}\,\mathrm{M}_\odot$ halos. $\Sigma_\mathrm{hot}$ is computed from the hot gas density profile shown in Figure 10 (Singh \etal\ 2018), assuming a unity volume filling fraction for the hot gas. The expected spatial mass profile of hot gas is more spatially extended than the observed $\Sigma_\mathrm{cool}$ profile (see also Liang \etal\ 2016). Within $d=160$ kpc from LRGs, we infer a total hot gas mass of  $\mathrm{\mathit{M}_{hot}(<160\,kpc)} = (1.7\pm0.5) \times10^{11}\,\mathrm{M_\odot}$. Comparing our estimate of $\mathrm{\mathit{M}_{cool}}$ with the inferred $\mathrm{\mathit{M}_{hot}}$, the cool-to-hot gas mass ratio in the CGM of LRGs is
\begin{equation}
X_\mathrm{cool}\equiv\mathrm{\mathit{M}_{cool}/\mathit{M}_{hot}}\approx0.06-0.13
\end{equation}
at $d<160$ kpc,  which is comparable to the inferred $X_\mathrm{cool}$ in the interstellar medium (ISM) of one of these massive ellipticals (Zahedy \etal\ 2017a). 

Furthermore, our data also hint at a declining $X_\mathrm{cool}$ with increasing projected distance from LRGs, from $X_\mathrm{cool}\sim0.1 - 0.2$ at $d<100$ kpc, to no more than  $X_\mathrm{cool}\sim0.01-0.03$ at $d=100-160$ kpc. The declining $X_\mathrm{cool}$ with increasing projected distance implies that the volume filling factor of cool gas is significantly lower in the outer CGM, at  galacto-centric radius $r\gtrsim100$ kpc, than it is in the inner CGM, at $r\lesssim100$ kpc.
The mean volume filling factor of cool gas can be estimated using line-of-sight observables according to the following expression (e.g., McCourt \etal\ 2018),  
\begin{equation}
\langle f_\mathrm{V} \rangle = \langle N_\mathrm{cl} \rangle \times \frac{ l}{L},
\end{equation}
where $l \equiv N_\mathrm{H}/n_\mathrm{H}$ is the clump thickness along the line of sight, $L$ is the path length through the CGM, and $\langle N_\mathrm{cl} \rangle$ is the mean number of clumps per line of sight. 
As shown in the right panel of Figure 9, the inferred clump sizes in COS-LRG range from 10 pc to 10 kpc, with a median value and mode of $l \sim100$ pc. For the purpose of this calculation, we approximate $\langle N_\mathrm{cl} \rangle$ to be the average number of discrete components identified per sightline. Based on our data, there are on average $\langle N_\mathrm{cl} \rangle_{<100}=3.7^{+0.6}_{-0.4}$ discrete components at $d<100$ kpc, which subsequently declines to $\langle N_\mathrm{cl} \rangle_{>100}=2.2^{+0.9}_{-0.5}$ at $d>100$ kpc. Given that $\langle N_\mathrm{cl} \rangle_{<100}$ has contributions from both the inner and outer parts of the halo, we can solve for $\langle f_\mathrm{V} \rangle$ in the inner and outer CGM separately using the following approximations:
\begin{align}
\langle f_\mathrm{V} \rangle_\mathrm{inner}&\approx(\langle N_\mathrm{cl} \rangle_{<100}-\langle N_\mathrm{cl} \rangle_{>100}) \times\frac{ l}{L_\mathrm{inner}}\\
\langle f_\mathrm{V} \rangle_\mathrm{outer}&\approx\langle N_\mathrm{cl} \rangle_{>100} \times\frac{ l}{L_\mathrm{outer}}.
\end{align}
By plugging the different quantities above to equations (6) and (7) and adopting $L_\mathrm{inner}=100$ kpc and  $L_\mathrm{inner}=500$ kpc, 
we estimate that the mean volume filling factor for typical clumps with $l=100$ pc is
$\langle f_\mathrm{V} \rangle_\mathrm{inner} \sim 2\times10^{-3}$ in the inner halo ($r\lesssim100$ kpc), and 
$\langle f_\mathrm{V} \rangle_\mathrm{outer} \sim 4\times10^{-4}$ in the outer ($r\gtrsim100$ kpc) halo of LRGs. This exercise illustrates that while the  cool gas covering fraction in the CGM of LRGs is high, the volume filling factor can remain very low (for possible theoretical explanations, see e.g., McCourt \etal\ 2018; Liang \& Remming 2018)

\subsection {On the origin and fate of cool gas in LRG halos}

In the previous section, we show that despite their ``red and dead" nature, LRGs at $z\sim0.4$ harbor as much as $\sim10^{10}\,\mathrm{M_\odot}$ of photoionized $T\sim10^4$ K gas in their extended halos. This massive reservoir of cool gas appears to consist of compact clumps with a characteristic size of $\sim100$ pc (\S 4.3.2), which are pressure confined by the hot gaseous halo that is expected to be ubiquitous around LRGs.

To gain a better understanding of the nature of the cool gas around LRGs, we now consider our observational results in the larger context of a multiphase gaseous halo around LRGs.

The physical formalism for a two-phase CGM was first explored by Mo \& Miralda Escude (1996), who argued that QSO absorption systems in the vicinity of galaxies originate in cool clouds which are in thermal pressure equilibrium with the hot halo. To explain the formation of cool clumps within an otherwise hot corona, Maller \& Bullock (2004) elaborated on this simple model by incorporating multiphase cooling in the halo. In their analytic model, cool clumps originate from condensation in a hydrostatically stable hot halo, triggered by thermal instabilities which develop locally when the cooling time ($\tau_\mathrm{cool}$) is comparable to the the dynamical timescale ($\tau_\mathrm{ff}$) of the gas. Building on these earlier works, more recent numerical simulations have shown that a multiphase halo can form as soon as $\tau_\mathrm{cool}/\tau_\mathrm{ff}\lesssim10$ (e.g., Sharma \etal\ 2012; McCourt \etal\ 2012), which is consistent with observations of multiphase gas in a number of nearby galaxy clusters and elliptical galaxies (e.g., Voit \etal\ 2015a, b).

Under the multiphase-cooling paradigm, cool clumps form within the cooling radius, $R_c$ inside the halo, where thermal instability is prone to develop. For LRG-sized halos, $R_c$ is estimated to be between $100$ and $200$ kpc (Maller \& Bullock 2004, equation 18), which is qualitatively consistent with a number of COS-LRG findings, including the observed decline in \ion{H}{I} covering fraction with $d$ (Paper I and \S 4.1) and the steep drop in inferred cool gas surface mass density and volume filling factor at $d\gtrsim100$ kpc (\S 5.1). 

In the absence of vigorous star formation activity capable of driving large scale outflows, circumgalactic cool gas is likely falling toward the center of the halo. The infall interpretation is supported by the observed line-of-sight velocity dispersion of individual absorbing components in the COS-LRG sample, $\sigma_\mathrm{gas}\approx150$ \kms, which is merely $\sim 60$ percent of what is expected from virial motion (see also Huang \etal\ 2016; Lan \etal\ 2018). The observed narrow distribution of line-of-sight velocities indicates that dissipative processes are effective in slowing down cool clumps as they undergo orbital motions in the halo. By attributing the observed deceleration as due to ram pressure drag exerted by the hot halo, Huang \etal\ (2016) calculated an upper limit on the cool clump mass of $m_{\rm cl}\lesssim10^4\,\mathrm{M}_\odot$ in LRG halos. This dissipative interaction with the hot gas would lead to orbital decay, causing cool clumps to fall toward the galaxy. {\it But does the cool gas survive this inward journey ?}

The survival of cool clumps depends on whether the infall time is sufficiently short compared to the timescales of cloud disruption processes acting on them. Cloud destruction is driven predominantly by thermal conduction between cool clumps and and the surrounding hot gas. If cool clumps are not sufficiently massive, they will not only decelerate due to ram pressure drag, but also evaporate due to thermal conduction before reaching the LRG at the center of the halo. 

We expect cool clumps to eventually reach terminal speed when the ram pressure drag force exerted by the hot gas is balanced by the gravitational pull of the halo on the clump. By identifying this terminal speed with the observed $\sigma_\mathrm{gas}\sim 0.6\, \sigma_\mathrm{vir}$ in LRG absorbers, we can compute the typical cool clump mass (Maller \& Bullock 2004, equation 39),
\begin{equation}
m_{\rm cl}\approx 7.7\times 10^2\,T_6^{-3/8}(\Lambda_Z\,t_8)^{1/2}\,\mathrm{M_\odot},
\end{equation}
where $T_6=T/10^6\,{\mathrm K}$ is the temperature of the hot corona, $\Lambda_Z$ is a cooling parameter which depends on the gas metallicity and $t_8 = t_f/8\, {\rm Gyr}$ is the halo formation timescale. For typical LRG halos in our sample, $T\sim 6\times10^6\,\rm K$ assuming an isothermal gas, and $t_f \sim 9\, {\rm Gyr}$ assuming $t_f$ is comparable to the age of the Universe at $z\sim0.4$. Using equation (8), we find that $m_{\rm cl}=(2-8)\times 10^2\,\mathrm{M}_\odot$ for metallicities of between 0.01 solar and solar, respectively.
This kinematics-based mass estimate can be compared to the cool clump mass independently constrained from our ionization analysis. Based on a combined bootstrap and Monte-Carlo resampling of the full range of inferred cool gas densities and characteristic clump sizes (\S 4.3.2), we estimate that the characteristic clump mass has a median value of 
$\langle m_{\rm cl}\rangle=50-1000\,\mathrm{M}_\odot$, which is consistent with the mass range estimated using equation (8). 

Given a mass of cool clump $m_{\rm cl}$, the characteristic   for cloud evaporation due to thermal conduction is given by (Maller \& Bullock 2004, equation 35),
\begin{equation}
\tau_{\rm evap}\approx 1.6\,  m_{\rm cl}^{2/3}\, T_6^{-3/2}(\Lambda_Z\,t_8)^{-1/3} \, \mathrm{Myr}.
\end{equation}
For typical $m_{\rm cl}\sim10^2-10^3 \,\mathrm{M}_\odot$ and a metallicity of between 0.01 solar and solar, we find that the evaporation timescale is 
$\tau_{\rm evap}\sim 1-20$ Myr. The expected evaporation time for typical cool clumps is vastly shorter than the minimum infall time of $\tau_{\rm infall}\sim 200-500$ Myr estimated for cool clumps which condense from the hot gas at $R_c\sim100-200$ kpc. The evaporation time is still significantly less than infall time even for clumps as massive as $m_{\rm cl}\sim10^4 \,\mathrm{M}_\odot$ (Huang \etal\ 2016). This exercise suggests that cool clumps travel only a relatively small distance in the halo during their lifetimes, and a majority of clumps originating at large distances will evaporate before reaching the center of the halo. 

The implication that a majority of cool clumps in the gaseous halo of LRGs never reaches the central galaxy could explain a number of observational findings that the cool ISM mass in massive quiescent galaxies remains low, $\mathit{M}\mathrm{_{cool}(ISM)}\sim 10^{8-9}\,\mathrm{M_\odot}$ (e.g., Serra \etal\ 2012; Zahedy \etal\ 2017a; Young \etal\ 2018), despite the existence of a much larger reservoir of cool gas in the halo, $\mathit{M}\mathrm{_{cool}(CGM)}\sim 10^{10}\,\mathrm{M_\odot}$. At the same time, the fact that cool gas is routinely observed in the gaseous halo of LRGs suggests that cool clumps are continuously formed and destroyed in the predominantly hot gaseous halo. In this quasi steady state, $\sim5-10$ percent of the CGM gas by mass reside in cool, $\sim10^4\,$K phase at any given time, a balance which is most likely determined by the amount of additional heating available to offset the increased cooling rate from the cool gas. 

Finally, we note although our discussion above is based entirely on considering our observations in the context of thermal instability in a multiphase CGM, it does not exclude the possibility that cool gas in LRG halos is also generated by other physical processes. These additional mechanisms include cool gas recently accreted from the IGM along filaments (e.g., Churchill \etal\ 2012; Huang \etal\ 2016), gas originating in and/or stripped from the CGM or ISM of satellite galaxies (e.g. Gauthier \etal\ 2010; Huang \etal\ 2016), and gas ejected by SNe Ia (e.g., Zahedy \etal\ 2016; 2017b).  Indeed, our finding that gas density and metallicity can vary by more than a factor of ten within individual LRG halos indicates that the CGM is a multiphase mixture of gas with different chemical enrichment histories, which hints at multiple origins of the cool gas. However, cool clumps in LRG halos are subject to the same interactions with the hot gas regardless of their physical origin. Therefore, our conclusion above can be applied generally on the nature of cool gas in massive quiescent halos.

\subsection {The nature of \ion{O}{\uppercase{VI}} absorbers in the CGM: insight from massive halos}

A significant finding in CGM studies over the past decade is the ubiquitous presence of strong \ion{O}{VI} absorption with log\,$N\mathrm{(\ion{O}{VI})/\cmjj}\sim14.5$ around $\sim L^*$ star-forming galaxies (e.g., Tumlinson \etal\ 2011). At the same time, \ion{O}{VI}-bearing is found to be less prevalent in the gaseous halos of passive galaxies (e.g., Chen \& Mulchaey 2009; Tumlinson \etal\ 2011; Johnson \etal\ 2015). The apparent dichotomy between \ion{O}{VI} absorption properties around late-type and early-type galaxies is often ascribed to a direct link between star-formation and the observed warm gas properties: recent star formation drive powerful outflows that eject metals to large distances in the CGM. Alternatively, the lower incidence of strong  \ion{O}{VI} absorption in passive galaxies has been attributed to further ionization of oxygen to higher states (e.g., \ion{O}{VII} and \ion{O}{VIII}) in the more massive and hotter halos of passive galaxies (e.g., Oppenheimer \etal\ 2016). 

To gain new insights into the nature of \ion{O}{VI} absorbers around galaxies, it is necessary to compare the observed \ion{O}{VI} absorption properties around galaxies of different masses. In Figure 13, we present current observational constraints on circumgalactic \ion{O}{VI} absorption spanning over more than three decades in galaxy stellar mass, from log\,$M_\mathrm{star}/\mathrm{M}_\odot\sim8$ to log\,$M_\mathrm{star}/\mathrm{M}_\odot>11$.
 \ion{O}{VI} measurements for massive quiescent galaxies are from COS-LRG sample, whereas constraints for $\sim L^*$ star-forming galaxies are from Johnson \etal\ (2015) and those for star-forming dwarf galaxies are adopted from Johnson \etal\ (2017). The mean covering fraction of \ion{O}{VI} at $d<160$ kpc is plotted versus $M_\mathrm{star}$ in the top panel of Figure 13, for a column density threshold of log\,$N\mathrm{(\ion{O}{VI})/\cmjj}>13.5$. In contrast to the near-unity covering fraction of \ion{O}{VI} absorbers around $L^*$ star-forming galaxies, passive LRGs and star-forming dwarf galaxies exhibit lower covering fraction of \ion{O}{VI} gas, at $\sim50-60$ percent. While star-formation driven winds is an attractive scenario to account for the ubiquity of \ion{O}{VI} absorption around $\sim L^*$ star-forming galaxies, it does not explain the lower \ion{O}{VI} covering fraction around star-forming dwarfs. Furthermore, despite the likely absence of strong outflows in LRGs, they still exhibit a significant incidence of \ion{O}{VI}, not to mention comparable covering fractions of lower-ionization metals to what have been observed around star-forming galaxies (Paper I).
 
\begin{figure}
\hspace{0.em}
\includegraphics[scale=1.08]{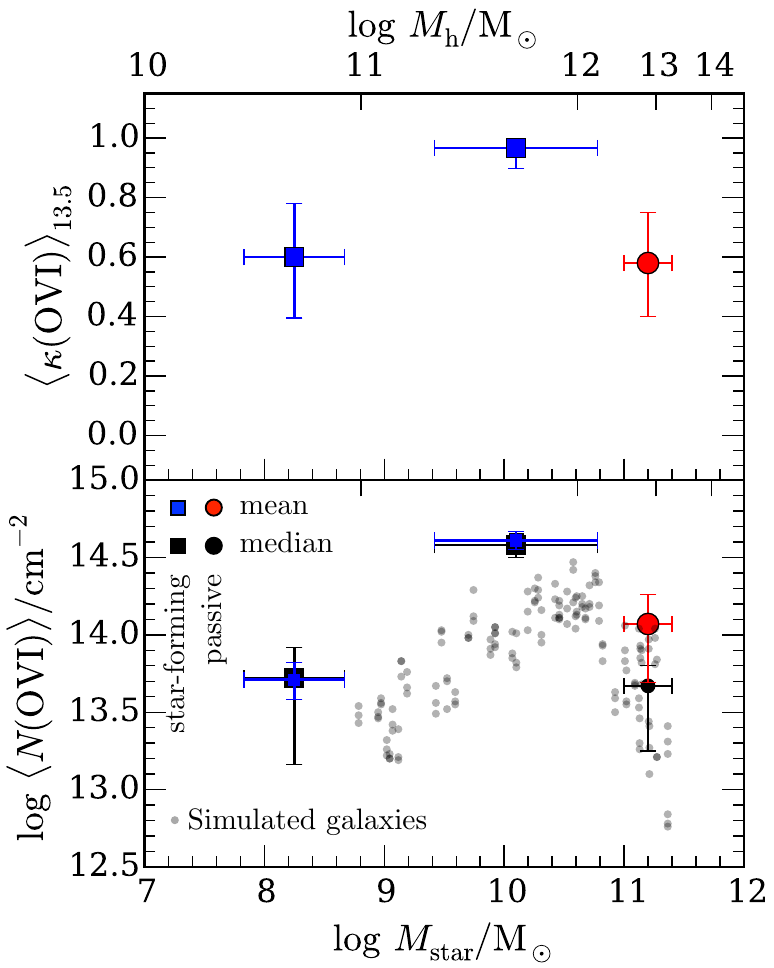}
\vspace{-1.2em}

\caption{Observational constraints on CGM \ion{O}{VI} absorption spanning more than three decades in galaxy stellar mass. Constraints for massive quiescent galaxies are from COS-LRG. Constraints for $\sim L^*$ star-forming galaxies are from Johnson \etal\ (2015), whereas those for star-forming dwarf galaxies are adopted from Johnson \etal\ (2017). {\it Top}:  The mean covering fraction of \ion{O}{VI} plotted versus $M_\mathrm{star}$ at $d<160$ kpc, for a column density threshold of log\,$N\mathrm{(\ion{O}{VI})/\cmjj}>13.5$. For each sample, the median stellar mass is plotted, with the horizontal error bars showing the sample dispersion. The corresponding halo mass for each point is indicated as well, based on the Kravtsov \etal\ (2018) stellar-to-halo-mass relation. The vertical error bars are calculated assuming binomial statistics. 
{\it Bottom}:  the mean and median $N\mathrm{(\ion{O}{VI})}$  at $d<160$ kpc plotted as a function of $M_\mathrm{star}$. The vertical error bars represent the 68\% confidence intervals for the mean and median $N\mathrm{(\ion{O}{VI})}$, calculated using a combined bootstrap and Monte-Carlo resampling. For comparison, in gray circles we plot the mean $N\mathrm{(\ion{O}{VI})}$ within $d<150$ kpc from simulated galaxies in \textsc{eagle} zoom simulations (Oppenheimer \etal\ 2017). 
}
\label{figure:ions}
\end{figure}

\begin{figure*}
\includegraphics[scale=0.77]{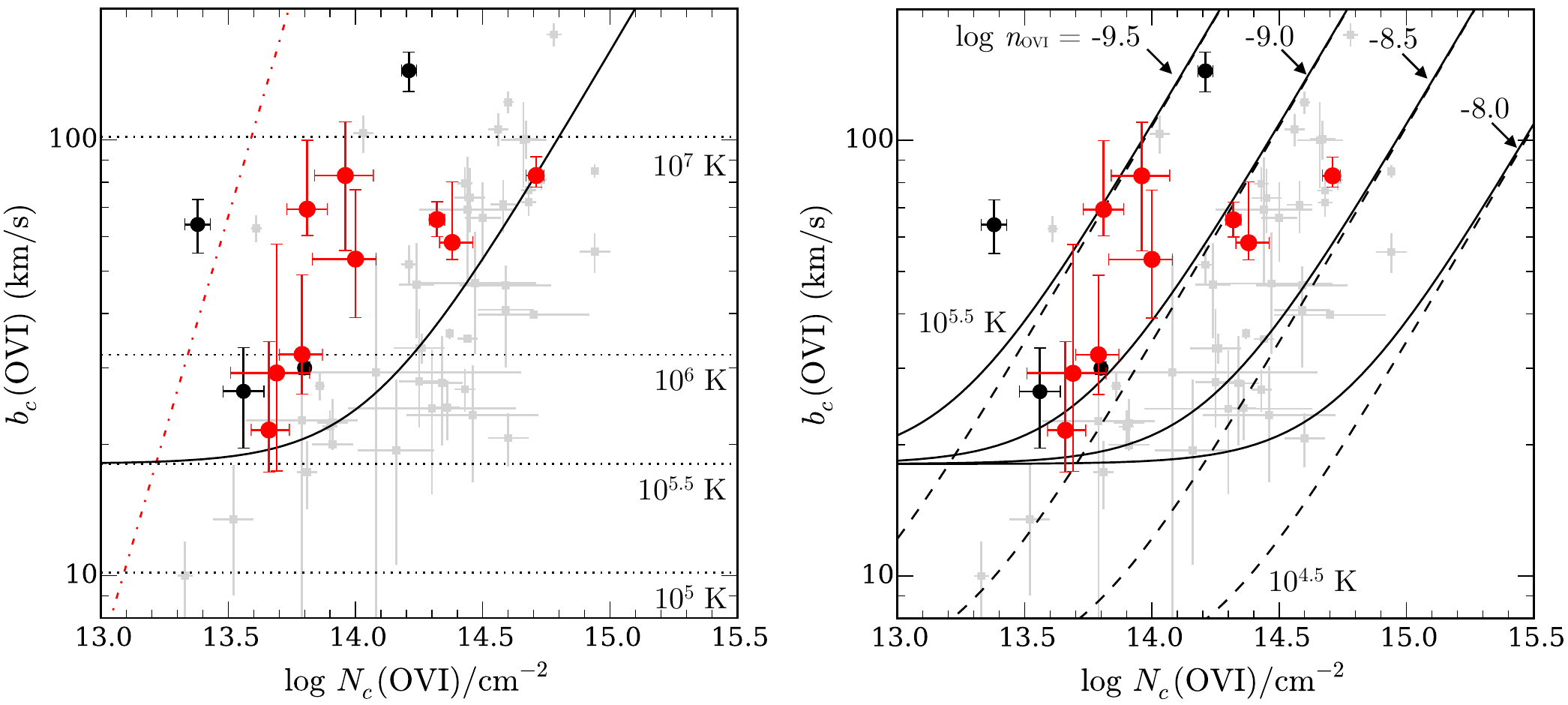}
\vspace{-0.5em}
\caption{\ion{O}{VI} component linewidth versus column density in the COS-LRG sample (red circles). The dashed-dotted line in the left panel represents the typical $3\,\sigma$ detection limit in the COS spectra for our sample. \ion{O}{VI} absorbers detected in the vicinity of massive quiescent galaxies (log\,$M_\mathrm{star}/\mathrm{M}_\odot>11$) in Johnson \etal\ (2015) are shown in black circles, whereas absorbers around $\sim L^*$  star-forming galaxies in in Johnson \etal\ (2015)  are shown in pale gray squares. The solid curve in the left panel shows the expected behavior
for a radiatively cooling, collisionally ionized gas at $T=10^{5.5}$ K  (Heckman \etal\ 2002), whereas the dotted horizontal lines indicated the expected thermal broadening at different temperatures. In the right panel, the curves show expected behaviors for a gravitationally broadened \ion{O}{VI}-bearing gas (Stern \etal\ 2018) at different O$^\mathrm{5+}$ volume densities, calculated for a $T=10^{4.5}$ K gas in dashed curves and $T=10^{5.5}$ K in solid curves.}
\label{figure:ions}
\end{figure*}
 
In the bottom panel of Figure 13, the mean and median $N\mathrm{(\ion{O}{VI})}$  at $d<160$ kpc are plotted versus $M_\mathrm{star}$ for the three galaxy samples. It is clear that CGM \ion{O}{VI} absorption strength peaks in $\sim L^*$ star-forming galaxies (where $M_\mathrm{h}=10^{11.3-12.1}\,\mathrm{M}_\odot$). The mean $N\mathrm{(\ion{O}{VI})}$ declines toward both lower and higher-mass halos, where a majority of \ion{O}{VI} absorbers around LRGs and star-forming dwarf galaxies have $N\mathrm{(\ion{O}{VI})}$ which are 0.5 to 1 dex lower than typical strong \ion{O}{VI} absorbers around $\sim L^*$ galaxies. We note that a similar trend of $N\mathrm{(\ion{O}{VI})}$ with galaxy mass (or halo mass) is also present in simulated galaxies from the \textsc{eagle} zoom simulations (gray points; Oppenheimer \etal\ 2017), despite the fact that the predicted $N\mathrm{(\ion{O}{VI})}$ are systematically lower than observations over the range of galaxy masses considered (see also Nelson \etal\ 2018). 

The strong dependence of circumgalactic $N\mathrm{(\ion{O}{VI})}$ on stellar mass, which in turn correlates with the total halo mass including dark matter, hints at a connection between the dominant ionization state of oxygen and the virial temperature of the halo. In particular, the inferred halo virial temperature for the $\sim L^*$  star-forming galaxy sample is $T_\mathrm{vir}=10^{5.3-5.8}\,$K, which is coincident with the narrow range of temperatures where the fractional abundance of \ion{O}{VI} is at a maximum in collisionally ionized gas (e.g., Heckman \etal\ 2002; Gnat \& Sternberg 2007; Oppenheimer \& Schaye 2013). In contrast, the expected virial temperatures for LRGs ($M_\mathrm{h}=10^{12.6-13.5}\,\mathrm{M}_\odot$) and dwarf galaxies ($M_\mathrm{h}<10^{11}\,\mathrm{M}_\odot$) are $T_\mathrm{vir}=10^{6.5-7.0}\,$K and $T_\mathrm{vir}\lesssim10^{5}\,$K, respectively. At these temperatures, the expected \ion{O}{VI} ionization fractions are very small ($<0.01$) under collisional ionization models. The observed peak of $N\mathrm{(\ion{O}{VI})}$ versus galaxy mass relation in $\sim L^*$ star-forming halos supports the interpretation that the high column of \ion{O}{VI} around these galaxies originate in collisionally ionized gas at $T\sim T_\mathrm{vir}$ (e.g., Oppenheimer \etal\ 2016; Werk \etal\ 2016) or, perhaps more realistically, a gas that follows a temperature distribution centered at $T_\mathrm{vir}$ (McQuinn \& Werk 2018).

For \ion{O}{VI} absorbers originating in a radiatively cooling flow of coronal ($T\sim10^{5.5}$ K) gas, $N\mathrm{(\ion{O}{VI})}$ is expected to be related to the flow velocity (e.g., Edgar \& Chevalier 1986; Heckman \etal\ 2002; Bordoloi \etal\ 2017). A cooling flow develops because as OVI-bearing gas cools in the halo, its density must increase to maintain pressure equilibrium. Consequently, the cooling gas sinks and flow inward. 
Because bulk motion in the gas broadens its line profile, the observed O\,VI linewidth is a combination of pure thermal broadening and additional broadening due to  cooling-flow velocity. By investigating \ion{O}{VI} absorbers  in a wide range of environments (Galactic disk and high velocity clouds, the Large and Small Magellanic Clouds, nearby starburst galaxies, and the IGM), Heckman \etal\ (2002) found a correlation between \ion{O}{VI}  column density and linewidth that is consistent with the theoretical prediction from the radiative cooling flow model. Later studies have also reported similar trends at both low and high redshifts (e.g., Tripp \etal\ 2008; Lehner \etal\ 2014; Werk \etal\ 2016). 

To investigate whether \ion{O}{VI} absorbers around LRGs can be explained by a radiatively cooling flow, we plot the observed Doppler linewidth versus column density for COS-LRG \ion{O}{VI} absorbers in the left panel of Figure 14 (red circles). Additional \ion{O}{VI} absorbers detected in the vicinity of massive quiescent galaxies (log\,$M_\mathrm{star}/\mathrm{M}_\odot>11$) in Johnson \etal\ (2015) are shown in black circles. For comparison, \ion{O}{VI} measurements around $\sim L^*$  star-forming galaxies from Johnson \etal\ (2015) are shown in pale gray squares. Finally, the predicted relationship between \ion{O}{VI} linewidth and column density for a radiatively cooling flow is shown in solid curve, for temperature $T_\mathrm{\ion{O}{VI}}=10^{5.5}\,$K. We note that if the gas is radiatively cooling at a higher/lower temperature, the effect is to shift the prediction curve upward/downward in the parameter space (see e.g., Bordoloi \etal\ 2017).

It is apparent from the left panel of Figure 14 that \ion{O}{VI} absorbers around $\sim L^*$ star-forming galaxies follow the trend predicted by the cooling flow model. This is consistent with the finding of Werk \etal\ (2016), who reported a statistically significant correlation between $b_c\mathrm{(\ion{O}{VI})}$ and $N_c\mathrm{(\ion{O}{VI})}$. For COS-LRG \ion{O}{VI} absorbers, a Spearman test on the sample indicates a $2.3\,\sigma$ correlation between $b_c\mathrm{(\ion{O}{VI})}$ and $N_c\mathrm{(\ion{O}{VI})}$, with a coefficient of $r=0.73$. While this marginal correlation is suggestive a cooling flow, note that most \ion{O}{VI} absorbers around LRGs are situated above the prediction curve for a $T_\mathrm{\ion{O}{VI}}=10^{5.5}\,$K cooling flow. These vertical displacements imply that if \ion{O}{VI} absorbers in LRG halos trace collisionally ionized gas in a radiatively cooling flow, the gas has to be significantly hotter with $T_\mathrm{\ion{O}{VI}}\approx10^{6}\,$K (Bordoloi \etal\ 2017). 

On a superficial level, the existence of a $10^{6}\,$K cooling gas may not be that surprising given the expectation that LRGs are surrounded by a hot gaseous halo with $T\sim T_\mathrm{vir}$. However, the expected \ion{O}{VI} ionization fraction in a $10^{6}\,$K gas is very low ($\sim10^{-3}$) under collisional ionization models (e.g., Gnat \& Sternberg 2007; Oppenheimer \& Schaye 2013). For a solar metallicity gas at $T=10^{6}\,$K and density of log\,$n_\mathrm{H}/\cmjjj=-4$, the implied cloud thickness for a log\,$N\mathrm{(\ion{O}{VI})/\cmjj}=14$ absorber is in excess of 200 kpc. The absorber size would be even larger for a lower metallicity and/or lower density gas, exceeding the size of typical LRG halos. For that reason, we consider it unlikely that \ion{O}{VI} absorbers in LRG halos originate in a $10^{6}\,$K cooling gas.

Alternatively, we consider the possibility that \ion{O}{VI} absorbers around LRGs trace cooler, photoionized gas. The expected \ion{O}{VI} thermal linewidth for a $T\approx10^{4.5}\,$K gas is 6\,\kms, which is significantly smaller than the observed O\,VI linewidths in COS-LRG, $b_c\,\mathrm{(\ion{O}{VI})}=20-100$\,\kms (Figure 14). If these \ion{O}{VI} absorbers originate in a photoionized gas, then their broad line profiles are predominantly due to non-thermal motions. At the same time, the implied non-thermal broadening of $b_\mathrm{nt}= 20-100$\,\kms\ for O\,VI gas is significantly higher than the modest non-thermal line broadening seen in lower-ionization gas around LRGs, $\langle b_\mathrm{nt} \rangle =7\pm5\,$\kms (\S\ 4.2). Because of this large discrepancy in the implied non-thermal motion {\it and} observed kinematic misalignments between \ion{O}{VI} and \ion{H}{I} as well as lower-ionization species (\S 4.2), we conclude that any photoionized \ion{O}{VI} gas has a different physical origin from cool gas traced by \ion{H}{I} and lower ions (\S 5.2). 

In a recent study, Stern \etal\ (2018) considered the possibility that circumgalactic O\,VI absorbers trace infalling cool gas which has yet to be virially shocked by the halo. Assuming photoionization and thermal equilibrium with the UVB, the implied absorber size is $\sim\mathrm{a\,few}\times10$ kpc for a gas with log\,$N\mathrm{(\ion{O}{VI})/\cmjj}=14$ and a metallicity between $0.1-1$ solar (e.g, Oppenheimer \& Schaye \etal\ 2013). Because of the substantial size of the absorber, bulk gravitational infall will produce a velocity shear which broadens the \ion{O}{VI} line profile. This gravitational line broadening is expected to grow with increasing absorber size. Because absorber size is proportional to column density for a fixed gas density, a correlation between O\,VI linewidth and column density is naturally expected. In the right panel of Figure 14, we show the relationships between the two variables as predicted by Stern \etal\ (2018) for different densities of $\mathrm{O^{+5}}$ ions, $n_\mathrm{\ion{O}{VI}}$.

Under the gravitational broadening scenario, the observed correlation between $b_c\mathrm{(\ion{O}{VI})}$ and $N_c\mathrm{(\ion{O}{VI})}$ implies that  \ion{O}{VI} absorbers around $L^*$  star-forming galaxies have densities of $-9\lesssim$ log\,$n_\mathrm{\ion{O}{VI}}/\cmjjj \lesssim -8.5$. In contrast, the implied \ion{O}{VI} volume density is significantly lower for most \ion{O}{VI} absorbers detected around massive quiescent galaxies in both COS-LRG (red circles) and Johnson \etal\ (2015, black circles) samples, log\,$n_\mathrm{\ion{O}{VI}}/\cmjjj \lesssim -9$.\footnote{It is possible that some of the broad \ion{O}{VI} absorbers shown in Figure 14 are the result of unresolved blending of multiple, narrow \ion{O}{VI} components. If unresolved components were present, they would naturally have narrower $b_c$ and lower $N_c\mathrm{(\ion{O}{VI})}$ than the measurements shown, and as a consequence the data points in Figure 14 would move downward and leftward.} 
For a photoionized \ion{O}{VI}-bearing gas with a metallicity of $\mathrm{[M/H]}=-0.7$, which is the median metallicity of lower-ionization gas in COS-LRG, the implied upper limit on $n_\mathrm{\ion{O}{VI}}$ corresponds to an upper limit on gas density of log\,$n_\mathrm{H} / \cmjjj\lesssim -4.3$.
If one assumes that gas density monotonically declines with increasing galacto-centric distance, the lower \ion{O}{VI} volume densities around LRGs suggest that these absorbers trace gas which resides at larger distances than typical \ion{O}{VI} absorbers around $L^*$  star-forming galaxies. This interpretation is consistent with our understanding that LRG halos are roughly twice the size of $L^*$ star-forming halos, and that stable accretion shocks in LRG halos are expected to be situated further out from the galaxies than accretion shocks in lower-mass halos. 

The inferred low density of \ion{O}{VI}-bearing gas is also consistent with the lack of detection of \ion{N}{V} absorption associated with \ion{O}{VI} absorbers in COS-LRG. Coverage of the \ion{N}{V} doublet is available for seven out of nine high-ionization absorption components detected in \ion{O}{VI}. We do not detect \ion{N}{V} absorption associated with any of these \ion{O}{VI} absorbers. The typical upper limit on the \ion{N}{V} to \ion{O}{VI} column density ratio in COS-LRG is log\,$N_c\mathrm{(\ion{N}{V})}/N_c\mathrm{(\ion{O}{VI})<-0.4}$, estimated from the error array by assuming that \ion{N}{V} has the same linewidth as \ion{O}{VI}. For a photoionized gas with solar $\mathrm{N/O}$ elemental abundance ratio, this upper limit constrains the gas density of \ion{O}{VI}-bearing gas to log\,$n_\mathrm{H} / \cmjjj< -4.1$ under the HM05 UVB.\footnote{Zahedy \etal\ (2017b) reported that the outer CGM of quiescent galaxies exhibit $\alpha$-element enhanced abundance pattern that is similar to what have been observed in the high-redshift IGM and damped Ly$\alpha$ absorbers (DLAs). Sub-solar $\mathrm{N/O}$ values of $[\mathrm{N/O}]\lesssim-1$ have been reported in high-redshift DLAs (e.g., Petitjean \etal\ 2008). Therefore, it is possible that the outskirts of LRG halos have similarly sub-solar $\mathrm{N/O}$ ratios. If \ion{O}{VI} absorbers in COS-LRG arise in gas with low $[\mathrm{N/O}]\approx -1$, the resulting constraint on gas density from the lack of \ion{N}{V} would be less sensitive, log\,$n_\mathrm{H} / \cmjjj\lesssim -3$.} At these low densities, the gas is highly ionized and little associated absorption is expected from low-ionization states. For instance, in a photoionized gas with log\,$n_\mathrm{H} / \cmjjj \approx -5$ and a typical \ion{O}{VI} column of of log\,$N\mathrm{(\ion{O}{VI})/\cmjj}=14$, the expected column densities in \ion{C}{III} and \ion{Si}{III} are very low, log\,$N/\cmjj<12$, which is consistent with the lack of lower ionization gas observed to be associated with \ion{O}{VI} absorbers in COS-LRG. Therefore, while the current sample of \ion{O}{VI} absorbers around LRGs is still small, our observations are suggestive of a physical picture where \ion{O}{VI} absorbers around LRGs trace photoionized and low-density gas at large distances from the galaxy (see also Voit \etal\ in prep).

\section{Summary and Conclusions}

We carried out a systematic investigation of the physical conditions and elemental abundances in the CGM within $d<160$ kpc from LRGs. The COS-LRG sample comprises 16 LRGs with log\,$M_\mathrm{star}/\mathrm{M}_\odot>11$ at $z=0.21-0.55$, which were selected without prior knowledge of the presence or absence of any CGM absorption features. The primary objectives of the COS-LRG program are: (1) to probe the bulk of cool gas in LRG halos by obtaining accurate measurements of $N\mathrm{(\ion{H}{I})}$; and 
(2) to constrain the physical properties and chemical enrichment in massive quiescent halos by observing different ionic metal transitions
that probe a wide range of ionization states. In Paper I, we presented the $N\mathrm{(\ion{H}{I})}$ measurements for the sample and reported that 
LRGs contains widespread chemically enriched gas traced by various metal ions. In this paper, we expanded our investigation with a detailed ionization analysis based on resolved component structures of a suite of absorption transitions, including the full \ion{H}{I} Lyman series and multiple
low-, intermediate-, and high-ionization metal transitions. Resolving the component structures of the various absorption lines was made possible by the high-resolution {\it HST} /COS FUV spectra and ground-based echelle optical spectra of the background QSOs. Our main findings are summarized below. \\

(1) LRGs exhibit enhanced absorption in \ion{H}{I}, low-ionization (\ion{Mg}{II}), and intermediate-ionization (\ion{Si}{III} and \ion{C}{III}) metals at projected distances $d\lesssim100$ kpc, compared to absorption at larger $d$ (Figure 2). 

(2) \ion{H}{I}-bearing gas detected around LRGs is predominantly cool, with temperatures of $T<10^5$ K inferred from the \ion{H}{I} linewidths. Using the ratios of Doppler linewidths for matched \ion{H}{I} and \ion{Mg}{II} components, we find that the gas has a mean temperature and dispersion of $\langle T \rangle =2.0\times10^4\,$K and $\sigma_T =1.4\times10^4\,$K, with a modest inferred non-thermal broadening of $\langle b_\mathrm{nt} \rangle =7\pm5\,$\kms (Figure 4). 

(3) The line-of-sight velocity distribution of individual absorption components relative to the systemic redshift of LRGs can be characterized by a mean and dispersion of $\langle \Delta v_\mathrm{gas-galaxy} \rangle =17$ \kms and $\sigma_{\Delta v_\mathrm{gas-galaxy}}=147$ \kms (Figure 3). The observed radial velocity dispersion is consistent with what have been observed in \ion{Mg}{II} absorbers around LRGs using low-resolution data (e.g., Huang \etal\ 2016), but it is only $\sim 60$ percent of what is expected from virial motion. 

(4) By considering matched absorbing components and comparing the relative abundances of different ions for each component, we find that
the underlying gas density and metallicity can vary by more than a factor of 10 within the gaseous halo of an LRG (left panels of Figures 8 and 9).  Such large variations in gas density and metallicity within individual sightlines highlight a complex multiphase structure and poor chemical mixing in the gaseous halos of LRGs. Moreover, they underscore the importance of resolving the component structures of CGM absorbers using high-resolution absorption spectra, because any information on variations in gas metallicity and density within individual halos is lost in ionization studies utilizing only the integrated \ion{H}{I} and metal column densities along individual sightlines. 

(5) Over the full sample, the median metallicity of absorbing components is $\mathrm{\langle [M/H] \rangle}=-0.7\pm0.2$, with an estimated 16-84 percentile range of $\mathrm{[M/H]}=(-1.6,-0.1)$.  Metal-poor components with $<1/10$ solar metallicity are seen in 50\% of the LRG halos, while gas with near-solar and super-solar metallicity is also common (Figure 8). Furthermore, we find a significant incidence of optically thick components with very low metallicities: $43^{+25}_{-22}$ percent of LLSs in the gaseous halos of LRGs have metallicities lower than a few percent solar. 

(6) The median gas density for individual components in the COS-LRG sample is log\,$\langle n_\mathrm{H}\rangle/ \cmjjj=-2.4\pm0.1$, with an estimated 16-84 percentile range of log\,$n_\mathrm{H}/ \cmjjj=(-3.0,-1.8)\, \cmjjj$. The inferred median gas density implies a median ionization parameter of $\mathrm{log}\,\langle U \rangle=-3.0\pm0.1$ under the HM05 UVB. The data points exhibit a trend of rising gas density with increasing \ion{H}{I} column density (Figure 9, right panel). 

(7) We infer a density contrast of $\sim100$ between optically thick components and the expected gas densities in the hot CGM (Figure 10). The inferred density contrast indicates that optically thick gas in the CGM of LRGs is roughly in thermal pressure equilibrium with the hot halo at galactocentric radius $r\sim d$. In contrast, only $\sim40$ percent of optically thin components have densities consistent with thermal pressure equilibrium with the hot halo at $r\sim d$, which implies that a majority of optically thin absorbers occur at larger radii, $r>d$. 

(8) Cool clumps in LRG halos are compact. The inferred clump sizes are between 10 pc and $\sim1$ kpc thick, with a mode of $\sim100$ pc (Figure 9, right panel). The estimated median clump size for the sample is $\langle l \rangle =120^{+80}_{-40}$ pc.

(9) We find that high-ionization \ion{O}{VI} and low-ionization species (low-ionization metals and  \ion{H}{I}) exhibit distinct kinematic structures. The median absolute difference in centroid velocity between \ion{O}{VI} components and the nearest low-ionization metal and \ion{H}{I} components is $24$ \kms, with a full range of from $|\Delta v|=4$ to  $|\Delta v|=71$ \kms (Figure 5). Furthermore, the implied non-thermal line broadening for \ion{O}{VI} gas is high, $b_\mathrm{nt}= 20-100$\,\kms, significantly higher than the modest non-thermal broadening inferred for lower-ionization gas. Such kinematic mismatches highlight different physical origins between high-ionization gas traced by \ion{O}{VI} and lower-ionization gas traced by other metal ions. Based on the observed relation between \ion{O}{VI} column density and linewidth, our data suggest that \ion{O}{VI} absorbers around LRGs trace photoionized, low-density gas at large distances from the galaxy (Figure 14). 

(10) We calculate the total surface mass density of cool ($T\sim10^4$ K) gas in the LRG halos, $\Sigma_\mathrm{cool}$, by applying estimated ionization fraction corrections to the observed \ion{H}{I} column densities. The spatial profile of $\Sigma_\mathrm{cool}$ is equally well-described by an exponential profile in 2D, $\Sigma_\mathrm{cool}  = (4.1\pm1.4)\,e^{-(d/27\pm4\,\mathrm{kpc})}\,\mathrm{M_\odot\,pc^{-2}}$, and a steep projected Einasto profile with shape parameter $\alpha=1.0^{+0.6}_{-0.2}$ and scale radius $r_s=48^{+19}_{-8}\,$kpc, consistent with a true exponential profile in 3D (Figure 11). On the other hand, a projected NFW profile or shallow Einasto profile with $\alpha<0.3$ are ruled out because they cannot reproduce the steep decline of $\Sigma_\mathrm{cool}$ with $d$. We conclude that the mass distribution of cool gas in the CGM of LRGs is different from the expected mass distribution of the underlying dark matter halo. 

(11) We estimate that typical LRGs at $z\sim0.4$ harbor at least $\mathrm{\mathit{M}_{cool} =  (1-2)\times10^{10}\,\mathrm{M_\odot}}$ of photoionized $T\sim10^4$ K gas at $d<160\,$kpc in their halos (or as much as $\mathrm{\mathit{M}_{cool} \approx 4\times10^{10}\,\mathrm{M_\odot}}$ at $d<500$ kpc), which is comparable to the estimated cool CGM mass of star-forming $L^*$ galaxies. The inferred cool CGM mass is about $\sim6-13$ percent of the expected gas mass in the hot phase of the CGM (Figure 12). \\

Considering our observations in the context of a multiphase gaseous halo surrounding LRGs, our findings are consistent with a scenario in which cool clumps condense from the hot halo due to local thermal instabilities. The observed distribution of line-of-sight velocities indicates that ram pressure drag exerted by the hot halo is effective at dissipating the kinetic energy of cool clumps, causing them to fall toward the galaxy. It is likely that a large majority of cool clumps in the CGM of LRGs are destroyed before reaching the central galaxy, thereby explaining the continuing lack of star-formation activity in these galaxies despite the existence of a large reservoir of cool gas in the CGM. Interactions with the hot gas (such as thermal conduction) and/or some form of energetic feedback from the galaxy itself (e.g., heating from stellar winds, SNe Ia, or an active nucleus) likely play an active role in preventing an accumulation of cool gas in the ISM the LRG. Moving forward, a systematic study of the incidence and physical properties of the cool ISM ($d\lesssim10$\,kpc; see e.g., Zahedy \etal\ 2017a) of LRGs is necessary to connect the observed plethora of cool gas at $d\sim100$ kpc scales in the CGM with the continuing ``red and dead" nature of LRGs over cosmic time.

\section*{Acknowledgments}

The authors thank Jonathan Stern, Mark Voit, Ann Zabludoff, and Rongmon Bordoloi for illuminating discussions. We are grateful to Ben Oppenheimer for providing \ion{O}{VI} data points for galaxies in the \textsc{EAGLE} zoom simulations. This work is based on data gathered under the HST-GO-14145.01A observing program using the NASA/ESA {\it Hubble Space Telescope} operated by the Space Telescope Science Institute and the Association of Universities for Research in Astronomy, Inc., under NASA contract NAS 5-26555. FSZ and HWC acknowledge partial support from NSF grant AST-1715692. SDJ is supported by a NASA Hubble Fellowship ({\it HST}-HF2-51375.001-A). This research has made use of the Keck Observatory Archive (KOA), which is operated by the W. M. Keck Observatory and the NASA Exoplanet Science Institute (NExScI), under contract with the National Aeronautics and Space Administration.

\appendix
\section{Description of Individual LRG Halos}

Here we describe the observed absorption properties and discuss the physical conditions (density and temperature) and chemical abundances 
of individual halos in the COS-LRG sample. The 16 QSO sightlines in COS-LRG are ordered by increasing projected distance from each LRG.  Appendix A1 is shown below for the closest QSO sightline in the sample, whereas sections A2 to A16 are published as online material.

\subsection{SDSS\, J0946$+$5123 at $d=42$ kpc}

\begin{subfigures}
\begin{figure*}
\includegraphics[scale=1.08]{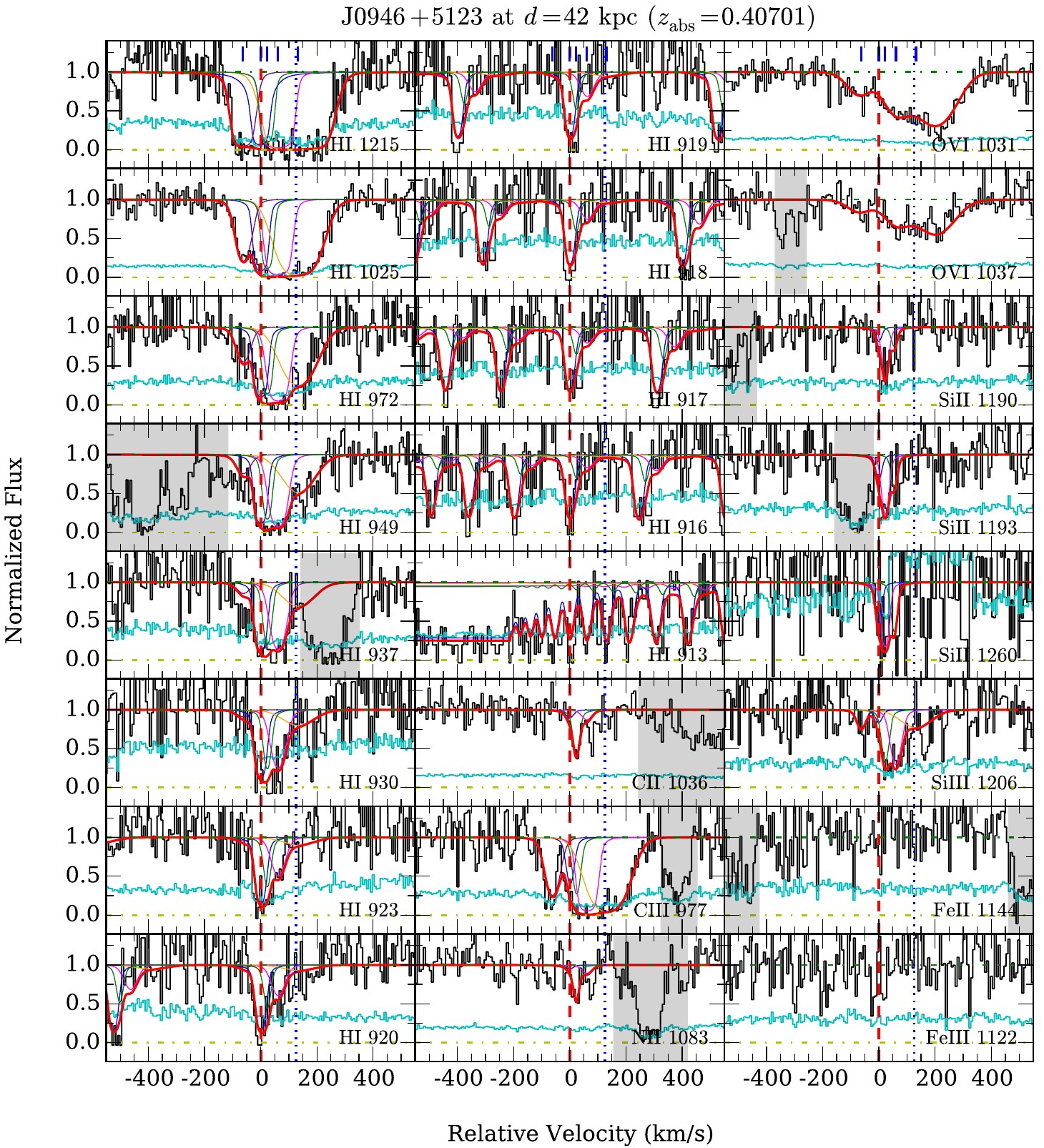}
\vspace{-0.75em}
\caption{Continuum normalized absorption profiles of different transitions along QSO sightline SDSS\,J0946$+$5123 at $d=42$ kpc
 from the LRG. The absorption transition is identified in the bottom-right corner of each panel. 
Zero velocity marks the redshift of the strongest \ion{H}{I} absorption component identified in the Voigt profile analysis, $z_\mathrm{abs}=0.40701$.
The systemic redshift of the LRG is indicated with a blue dotted line. The 1-$\sigma$ error spectrum is included in cyan, above the zero-flux level.
Contaminating features have been grayed out for clarity. The best-fit Voigt profiles for each individual transition detected are plotted, 
both for the sum of all components (red curve) and for individual components (different-colored curves). The centroid of each absorption component is marked by a blue tick mark at the top of panels in the first row.}
\label{figure:ions}
\end{figure*}

\begin{figure}
\hspace{-0.8em}
\includegraphics[scale=1.53]{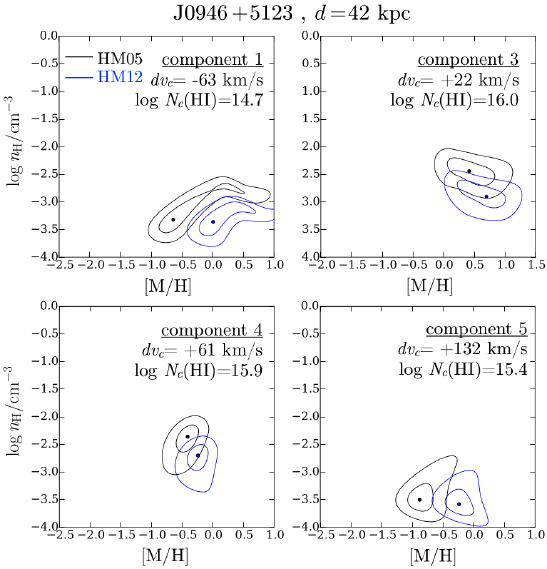}
\vspace{-0.2em}
\caption{Probability distribution of gas metallicity and density for the individual absorption components detected along QSO sightline
SDSS\,J0946$+$5123, at $d=42$ kpc from the LRG. Each component shown here has at least two ionic metal species detected in absorption.   
The contour levels indicate areas enclosing estimated 68\% and 95\% probabilities of the model parameters, shown in black for models assuming the HM05 UVB and in blue for the HM12 UVB (see \S 3.2). Not shown here is component 2 at $dv_c=0\,$\kms, which has the strongest \ion{H}{I} absorption in the absorber, with log\,$N_c\mathrm{(\ion{H}{I})/\cmjj}=17.3$, yet shows little metal absorption. The weak \ion{C}{III} absorption seen in component 2, along with upper limits on the column density of other ions still allow us to constrain the gas metallicity to $\mathrm{[M/H]}\lesssim-1.5$.}
\label{figure:ions}
\end{figure}
\end{subfigures}

This LRG is at redshift $z_\mathrm{LRG}=0.4076$. A LLS with a total $N\mathrm{(\ion{H}{I})}$ of log\,$N\mathrm{(\ion{H}{I})/\cmjj}=17.34\pm0.01$ is present near the redshift of the galaxy (Figure A1a). In addition, the following ionic metal species are also detected: \ion{C}{II}, \ion{C}{III}, \ion{N}{II}, \ion{O}{VI}, \ion{Si}{II}, and \ion{Si}{III}. 

Based on a combined Voigt profile analysis of \ion{H}{I} and the corresponding metal absorption profiles, we identify five components in the absorption system (Figure A1a and Table A1a). The observed velocity spread of the absorber is $\Delta v\approx200$ \kms\ from the bluest to the reddest component. Most (90 percent) of the \ion{H}{I} column density is in component 2 at $z_\mathrm{abs}=0.40701$, or $126$ \kms\ blueward of the LRG. Two other components have log\,$N_c\mathrm{(\ion{H}{I})/\cmjj}\sim16$, components 3 and 4 at $dv_c=+22$ and $+61$ \kms\ from the strongest component, respectively. While the bulk of the neutral hydrogen content is in component 2, little metal absorption is associated with it. In contrast, both low-ionization (e.g., \ion{C}{II} and \ion{Si}{II}) and intermediate-ionization (e.g., \ion{C}{III}) absorption are very prominent in components 3 and 4. This particular characteristic of the absorber suggests a large variation in chemical abundances across different components.

The observed Doppler linewidths of individual \ion{H}{I} components ($b_c\mathrm{(\ion{H}{I})}\lesssim25$\,\kms\ for all but one components) impose a temperature upper limit of $T\lesssim4\times10^4$ K for the gas, under a purely thermal broadening assumption. The other component, component 5, has a very broad \ion{H}{I} linewidth of $b_c\mathrm{(\ion{H}{I})}=71$ \kms . However, similar linewidths are observed for the corresponding \ion{Si}{III} and \ion{C}{III} absorption in component, which implies that the gas is cool ($T\sim10^4$ K) and the broad line profile is primarily due to non-thermal motion (e.g., turbulence) or the presence of blended narrow components.

As shown in Figure A1b and Table A1b, our ionization analysis separates the absorbing gas into two different regimes of gas density. For components 2, 3, and 4, which have log\,$N_c\mathrm{(\ion{H}{I})/\cmjj}\sim16$, good agreements between observations and models are achieved for a gas density range of from log\,$n_\mathrm{H}/ \cmjjj \approx -2.4$ to  log\,$n_\mathrm{H}/ \cmjjj \approx -2.1$ under HM05, and from log\,$n_\mathrm{H}/ \cmjjj \approx -2.7$ to  log\,$n_\mathrm{H}/ \cmjjj \approx -2.5$ under HM12. On the other hand, models for lower $N_c\mathrm{(\ion{H}{I})}$ components 1 and 5 require lower gas densities to match the data: between log\,$n_\mathrm{H}/ \cmjjj \approx -3.6$ and log\,$n_\mathrm{H}/ \cmjjj \approx -3.0$ under both HM05 and HM12 UVBs. 

Similarly, the \textsc{Cloudy} photoionization models indicate a large variation in metallicities ($>1$ dex) across different components. For component 2 at $dv_c=0$ \kms, which has the highest \ion{H}{I} column density in the absorber (log\,$N_c\mathrm{(\ion{H}{I})/\cmjj}=17.3$) but exhibits little associated metals, the inferred metallicity is very low with an upper limit of $\mathrm{[M/H]}\lesssim-1.5$ under both HM05 and HM12 UVBs. In contrast, the observed ionic column densities in components 1,4, and 5 are consistent with the gas having sub-solar metallicities of between $\mathrm{[M/H]}\approx-0.8$ and $\mathrm{[M/H]}\approx-0.4$ under HM05, and between $\mathrm{[M/H]}\approx-0.3$ and $\mathrm{[M/H]}\approx0$ under HM05. Finally, for component 3, which has log\,$N\mathrm{(\ion{H}{I})/\cmjj}\approx16$ yet shows the strongest metal absorption, solar or super-solar metallicities are required to match the data, $\mathrm{[M/H]}=0.4\pm0.4$ under HM05 and $\mathrm{[M/H]}=0.7\pm0.4$ under HM12. 

This absorption system is also noteworthy because it is the strongest \ion{O}{VI} absorber in the COS-LRG sample. The \ion{O}{VI} absorption profile is kinematically complex, comprising three components that extend over $\sim300$ \kms\ in line-of-sight velocity. The measured total \ion{O}{VI} column density is log\,$N\mathrm{(\ion{O}{VI})/\cmjj}=14.93\pm0.02$, which is the highest yet detected in the vicinity of a passive galaxy (cf., Tumlinson \etal\ 2011; Johnson \etal\ 2015), and among the highest $N\mathrm{(\ion{O}{VI})}$ seen in both star-forming and passive galaxies. The broad and asymmetric \ion{O}{VI} absorption profile is in stark contrast to the narrower absorption profiles the lower-ionization metals, which indicates different physical origins between the low- and high-ionization species.

\begin{subtables}
\begin{table}
\begin{center}
\caption{Absorption properties along QSO sightline SDSS\,J0946$+$5123 at $d=42$ kpc from the LRG}
\hspace{-2.5em}
\vspace{-0.5em}
\label{tab:Imaging}
\resizebox{3.5in}{!}{
\begin{tabular}{clrrr}\hline
Component	&	Species		&\multicolumn{1}{c}{$dv_c$} 		& \multicolumn{1}{c}{log\,$N_c$}	&\multicolumn{1}{c}{$b_c$}		\\	
 			&				&\multicolumn{1}{c}{(km\,s$^{-1}$)}	&   		   					& \multicolumn{1}{c}{(km\,s$^{-1}$)}  \\ \hline \hline

all	& \ion{H}{I}	&	$...$					& $17.34\pm0.01$& $...$ \\
	& \ion{C}{II}	&	$...$					& $14.11\pm0.16$		& $...$	\\	
	& \ion{C}{III}	&	$...$					& $>14.73$			& $...$\\
	& \ion{N}{II}	&	$...$					& $13.90\pm0.22$		& $...$ \\
	& \ion{N}{V}	&	$...$					& $<13.97$			& $...$ \\
	& \ion{O}{VI}	&	$...$					& $14.93\pm0.02$		& $...$	\\
	& \ion{Si}{II}	&	$...$					& $13.87^{+0.21}_{-0.14}$& $...$	\\	
	& \ion{Si}{III}	&	$...$					& $13.48^{+0.19}_{-0.06}$& $...$	\\	
	& \ion{Fe}{II}	&	$...$					& $<14.04$			& $...$	\\ 	
	& \ion{Fe}{III}	&	$...$					& $<14.14$			& $...$	\\ \hline	

1	& \ion{H}{I}	&	$-62.8^{+2.8}_{-3.0}$	& $14.70^{+0.10}_{-0.09}$& $18.2\pm1.7$ \\
	& \ion{C}{II}	&	$-62.8$				& $<13.24$			& 10	\\	
	& \ion{C}{III}	&	$-62.8$				& $13.67^{+0.13}_{-0.11}$	& $27.3\pm8.9$ 	\\
	& \ion{N}{II}	&	$-62.8$				& $<13.29$			& 10	\\
	& \ion{Si}{II}	&	$-62.8$				& $<12.88$			& 10	\\	
	& \ion{Si}{III}	&	$-62.8$				& $12.35^{+0.20}_{-0.23}$& $17.3^{+8.6}_{-6.1}$	\\	
	& \ion{Fe}{II}	&	$-62.8$				& $<13.48$			& 10	\\ 
	& \ion{Fe}{III}	&	$-62.8$				& $<13.50$			& 10	\\ \hline

2	& \ion{H}{I}	&	$0.0\pm0.2$			& $17.30\pm0.02$		& $10.6^{+0.7}_{-0.6}$ \\
	& \ion{C}{II}	&	$0.0$				& $<13.22$			& 10	\\	
	& \ion{C}{III}	&	$0.0$				& $13.20^{+0.28}_{-0.26}$& $13.4^{+5.6}_{-2.0}$ 	\\
	& \ion{N}{II}	&	$0.0$				& $<13.40$			& 10	\\
	& \ion{Si}{II}	&	$0.0$				& $<12.92$			& 10	\\	
	& \ion{Si}{III}	&	$0.0$				& $<12.08$			& 10	\\	
	& \ion{Fe}{II}	&	$0.0$				& $<13.47$			& 10	\\ 
	& \ion{Fe}{III}	&	$0.0$				& $<13.55$			& 10	\\ \hline

3	& \ion{H}{I}	&	$+22.1^{+2.2}_{-2.8}$	& $15.97^{+0.30}_{-0.46}$& $10.8^{+1.0}_{-0.9}$ \\
	& \ion{C}{II}	&	$+22.1$				& $14.11\pm0.16$		& $10.4^{+5.0}_{-2.5}$ \\		
	& \ion{C}{III}	&	$+22.1$				& $>13.60$			& $<22.5$ \\
	& \ion{N}{II}	&	$+22.1$				& $13.90\pm0.22$		& $9.1^{+5.8}_{-2.4}$	\\
	& \ion{Si}{II}	&	$+22.1$				& $13.80^{+0.22}_{-0.20}$& $9.9\pm4.7$ \\	
	& \ion{Si}{III}	&	$+22.1$				& $12.97^{+0.32}_{-0.26}$& $12.0^{+9.5}_{-2.3}$ \\	
	& \ion{Fe}{II}	&	$+22.1$				& $<13.52$			& 10	\\ 
	& \ion{Fe}{III}	&	$+22.1$				& $<13.49$			& 10	\\ \hline

4	& \ion{H}{I}	&	$+60.8^{+3.9}_{-4.9}$	& $15.90^{+0.13}_{-0.12}$& $26.9^{+2.6}_{-1.9}$ \\
	& \ion{C}{II}	&	$+60.8$				& $<13.31$			& 10 \\		
	& \ion{C}{III}	&	$+60.8$				& $>13.68$			& $<35.7$ \\
	& \ion{N}{II}	&	$+60.8$				& $<13.27$			& 10	\\
	& \ion{Si}{II}	&	$+60.8$				& $13.07^{+0.20}_{-0.31}$& $9.7^{+6.4}_{-3.1}$ \\	
	& \ion{Si}{III}	&	$+60.8$				& $13.09^{+0.29}_{-0.19}$& $20.0^{+8.3}_{-4.4}$ \\	
	& \ion{Fe}{II}	&	$+60.8$				& $<13.51$			& 10	\\ 
	& \ion{Fe}{III}	&	$+60.8$				& $<13.61$			& 10	\\ \hline

5	& \ion{H}{I}	&	$+132.3^{+5.2}_{-6.6}$	& $15.41^{+0.09}_{-0.07}$& $71.3^{+3.7}_{-3.1}$ \\
	& \ion{C}{II}	&	$+132.3$				& $<13.16$			& 10 \\		
	& \ion{C}{III}	&	$+132.3$				& $14.34^{+0.22}_{-0.07}$& $69.8^{+5.3}_{-11.8}$ \\
	& \ion{N}{II}	&	$+132.3$				& $<13.37$			& 10	\\
	& \ion{Si}{II}	&	$+132.3$				& $<12.73	$			& 10 \\	
	& \ion{Si}{III}	&	$+132.3$				& $12.81\pm0.23$		& $70.0^{+35.1}_{-19.1}$ \\	
	& \ion{Fe}{II}	&	$+132.3$				& $<13.49$			& 10	\\ 
	& \ion{Fe}{III}	&	$+132.3$				& $<13.59$			& 10	\\ \hline

high-1	& \ion{O}{VI}	&	$-69.7\pm12.7$		& $14.00^{+0.08}_{-0.17}$& $53.3^{+23.6}_{-14.2}$ \\
		& \ion{N}{V}	&	$-69.7$			& $<13.67$			& 				 \\
high-2	& \ion{O}{VI}	&	$+64.3\pm11.0$	& $14.38^{+0.08}_{-0.05}$& $58.2^{+22.0}_{-5.0}$ \\
		& \ion{N}{V}	&	$+64.3$			& $<13.67$			& \\
high-3	& \ion{O}{VI}	&	$+202.6\pm11.0$	& $14.71^{+0.03}_{-0.04}$& $82.9^{+8.6}_{-4.9}$ \\
		& \ion{N}{V}	&	$+202.6$			& $<13.82$			& \\
\hline
\end{tabular}}
\end{center}
\end{table}

\begin{table}
\begin{center}
\caption{Ionization modeling results for the absorber along SDSS\,J0946$+$5123 at $d=42$ kpc from the LRG}
\hspace{-2.5em}
\label{tab:Imaging}
\resizebox{3.5in}{!}{
\begin{tabular}{@{\extracolsep{3pt}}ccrrrr@{}}\hline
Component	&$N_\mathrm{metal}$& \multicolumn{2}{c}{$\mathrm{[M/H]}$} 	& \multicolumn{2}{c}{$\mathrm{log\,}n_\mathrm{H}/\cmjjj$}		\\
\cline{3-4} \cline {5-6}
	& &\multicolumn{1}{c}{HM05}&	\multicolumn{1}{c}{HM12}	&\multicolumn{1}{c}{HM05} 	& 	\multicolumn{1}{c}{HM12}			\\	\hline \hline

SC	&5& $-0.84\pm0.16$			& $-0.73^{+0.16}_{-0.13}$	& $-1.90^{+0.08}_{-0.20}$	& $-2.28^{+0.10}_{-0.18}$ \\ \hline
1	&2& $-0.61^{+0.91}_{-0.12}$	& $0.04^{+0.47}_{-0.17}$	& $-2.84^{+0.04}_{-0.60}$	& $-3.20^{+0.08}_{-0.34}$ \\
2	&1& $<-1.53$				& $<-1.41$			& $-2.08^{+0.52}_{-0.18}$	& $-2.52^{+0.60}_{-0.20}$	\\	
3	&5& $0.42^{+0.38}_{-0.29}$	& $0.70^{+0.26}_{-0.40}$	& $-2.44^{+0.16}_{-0.24}$	& $-2.84^{+0.14}_{-0.28}$ \\
4	&3& $-0.42^{+0.15}_{-0.24}$	& $-0.23^{+0.15}_{-0.20}$	& $-2.38^{+0.14}_{-0.46}$	& $-2.72^{+0.12}_{-0.40}$	\\	
5	&2& $-0.81^{+0.28}_{-0.19}$	& $-0.20^{+0.20}_{-0.22}$	& $-3.50^{+0.42}_{-0.16}$	& $-3.58^{+0.30}_{-0.16}$	\\

\hline
\end{tabular}}
\end{center}
\end{table}
\end{subtables}

\subsection{SDSS\, J1406$+$2509 at $d=47$ kpc}

\begin{subfigures}
\begin{figure*}
\includegraphics[scale=1.08]{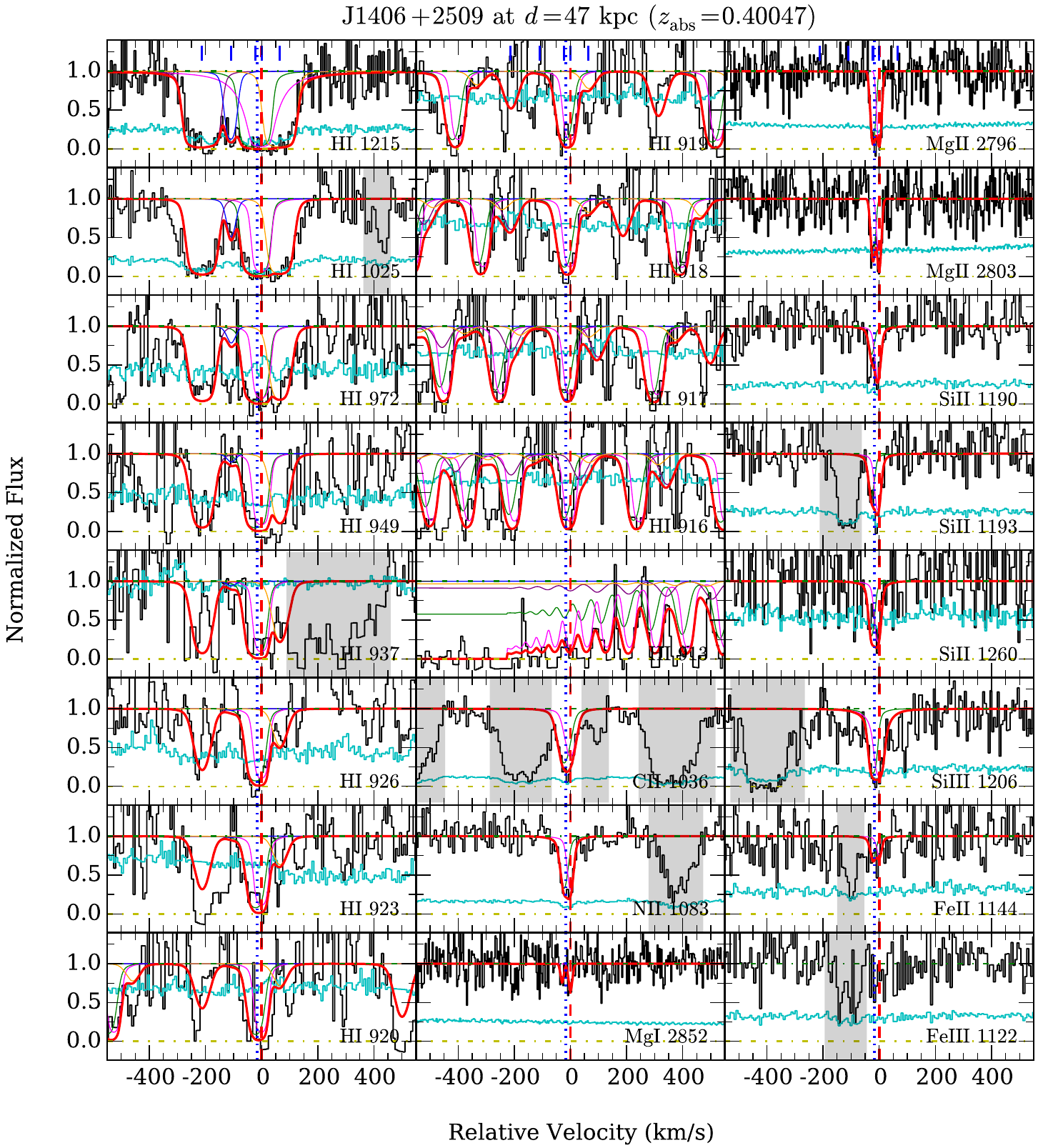}
\vspace{-0.75em}
\caption{Similar to Figure A1a, but for SDSS\,J1406$+$2509 at $d=47$ kpc from the LRG.}
\label{figure:ions}
\end{figure*}

\begin{figure}
\hspace{-0.8em}
\includegraphics[scale=0.62]{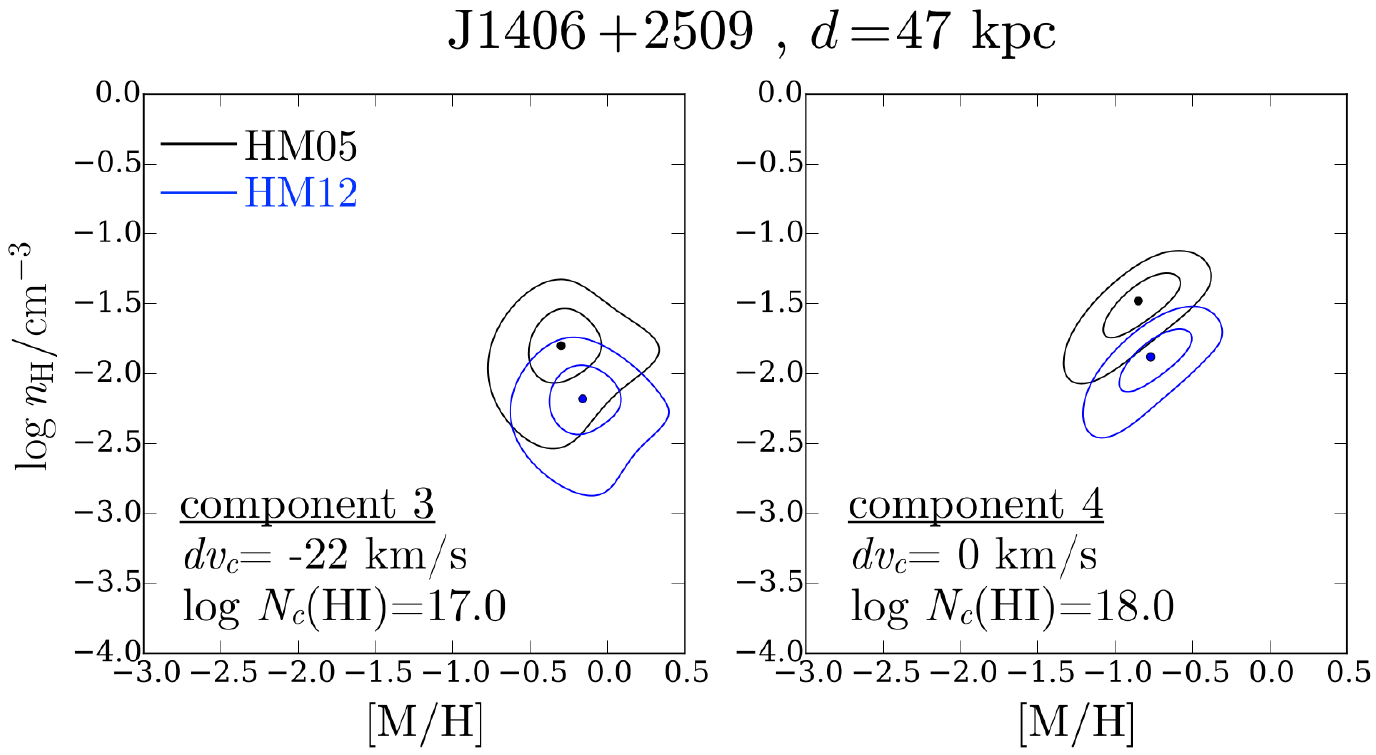}
\vspace{-1.5em}
\caption{Probability distribution contours of gas metallicity and density for optically thick individual absorption components identified along SDSS\,J1406$+$2509, at $d=47$ kpc from the LRG. Contour levels are the same as in Figure A1b. Note that optically thin components 1,2 and 5 (log\,$N_c\mathrm{(\ion{H}{I})/\cmjj}<16$) are not shown here, because the absence of metal detections for these three components result in a lack of strong constraints on the metallicity and density of the gas (see Table A2b).}
\label{figure:ions}
\end{figure}
\end{subfigures}

This LRG occurs at $z_\mathrm{LRG}=0.4004$. As shown in Figure A2a, a strong LLS is present at the LRG redshift, with a total $N\mathrm{(\ion{H}{I})}$ of log\,$N\mathrm{(\ion{H}{I})/\cmjj}=18.03^{+0.20}_{-0.03}$. The HI absorption is accompanied by associated detections of the following metal species: \ion{C}{II}, \ion{C}{III}, \ion{N}{II}, \ion{Mg}{I} \ion{Mg}{II}, \ion{Si}{II}, \ion{Si}{III}, and \ion{Fe}{II}. We note that two intervening absorbers at higher redshifts, a LLS at $z=0.578$ and a pLLS at $z=0.516$, absorb a majority of the QSO continuum flux below rest-frame $\lambda \approx 990$ \AA\ for the LRG absorber. Because of the significantly reduced S/N in this spectral regime, we cannot reliably constrain the absorption properties of \ion{C}{III} $\lambda 977$ transition, and we chose to exclude \ion{C}{III} from our subsequent analysis of the absorber.  

We identify five distinct components in the absorption system, based on the Voigt profile analysis (Figure A2a and Table A2a). The observed velocity spread of the absorber is $\Delta v\approx210$ \kms\ from the bluest to the reddest component. Two components are optically thick with log\,$N_c\mathrm{(\ion{H}{I})/\cmjj}\gtrsim17$. The strongest \ion{H}{I} absorption occurs in component 4 at $z_\mathrm{abs}=0.40047$, or $16$ \kms\ redward of the LRG. Component 4 comprises most (90 percent) of the total $N\mathrm{(\ion{H}{I})}$ of the absorption system. Most of the remaining neutral hydrogen is found in component 3, which is at $dv_c=-22$ \kms\ from the strongest component. These two strong components are also where all of the ionic metals in the absorber are detected. In contrast, the remaining components (components 1 to 3) are optically thin with log\,$N_c\mathrm{(\ion{H}{I})/\cmjj}<16$, and are not detected in associated metal absorption lines. 

We find a consistent velocity structure in optically thick \ion{H}{I} components (components 3 and 4) and metal species (e.g., \ion{Mg}{II} and \ion{Mg}{I}, see Figure A2a). Based on the observed linewidths of \ion{H}{I} and \ion{Mg}{I}, the implied temperatures of the absorbing gas for components 3 and 4 are $T\approx 2.5\times10^4 $ K and $10^4$ K, respectively. The observed linewidths also indicate little non-thermal broadening in the optically thick gas, with $b_\mathrm{nt}\approx3-4$\,\kms. For optically thin components 1 to 3, the observed \ion{H}{I} linewidths place an upper limit on the gas temperature of $T\lesssim5\times10^4$ K.

Our ionization analysis finds a modest difference ($\sim0.3$ dex) in gas densities between the two optically thick components 3 and 4 (Figure A2b and Table A2b). For both components, the observed absorption profile can be reproduced by the models over a gas density range of from log\,$n_\mathrm{H}/ \cmjjj\approx-1.8$ to log\,$n_\mathrm{H}/ \cmjjj\approx-1.5$ under the HM05 UVB, and from log\,$n_\mathrm{H}/ \cmjjj\approx-2.1$ to log\,$n_\mathrm{H}/ \cmjjj\approx-1.8$ under the HM12 ionizing spectrum. The inferred metallicities are sub-solar, ranging from $\mathrm{[M/H]}=-0.9^{+0.2}_{-0.3}$ (component 4) to $\mathrm{[M/H]}=-0.3^{+0.3}_{-0.2}$ (component 3) under HM05, and from $\mathrm{[M/H]}=-0.8\pm0.2$ (component 4) to $\mathrm{[M/H]}=-0.2\pm0.2$ (component 3) under HM12. Finally, we note that the inferred Fe and Mg abundance ratio of the gas is consistent with solar value, $\mathrm{[Fe/\alpha]}\sim0$, to within uncertainties. 

\begin{subtables}
\begin{table}
\begin{center}
\caption{Absorption properties along QSO sightline SDSS\,J1406$+$2509 at $d=47$ kpc from the LRG}
\hspace{-2.5em}
\vspace{-0.5em}
\label{tab:Imaging}
\resizebox{3.5in}{!}{
\begin{tabular}{clrrr}\hline
Component	&	Species		&\multicolumn{1}{c}{$dv_c$} 		& \multicolumn{1}{c}{log\,$N_c$}	&\multicolumn{1}{c}{$b_c$}		\\	
 			&				&\multicolumn{1}{c}{(km\,s$^{-1}$)}	&   		   					& \multicolumn{1}{c}{(km\,s$^{-1}$)}  \\ \hline \hline

all	& \ion{H}{I}	&	$...$					& $18.03^{+0.20}_{-0.03}$& $...$ \\
	& \ion{C}{II}	&	$...$					& $14.58^{+0.16}_{-0.04}$& $...$	\\
	& \ion{N}{II}	&	$...$					& $>14.47	$			& $...$	\\	
	& \ion{Mg}{I}	&	$...$					& $11.92^{+0.22}_{-0.17}$	& $...$	\\
	& \ion{Mg}{II}	&	$...$					& $>13.45$			& $...$\\
	& \ion{Si}{II}	&	$...$					& $14.29^{+0.47}_{-0.44}$& $...$	\\	
	& \ion{Si}{III}	&	$...$					& $>13.45$			& $...$	\\	
	& \ion{Fe}{II}	&	$...$					& $13.81^{+0.27}_{-0.24}$& $...$	\\
	& \ion{Fe}{III}	&	$...$					& $<14.18$			& $...$	\\ \hline	
	
1	& \ion{H}{I}	&	$-212.9^{+3.2}_{-3.0}$	& $16.16^{+0.26}_{-0.22}$& $28.5^{+1.9}_{-2.0}$ \\
	& \ion{N}{II}	&	$-212.9$				& $<13.25$			& 10	\\	
	& \ion{Mg}{I}	&	$-212.9$				& $<11.61$			& 10	\\
	& \ion{Mg}{II}	&	$-212.9$				& $<12.14$			& 10	\\
	& \ion{Si}{II}	&	$-212.9$				& $<12.97$			& 10	\\	
	& \ion{Si}{III}	&	$-212.9$				& $<12.11$			& 10	\\	
	& \ion{Fe}{II}	&	$-212.9$				& $<13.43$			& 10	\\
	& \ion{Fe}{III}	&	$-212.9$				& $<13.63$			& 10	\\ \hline	

2	& \ion{H}{I}	&	$-109.9^{+5.1}_{-4.7}$	& $14.15^{+0.14}_{-0.17}$& $16.7^{+1.6}_{-1.2}$ \\
	& \ion{N}{II}	&	$-109.9$				& $<13.22$			& 10	\\
	& \ion{Mg}{I}	&	$-109.9$				& $<11.58$			& 10	\\
	& \ion{Mg}{II}	&	$-109.9$				& $<12.13$			& 10 \\	
	& \ion{Si}{II}	&	$-109.9$				& $<12.96$			& 10 \\	
	& \ion{Si}{III}	&	$-109.9$				& $<12.05$			& 10 \\ \hline

3	& \ion{H}{I}	&	$-22.1\pm0.7$			& $16.98^{+0.50}_{-0.40}$& $21.5^{+1.5}_{-1.6}$ \\
	& \ion{C}{II}	&	$-22.1$				& $14.34^{+0.18}_{-0.06}$& $22.1^{+7.5}_{-4.6}$ \\
	& \ion{N}{II}	&	$-22.1$				& $>14.01$			& $<18.4$ \\
	& \ion{Mg}{I}	&	$-29.7\pm4.2$			& $11.47^{+0.25}_{-0.29}$	& $4.8^{+3.7}_{-1.8}$ \\
	& \ion{Mg}{II}	&	$-22.1\pm2.5$			& $>12.96$			& $<10.1$ \\
	& \ion{Si}{II}	&	$-22.1$				& $13.27\pm0.36$		& $11.6^{+4.7}_{-5.1}$  \\
	& \ion{Si}{III}	&	$-22.1$				& $>12.90$			& $<26.9$ \\	
	& \ion{Fe}{II}	&	$-22.1$				& $13.55^{+0.32}_{-0.21}$& $12.7^{+5.0}_{-5.6}$ \\
	& \ion{Fe}{III}	&	$-22.1$				& $<13.58$			& 10 \\ \hline

4	& \ion{H}{I}	&	$0.0^{+1.0}_{-1.3}$		& $17.98^{+0.19}_{-0.06}$& $11.4^{+1.0}_{-1.2}$ \\
	& \ion{C}{II}	&	$0.0$				& $14.20^{+0.19}_{-0.12}$& $19.7^{+5.0}_{-7.0}$ \\	
	& \ion{N}{II}	&	$0.0$				& $>13.98	$			& $<17.6$ \\	
	& \ion{Mg}{I}	&	$-2.2\pm2.6$			& $11.73^{+0.25}_{-0.28}$	& $5.1^{+3.3}_{-2.3}$ \\	
	& \ion{Mg}{II}	&	$0.0\pm2.2$			& $>13.08$			& $<9.0$ \\
	& \ion{Si}{II}	&	$0.0	$				& $14.25^{+0.49}_{-0.58}$& $6.0^{+3.9}_{-0.5}$ \\		
	& \ion{Si}{III}	&	$0.0	$				& $>12.92	$			& $<25.8$ \\	
	& \ion{Fe}{II}	&	$0.0	$				& $13.46\pm0.31$		& $11.0^{+2.4}_{-4.7}$\\	
	& \ion{Fe}{III}	&	$0.0	$				& $<13.55	$			& 10 \\ \hline

5	& \ion{H}{I}	&	$+64.5^{+5.9}_{-8.2}$	& $15.72^{+0.31}_{-0.22}$& $27.5^{+2.8}_{-2.2}$  \\
	& \ion{N}{II}	&	$+64.5$				& $<13.16$			& 10	\\
	& \ion{Mg}{I}	&	$+64.5$				& $<11.57$			& 10	\\
	& \ion{Mg}{II}	&	$+64.5$				& $<12.11$			& 10 \\
	& \ion{Si}{II}	&	$+64.5$				& $<12.71	$			& 10	\\	
	& \ion{Si}{III}	&	$+64.5$				& $<12.04$			& 10 \\	
	& \ion{Fe}{II}	&	$+64.5$				& $<13.43	$			& 10	\\
	& \ion{Fe}{III}	&	$+64.5$			 	& $<13.52$			& 10	\\ \hline

\end{tabular}}
\end{center}
\end{table}

\begin{table}
\begin{center}
\caption{Ionization modeling results for the absorber along SDSS\,J1406$+$2509 at $d=47$ kpc from the LRG}
\hspace{-2.5em}
\label{tab:Imaging}
\resizebox{3.5in}{!}{
\begin{tabular}{@{\extracolsep{3pt}}ccrrrr@{}}\hline
Component	&$N_\mathrm{metal}$& \multicolumn{2}{c}{$\mathrm{[M/H]}$} 	& \multicolumn{2}{c}{$\mathrm{log\,}n_\mathrm{H}/\cmjjj$}		\\
\cline{3-4} \cline {5-6}
	& &\multicolumn{1}{c}{HM05}&	\multicolumn{1}{c}{HM12}	&\multicolumn{1}{c}{HM05} 	& 	\multicolumn{1}{c}{HM12}			\\	\hline \hline

SC	&7& $-0.64^{+0.17}_{-0.21}$	& $-0.54^{+0.16}_{-0.19}$	& $-1.76^{+0.12}_{-0.32}$	& $-2.14^{+0.12}_{-0.34}$ \\ \hline
1	&0& $<0.06$				& $<0.14$				& $>-4.84$			& $>-4.82$ \\
2	&0& $<0.75$				& $<0.78$				& $>-4.72$			& $>-4.72$	\\	
3	&7& $-0.30^{+0.26}_{-0.23}$	& $-0.16^{+0.24}_{-0.23}$	& $-1.80^{+0.16}_{-0.42}$	& $-2.18^{+0.16}_{-0.38}$	\\
4	&7& $-0.85^{+0.21}_{-0.26}$	& $-0.77^{+0.21}_{-0.24}$	& $-1.48^{+0.12}_{-0.32}$	& $-1.88^{+0.14}_{-0.32}$	\\
5	&0& $<0.34$				& $<0.43$				& $>4.82$				& $>-4.80$	\\

\hline
\end{tabular}}
\end{center}
\end{table}
\end{subtables}

\subsection{SDSS\, J1111$+$5547 at $d=77$ kpc}

\begin{subfigures}
\begin{figure*}
\includegraphics[scale=1.08]{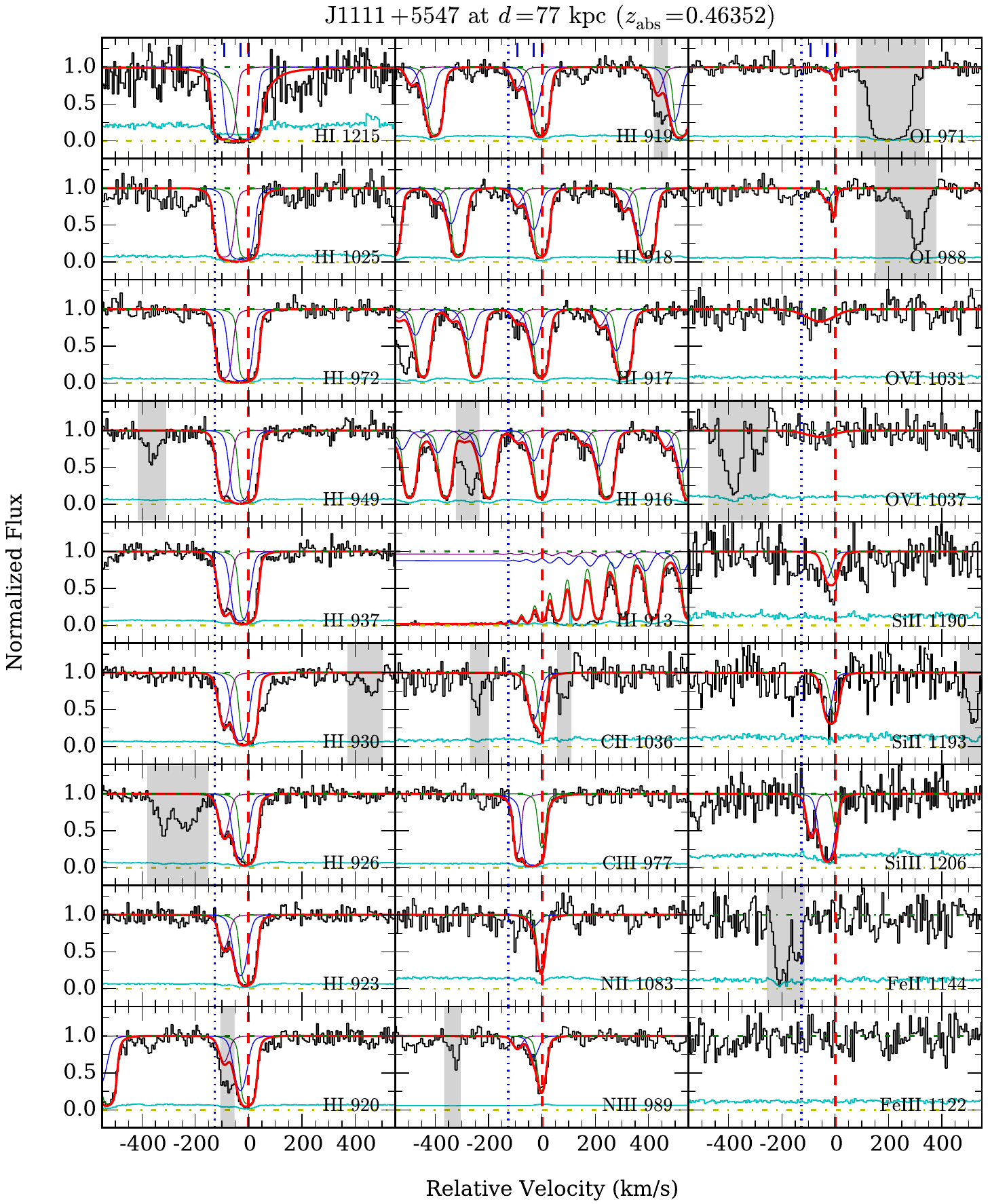}
\vspace{-0.75em}
\caption{Similar to Figure A1a, but for SDSS\, J1111$+$5547 at $d=77$ kpc from the LRG. For the \ion{N}{III} $\lambda 989$ transition, contamination from the adjacent \ion{Si}{II} $\lambda989$ line has been removed by dividing the observed absorption profile by a model profile of \ion{Si}{II} $\lambda989$ line, which is constrained by fitting other \ion{Si}{II} transitions in the spectrum.}
\label{figure:ions}
\end{figure*}

\begin{figure}
\hspace{-0.8em}
\includegraphics[scale=0.62]{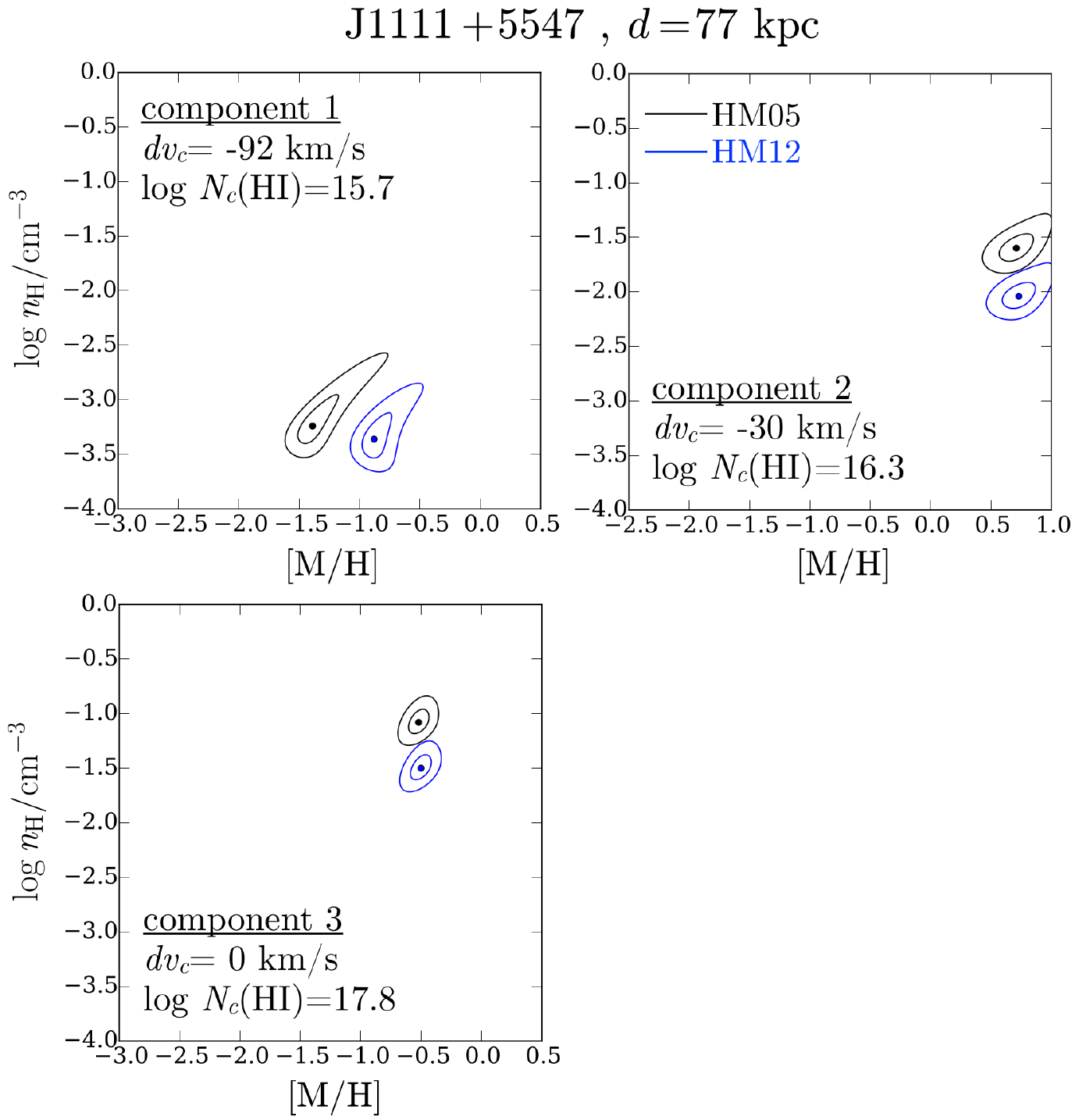}
\vspace{-1.5em}
\caption{Probability distribution contours of gas metallicity and density for individual absorption components identified along SDSS\, J1111$+$5547, at $d=77$ kpc from the LRG. Contour levels are the same as in Figure A1b.}
\label{figure:ions}
\end{figure}
\end{subfigures}

The LRG is located at $z_\mathrm{LRG}=0.4629$. A LLS with a total $N\mathrm{(\ion{H}{I})}$ of log\,$N\mathrm{(\ion{H}{I})/\cmjj}=17.82
\pm$0.01 is found near the galaxy redshift (Figure A3a). In addition, the following ionic metal species are detected: \ion{C}{II}, \ion{C}{III}, \ion{N}{II}, \ion{N}{III}, \ion{O}{I}, \ion{O}{VI}, \ion{Si}{II}, and \ion{Si}{III}.  

We identify three components in this absorber based on a combined Voigt profile analysis of \ion{H}{I} and the metal absorption profiles (Figure A3b and Table A3a). The observed velocity spread is $\Delta v\approx90$ \kms\ from the bluest to the reddest component. The bulk (95 percent) of the \ion{H}{I} column density is concentrated in component 3 at $z_\mathrm{abs}=0.46352$, or $127$ \kms\ redward of the LRG. One other component has log\,$N_c\mathrm{(\ion{H}{I})/\cmjj}>16$, component 2 at $dv_c=-30$ \kms. The observed \ion{H}{I} linewidths for all three components constrain the gas temperature to $T\lesssim 3\times10^4$ K, consistent with the expectation for a photoionized gas. 

Our ionization analysis finds a large variation in densities and metallicities among different components (Figure A3b and Table A3b). For component 1, the observed metal column densities are well-reproduced by a low density and metal-poor gas, with a density between log\,$n_\mathrm{H}/ \cmjjj\approx-3.4$ (HM12) and log\,$n_\mathrm{H}/ \cmjjj\approx-3.2$ (HM05), and a metallicity between $\mathrm{[M/H]}=-1.4^{+0.3}_{-0.1}$ (HM05) and $\mathrm{[M/H]}=-0.9^{+0.2}_{-0.1}$ (HM12). 

 In contrast, component 3, which contains most of the neutral hydrogen in the system, has an inferred density of between log\,$n_\mathrm{H}/ \cmjjj\approx-1.5$ (HM12) and log\,$n_\mathrm{H}/ \cmjjj\approx-1.1$ (HM05), and a sub-solar metallicity of $\mathrm{[M/H]}=-0.5\pm0.1$ under both HM05 and HM12. The close agreement in the derived metallicities between HM05 and HM12 for component 3 is due to the well-constrained \ion{O}{I} column density,  which scales proportionally with metallicity but is insensitive to different ionizing radiation fields. We note that the non-detection of \ion{Fe}{II} in this component is consistent with the gas having an $\alpha-$element enhanced abundance pattern, with $\mathrm{[Fe/\alpha]}\lesssim-0.3$.

For component 2, the presence of \ion{O}{I} absorption in a component with a relatively low $N\mathrm{(\ion{H}{I})}$ of log\,$N\mathrm{(\ion{H}{I})/\cmjj}=16.3$ indicates that the gas has been highly enriched by heavy elements. Indeed, our ionization analysis finds a super-solar metallicity of $0.7\pm0.2$ under both HM05 and HM12 UVBs. The gas density is found to be between log\,$n_\mathrm{H}/ \cmjjj\approx-2.0$ (HM12) and log\,$n_\mathrm{H}/ \cmjjj\approx-1.6$ (HM05). At the adopted metallicity, the non-detection of \ion{Fe}{II} in component 2 is consistent with the gas having an $\alpha-$element enhanced abundance pattern, with $\mathrm{[Fe/\alpha]}\lesssim-0.2$. Finally, we note that over the entire range of allowed gas densities and metallicities, \textsc{Cloudy} under-predicts the column densities of all intermediate ions (\ion{C}{III}, \ion{N}{III}, and \ion{Si}{III}) in component 2 by $\approx1$ dex, which implies that most of the absorption from intermediate ions arises from a lower density phase. 

A weak \ion{O}{VI} absorption is present in this system, centered at $dv_c=-58$ \kms. The \ion{O}{VI} absorption consists of a single component with log\,$N\mathrm{(\ion{O}{VI})/\cmjj}=14.0\pm0.1$ and a broad Doppler $b$ linewidth of of 69 \kms. No low- or intermediate-ionization metal or \ion{H}{I} component is found to match the \ion{O}{VI} absorption (Figure A3a) in velocity space, which occurs at  $\Delta v=-28$ \kms\ from the nearest low-ionization component.

\begin{subtables}
\begin{table}
\begin{center}
\caption{Absorption properties along QSO sightline SDSS\,J1111$+$5547 at $d=77$ kpc from the LRG}
\hspace{-2.5em}
\vspace{-0.5em}
\label{tab:Imaging}
\resizebox{3.5in}{!}{
\begin{tabular}{clrrr}\hline
Component	&	Species		&\multicolumn{1}{c}{$dv_c$} 		& \multicolumn{1}{c}{log\,$N_c$}	&\multicolumn{1}{c}{$b_c$}		\\	
 			&				&\multicolumn{1}{c}{(km\,s$^{-1}$)}	&   		   					& \multicolumn{1}{c}{(km\,s$^{-1}$)}  \\ \hline \hline

all	& \ion{H}{I}	&	$...$					& $17.82\pm0.01$& $...$ \\
	& \ion{C}{II}	&	$...$					& $14.64^{+0.15}_{-0.10}$& $...$	\\	
	& \ion{C}{III}	&	$...$					& $>14.76$			& $...$\\
	& \ion{N}{II}	&	$...$					& $14.55^{+0.26}_{-0.18}$& $...$ \\
	& \ion{N}{III}	&	$...$					& $14.51^{+0.09}_{-0.05}$& $...$ \\
	& \ion{O}{I}	&	$...$					& $14.27^{+0.04}_{-0.06}$& $...$	\\
	& \ion{O}{VI}	&	$...$					& $13.81\pm0.08$		& $...$	\\
	& \ion{Si}{II}	&	$...$					& $13.71^{+0.06}_{-0.03}$& $...$	\\	
	& \ion{Si}{III}	&	$...$					& $13.61^{+0.27}_{-0.07}$& $...$	\\	
	& \ion{Fe}{II}	&	$...$					& $<13.46$			& $...$	\\ 	
	& \ion{Fe}{III}	&	$...$					& $<14.57$			& $...$	\\ \hline	

1	& \ion{H}{I}	&	$-92.1^{+1.6}_{-1.3}$	& $15.73\pm0.03$		& $20.1^{+1.1}_{-1.0}$ \\
	& \ion{C}{II}	&	$-92.1$				& $<12.97$			& 10	\\	
	& \ion{C}{III}	&	$-92.1$				& $>13.55$			& $<13.9$ 	\\
	& \ion{N}{III}	&	$-92.1$				& $13.48\pm0.12$		& $20.0^{+7.7}_{-5.3}$ 	\\
	& \ion{O}{I}	&	$-92.1$				& $<13.18$			& 10	\\
	& \ion{Si}{II}	&	$-92.1$				& $<12.35$			& 10	\\	
	& \ion{Si}{III}	&	$-92.1$				& $12.78\pm0.13$		&$15.7^{+5.3}_{-5.4}$	\\	
	& \ion{Fe}{II}	&	$-92.1$				& $<13.07$			& 10	\\ 
	& \ion{Fe}{III}	&	$-92.1$				& $<13.19$			& 10	\\ \hline

2	& \ion{H}{I}	&	$-30.4^{+2.5}_{-3.0}$	& $16.32^{+0.06}_{-0.08}$& $23.9^{+1.7}_{-2.0}$ \\
	& \ion{C}{II}	&	$-30.4$				& $14.22^{+0.07}_{-0.06}$& $21.7^{+4.0}_{-3.3}$	\\	
	& \ion{C}{III}	&	$-30.4$				& $>14.72$			& $<25.8$ 	\\
	& \ion{N}{II}	&	$-30.4$				& $13.44^{+0.20}_{-0.31}$& $22.0^{+13.4}_{-7.7}$	\\
	& \ion{N}{III}	&	$-30.4$				& $13.75\pm0.11$		& $19.9^{+9.2}_{-3.9}$	\\
	& \ion{O}{I}	&	$-30.4$				& $13.83^{+0.09}_{-0.14}$& $18.3^{+9.8}_{-3.6}$	\\
	& \ion{Si}{II}	&	$-30.4$				& $13.42\pm0.09$		& $19.9^{+8.2}_{-5.2}$	\\	
	& \ion{Si}{III}	&	$-30.4$				& $13.50^{+0.31}_{-0.11}$& $25.1^{+4.5}_{-4.8}$	\\	
	& \ion{Fe}{II}	&	$-30.4$				& $<13.05$			& 10	\\ 
	& \ion{Fe}{III}	&	$-30.4$				& $<13.18$			& 10	\\ \hline

3	& \ion{H}{I}	&	$0.0\pm0.1$			& $17.80\pm0.01$		& $12.8^{+0.4}_{-0.3}$ \\
	& \ion{C}{II}	&	$0.0$				& $14.43^{+0.22}_{-0.17}$& $11.2^{+2.8}_{-1.9}$	\\	
	& \ion{C}{III}	&	$0.0$				& $13.50^{+0.41}_{-1.26}$& $11.2^{+15.1}_{-4.8}$ 	\\
	& \ion{N}{II}	&	$0.0$				& $14.52^{+0.28}_{-0.20}$	& $11.5^{+3.7}_{-2.0}$	\\
	& \ion{N}{III}	&	$0.0$				& $14.37^{+0.13}_{-0.07}$	& $15.3^{+1.6}_{-2.5}$	\\
	& \ion{O}{I}	&	$0.0$				& $14.08^{+0.07}_{-0.10}$	& $5.0^{+3.0}_{-0.5}$	\\
	& \ion{Si}{II}	&	$0.0$				& $13.40^{+0.14}_{-0.07}$	& $18.9^{+5.0}_{-9.9}$	\\	
	& \ion{Si}{III}	&	$0.0$				& $12.51^{+0.35}_{-0.33}$& $8.0^{+6.5}_{-1.9}$	\\	
	& \ion{Fe}{II}	&	$0.0$				& $<13.48$			& 10	\\ 
	& \ion{Fe}{III}	&	$0.0$				& $<13.50$			& 10	\\ \hline

high-1	& \ion{O}{VI}	&	$-58.0\pm12.6$		& $13.81\pm0.08$		& $69.4^{+30.5}_{-9.0}$ \\

\hline
\end{tabular}}
\end{center}
\end{table}

\begin{table}
\begin{center}
\caption{Ionization modeling results for the absorber along SDSS\,J1111$+$5547 at $d=77$ kpc from the LRG}
\hspace{-2.5em}
\label{tab:Imaging}
\resizebox{3.5in}{!}{
\begin{tabular}{@{\extracolsep{3pt}}ccrrrr@{}}\hline
Component	&$N_\mathrm{metal}$& \multicolumn{2}{c}{$\mathrm{[M/H]}$} 	& \multicolumn{2}{c}{$\mathrm{log\,}n_\mathrm{H}/\cmjjj$}		\\
\cline{3-4} \cline {5-6}
	& &\multicolumn{1}{c}{HM05}&	\multicolumn{1}{c}{HM12}	&\multicolumn{1}{c}{HM05} 	& 	\multicolumn{1}{c}{HM12}			\\	\hline \hline

SC	&6& $-0.33\pm0.06$			& $-0.32^{+0.05}_{-0.07}$	& $-1.02^{+0.08}_{-0.10}$	& $-1.48\pm0.10$ \\ \hline
1	&3& $-1.38^{+0.30}_{-0.08}$	& $-0.85^{+0.16}_{-0.09}$	& $-3.24^{+0.38}_{-0.10}$	& $-3.36^{+0.26}_{-0.14}$ \\
2	&4& $0.72^{+0.13}_{-0.14}$	& $0.74^{+0.12}_{-0.14}$	& $-1.60^{+0.12}_{-0.10}$	& $-2.04^{+0.12}_{-0.10}$	\\	
3	&6& $-0.52^{+0.07}_{-0.08}$	& $-0.50^{+0.07}_{-0.09}$	& $-1.08\pm0.10$		& $-1.50\pm0.10$	\\	

\hline
\end{tabular}}
\end{center}
\end{table}
\end{subtables}

\subsection{SDSS\, J0803$+$4332 at $d=79$ kpc}

\begin{subfigures}
\begin{figure*}
\includegraphics[scale=1.08]{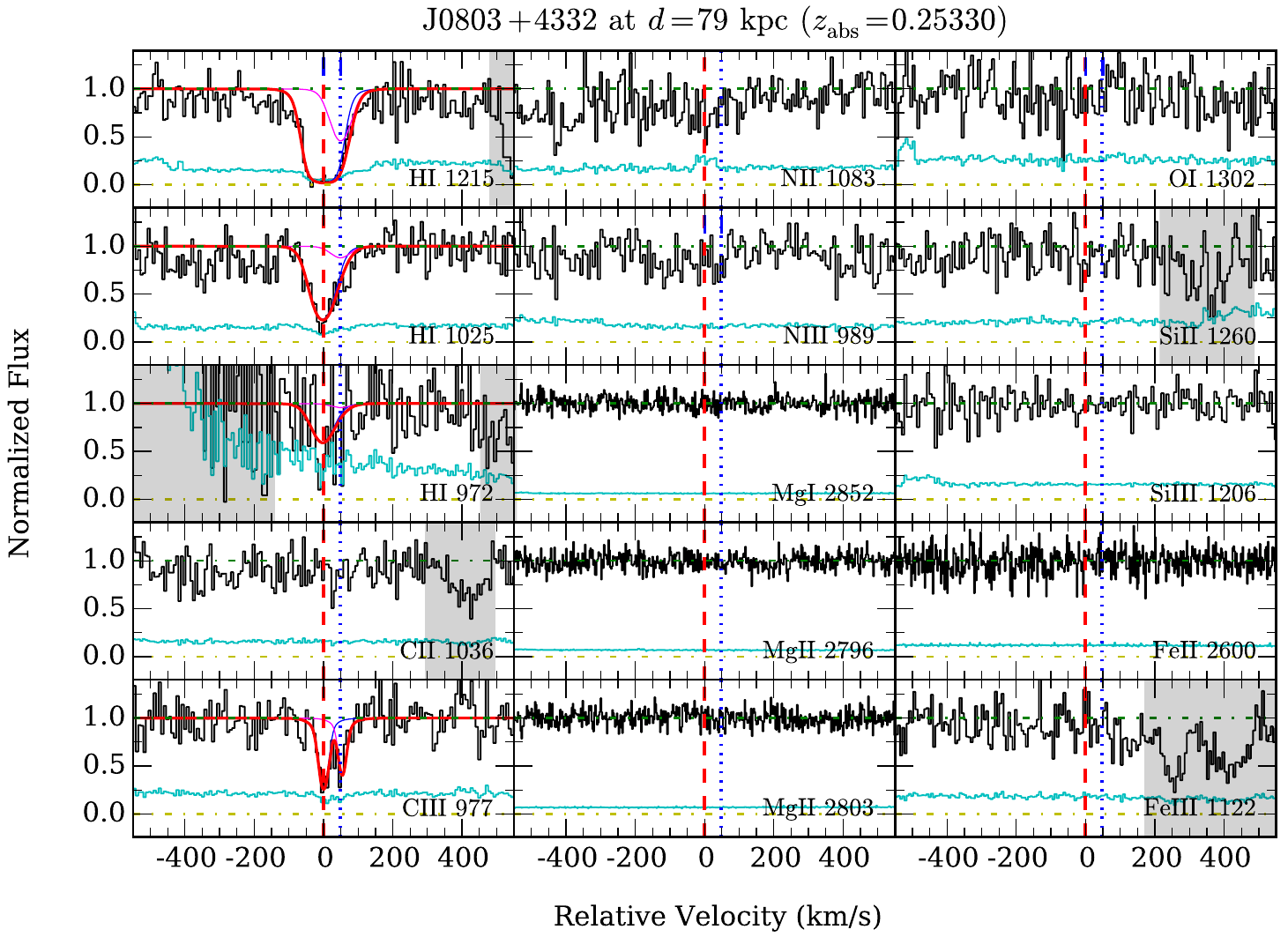}
\vspace{-0.75em}
\caption{Similar to Figure A1a, but for SDSS\,J0803$+$4332 at $d=79$ kpc from the LRG.}
\label{figure:ions}
\end{figure*}

\end{subfigures}

This LRG occurs at $z_\mathrm{LRG}=0.2535$. As shown in Figure A4a, an \ion{H}{I} absorber is present near the LRG redshift, with a total $N\mathrm{(\ion{H}{I})}$ of log\,$N\mathrm{(\ion{H}{I})/\cmjj}=14.78\pm0.05$. The only metal species detected is \ion{C}{III}, which exhibits excellent kinematic alignment with HI. 

Our Voigt profile analysis shows that the \ion{H}{I} and \ion{C}{III} absorption profile can be decomposed into two individual components (Figure A4a and Table A4a). 
The velocity spread of the absorbing gas is $\Delta v\approx50$ \kms\ between the two components. The stronger of the two components (component 1) occurs at $-48$ \kms\ from the LRG, at $z_\mathrm{abs}=0.25330$, whereas the weaker component occurs at the systemic redshift of the LRG. For both components, the measured ratios of \ion{H}{I} and \ion{C}{III} linewidths are consistent with expectations from pure thermal broadening. The inferred temperature of the gas is $T\sim(8-9)\times10^4$ K for both components. 

For both components, the observed $N_c\mathrm{(\ion{C}{III})}$ and upper limits on other ionic column densities constrain the gas to be low density, log\,$n_\mathrm{H}/ \cmjjj\lesssim-3$, under both HM05 and HM12 UVBs (Table A4b). Because \ion{C}{III} is the only metal species detected, the inferred chemical abundance of the gas is subject to large uncertainties. For the stronger component 1, the inferred metallicity is $\mathrm{[M/H]}=-1.2^{+0.5}_{-0.2}$ under HM05, and $\mathrm{[M/H]}=-0.4^{+0.7}_{-0.2}$ under HM12. For component 2, which has comparable $N_c\mathrm{(\ion{C}{III})}$ to component 1 despite a much lower $N_c\mathrm{(\ion{H}{I})}$, a higher metallicity is required to match the data, ranging from a sub-solar $\mathrm{[M/H]}=-0.2^{+0.5}_{-0.6}$  under HM05, to possibly super-solar $\mathrm{[M/H]}=0.5^{+0.3}_{-0.7}$ under HM12.

This absorber was also studied in the COS-Halos survey. A major difference between our analysis and that of the COS-Halos survey (Werk \etal\ 2014; Prochaska \etal\ 2017) is that the COS-Halos ionization analysis utilized only the integrated \ion{H}{I} and metal column densities summed over all components in each system, which is the same as imposing a single-clump model (see \S 3.2) where different components have the same density and metallicity. Using the Haardt \& Madau (2001) UVB (which is similar of the HM05 UVB at energies $\lesssim2$ Ryd), Werk \etal\ (2014) inferred a mean metallicity of $\mathrm{[M/H]}=-0.9^{+0.9}_{-0.8}$ and density of log\,$n_\mathrm{H}/\cmjjj<-3.7$ under the Haardt \& Madau (2001) UVB. In an updated analysis using the HM12 UVB, Prochaska \etal\ (2017) found a mean metallicity of $\mathrm{[M/H]}=0.1^{+0.7}_{-0.6}$ and density of  log\,$n_\mathrm{H}/ \cmjjj=-3.0^{+0.3}_{-0.8}$. The $\mathrm{[M/H]}$ and $n_\mathrm{H}$ values from COS-Halos are consistent what we find in our analysis assuming the single-clump model (Table A4b).

\begin{subtables}
\begin{table}
\begin{center}
\caption{Absorption properties along QSO sightline SDSS\,J0803$+$4332, $d=79$ kpc from the LRG}
\hspace{-2.5em}
\vspace{-0.5em}
\label{tab:Imaging}
\resizebox{3.5in}{!}{
\begin{tabular}{clrrr}\hline
Component	&	Species		&\multicolumn{1}{c}{$dv_c$} 		& \multicolumn{1}{c}{log\,$N_c$}	&\multicolumn{1}{c}{$b_c$}		\\	
 			&				&\multicolumn{1}{c}{(km\,s$^{-1}$)}	&   		   					& \multicolumn{1}{c}{(km\,s$^{-1}$)}  \\ \hline \hline

all	& \ion{H}{I}	&	$...$					& $14.78\pm0.05$		& $...$ \\
	& \ion{C}{II}	&	$...$					& $<13.54$			& $...$	\\	
	& \ion{C}{III}	&	$...$					& $13.78^{+0.20}_{-0.08}$& $...$\\
	& \ion{N}{II}	&	$...$					& $<13.70$			& $...$ \\
	& \ion{N}{III}	&	$...$					& $<13.65$			& $...$ \\
	& \ion{N}{V}	&	$...$					& $<13.28$			& $...$ \\
	& \ion{O}{I}	&	$...$					& $<14.07$			& $...$ \\
	& \ion{Mg}{I}	&	$...$					& $<11.31$			& $...$	\\
	& \ion{Mg}{II}	&	$...$					& $<11.84$			& $...$\\
	& \ion{Si}{II}	&	$...$					& $<12.68$			& $...$	\\	
	& \ion{Si}{III}	&	$...$					& $<12.35$			& $...$	\\
	& \ion{Si}{IV}	&	$...$					& $<12.82$			& $...$	\\	
	& \ion{Fe}{II}	&	$...$					& $<12.57$			& $...$	\\
	& \ion{Fe}{III}	&	$...$					& $<13.84$			& $...$	\\ \hline	
	
1	& \ion{H}{I}	&	$0.0^{+2.9}_{-2.3}$		& $14.75^{+0.06}_{-0.05}$& $38.7^{+6.0}_{-1.7}$ \\
	& \ion{C}{II}	&	$0.0$				& $<13.16$			& 10	\\	
	& \ion{C}{III}	&	$-1.2	\pm2.9$			& $13.61^{+0.23}_{-0.15}$& $11.9^{+7.3}_{-2.1}$	\\
	& \ion{N}{II}	&	$0.0$				& $<13.41$			& 10	\\
	& \ion{N}{III}	&	$0.0$				& $<13.27$			& 10	\\
	& \ion{O}{I}	&	$0.0$				& $<13.67$			& 10	\\
	& \ion{Mg}{I}	&	$0.0$				& $<10.91$			& 10	\\
	& \ion{Mg}{II}	&	$0.0$				& $<11.44$			& 10	\\
	& \ion{Si}{II}	&	$0.0$				& $<12.26$			& 10	\\	
	& \ion{Si}{III}	&	$0.0$				& $<11.93$			& 10	\\
	& \ion{Si}{IV}	&	$0.0$				& $<12.42$			& 10	\\	
	& \ion{Fe}{II}	&	$0.0$				& $<12.17$			& 10	\\
	& \ion{Fe}{III}	&	$0.0$				& $<13.47$			& 10	\\ \hline	

2	& \ion{H}{I}	&	$+50.3^{+7.2}_{-6.4}$	& $13.64^{+0.09}_{-0.48}$& $35.0^{+11.0}_{-6.2}$ \\
	& \ion{C}{II}	&	$+50.3$				& $<13.14$			& 10 \\		
	& \ion{C}{III}	&	$+50.2\pm3.5$			& $13.30^{+0.26}_{-0.09}$& $10.0^{+6.3}_{-3.5}$ \\
	& \ion{N}{II}	&	$+50.3$				& $<13.23$			& 10	\\
	& \ion{N}{III}	&	$+50.3$				& $<13.26$			& 10	\\
	& \ion{O}{I}	&	$+50.3$				& $<13.69$			& 10	\\
	& \ion{Mg}{I}	&	$+50.3$				& $<10.92$			& 10	\\
	& \ion{Mg}{II}	&	$+50.3$				& $<11.45$			& 10 \\	
	& \ion{Si}{II}	&	$+50.3$				& $<12.25	$			& 10 \\	
	& \ion{Si}{III}	&	$+50.3$				& $<11.98$			& 10 \\	
	& \ion{Si}{IV}	&	$+50.3$				& $<12.41$			& 10 \\	
	& \ion{Fe}{II}	&	$+50.3$				& $<12.18$			& 10	\\
	& \ion{Fe}{III}	&	$+50.3$				& $<13.44$			& 10	\\
\hline
\end{tabular}}
\end{center}
\end{table}

\begin{table}
\begin{center}
\caption{Ionization modeling results for the absorber along SDSS\,J0803$+$4332, at $d=79$ kpc from the LRG}
\hspace{-2.5em}
\label{tab:Imaging}
\resizebox{3.5in}{!}{
\begin{tabular}{@{\extracolsep{3pt}}ccrrrr@{}}\hline
Component	&$N_\mathrm{metal}$& \multicolumn{2}{c}{$\mathrm{[M/H]}$} 	& \multicolumn{2}{c}{$\mathrm{log\,}n_\mathrm{H}/\cmjjj$}		\\
\cline{3-4} \cline {5-6}
	& &\multicolumn{1}{c}{HM05}&	\multicolumn{1}{c}{HM12}	&\multicolumn{1}{c}{HM05} 	& 	\multicolumn{1}{c}{HM12}			\\	\hline \hline

SC	&1& $-0.91^{+0.48}_{-0.23}$	& $-0.10^{+0.52}_{-0.22}$	& $<-3.18$		& $<-3.16$ \\ \hline
1	&1& $-1.17^{+0.54}_{-0.22}$	& $-0.36^{+0.65}_{-0.21}$& $<-3.50$		& $<-3.58$ \\
2	&1& $-0.16^{+0.52}_{-0.64}$	& $+0.52^{+0.25}_{-0.67}$& $<-2.84$		& $<-3.08$	\\

\hline
\end{tabular}}
\end{center}
\end{table}
\end{subtables}

\subsection{SDSS\, J0925$+$4004 at $d=84$ kpc}

\begin{subfigures}
\begin{figure*}
\includegraphics[scale=1.08]{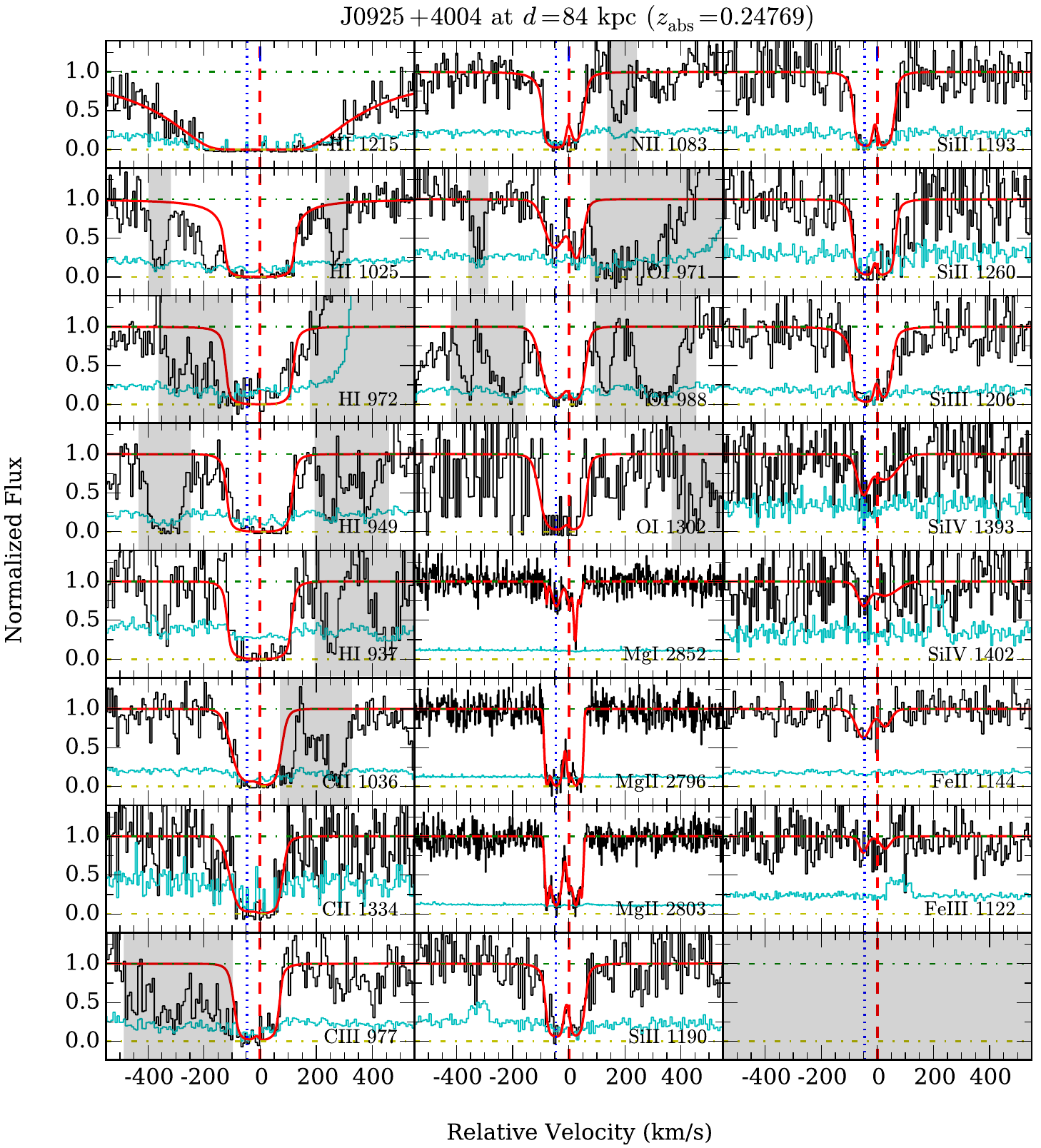}
\vspace{-0.75em}
\caption{Similar to Figure A1a, but for SDSS\,J0925$+$4004 at $d=84$ kpc from the LRG. Note that while the saturated \ion{C}{III} $\lambda977$ transition is partially contaminated on the blue side, we can still robustly constrain a lower limit on the \ion{C}{III} column density.}
\label{figure:ions}
\end{figure*}

\begin{figure}
\centering
\hspace{-0.8em}
\includegraphics[scale=0.68]{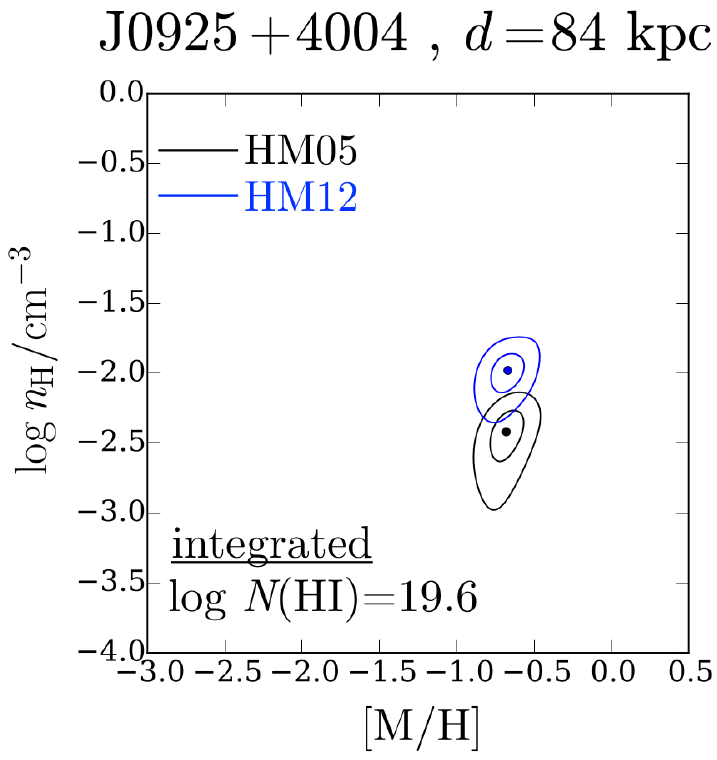}
\vspace{-1em}
\caption{Probability distribution contours of gas metallicity and density for the integrated absorption near the LRG redshift, seen at $d=84$ kpc from the LRG along sightline SDSS\,J0925$+$4004. Contour levels are the same as in Figure A1b. Note the close agreement between the metallicities derived using the HM05 and HM12 UVBs, which can be attributed to the well-constrained \ion{O}{I} column density in this system (see \S A5).
}
\label{figure:ions}
\end{figure}
\end{subfigures}

This LRG occurs at $z_\mathrm{LRG}=0.2475$. As shown in Figure A5a, a sub-damped Ly$\alpha$ absorber (sub-DLA) with a total $N\mathrm{(\ion{H}{I})}$ of log\,$N\mathrm{(\ion{H}{I})/\cmjj}=19.58\pm0.02$ is present at $z_\mathrm{abs}=0.24769$, or $46$ \kms\ redward of the LRG. This absorber has the highest \ion{H}{I} column density in the COS-LRG sample. In addition, the following ionic metal species are also detected: \ion{C}{II}, \ion{C}{III}, \ion{N}{II}, \ion{O}{I}, \ion{Mg}{I}, \ion{Mg}{II}, \ion{Si}{II}, \ion{Si}{III}, \ion{Si}{IV}, and \ion{Fe}{II}.  

Our Voigt profile analysis of the echelle spectrum of the \ion{Mg}{I} and \ion{Mg}{II} metal absorption has identified five absorption components which are tightly separated ($\lesssim 20$ \kms) in velocity space. However, because of the lower resolution of the FUV COS spectrum and the fact that all \ion{H}{I} and many metal transitions are saturated, we cannot reliably constrain the \ion{H}{I} and metal column densities of individual components in this system. For that reason, we forgo component-by-component ionization analysis for this absorber and proceed by modeling the mean physical properties of the gas using the integrated column densities of \ion{H}{I} and metal species (Table A5a).

Our ionization analysis constrains the gas to a density of between log\,$n_\mathrm{H}/ \cmjjj=-2.0^{+0.1}_{-0.2}$ under HM05 and
log\,$n_\mathrm{H}/ \cmjjj=-2.4^{+0.1}_{-0.4}$ under HM12 (Figure A5b and Table A5b). The gas density is well-constrained due to a combination of lower limits on the column density for saturated \ion{C}{III} and \ion{Si}{III} absorption, which set an upper bound on gas density, and the upper limit on \ion{Fe}{III} column density, which imposes a lower bound on gas density. At the adopted gas density, \textsc{Cloudy} under-predicts the observed \ion{Si}{IV} column density by almost 0.5 dex, suggesting that some of the observed \ion{Si}{IV} absorption likely arises from a lower-density gas phase. 
 
The inferred gas metallicity is $\mathrm{[M/H]}=-0.7\pm0.1$ under both HM05 and HM12 UVBs (Figure A5b and Table A5b). The close agreement in the derived metallicities between HM05 and HM12 is due to the well-constrained \ion{O}{I} column density, which is insensitive to different ionizing background radiation fields in high $N\mathrm{(\ion{H}{I})}$ regimes (\S 4.3.1). We note that at this metallicity, the models over-predict the observed \ion{Fe}{II} and \ion{Mg}{II} column densities by 0.4 to 0.5 dex. The inferred depletions in {\it both} the observed Fe and Mg abundances are comparable to what have been observed in the Galactic halo (e.g., Savage \& Sembach 1996; de Cia \etal\ 2016), and indicate the likely presence of dust grains in the gas (see also Zahedy \etal\ 2017). 
 
This absorber was also studied in the COS-Halos survey. Using the  Haardt \& Madau (2001) UVB (which is similar of the HM05 UVB at energies $\lesssim2$ Ryd), Werk \etal\ (2014) inferred a mean metallicity of $\mathrm{[M/H]}=-0.7\pm0.2$ under the Haardt \& Madau (2001) UVB. In an updated analysis using the HM12 UVB, Prochaska \etal\ (2017) found a mean metallicity of $\mathrm{[M/H]}=-0.81^{+0.15}_{-0.14}$. The chemical abundance estimated from these COS-Halos studies is consistent with what we find in our analysis.

\begin{subtables}
\begin{table}
\begin{center}
\caption{Absorption properties along QSO sightline SDSS\,J0925$+$4004 at $d=84$ kpc from the LRG}
\hspace{-2.5em}
\vspace{-0.5em}
\label{tab:Imaging}
\resizebox{3.5in}{!}{
\begin{tabular}{clrrr}\hline
Component	&	Species		&\multicolumn{1}{c}{$dv_c$} 		& \multicolumn{1}{c}{log\,$N_c$}	&\multicolumn{1}{c}{$b_c$}		\\	
 			&				&\multicolumn{1}{c}{(km\,s$^{-1}$)}	&   		   					& \multicolumn{1}{c}{(km\,s$^{-1}$)}  \\ \hline \hline

all	& \ion{H}{I}	&	$0.0\pm1.7$			& $19.58\pm0.02$		& $36.2^{+0.7}_{-1.6}$ \\
	& \ion{C}{II}	&	$...$					& $>15.17$			& $...$	\\	
	& \ion{C}{III}	&	$...$					& $>14.48$			& $...$\\
	& \ion{N}{II}	&	$...$					& $>15.15$			& $...$ \\
	& \ion{N}{V}	&	$...$					& $<13.31$			& $...$ \\
	& \ion{O}{I}	&	$...$					& $15.63^{+0.11}_{-0.07}$& $...$ \\
	& \ion{Mg}{I}	&	$...$					& $12.51^{+0.05}_{-0.04}$& $...$	\\
	& \ion{Mg}{II}	&	$...$					& $13.95^{+0.05}_{-0.04}$& $...$\\
	& \ion{Si}{II}	&	$...$					& $>14.60$			& $...$	\\	
	& \ion{Si}{III}	&	$...$					& $>14.10$			& $...$	\\
	& \ion{Si}{IV}	&	$...$					& $13.61^{+0.07}_{-0.12}$& $...$	\\	
	& \ion{Fe}{II}	&	$...$					& $14.12^{+0.08}_{-0.10}$& $...$	\\
	& \ion{Fe}{III}	&	$...$					& $<14.05$			& $...$	\\ 
	
\hline
\end{tabular}}
\end{center}
\end{table}

\begin{table}
\begin{center}
\caption{Ionization modeling results for the absorber along SDSS\,J0925$+$4004 at $d=84$ kpc from the LRG}
\hspace{-2.5em}
\label{tab:Imaging}
\resizebox{3.5in}{!}{
\begin{tabular}{@{\extracolsep{3pt}}ccrrrr@{}}\hline
Component	&$N_\mathrm{metal}$& \multicolumn{2}{c}{$\mathrm{[M/H]}$} 	& \multicolumn{2}{c}{$\mathrm{log\,}n_\mathrm{H}/\cmjjj$}		\\
\cline{3-4} \cline {5-6}
	& &\multicolumn{1}{c}{HM05}&	\multicolumn{1}{c}{HM12}	&\multicolumn{1}{c}{HM05} 	& 	\multicolumn{1}{c}{HM12}			\\	\hline \hline

SC	&10& $-0.68^{+0.10}_{-0.11}$	& $-0.69^{+0.10}_{-0.11}$	& $-1.98^{+0.08}_{-0.22}$		& $-2.42^{+0.08}_{-0.36}$ \\ 

\hline
\end{tabular}}
\end{center}
\end{table}
\end{subtables}

\subsection{SDSS\, J0950$+$4831 at $d=94$ kpc}

\begin{subfigures}
\begin{figure*}
\includegraphics[scale=2.75]{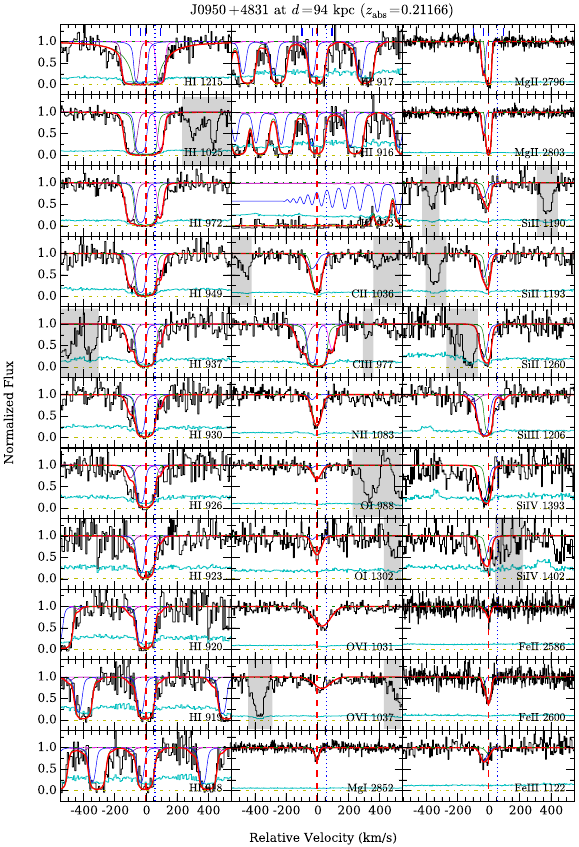}
\vspace{-0.75em}
\caption{Similar to Figure A1a, but for SDSS\,J0950$+$4831 at $d=94$ kpc from the LRG.}
\label{figure:ions}
\end{figure*}

\begin{figure}
\hspace{-0.8em}
\includegraphics[scale=0.62]{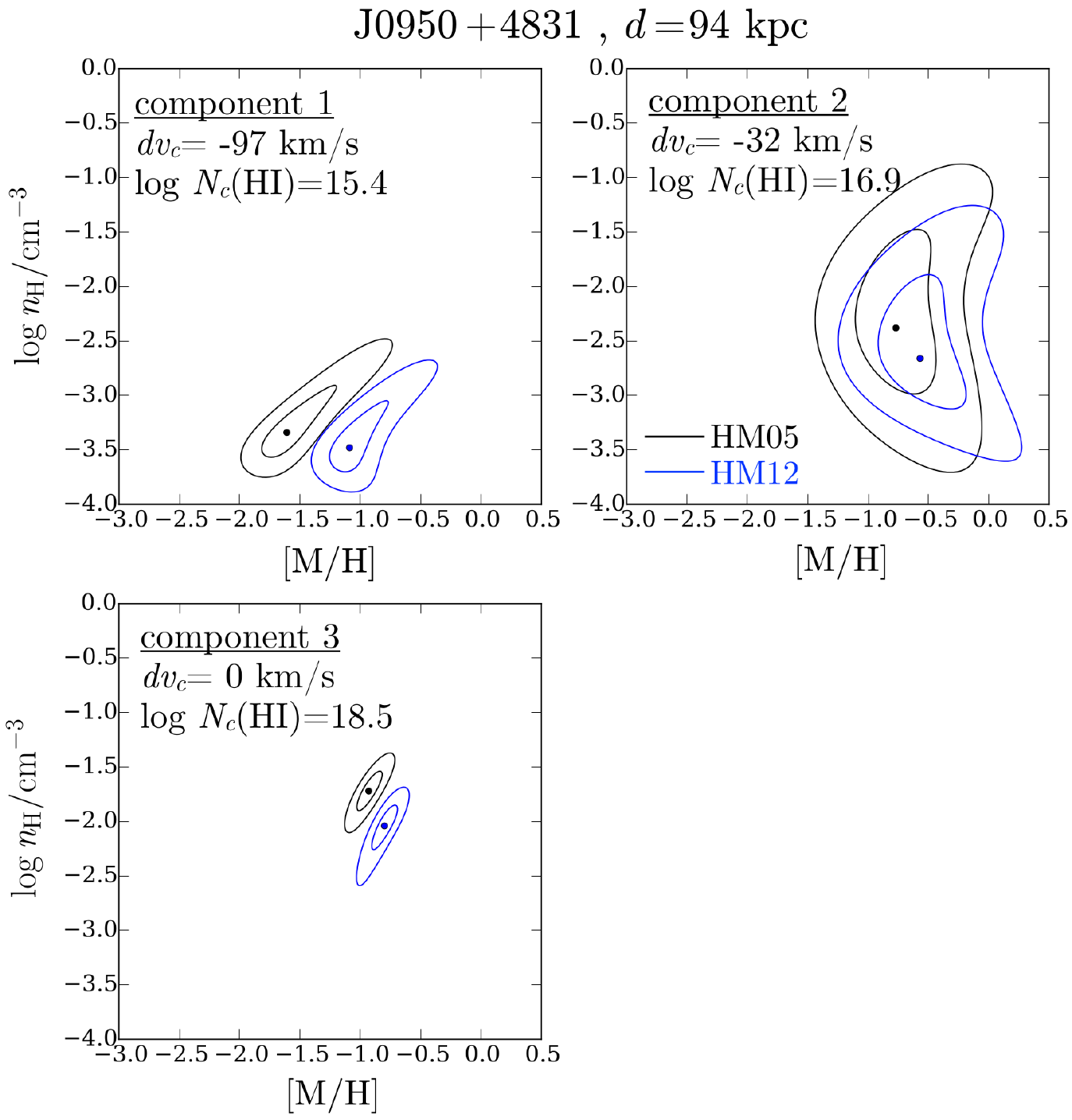}
\vspace{-1.5em}
\caption{Probability distribution contours of gas metallicity and density for individual absorption components identified along SDSS\,J0950$+$4831, at $d=94$ kpc from the LRG. Contour levels are the same as in Figure A1b. Not plotted here is component 4 at $dv_c=+96$ \kms, which has the lowest $N\mathrm{(\ion{H}{I})}$ in the absorber (log\,$N_c\mathrm{(\ion{H}{I})/\cmjj}=15.1$) and exhibits associated metal absorption in only \ion{C}{III}, resulting in weak constraints on the gas metallicity and density (see Table A6b).}
\label{figure:ions}
\end{figure}
\end{subfigures}

This LRG is at redshift $z_\mathrm{LRG}=0.2119$. A strong LLS with a total $N\mathrm{(\ion{H}{I})}$ of log\,$N\mathrm{(\ion{H}{I})/\cmjj}=18.51^{+0.05}_{-0.10}$ is present near the galaxy redshift (Figure A6a). In addition, the following ionic metal species are detected: \ion{C}{II}, \ion{C}{III}, \ion{N}{II}, \ion{O}{I}, \ion{O}{VI}, 
\ion{Mg}{I}, \ion{Mg}{II}, \ion{Si}{II}, \ion{Si}{III}, \ion{Si}{IV}, \ion{Fe}{II}, and \ion{Fe}{III}.  

We identify four components in the absorber, based on a combined Voigt profile analysis of \ion{H}{I} and the corresponding metal absorption profiles (Figure A6a and Table A6a). The \ion{H}{I} component structure is in very good agreement with those of the low- and intermediate-ionization metals. The observed velocity spread is $\Delta v\approx200$ \kms\ from the bluest to the reddest component. Most (98 percent) of the \ion{H}{I} column density is concentrated in component 3 at $z_\mathrm{abs}=0.21166$, or $59$ \kms\ blueward of the LRG. One other component has log\,$N_c\mathrm{(\ion{H}{I})/\cmjj}>16$, component 2 at $dv_c=-32$ \kms\ . These two high $N\mathrm{(\ion{H}{I})}$ components exhibit associated absorption of multiple metal species, whereas optically thin components 1 and 4 at $dv_c=+22$ and $+61$ \kms\ show metal absorption only in intermediate ions (e.g., \ion{C}{III}).  

Comparing the \ion{H}{I} and \ion{Mg}{II} linewidths of the optically thick components 2 and 3, the implied gas temperature is $T\approx (1-3)\times10^4$ K for both components, with $b_\mathrm{nt}\approx10-12$\,\kms\ of non-thermal line broadening. For components 1 and 4, the observed \ion{H}{I} linewidths constrain the gas temperature to $T\lesssim 4\times10^4$ K, consistent with those of the optically thick components. 

Our ionization analysis shows that the absorbing gas separates into two different regimes of gas density, as shown in Figure A6b and Table A6b. The inferred densities for the lower $N\mathrm{(\ion{H}{I})}$ gas in components 1 and 4 are low, log\,$n_\mathrm{H}/ \cmjjj\lesssim-3$ under both HM05 and HM12 UVBs. In contrast, models for the optically thick (log\,$N\mathrm{(\ion{H}{I})/\cmjj}\gtrsim17$) components 2 and 3 require higher densities to match the observations, from log\,$n_\mathrm{H}/ \cmjjj=-2.7^{+0.8}_{-0.4}$ (HM12) to  log\,$n_\mathrm{H}/ \cmjjj=-2.4^{+0.8}_{-0.6}$ (HM05) for component 2, and from 
log\,$n_\mathrm{H}/ \cmjjj=-2.0^{+0.1}_{-0.3}$ (HM12) to  log\,$n_\mathrm{H}/ \cmjjj=-1.7\pm0.2$ (HM05) for component 3. A caveat from the analysis is that over the allowed gas densities, \textsc{Cloudy} under-predicts the absorption column density of \ion{Si}{IV} by more than 1 dex for optically thick components 2 and 3, indicating that most of the Si\,IV absorption likely arises from a lower-density gas phase. 

Similarly, our \textsc{Cloudy} photoionization models indicate that the optically thin and thick gases are well-separated in metallicity space (Figure A6b and Table A6b). The observed ionic column densities in the lower $N\mathrm{(\ion{H}{I})}$ components 1 and 4 are consistent with a low-metallicity gas of $\mathrm{[M/H]}\sim-1.6$ under HM05 or $\mathrm{[M/H]}\sim-1.0$ under HM12, for both components. In contrast, the inferred metallicities for component 2 and 3 are sub-solar, from $\mathrm{[M/H]}=-0.8\pm0.3$ (HM05) to $\mathrm{[M/H]}=-0.6\pm0.3$ (HM12) for component 2, and from $\mathrm{[M/H]}=-0.9\pm0.1$ (HM05) to $\mathrm{[M/H]}=-0.8\pm0.1$ (HM12). The close agreement between HM05- and HM12-derived metallicities for component 3 is due to the well-constrained \ion{O}{I} column density, which scales proportionally with metallicity but is insensitive to different ionizing radiation fields. Finally, we note that the bulk of the absorption (component 3) is consistent with arising from a gas with an $\alpha-$element enhanced abundance pattern of $\mathrm{[Fe/\alpha]}\approx-0.2\pm0.1$.

This LLS was previously studied in the COS-Halos survey. Using the  Haardt \& Madau (2001) UVB (which is similar of the HM05 UVB at energies $\lesssim2$ Ryd), Werk \etal\ (2014) inferred a mean metallicity of $\mathrm{[M/H]}=-1.0^{+1.0}_{-0.5}$ and density of between log\,$n_\mathrm{H}/ \cmjjj=-3.8$ and  log\,$n_\mathrm{H}/ \cmjjj=-3.2$ under the Haardt \& Madau (2001) UVB. In an updated analysis using the HM12 UVB, Prochaska \etal\ (2017) found a mean metallicity of $\mathrm{[M/H]}=-0.91^{+0.14}_{-0.10}$ and density of  log\,$n_\mathrm{H}/ \cmjjj=-2.8\pm0.2$. The Prochaska \etal\ (2017) values are consistent with what we find in our analysis using the single-clump model, after accounting for the difference in the adopted redshift of the HM12 UVB spectrum used in both studies.

The strong \ion{O}{VI} absorption in this system can be fitted by a single component centered at $dv_c=+31$ \kms. The \ion{O}{VI} absorption profile is broad with a Doppler linewidth of 66 \kms, and a column density of log\,$N\mathrm{(\ion{O}{VI})/\cmjj}=14.32\pm0.03$. No low- or intermediate-ionization metal or \ion{H}{I} component is found to correspond with the \ion{O}{VI} absorption in velocity space (Figure A6a), with the \ion{O}{VI} doublet situated at  $\Delta v=+31$ \kms\ from the nearest low-ionization component.

\begin{subtables}
\begin{table}
\begin{center}
\caption{Absorption properties along QSO sightline SDSS\,J0950$+$4831 at $d=94$ kpc from the LRG}
\hspace{-2.5em}
\vspace{-0.5em}
\label{tab:Imaging}
\resizebox{3.5in}{!}{
\begin{tabular}{clrrr}\hline
Component	&	Species		&\multicolumn{1}{c}{$dv_c$} 		& \multicolumn{1}{c}{log\,$N_c$}	&\multicolumn{1}{c}{$b_c$}		\\	
 			&				&\multicolumn{1}{c}{(km\,s$^{-1}$)}	&   		   					& \multicolumn{1}{c}{(km\,s$^{-1}$)}  \\ \hline \hline

all	& \ion{H}{I}	&	$...$					& $18.51^{+0.05}_{-0.10}$& $...$ \\
	& \ion{C}{II}	&	$...$					& $14.76^{+0.13}_{-0.06}$& $...$	\\	
	& \ion{C}{III}	&	$...$					& $>14.60$			& $...$\\
	& \ion{N}{II}	&	$...$					& $14.47^{+0.07}_{-0.04}$& $...$ \\
	& \ion{N}{V}	&	$...$					& $<13.55$			& $...$	\\
	& \ion{O}{I}	&	$...$					& $14.35\pm0.05$& $...$	\\
	& \ion{O}{VI}	&	$...$					& $14.32\pm0.03$		& $...$	\\
	& \ion{Mg}{I}	&	$...$					& $11.86\pm0.06$		& $...$	\\	
	& \ion{Mg}{II}	&	$...$					& $13.80^{+0.06}_{-0.04}$& $...$	\\	
	& \ion{Si}{II}	&	$...$					& $13.85^{+0.06}_{-0.05}$& $...$	\\	
	& \ion{Si}{III}	&	$...$					& $>13.96$			& $...$	\\	
	& \ion{Si}{IV}	&	$...$					& $14.00^{+0.22}_{-0.07}$& $...$	\\	
	& \ion{Fe}{II}	&	$...$					& $13.44\pm0.03$		& $...$	\\ 	
	& \ion{Fe}{III}	&	$...$					& $14.06\pm0.14$		& $...$	\\ \hline	

1	& \ion{H}{I}	&	$-96.9^{+7.5}_{-3.5}$	& $15.43^{+0.11}_{-0.07}$& $25.8^{+5.1}_{-2.4}$ \\
	& \ion{C}{II}	&	$-96.9$				& $<12.91$			& 10	\\	
	& \ion{C}{III}	&	$-96.9$				& $13.57^{+0.06}_{-0.11}$	& $36.1^{+8.7}_{-7.1}$ 	\\
	& \ion{N}{II}	&	$-96.9$				& $<13.07$			& 10	\\
	& \ion{O}{I}	&	$-96.9$				& $<13.47$			& 10	\\
	& \ion{Mg}{I}	&	$-96.9$				& $<10.90$			& 10	\\	
	& \ion{Mg}{II}	&	$-96.9$				& $<11.43$			& 10	\\	
	& \ion{Si}{II}	&	$-96.9$				& $<12.40$			& 10	\\	
	& \ion{Si}{III}	&	$-96.9$				& $12.25^{+0.29}_{-0.21}$& $39.3^{+6.6}_{-19.1}$	\\
	& \ion{Si}{IV}	&	$-96.9$				& $<12.49$			& 10	\\	
	& \ion{Fe}{II}	&	$-96.9$				& $<12.12$			& 10	\\ 
	& \ion{Fe}{III}	&	$-96.9$				& $<13.21$			& 10	\\ \hline

2	& \ion{H}{I}	&	$-32.1$				& $16.94^{+0.35}_{-0.74}$& $15.0^{+2.6}_{-6.5}$ \\
	& \ion{C}{II}	&	$-32.1$				& $13.99^{+0.09}_{-0.14}$& $36.9^{+13.8}_{-5.7}$	\\	
	& \ion{C}{III}	&	$-32.1$				& $>13.72$			& $<27.5$ 	\\
	& \ion{N}{II}	&	$-32.1$				& $<13.27$			& 10	\\
	& \ion{O}{I}	&	$-32.1$				& $<13.45$			& 10	\\
	& \ion{Mg}{I}	&	$-32.1$				& $<10.85$			& 10	\\	
	& \ion{Mg}{II}	&	$-32.1^{+1.7}_{-0.8}.$	& $12.87^{+0.07}_{-0.02}$& $12.5^{+2.1}_{-0.5}$	\\	
	& \ion{Si}{II}	&	$-32.1$				& $13.45^{+0.09}_{-0.06}$& $23.6^{+7.1}_{-4.1}$	\\	
	& \ion{Si}{III}	&	$-32.1$				& $>13.38$			& $<40.1$	\\
	& \ion{Si}{IV}	&	$-32.1$				& $13.75^{+0.16}_{-0.09}$& $27.5^{+9.0}_{-6.8}$	\\	
	& \ion{Fe}{II}	&	$-32.1$				& $12.58\pm0.15$		& $13.1\pm5.4$	\\ 
	& \ion{Fe}{III}	&	$-32.1$				& $13.92\pm0.12$		& $23.4\pm9.2$	\\ \hline

3	& \ion{H}{I}	&	$0.0\pm0.9$			& $18.50^{+0.05}_{-0.10}$& $24.7^{+0.9}_{-1.1}$ \\
	& \ion{C}{II}	&	$0.0$				& $14.68^{+0.15}_{-0.06}$& $27.2^{+3.4}_{-4.1}$ \\	
	& \ion{C}{III}	&	$0.0$				& $>14.45$			& $<36.8$ 	\\
	& \ion{N}{II}	&	$0.0$				& $14.47^{+0.07}_{-0.04}$& $29.7^{+5.1}_{-2.7}$	\\
	& \ion{O}{I}	&	$0.0$				& $14.35\pm0.05$& $30.2^{+10.3}_{-3.7}$	\\
	& \ion{Mg}{I}	&	$0.0$				& $11.86\pm0.06$		& $18.1\pm2.5$	\\	
	& \ion{Mg}{II}	&	$0.0^{+0.7}_{-0.3}.$		& $13.75\pm0.05$		& $13.0^{+0.4}_{-0.8}$	\\	
	& \ion{Si}{II}	&	$0.0$				& $13.63^{+0.09}_{-0.10}$& $16.0^{+2.7}_{-3.1}$	\\	
	& \ion{Si}{III}	&	$0.0$				& $>13.59$			& $<23.3$	\\
	& \ion{Si}{IV}	&	$0.0$				& $13.63^{+0.32}_{-0.21}$& $18.8^{+7.3}_{-7.2}$	\\	
	& \ion{Fe}{II}	&	$+0.4\pm1.3$			& $13.37\pm0.03$		& $19.5^{+1.9}_{-1.2}$	\\ 
	& \ion{Fe}{III}	&	$0.0$				& $13.49\pm0.28$		& $20.2\pm19.5$	\\ \hline

4	& \ion{H}{I}	&	$+95.9^{+2.4}_{-6.0}$	& $15.06^{+0.13}_{-0.08}$& $14.4^{+4.1}_{-1.6}$ \\
	& \ion{C}{II}	&	$+95.9$				& $<12.89$			& 10	\\	
	& \ion{C}{III}	&	$+95.9$				& $13.40^{+0.10}_{-0.11}$	& $19.3^{+4.3}_{-5.3}$ 	\\
	& \ion{N}{II}	&	$+95.9$				& $<13.06$			& 10	\\
	& \ion{O}{I}	&	$+95.9$				& $<13.51$			& 10	\\
	& \ion{Mg}{I}	&	$+95.9$				& $<10.92$			& 10	\\	
	& \ion{Mg}{II}	&	$+95.9$				& $<11.40$			& 10	\\	
	& \ion{Si}{II}	&	$+95.9$				& $<12.23$			& 10	\\	
	& \ion{Si}{III}	&	$+95.9$				& $<12.10$			& 10	\\
	& \ion{Si}{IV}	&	$+95.9$				& $<12.63$			& 10	\\	
	& \ion{Fe}{II}	&	$+95.9$				& $<12.14$			& 10	\\ 
	& \ion{Fe}{III}	&	$+95.9$				& $<13.19$			& 10	\\ \hline

high-1	& \ion{O}{VI}	&	$+31.1\pm3.5$		& $14.32\pm0.03$		& $65.7^{+6.5}_{-5.7}$ \\
		& \ion{N}{V}	&	$+31.1$			& $<13.55$			& \\

\hline
\end{tabular}}
\end{center}
\end{table}

\begin{table}
\begin{center}
\caption{Ionization modeling results for the absorber along SDSS\,J0950$+$4831 at $d=94$ kpc from the LRG}
\hspace{-2.5em}
\label{tab:Imaging}
\resizebox{3.5in}{!}{
\begin{tabular}{@{\extracolsep{3pt}}ccrrrr@{}}\hline
Component	&$N_\mathrm{metal}$& \multicolumn{2}{c}{$\mathrm{[M/H]}$} 	& \multicolumn{2}{c}{$\mathrm{log\,}n_\mathrm{H}/\cmjjj$}		\\
\cline{3-4} \cline {5-6}
	& &\multicolumn{1}{c}{HM05}&	\multicolumn{1}{c}{HM12}	&\multicolumn{1}{c}{HM05} 	& 	\multicolumn{1}{c}{HM12}			\\	\hline \hline

SC	&10& $-0.97\pm0.08$		& $-0.92^{+0.09}_{-0.08}$		& $-1.96^{+0.12}_{-0.18}$	& $-2.46^{+0.16}_{-0.26}$ \\ \hline
1	&2& $-1.59^{+0.47}_{-0.15}$	& $-1.05^{+0.32}_{-0.14}$		& $-3.34^{+0.50}_{-0.16}$	& $-3.48^{+0.46}_{-0.16}$ \\
2	&7& $-0.76^{+0.36}_{-0.34}$	& $-0.57^{+0.37}_{-0.34}$		& $-2.42^{+0.86}_{-0.58}$	& $-2.68^{+0.82}_{-0.36}$	\\	
3	&10& $-0.93\pm0.10$			& $-0.81^{+0.09}_{-0.12}$	& $-1.72^{+0.16}_{-0.20}$	& $-2.02^{+0.12}_{-0.32}$	\\	
4	&1& $-1.66^{+0.56}_{-0.17}$	& $-0.90^{+0.58}_{-0.15}$		& $<-3.10$			& $<-3.20$	\\	

\hline
\end{tabular}}
\end{center}
\end{table}
\end{subtables}

\subsection{SDSS\, J1127$+$1154 at $d=99$ kpc}

\begin{subfigures}
\begin{figure*}
\includegraphics[scale=1.08]{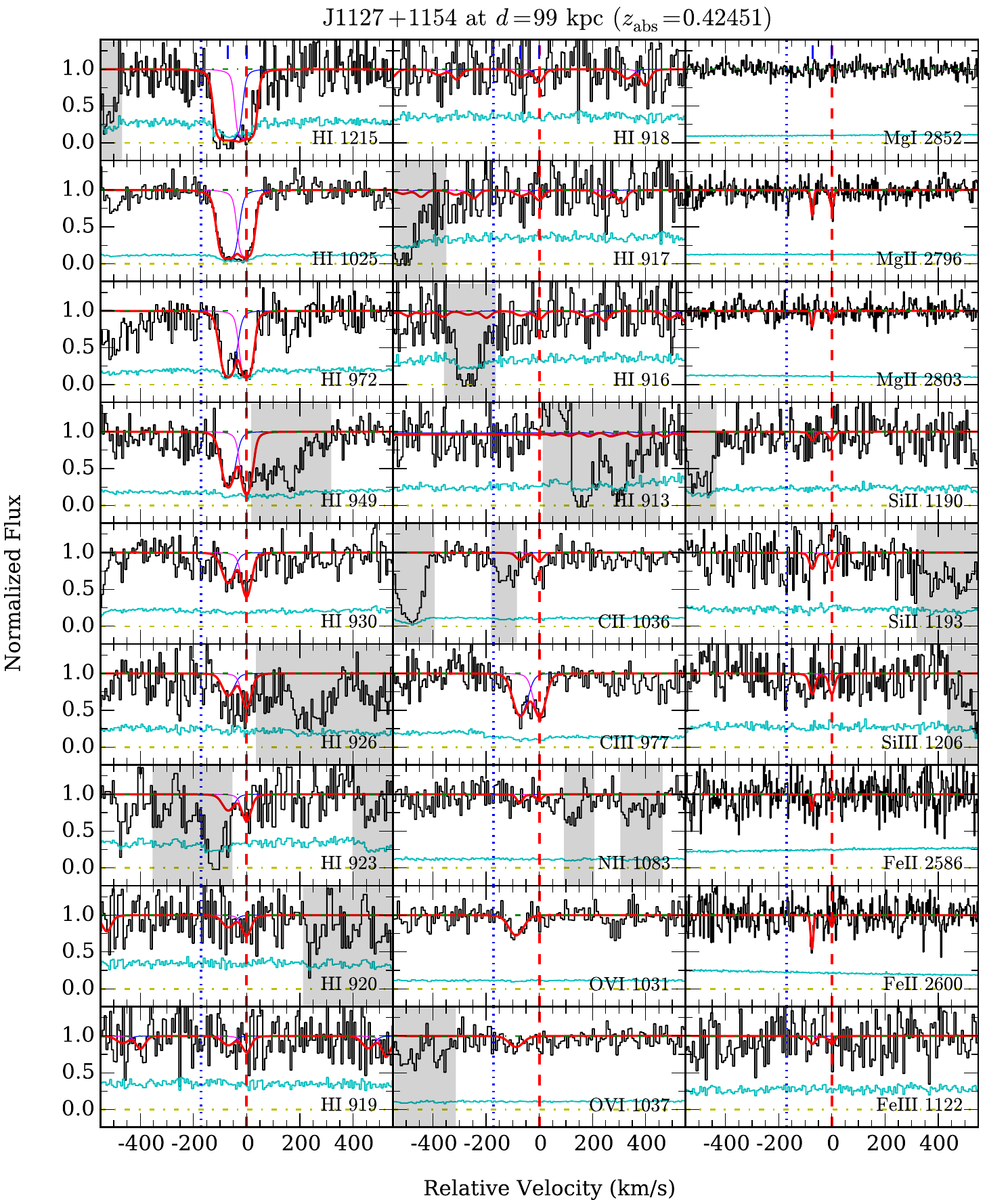}
\vspace{-0.75em}
\caption{Similar to Figure A1a, but for SDSS\,J1127$+$1154 at $d=99$ kpc from the LRG.}
\label{figure:ions}
\end{figure*}

\begin{figure}
\hspace{-0.5em}
\includegraphics[scale=0.62]{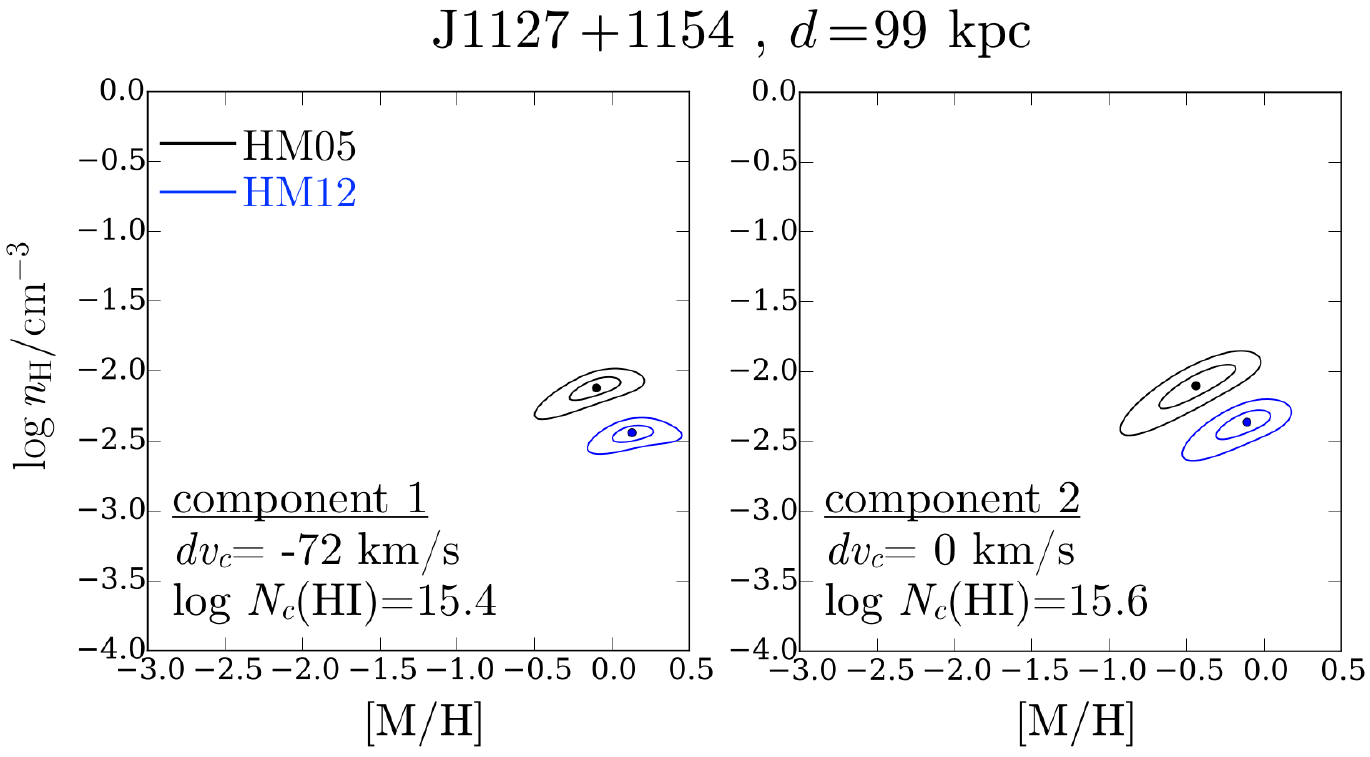}
\vspace{-1.5em}
\caption{Probability distribution contours of gas metallicity and density for individual absorption components identified along SDSS\,J1127$+$1154, at $d=99$ kpc from the LRG. Contour levels are the same as in Figure A1b.}
\label{figure:ions}
\end{figure}
\end{subfigures}

This LRG is found at $z_\mathrm{LRG}=0.4237$. As shown in Figure A7a, an \ion{H}{I} absorber with a total $N\mathrm{(\ion{H}{I})}$ of log\,$N\mathrm{(\ion{H}{I})/\cmjj}=15.81\pm0.06$ is present near the LRG redshift. The HI absorption is accompanied by detections of \ion{C}{III}, \ion{Mg}{II}, \ion{Fe}{II}, and \ion{O}{VI} metal species.

The HI absorption profile can be decomposed into two individual components of comparable strengths, log\,$N_c\mathrm{(\ion{H}{I})/\cmjj}=15.4-15.6$ (Figure A7a and Table A7a). The stronger \ion{H}{I} absorption component (component 2) occurs $171$ \kms\ redward of the LRG at $z_\mathrm{abs}=0.42451$, whereas the weaker component 1 occurs at  $dv_c=-72$ \kms\ from this component. A curious feature of this absorption system is while the \ion{Mg}{II} and \ion{C}{III} absorption are of comparable strengths in both components, \ion{Fe}{II} is detected only in component 1 (Figure A7a), which suggests a difference in $\mathrm{[Fe/Mg]}$ abundance ratio between the two components. 

The velocity spread of the absorbing gas is $\Delta v\approx75$ \kms\ from the bluer to redder components. The observed Doppler $b$ parameters of individual \ion{H}{I} and \ion{Mg}{II} components indicate that the absorbing gas is relatively cool, $T\sim(2-4)\times10^4$ K, with only a modest amount of non-thermal broadening in the gas, with $b_\mathrm{nt}\lesssim6$\,\kms.

The matching component structure between \ion{H}{I} and low- to intermediate-ionization metals in the absorber justifies a single-phase photoionization model for the gas. As shown in Figure A7b and Table A7b, our ionization analysis finds little variation in gas densities between the two components, where log\,$n_\mathrm{H}/ \cmjjj\approx-2.1$ is inferred under the HM05 UVB and  log\,$n_\mathrm{H}/ \cmjjj\approx-2.4$ is inferred under HM12 UVB. In contrast, our models indicate a larger variation in chemical abundance. For component 2, we find a sub-solar metallicity of $\mathrm{[M/H]}=-0.4\pm0.2$ under HM05 and $\mathrm{[M/H]}=-0.1\pm0.2$ under HM12. A high chemical abundance, consistent with solar value, is inferred for component 1, where the \textsc{Cloudy} models constrain the metallicity to $\mathrm{[M/H]}=-0.1\pm0.2$  under HM05 or $\mathrm{[M/H]}=+0.1\pm0.2$ under HM12. Furthermore, the unusual strength of \ion{Fe}{II} absorption  relative to that of \ion{Mg}{II} in component 1 suggests that the gas is particularly iron-rich. The inferred $\mathrm{[Fe/\alpha]}$ for this component is $\mathrm{[Fe/\alpha]}\approx +1.0\pm0.3$, which is among the highest known in the literature (e.g, Narayanan \etal\ 2008; Zahedy \etal\ 2016; 2017b). 

Finally, we note that a modest absorption of highly ionized \ion{O}{VI} gas is detected at $dv_c=-86$ \kms. The \ion{O}{VI} absorption consists of a single component with log\,$N\mathrm{(\ion{O}{VI})/\cmjj}=13.79^{+0.08}_{-0.09}$ and a Doppler $b$ value of of 32 \kms. No low- or intermediate-ionization metal or \ion{H}{I} component is found to correspond with the \ion{O}{VI} absorption in velocity space (Figure A7a), with the \ion{O}{VI} doublet situated at $\Delta v= -15$ \kms\ from the nearest component.

\begin{subtables}
\begin{table}
\begin{center}
\caption{Absorption properties along QSO sightline SDSS\,J1127$+$1154 at $d=99$ kpc from  the LRG}
\hspace{-2.5em}
\vspace{-0.5em}
\label{tab:Imaging}
\resizebox{3.5in}{!}{
\begin{tabular}{clrrr}\hline
Component	&	Species		&\multicolumn{1}{c}{$dv_c$} 		& \multicolumn{1}{c}{log\,$N_c$}	&\multicolumn{1}{c}{$b_c$}		\\	
 			&				&\multicolumn{1}{c}{(km\,s$^{-1}$)}	&   		   					& \multicolumn{1}{c}{(km\,s$^{-1}$)}  \\ \hline \hline

all	& \ion{H}{I}	&	$...$					& $15.81\pm0.06$		& $...$ \\
	& \ion{C}{II}	&	$...$					& $<13.43$			& $...$	\\	
	& \ion{C}{III}	&	$...$					& $13.73\pm0.04$		& $...$\\
	& \ion{N}{II}	&	$...$					& $<13.52$			& $...$ \\
	& \ion{N}{V}	&	$...$					& $<13.64$			& $...$	\\
	& \ion{O}{I}	&	$...$					& $<13.97$			& $...$ \\
	& \ion{O}{VI}	&	$...$					& $13.79^{+0.08}_{-0.09}$& $...$	\\
	& \ion{Mg}{I}	&	$...$					& $<11.64$			& $...$	\\
	& \ion{Mg}{II}	&	$...$					& $12.39^{+0.14}_{-0.12}$& $...$\\
	& \ion{Si}{II}	&	$...$					& $<13.08$			& $...$	\\	
	& \ion{Si}{III}	&	$...$					& $<12.61$			& $...$	\\	
	& \ion{Fe}{II}	&	$...$					& $12.88^{+0.21}_{-0.29}$& $...$	\\
	& \ion{Fe}{III}	&	$...$					& $<14.00$			& $...$	\\ \hline	
	
1	& \ion{H}{I}	&	$-71.5^{+3.0}_{-1.6}$	& $15.42^{+0.07}_{-0.06}$& $25.6^{+2.3}_{-0.9}$ \\
	& \ion{C}{II}	&	$-71.5$				& $<13.22$			& 10	\\	
	& \ion{C}{III}	&	$-71.5$				& $13.45\pm0.06$		& $30.5^{+10.0}_{-5.2}$	\\
	& \ion{N}{II}	&	$-71.5$				& $<13.10$			& 10	\\
	& \ion{O}{I}	&	$-71.5$				& $<13.50$			& 10	\\
	& \ion{Mg}{I}	&	$-71.5$				& $<11.18$			& 10	\\
	& \ion{Mg}{II}	&	$-73.3^{+1.4}_{-1.5}$	& $12.15^{+0.21}_{-0.15}$&  $3.0^{+3.1}_{-1.6}$	\\
	& \ion{Si}{II}	&	$-71.5$				& $<12.70$			& 10	\\	
	& \ion{Si}{III}	&	$-71.5$				& $<12.26$			& 10	\\	
	& \ion{Fe}{II}	&	$-74.5^{+1.5}_{-1.5}$	& $12.88^{+0.21}_{-0.29}$& $2.7\pm2.6$	\\
	& \ion{Fe}{III}	&	$-71.5$				& $<13.50$			& 10	\\ \hline	

2	& \ion{H}{I}	&	$0.0^{+3.2}_{-2.0}$		& $15.59^{+0.06}_{-0.10}$& $18.4^{+1.3}_{-1.4}$ \\
	& \ion{C}{II}	&	$0.0$				& $<13.20$			& 10 \\		
	& \ion{C}{III}	&	$0.0$				& $13.41\pm0.07$		& $24.8^{+6.8}_{-4.6}$ \\
	& \ion{N}{II}	&	$0.0$				& $<13.05$			& 10	\\
	& \ion{O}{I}	&	$0.0$				& $<13.48$			& 10	\\
	& \ion{Mg}{I}	&	$0.0$				& $<11.19$			& 10	\\
	& \ion{Mg}{II}	&	$0.0^{+3.2}_{-3.1}$		& $12.02^{+0.09}_{-0.23}$& $7.4^{+6.4}_{-3.7}$ \\	
	& \ion{Si}{II}	&	$0.0$				& $<12.68	$			& 10 \\	
	& \ion{Si}{III}	&	$0.0$				& $<12.30$			& 10 \\	
	& \ion{Fe}{II}	&	$0.0$				& $<12.40$			& 10	\\
	& \ion{Fe}{III}	&	$0.0$				& $<13.50$			& 10	\\ \hline
	
high-1	& \ion{O}{VI}	&	$-86.4\pm5.3$		& $13.79^{+0.08}_{-0.09}$& $32.2^{+16.8}_{-6.1}$ \\
		& \ion{N}{V}	&	$-86.4$			& $<13.64$			&\\
\hline
\end{tabular}}
\end{center}
\end{table}

\begin{table}
\begin{center}
\caption{Ionization modeling results for the absorber along SDSS\,J1127$+$1154, at $d=99$ kpc from the LRG}
\hspace{-2.5em}
\label{tab:Imaging}
\resizebox{3.5in}{!}{
\begin{tabular}{@{\extracolsep{3pt}}ccrrrr@{}}\hline
Component	&$N_\mathrm{metal}$& \multicolumn{2}{c}{$\mathrm{[M/H]}$} 	& \multicolumn{2}{c}{$\mathrm{log\,}n_\mathrm{H}/\cmjjj$}		\\
\cline{3-4} \cline {5-6}
	& &\multicolumn{1}{c}{HM05}&	\multicolumn{1}{c}{HM12}	&\multicolumn{1}{c}{HM05} 	& 	\multicolumn{1}{c}{HM12}			\\	\hline \hline

SC	&3& $-0.22^{+0.11}_{-0.18}$	& $+0.07\pm0.12$		& $-2.10^{+0.04}_{-0.10}$		& $-2.40^{+0.02}_{-0.06}$ \\ \hline
1	&3& $-0.11^{+0.14}_{-0.21}$	& $+0.12^{+0.17}_{-0.14}$& $-2.12^{+0.04}_{-0.12}$	& $-2.46^{+0.04}_{-0.06}$ \\
2	&2& $-0.44^{+0.19}_{-0.26}$	& $-0.13^{+0.13}_{-0.22}$	& $-2.08^{+0.08}_{-0.20}$	& $-2.36^{+0.04}_{-0.16}$	\\

\hline
\end{tabular}}
\end{center}
\end{table}
\end{subtables}

\subsection{SDSS\, J1243$+$3539 at $d=102$ kpc}

The LRG is at $z_\mathrm{LRG}=0.3896$. As shown in Figure A8a, no \ion{H}{I} absorption is detected within the adopted search window of $\pm500$ \kms\ from the LRG redshift. We are able to place a sensitive 2-$\sigma$ column density upper limit of log\,$N\mathrm{(\ion{H}{I})/\cmjj}<12.7$, calculated for an \ion{H}{I} line with $b\mathrm{(\ion{H}{I})}=15$ \kms\ that is centered at the LRG redshift (Table A8a).

\begin{subfigures}
\begin{figure*}
\includegraphics[scale=1.08]{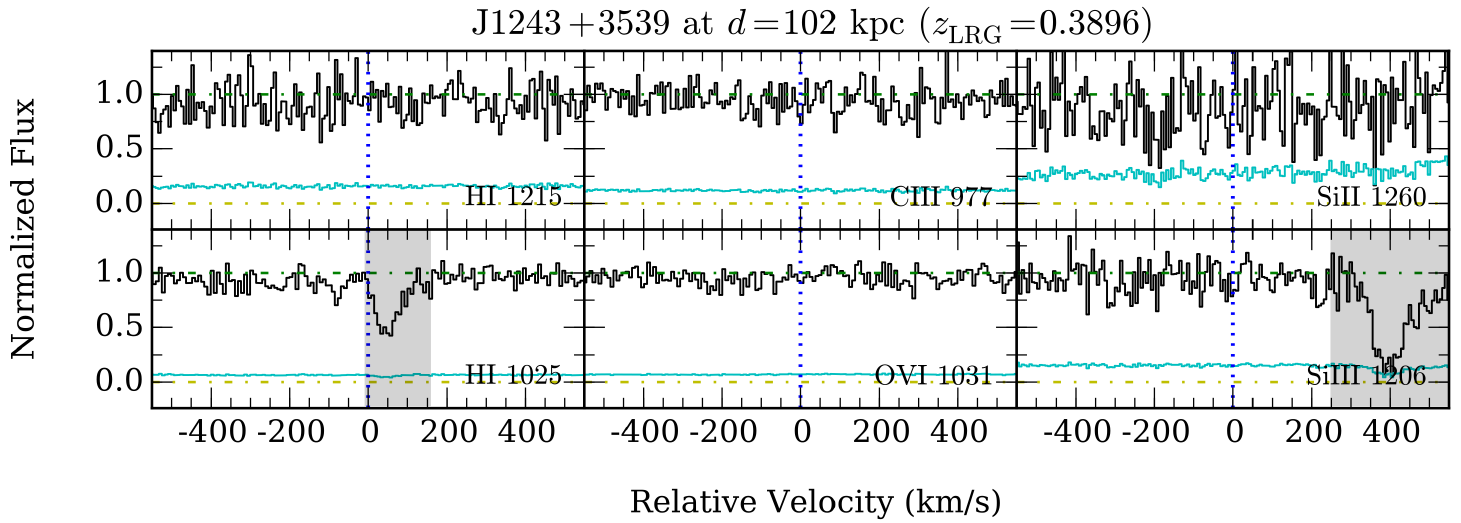}
\vspace{-0.75em}
\caption{Similar to Figure A1a, but for SDSS\,J1243$+$3539 at $d=102$ kpc from the LRG.}
\label{figure:ions}
\end{figure*}
\end{subfigures}

\begin{subtables}
\begin{table}
\begin{center}
\caption{Absorption properties along QSO sightline  SDSS\,J1243$+$3539 at $d=102$ kpc from the LRG}
\hspace{-2.5em}
\vspace{-0.5em}
\label{tab:Imaging}
\resizebox{3.3in}{!}{
\begin{tabular}{clcrc}\hline
Component	&	Species		&\multicolumn{1}{c}{$dv_c$} 		& \multicolumn{1}{c}{log\,$N_c$}	&\multicolumn{1}{c}{$b_c$}		\\	
 			&				&\multicolumn{1}{c}{(km\,s$^{-1}$)}	&   		   					& \multicolumn{1}{c}{(km\,s$^{-1}$)}  \\ \hline \hline

...	& \ion{H}{I}	&	$0.0$					& $<12.65$			& $15$ \\
	& \ion{C}{II}	&	$0.0$					& $<12.80$			& $10$	\\	
	& \ion{C}{III}	&	$0.0$					& $<12.24$			& $10$\\
	& \ion{N}{II}	&	$0.0$					& $<12.91$			& $10$ \\
	& \ion{N}{III}	&	$0.0$					& $<13.06$			& $10$ \\
	& \ion{N}{V}	&	$0.0$					& $<13.21$			& $30$ \\
	& \ion{O}{I}	&	$0.0$					& $<13.49$			& $10$ \\
	& \ion{O}{VI}	&	$0.0$					& $<13.06$			& $30$ \\
	& \ion{Si}{II}	&	$0.0$					& $<12.38$			& $10$	\\	
	& \ion{Si}{III}	&	$0.0$					& $<11.91$			& $10$	\\
	& \ion{Fe}{II}	&	$0.0$					& $<12.40$			& $10$	\\
	& \ion{Fe}{III}	&	$0.0$					& $<13.50$			& $10$	\\ 	

\hline
\end{tabular}}
\end{center}
\end{table}
\end{subtables}

\subsection{SDSS\, J1550$+$4001 at $d=107$ kpc}

\begin{subfigures}
\begin{figure*}
\includegraphics[scale=1.08]{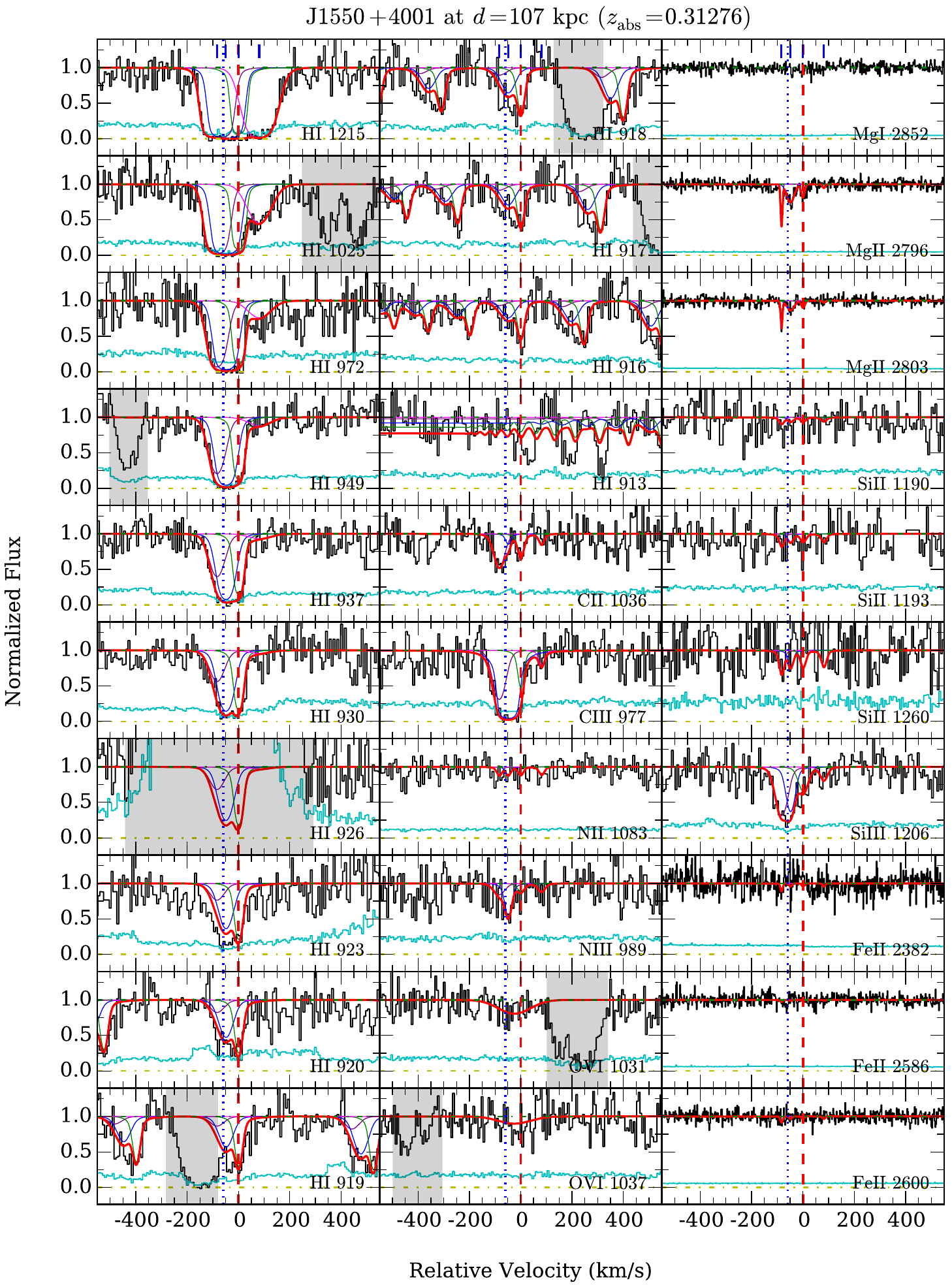}
\vspace{-0.75em}
\caption{Similar to Figure A1a, but for SDSS\,J1550$+$4001 at $d=107$ kpc from the LRG. The excess flux seen redward of the Lyman limit can be attributed to a second point source that falls within the same COS aperture as the background QSO (see Paper I)}
\label{figure:ions}
\end{figure*}

\begin{figure}
\hspace{-0.8em}
\includegraphics[scale=0.62]{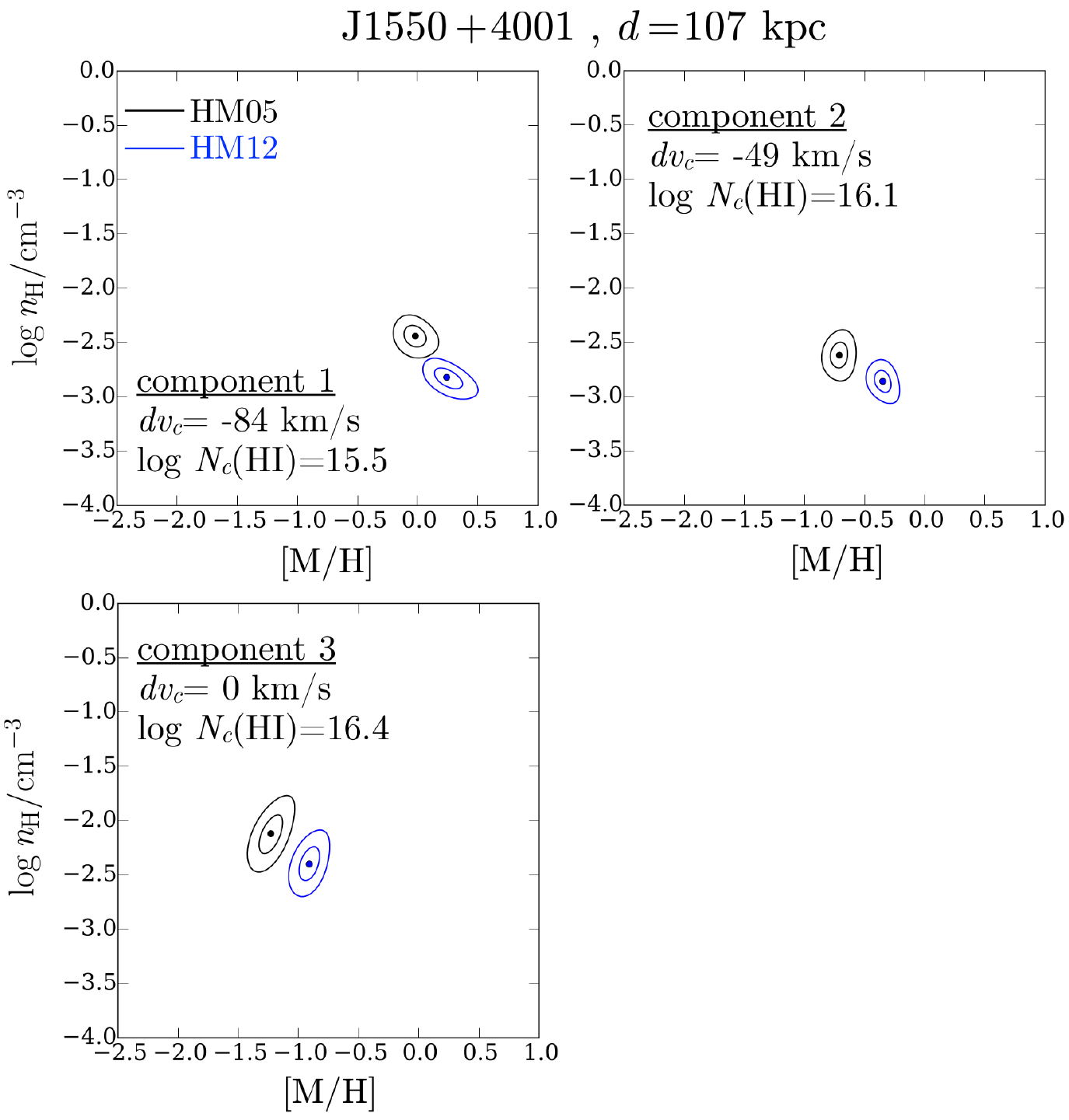}
\vspace{-1.5em}
\caption{Probability distribution contours of gas metallicity and density for individual absorption components identified along SDSS\,J1550$+$4001, at $d=107$ kpc from the LRG. Contour levels are the same as in Figure A1b. Not plotted here is component 4 at $dv_c=+81$ \kms, which has the lowest $N\mathrm{(\ion{H}{I})}$ in the absorber (log\,$N_c\mathrm{(\ion{H}{I})/\cmjj}=14.6$) and shows only \ion{C}{III} metal absorption, resulting in weak constraints on the gas metallicity and density (see Table A9b).}
\label{figure:ions}
\end{figure}
\end{subfigures}

The LRG, which is at $z_\mathrm{LRG}=0.3125$, has an \ion{H}{I} absorber with a total $N\mathrm{(\ion{H}{I})}$ of log\,$N\mathrm{(\ion{H}{I})/\cmjj}=16.61\pm0.04$. This pLLS also exhibits corresponding absorption of ionic species \ion{C}{II}, \ion{C}{III}, \ion{N}{III}, \ion{Mg}{II}, \ion{Si}{II}, \ion{Si}{III}, \ion{Fe}{II}, and \ion{O}{VI} (Figure A9a).

The HI absorption profile consists of four individual components (Figure A9a and Table A9a).  The strongest \ion{H}{I} absorption is in component 3 
at $z_\mathrm{abs}=0.31276$, or $58$ \kms\ redward of the LRG. This component contributes to $\sim60\%$ of the total $N\mathrm{(\ion{H}{I})}$ of the absorber. In contrast, the strongest metal absorption corresponds to weaker \ion{H}{I} components, component 1 for low ions (e.g., \ion{C}{II} and \ion{Mg}{II}) and component 2 for intermediate ions (e.g., \ion{C}{III}), suggesting a variation in chemical abundance among different components. 

The observed velocity spread of the absorbing gas is $\Delta v=164$ \kms\ from the bluest to reddest component. Our Voigt profile analysis shows that the absorption profiles of both low- and intermediate-ionization metals match the component structure of \ion{H}{I} very well (see Figure A4a). By comparing the Doppler $b$ linewidths of \ion{H}{I} and ionic metal components, we find that the bulk of the absorbing gas (component 3) is cool, $T\sim 10^4$ K, with a modest amount of non-thermal broadening, $b_\mathrm{nt}\approx8$\,\kms. The other three components are at higher temperatures, $T\sim(3-6)\times10^4$ K, with inferred non-thermal broadening that varies from $b_\mathrm{nt}\lesssim5$\,\kms\ in component 1 to $b_\mathrm{nt}\gtrsim20$ \,\kms\ in components 2 and 4.

As shown in Figure A9b and Table A9b, our ionization analysis finds a modest variation in gas densities across different components, which range from log\,$n_\mathrm{H}/ \cmjjj\approx-2.6$ to log\,$n_\mathrm{H}/ \cmjjj\approx-2.1$ under the HM05 UVB, and from log\,$n_\mathrm{H}/ \cmjjj\approx-2.9$ to  log\,$n_\mathrm{H}/ \cmjjj\approx-2.4$ under the HM12 UVB. To reproduce the observed metal column densities, the required chemical abundances range from sub-solar values of between $\mathrm{[M/H]}=-1.2\pm0.1$ (HM05) and $\mathrm{[M/H]}=-0.9\pm0.1$ (HM12) for the strongest component, to at least solar metallicity for component 1,   
$\mathrm{[M/H]}=0.0\pm0.1$ under HM05 and $\mathrm{[M/H]}=0.2\pm0.1$ under HM12.

This absorber was also studied in the COS-Halos survey. Using the  Haardt \& Madau (2001) UVB (which is similar of the HM05 UVB at energies $\lesssim2$ Ryd), Werk \etal\ (2014) inferred a mean metallicity of $\mathrm{[M/H]}=-0.8\pm0.2$ and density of between log\,$n_\mathrm{H}/ \cmjjj=-4.0$ and  log\,$n_\mathrm{H}/ \cmjjj=-3.5$ under the Haardt \& Madau (2001) UVB. In an updated analysis using the HM12 UVB, Prochaska \etal\ (2017) found a mean metallicity of $\mathrm{[M/H]}=-0.35^{+0.05}_{-0.05}$ and density of  log\,$n_\mathrm{H}/ \cmjjj=-2.75\pm0.2$, which are in agreement with what we find in our analysis using the single-clump model (Table A6b).

We detect a weak \ion{O}{VI} absorption at $dv_c=-25$ \kms. The \ion{O}{VI} absorption profile consists of a single component with log\,$N\mathrm{(\ion{O}{VI})/\cmjj}=14.0\pm0.1$ and a broad Doppler $b$ value of of 83 \kms. No low- or intermediate-ionization metal or \ion{H}{I} component is found to correspond with the \ion{O}{VI} absorption in velocity space (Figure A9a), with the \ion{O}{VI} doublet situated at  $\Delta v=-24$ \kms\ from the nearest low-ionization component.

\begin{subtables}
\begin{table}
\begin{center}
\caption{Absorption properties along QSO sightline SDSS\,J1550$+$4001 at $d=107$ kpc from the LRG}
\hspace{-2.5em}
\vspace{-0.5em}
\label{tab:Imaging}
\resizebox{3.5in}{!}{
\begin{tabular}{clrrr}\hline
Component	&	Species		&\multicolumn{1}{c}{$dv_c$} 		& \multicolumn{1}{c}{log\,$N_c$}	&\multicolumn{1}{c}{$b_c$}		\\	
 			&				&\multicolumn{1}{c}{(km\,s$^{-1}$)}	&   		   					& \multicolumn{1}{c}{(km\,s$^{-1}$)}  \\ \hline \hline

all	& \ion{H}{I}	&	$...$					& $16.61\pm0.04$		& $...$ \\
	& \ion{C}{II}	&	$...$					& $14.16^{+0.11}_{-0.07}$& $...$	\\	
	& \ion{C}{III}	&	$...$					& $>14.53$			& $...$\\
	& \ion{N}{II}	&	$...$					& $<13.59$			& $...$ \\
	& \ion{N}{III}	&	$...$					& $13.97^{+0.22}_{-0.19}$& $...$ \\
	& \ion{N}{V}	&	$...$					& $<13.62$			& $...$	\\
	& \ion{O}{I}	&	$...$					& $<14.26$			& $...$ \\
	& \ion{O}{VI}	&	$...$					& $13.96^{+0.11}_{-0.12}$& $...$	\\
	& \ion{Mg}{I}	&	$...$					& $<11.27$			& $...$	\\
	& \ion{Mg}{II}	&	$...$					& $12.70\pm0.02$		& $...$\\
	& \ion{Si}{II}	&	$...$					& $12.57\pm0.30$		& $...$	\\	
	& \ion{Si}{III}	&	$...$					& $13.42^{+0.12}_{-0.06}$& $...$	\\	
	& \ion{Fe}{II}	&	$...$					& $11.84\pm0.20		$& $...$	\\ \hline	
	
1	& \ion{H}{I}	&	$-84.1^{+3.1}_{-2.4}$	& $15.54^{+0.13}_{-0.17}$& $32.6^{+2.2}_{-1.7}$ \\
	& \ion{C}{II}	&	$-84.1$				& $14.01^{+0.12}_{-0.09}$& $23.4^{+10.8}_{-4.6}$	\\	
	& \ion{C}{III}	&	$-84.1$				& $13.58^{+0.44}_{-0.34}$& $26.0^{+10.1}_{-9.9}$	\\
	& \ion{N}{II}	&	$-84.1$				& $<13.09$			& 10	\\
	& \ion{N}{III}	&	$-84.1$				& $<13.42$			& 10	\\
	& \ion{O}{I}	&	$-84.1$				& $<13.78$			& 10	\\
	& \ion{Mg}{I}	&	$-84.1$				& $<10.76$			& 10	\\
	& \ion{Mg}{II}	&	$-84.1\pm0.3$			& $12.29^{+0.04}_{-0.03}$& $2.8\pm0.5$	\\
	& \ion{Si}{II}	&	$-84.1$				& $12.57\pm0.30$		& $8.0^{+7.1}_{-3.7}$	\\	
	& \ion{Si}{III}	&	$-84.1$				& $13.12^{+0.14}_{-0.10}$& $32.7^{+9.4}_{-8.3}$	\\	
	& \ion{Fe}{II}	&	$-84.1$				& $11.84\pm0.20$		& $4.4\pm4.2$	\\ \hline

2	& \ion{H}{I}	&	$-49.2^{+2.9}_{-2.4}$	& $16.12^{+0.08}_{-0.09}$& $30.7^{+4.6}_{-2.9}$ \\
	& \ion{C}{II}	&	$-49.2$				& $<13.17$			& 10 \\		
	& \ion{C}{III}	&	$-49.2$				& $>14.34	$			& $<30.8$ \\
	& \ion{N}{II}	&	$-49.2$				& $<13.09$			& 10	\\
	& \ion{N}{III}	&	$-49.2$				& $13.97^{+0.22}_{-0.19}$& $13.1^{+6.3}_{-2.0}$	\\
	& \ion{O}{I}	&	$-49.2$				& $<13.76$			& 10	\\
	& \ion{Mg}{I}	&	$-49.2$				& $<10.78$			& 10	\\
	& \ion{Mg}{II}	&	$-49.2\pm1.3$			& $12.36\pm0.03$		& $19.6^{+2.2}_{-1.7}$ \\	
	& \ion{Si}{II}	&	$-49.2$				& $<12.37	$			& 10 \\	
	& \ion{Si}{III}	&	$-49.2$				& $12.97^{+0.21}_{-0.25}$& $18.4^{+10.9}_{-6.8}$ \\	
	& \ion{Fe}{II}	&	$-49.2$				& $<11.60$			& 10	\\ \hline

3	& \ion{H}{I}	&	$0.0\pm0.3$			& $16.37^{+0.07}_{-0.08}$& $12.7^{+0.9}_{-1.0}$ \\
	& \ion{C}{II}	&	$0.0$				& $13.64^{+0.15}_{-0.19}$& $12.4^{+6.5}_{-4.5}$ \\		
	& \ion{C}{III}	&	$0.0$				& $13.20^{+0.50}_{-0.44}$& $10.0^{+10.3}_{-3.2}$ \\
	& \ion{N}{II}	&	$0.0$				& $<13.07$			& 10	\\
	& \ion{N}{III}	&	$0.0$				& $<13.40$			& 10	\\
	& \ion{O}{I}	&	$0.0$				& $<13.78$			& 10	\\
	& \ion{Mg}{I}	&	$0.0$				& $<10.78$			& 10	\\
	& \ion{Mg}{II}	&	$0.0\pm1.2$			& $11.89^{+0.05}_{-0.07}$& $8.6^{+2.1}_{-1.6}$ \\	
	& \ion{Si}{II}	&	$0.0$				& $<12.44$			& 10 \\	
	& \ion{Si}{III}	&	$0.0$				& $12.60^{+0.13}_{-0.22}$& $21.3^{+12.5}_{-7.0}$ \\	
	& \ion{Fe}{II}	&	$0.0$				& $<11.60$			& 10	\\ \hline

4	& \ion{H}{I}	&	$+80.7^{+5.1}_{-8.3}$	& $14.59^{+0.06}_{-0.05}$& $55.2^{+7.7}_{-4.9}$ \\
	& \ion{C}{II}	&	$+80.7$				& $<13.20$			& 10 \\		
	& \ion{C}{III}	&	$+80.7$				& $12.84^{+0.19}_{-0.42}$& $48.5^{+18.5}_{-26.1}$ \\
	& \ion{N}{II}	&	$+80.7$				& $<13.08$			& 10	\\
	& \ion{N}{III}	&	$+80.7$				& $<13.41$			& 10	\\
	& \ion{O}{I}	&	$+80.7$				& $<13.82$			& 10	\\
	& \ion{Mg}{I}	&	$+80.7$				& $<10.80$			& 10	\\
	& \ion{Mg}{II}	&	$+80.7$				& $<11.31$			& 10 \\	
	& \ion{Si}{II}	&	$+80.7$				& $<12.39	$			& 10 \\	
	& \ion{Si}{III}	&	$+80.7$				& $<12.20$			& 10 \\	
	& \ion{Fe}{II}	&	$+80.7$				& $<11.60$			& 10	\\ \hline
	
high-1	& \ion{O}{VI}	&	$-25.2\pm18.2$		& $13.96^{+0.11}_{-0.12}$& $82.9\pm27.1$ \\
		& \ion{N}{V}	&	$-25.2$			& $<13.62$			&  \\

\hline
\end{tabular}}
\end{center}
\end{table}

\begin{table}
\begin{center}
\caption{Ionization modeling results for the absorber along SDSS\,J1550$+$4001, at $d=107$ kpc from the LRG}
\hspace{-2.5em}
\label{tab:Imaging}
\resizebox{3.5in}{!}{
\begin{tabular}{@{\extracolsep{3pt}}ccrrrr@{}}\hline
Component	&$N_\mathrm{metal}$& \multicolumn{2}{c}{$\mathrm{[M/H]}$} 	& \multicolumn{2}{c}{$\mathrm{log\,}n_\mathrm{H}/\cmjjj$}		\\
\cline{3-4} \cline {5-6}
	& &\multicolumn{1}{c}{HM05}&	\multicolumn{1}{c}{HM12}	&\multicolumn{1}{c}{HM05} 	& 	\multicolumn{1}{c}{HM12}			\\	\hline \hline

SC	&6& $-0.76^{+0.03}_{-0.04}$	& $-0.40\pm0.04$		& $-2.34^{+0.06}_{-0.08}$	& $-2.64^{+0.06}_{-0.08}$ \\ \hline
1	&6& $-0.03^{+0.10}_{-0.08}$	& $+0.24^{+0.11}_{-0.10}$& $-2.44^{+0.06}_{-0.10}$	& $-2.84\pm0.08$ \\
2	&3& $-0.71^{+0.06}_{-0.07}$	& $-0.35^{+0.06}_{-0.07}$	& $-2.62^{+0.10}_{-0.12}$	& $-2.86^{+0.08}_{-0.10}$	\\	
3	&4& $-1.23^{+0.09}_{-0.10}$	& $-0.91\pm0.08$		& $-2.12^{+0.16}_{-0.18}$	& $-2.40\pm0.14$	\\	
4	&1& $-1.48^{+0.78}_{-0.50}$	& $-0.71^{+0.59}_{-0.55}$	& $<-2.26$			& $<-2.54$		\\

\hline
\end{tabular}}
\end{center}
\end{table}
\end{subtables}

\subsection{SDSS\, J0246$-$0059 at $d=109$ kpc}

\begin{subfigures}
\begin{figure*}
\includegraphics[scale=1.08]{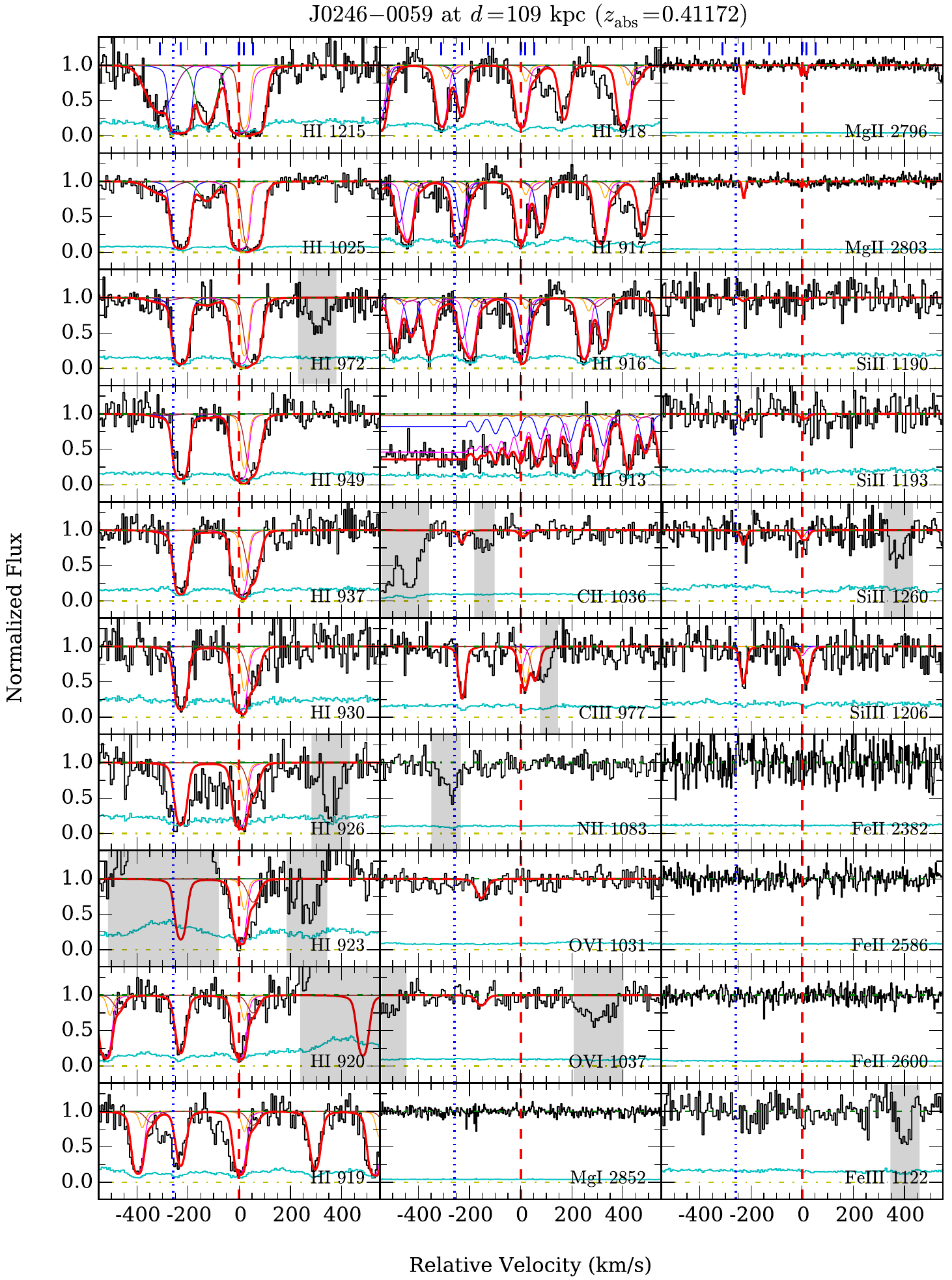}
\vspace{-0.75em}
\caption{Similar to Figure A1a, but for SDSS\,J0246$-$0059 at $d=109$ kpc from the LRG.}
\label{figure:ions}
\end{figure*}

\begin{figure}
\hspace{-0.8em}
\includegraphics[scale=0.62]{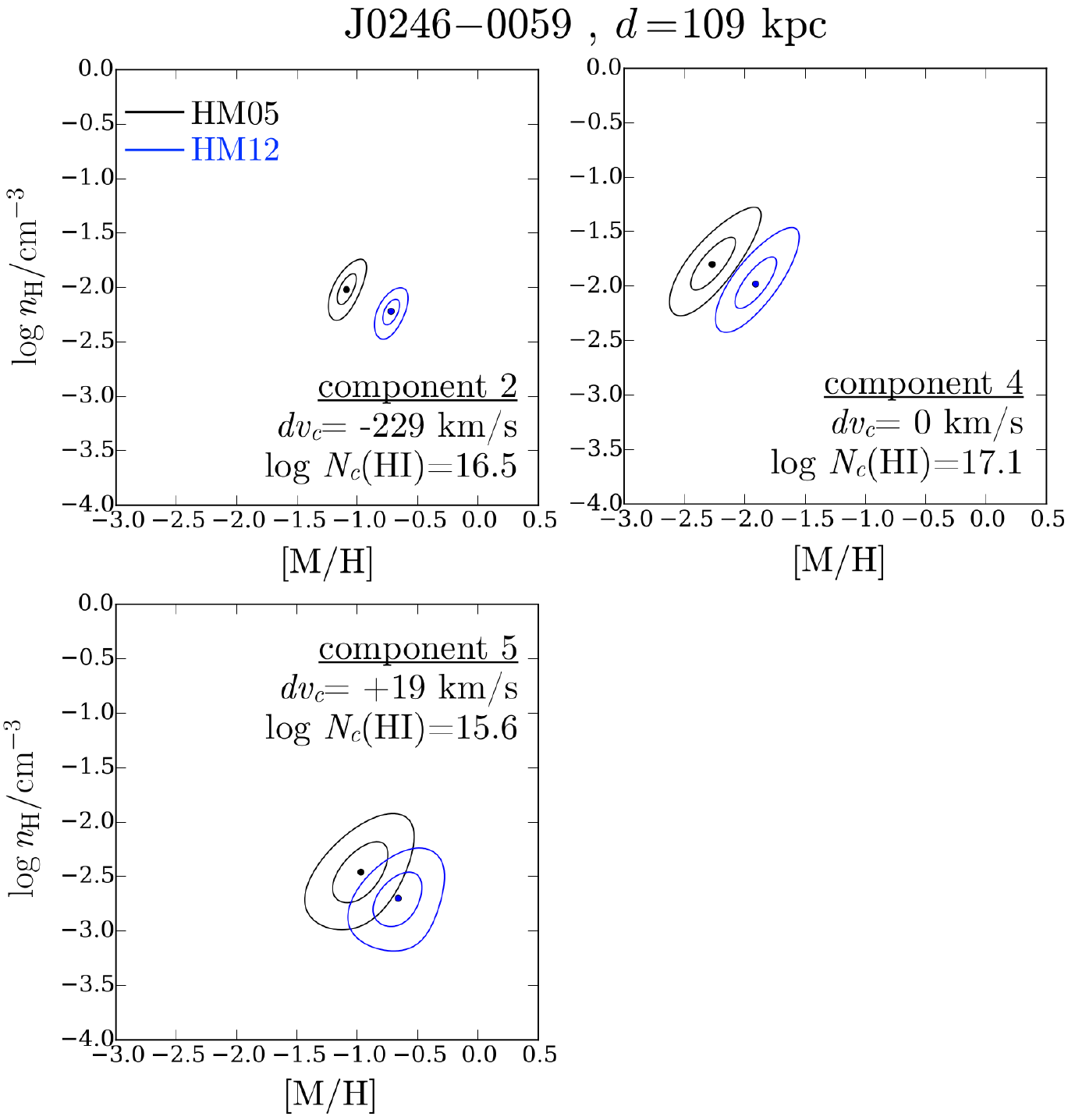}
\vspace{-1.5em}
\caption{Probability distribution contours of gas metallicity and density for individual absorption components identified along SDSS\,J0246$-$0059, at $d=109$ kpc from the LRG. Contour levels are the same as in Figure A1b. Components 1 and 3 show weaker \ion{H}{I} absorption (log\, $N_c\mathrm{(\ion{H}{I})}\lesssim14$) and are not shown here, because the absence of metal detections for these components result in a lack of strong constraints on the metallicity and density of the gas (see Table A10b). Also not plotted here is component 6 at $dv_c=+54$ \kms, which has log\,$N_c\mathrm{(\ion{H}{I})/\cmjj}=15.6$ yet shows little metal absorption. The  \ion{C}{III} absorption seen in component 6, along with upper limits on the column density of other ions, still allows us to constrain the gas density to $n_\mathrm{H}\lesssim10^{-3}\,$\cmjjj\ and the metallicity of $\mathrm{[M/H]}\lesssim-1$ (Table A10b).}
\end{figure}
\end{subfigures}

This LRG is at $z_\mathrm{LRG}=0.4105$. As shown in Figure A10a, a LLS with a total $N\mathrm{(\ion{H}{I})}$ of log\,$N\mathrm{(\ion{H}{I})/\cmjj}=17.21\pm0.01$ is present near the LRG redshift. The HI absorption is accompanied by ionic metal detections of \ion{C}{II}, \ion{C}{III}, \ion{Mg}{II}, \ion{Si}{II}, \ion{Si}{III}, and \ion{O}{VI}.

Our Voigt profile analysis indicates that the HI absorption profile can be decomposed into six components, with the bulk of the HI gas concentrated on two components with log\,$N_c\mathrm{(\ion{H}{I})/\cmjj}>16$ (Figure A10a and Table A10a). The strongest \ion{H}{I} absorption is in component 4 at $z_\mathrm{abs}=0.41172$ or $+259$ \kms\ from the LRG, which contributes to a majority (76\%) of the total $N\mathrm{(\ion{H}{I})}$. Most of the remaining \ion{H}{I} absorption is found in component 2 which is found at $dv_c=-229$ \kms\ from the strongest component. An interesting feature of this absorption system, which can be seen in Figure A10a, is that the strongest low- and intermediate-ionization metal absorption (e.g., \ion{Mg}{II} and \ion{Si}{III}) occurs in component 2, instead of in component 4 where the $N\mathrm{(\ion{H}{I})}$ is almost four times higher. The discrepancy between where the bulk of the \ion{H}{I} gas is located, and where most of the heavy metals are, suggests a significant variation in metallicities across different components. We also note that no corresponding absorption of metal ions is detected in the two components with the weakest \ion{H}{I} absorption, components 1 and 3.

The observed velocity spread of the absorbing gas is $\Delta v\approx360$ \kms\ from the bluest to reddest component. Based on the Doppler $b$ parameters of individual \ion{H}{I} components ($b_c\mathrm{(\ion{H}{I})}<60$\,\kms), we place a temperature upper limit of $T\lesssim2\times10^5$ K for all components. For two components with log\,$N\mathrm{(\ion{H}{I})/\cmjj}>16$, components 2 and 4, the best-fit line profiles are narrow with $b_c\mathrm{(\ion{H}{I})}=16-18$\,\kms, indicating that the absorption arises in relatively cool gas with $T\lesssim2\times10^4$ K. Moreover, these two components also have associated \ion{Mg}{II} detections in the high-resolution MIKE data. Comparing the \ion{Mg}{II} and \ion{H}{I} linewidths for these two components shows that there is little non-thermal broadening in the gas, with $b_\mathrm{nt}\lesssim3$\,\kms.

The well-matched component structure between \ion{H}{I} and low- to intermediate-ionization metals in this system justifies a single-phase photoionization model. 
As shown in Figure A10b and Table A10b, our ionization analysis indicates only a modest variation in gas densities across different components, where values between log\,$n_\mathrm{H}/ \cmjjj\approx-2.5$ and  log\,$n_\mathrm{H}/ \cmjjj\approx-2.0$ are found under the HM05 UVB. A similar spread in of $n_\mathrm{H}$ are found under the HM12 UVB albeit at lower ($\sim0.2$ dex) densities. 

In contrast to the relative uniformity seen in $n_\mathrm{H}$, a large variation in metallicities is seen in the gas. As previously mentioned, while the strongest \ion{H}{I} absorption is found in component 4 (log\,$N_c\mathrm{(\ion{H}{I})/\cmjj}=17.1$), that component exhibits weak low- and intermediate-ionization metal absorption, suggesting low gas-phase metallicity. Indeed, our analysis infers a very low metallicity for this component, from $\mathrm{[M/H]}=-2.3\pm0.2$ under HM05 to $\mathrm{[M/H]}=-1.9\pm0.2$ under HM12. For two other components with detection of multiple ionic metal species, components 2 and 5,  significantly higher chemical abundances are required to match the data, from $\mathrm{[M/H]}\approx-1$ under HM05 to $\mathrm{[M/H]}\approx-0.7$ under HM12. The large (more than a factor of 10) variations of metallicities seen over $\Delta v=360$ \kms\ in line-of-sight velocity indicates multiple physical origins for the halo gas of this LRG. 

Finally, we note that a modest absorption of highly ionized \ion{O}{VI} gas is seen in this system at $dv_c=-152$ \kms. The \ion{O}{VI} absorption consists of a single component with log\,$N\mathrm{(\ion{O}{VI})/\cmjj}=13.66^{+0.08}_{-0.07}$ and a relatively narrow $b$ parameter of 22 \kms. Although the \ion{O}{VI} absorption profile is narrow, no low- or intermediate-ionization metal or \ion{H}{I} component is found to match the \ion{O}{VI} absorption in velocity space (Figure A10a), with the \ion{O}{VI} doublet situated at  $\Delta v=-25$ \kms\ from the nearest low-ionization component. 

\begin{subtables}
\begin{table}
\begin{center}
\caption{Absorption properties along QSO sightline SDSS\,J0246$-$0059 at $d=108$ kpc from the LRG}
\hspace{-2.5em}
\vspace{-0.5em}
\label{tab:Imaging}
\resizebox{3.15in}{!}{
\begin{tabular}{clrrr}\hline
Component	&	Species		&\multicolumn{1}{c}{$dv_c$} 		& \multicolumn{1}{c}{log\,$N_c$}	&\multicolumn{1}{c}{$b_c$}		\\	
 			&				&\multicolumn{1}{c}{(km\,s$^{-1}$)}	&   		   					& \multicolumn{1}{c}{(km\,s$^{-1}$)}  \\ \hline \hline

all	& \ion{H}{I}	&	$...$					& $17.21\pm0.01$		& $...$ \\
	& \ion{C}{II}	&	$...$					& $13.31^{+0.12}_{-0.28}$& $...$	\\	
	& \ion{C}{III}	&	$...$					& $13.89^{+0.17}_{-0.16}$& $...$\\
	& \ion{N}{V}	&	$...$					& $<13.27$			& $...$	\\
	& \ion{O}{VI}	&	$...$					&$13.66^{+0.08}_{-0.07}$	& $...$	\\
	& \ion{Mg}{I}	&	$...$					& $<11.32$			& $...$	\\
	& \ion{Mg}{II}	&	$...$					& $12.43^{+0.05}_{-0.03}$& $...$\\
	& \ion{Si}{II}	&	$...$					& $12.31^{+0.16}_{-0.35}$& $...$	\\	
	& \ion{Si}{III}	&	$...$					& $13.04^{+0.11}_{-0.10}$	& $...$	\\	
	& \ion{Fe}{II}	&	$...$					& $<12.11$			& $...$	\\
	& \ion{Fe}{III}	&	$...$					& $<13.83$			& $...$	\\ \hline	
	
1	& \ion{H}{I}	&	$-309.6^{+5.0}_{-4.5}$	& $13.96\pm0.06$		& $59.0^{+2.6}_{-2.5}$ \\
	& \ion{C}{II}	&	$-309.6$				& $<12.94$			& 10	\\	
	& \ion{C}{III}	&	$-309.6$				& $<12.48$			& 10	\\
	& \ion{Mg}{I}	&	$-309.6$				& $<10.80$			& 10	\\
	& \ion{Mg}{II}	&	$-309.6$				& $<11.35$			& 10	\\
	& \ion{Si}{II}	&	$-309.6$				& $<12.24$			& 10	\\	
	& \ion{Si}{III}	&	$-309.6$				& $<12.02$			& 10	\\	
	& \ion{Fe}{II}	&	$-309.6$				& $<12.02$			& 10	\\
	& \ion{Fe}{III}	&	$-309.6$				& $<13.34$			& 10	\\ \hline	

2	& \ion{H}{I}	&	$-228.8\pm0.9$			& $16.49^{+0.06}_{-0.05}$& $17.4\pm0.4$ \\
	& \ion{C}{II}	&	$-228.8$				& $13.31^{+0.12}_{-0.28}$& $10.1^{+6.2}_{-3.5}$ \\		
	& \ion{C}{III}	&	$-228.8$				& $13.63^{+0.23}_{-0.14}$& $9.9\pm4.6$ \\
	& \ion{N}{II}	&	$-228.8$				& $<13.05$			& 10	\\
	& \ion{Mg}{I}	&	$-228.8$				& $<10.82$			& 10	\\
	& \ion{Mg}{II}	&	$-227.2\pm0.5$			& $12.25\pm0.04$		& $4.9^{+0.9}_{-1.2}$ \\	
	& \ion{Si}{II}	&	$-228.8$				& $12.31^{+0.16}_{-0.35}$& $12.2^{+4.3}_{-5.1}$ \\	
	& \ion{Si}{III}	&	$-228.8$				& $12.72^{+0.15}_{-0.19}$& $8.8^{+4.2}_{-3.4}$ \\	
	& \ion{Fe}{II}	&	$-228.8$				& $<12.01$			& 10	\\
	& \ion{Fe}{III}	&	$-228.8$				& $<13.28$			& 10	\\ \hline

3	& \ion{H}{I}	&	$-127.5\pm0.6$			& $14.04\pm0.05$		& $37.2^{+1.8}_{-1.5}$ \\
	& \ion{C}{II}	&	$-127.5$				& $<12.90$			& 10 \\
	& \ion{C}{III}	&	$-127.5$				& $<12.46$			& 10 \\
	& \ion{N}{II}	&	$-127.5$				& $<13.04$			& 10 \\
	& \ion{Mg}{I}	&	$-127.5$				& $<10.80$			& 10 \\
	& \ion{Mg}{II}	&	$-127.5$				& $<11.34$			& 10 \\
	& \ion{Si}{II}	&	$-127.5$				& $<12.15$			& 10 \\
	& \ion{Si}{III}	&	$-127.5$				& $<12.03$			& 10 \\	
	& \ion{Fe}{II}	&	$-127.5$				& $<12.04$			& 10 \\
	& \ion{Fe}{III}	&	$-127.5$				& $<13.32$			& 10 \\ \hline

4	& \ion{H}{I}	&	$0.0^{+0.8}_{-0.4}$		& $17.09^{+0.01}_{-0.02}$& $15.9^{+0.4}_{-0.3}$ \\
	& \ion{C}{II}	&	$0.0$				& $<12.76$			& 10 \\	
	& \ion{C}{III}	&	$0.0$				& $12.84^{+0.18}_{-0.33}$& $13.0^{+6.0}_{-5.1}$ \\
	& \ion{N}{II}	&	$0.0$				& $<13.05	$			& 10 \\	
	& \ion{Mg}{I}	&	$0.0$				& $<10.80$			& 10 \\	
	& \ion{Mg}{II}	&	$-1.3	\pm1.4$			& $11.67^{+0.10}_{-0.15}$	& $3.0^{+3.5}_{-1.9}$ \\
	& \ion{Si}{II}	&	$0.0	$				& $<12.03	$			& 10 \\		
	& \ion{Si}{III}	&	$0.0	$				& $<12.19	$			& 10 \\	
	& \ion{Fe}{II}	&	$0.0	$				& $<12.02	$			& 10 \\	
	& \ion{Fe}{III}	&	$0.0	$				& $<13.33	$			& 10 \\ \hline

5	& \ion{H}{I}	&	$+19.0^{+2.2}_{-0.7}$	& $15.61^{+0.24}_{-0.32}$& $10.0\pm0.4$ \\
	& \ion{C}{II}	&	$+19.0$				& $<12.84	$			& 10	\\	
	& \ion{C}{III}	&	$+19.0$				& $13.12^{+0.32}_{-0.20}$& $10.3^{+3.7}_{-3.8}$ \\
	& \ion{N}{II}	&	$+19.0$				& $<13.00$			& 10	\\
	& \ion{Mg}{I}	&	$+19.0$				& $<10.76$			& 10	\\
	& \ion{Mg}{II}	&	$+17.3\pm1.4$			& $11.67\pm0.11$		& $3.3^{+6.7}_{-1.3}$ \\
	& \ion{Si}{II}	&	$+19.0$				& $<12.00	$			& 10	\\	
	& \ion{Si}{III}	&	$+19.0$				& $12.75\pm0.15$		& $12.4\pm4.7$ \\	
	& \ion{Fe}{II}	&	$+19.0$				& $<12.02	$			& 10	\\
	& \ion{Fe}{III}	&	$+19.0$			 	& $<13.26$			& 10	\\ \hline

6	& \ion{H}{I}	&	$+54.0^{+2.6}_{-2.4}$		& $15.58\pm0.08$	 	& $24.2\pm0.9$ \\	
	& \ion{C}{II}	&	$+54.0$				& $<12.94$			& 10 \\	
	& \ion{C}{III}	&	$+54.0$				& $13.18^{+0.12}_{-0.13}$& $22.8^{+5.8}_{-9.3}$ \\
	& \ion{N}{II}	&	$+54.0$				& $<13.03	$			& 10 \\
	& \ion{Mg}{I}	&	$+54.0$				& $<10.80	$			& 10 \\
	& \ion{Mg}{II}	&	$+54.0$				& $<11.33	$			& 10 \\
	& \ion{Si}{II}	&	$+54.0$				& $<12.08	$			& 10 \\	
	& \ion{Si}{III}	&	$+54.0$				& $<12.02	$			& 10 \\	
	& \ion{Fe}{II}	&	$+54.0$				& $<12.00	$			& 10 \\
	& \ion{Fe}{III}	&	$+54.0$				& $<13.23	$			& 10 \\ \hline

high-1	& \ion{O}{VI}	&	$-152.3\pm4.1$			& $13.66^{+0.08}_{-0.07}$& $21.6^{+12.9}_{-4.3}$ \\
		& \ion{N}{V}	&	$-152.3$				& $<13.27$			&  \\

\hline
\end{tabular}}
\end{center}
\end{table}

\begin{table}
\begin{center}
\caption{Ionization modeling results for the absorber along SDSS\,J0246$-$0059, at $d=108$ kpc from the LRG}
\hspace{-2.5em}
\label{tab:Imaging}
\resizebox{3.5in}{!}{
\begin{tabular}{@{\extracolsep{3pt}}ccrrrr@{}}\hline
Component	&$N_\mathrm{metal}$& \multicolumn{2}{c}{$\mathrm{[M/H]}$} 	& \multicolumn{2}{c}{$\mathrm{log\,}n_\mathrm{H}/\cmjjj$}		\\
\cline{3-4} \cline {5-6}
	& &\multicolumn{1}{c}{HM05}&	\multicolumn{1}{c}{HM12}	&\multicolumn{1}{c}{HM05} 	& 	\multicolumn{1}{c}{HM12}			\\	\hline \hline

SC	&5& $-1.66^{+0.05}_{-0.06}$	& $-1.26^{+0.04}_{-0.06}$	&  $-2.10^{+0.06}_{-0.10}$& $-2.28^{+0.04}_{-0.10}$ \\ \hline
1	&0& $<0.69$				& $<0.71$				& $>-4.56$			& $>-4.66$ \\
2	&5& $-1.09\pm0.07$			& $-0.72^{+0.06}_{-0.07}$	& $-2.02^{+0.12}_{-0.14}$	& $-2.22^{+0.08}_{-0.14}$	\\	
3	&0& $<0.68$				& $<0.70$				& $>-4.54$			& $>-4.66$	\\
4	&2& $-2.27^{+0.19}_{-0.18}$	& $-1.91\pm0.17$		&  $-1.80^{+0.26}_{-0.22}$& $-1.98^{+0.26}_{-0.20}$	\\
5	&3& $-0.96^{+0.20}_{-0.23}$	& $-0.64^{+0.17}_{-0.21}$	& $-2.48\pm0.26$		& $-2.72^{+0.22}_{-0.24}$	\\
6	&1& $-2.47^{+0.78}_{-0.12}$	& $-1.71^{+0.63}_{-0.12}$	& $<-2.88$			& $<-3.00$\\

\hline
\end{tabular}}
\end{center}
\end{table}
\end{subtables}

\subsection{SDSS\, J1357$+$0435 at $d=126$ kpc}

\begin{subfigures}
\begin{figure*}
\includegraphics[scale=1.08]{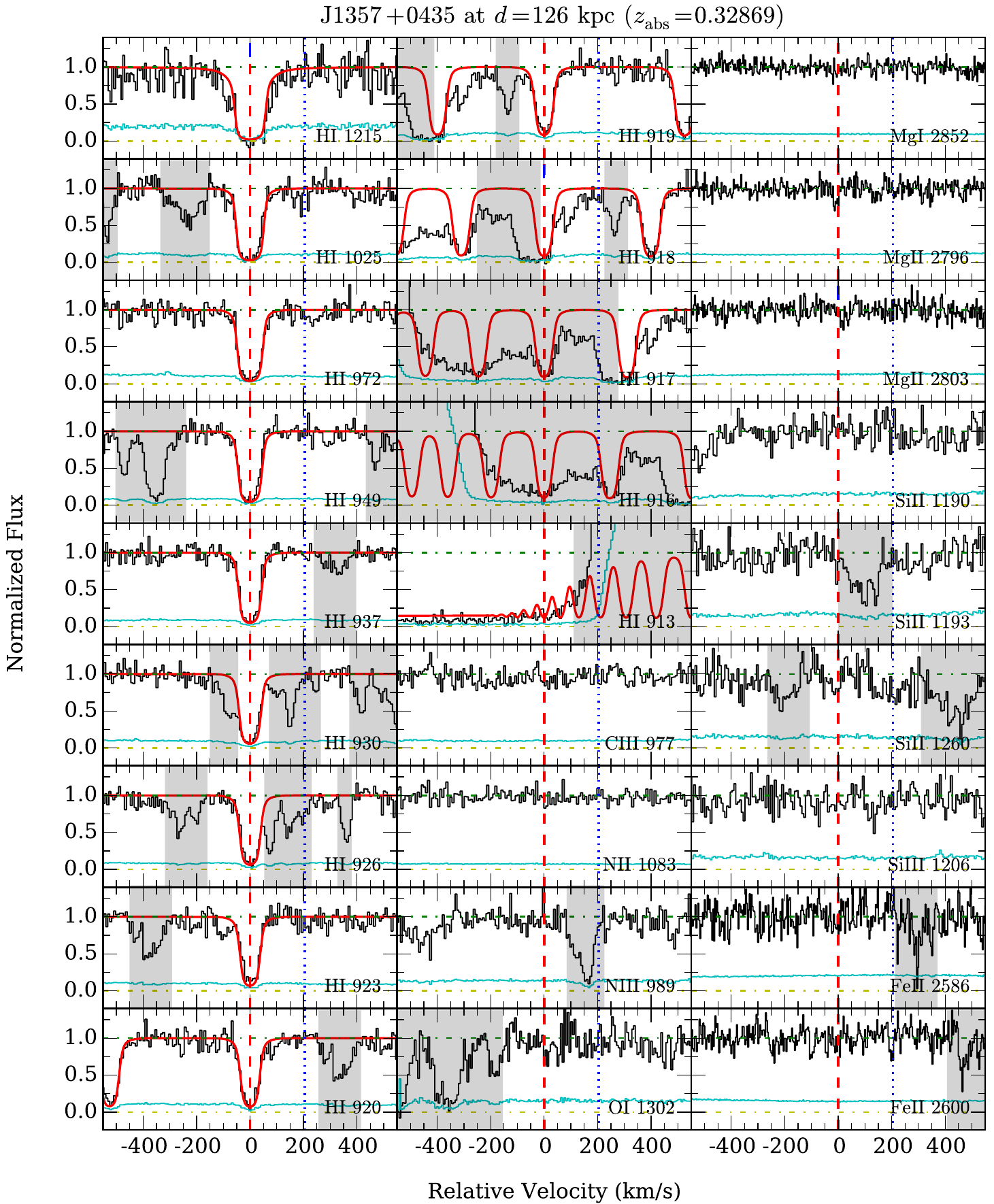}
\vspace{-0.75em}
\caption{Similar to Figure A1a, but for SDSS\,J1357$+$0435 at $d=126$ kpc from the LRG.}
\label{figure:ions}
\end{figure*}
\end{subfigures}

The LRG, which is at $z_\mathrm{LRG}=0.3296$, has a LLS with a total column density of log\,$N\mathrm{(\ion{H}{I})/\cmjj}=17.48\pm0.01$. Despite the strength of the \ion{H}{I} Lyman series absorption in this system, no trace of corresponding metal absorption is found down to very sensitive column density limits (see Figure A11a and Table A11a). The HI absorption profile consists of a single component centered at $z_\mathrm{abs}=0.32869$, or $-205$ \kms\ from the LRG systemic redshift. Our Voigt profile analysis finds a narrow Doppler parameter of  $b_c\mathrm{(\ion{H}{I})}=18$\,\kms, which constrains the gas temperature to $T\lesssim2\times10^4$ K.

The most intriguing aspect of this system is the absence of ionic metals in an optically thick absorber occurring at $d=126$ kpc from an massive quiescent galaxy. To put an upper limit on the metallicity given the observed $N\mathrm{(\ion{H}{I})}$ and upper limits on ionic column densities (Table A11a), we make a couple of assumptions about the physical conditions of the gas. First, we assume that the gas is predominantly ionized, $f_\mathrm{H^+}>0.9$, which is motivated by previous findings on the physical conditions of $z<1$ LLSs (e.g., Lehner \etal\ 2013). Furthermore, we impose an upper limit on the cloud length of $\sim 1$ kpc, based on observations of Galactic high-velocity clouds (e.g., Putman \etal\ 2012). Together, these assumptions limit the allowed gas density to $-2.5\lesssim$ log\,$n_\mathrm{H}/ \cmjjj\lesssim-1.1$ under the HM05 UVB, and $-2.3\lesssim$ log\,$n_\mathrm{H}/ \cmjjj\lesssim-1.5$ under the HM12 UVB. Using this prior on gas density, we estimate that the 95 percent upper limit on the metallicity of the gas is $\mathrm{[M/H]}<-2.3$ under HM05 and $\mathrm{[M/H]}<-2.2$ under HM12. The inferred low metallicity ($\lesssim 5 \times 10^{-3}$ solar) places this absorber among the lowest-metallicity LLSs known to be in the vicinity of $z<1$ luminous galaxies (c.f., Ribaudo \etal\ 2011; Prochaska \etal\ 2017).

\begin{subtables}
\begin{table}
\begin{center}
\caption{Absorption properties along QSO sightline SDSS\,J1357$+$0435 $d=126$ kpc from the LRG}
\hspace{-2.5em}
\vspace{-0.5em}
\label{tab:Imaging}
\resizebox{3.3in}{!}{
\begin{tabular}{clcrc}\hline
Component	&	Species		&\multicolumn{1}{c}{$dv_c$} 		& \multicolumn{1}{c}{log\,$N_c$}	&\multicolumn{1}{c}{$b_c$}		\\	
 			&				&\multicolumn{1}{c}{(km\,s$^{-1}$)}	&   		   					& \multicolumn{1}{c}{(km\,s$^{-1}$)}  \\ \hline \hline

1	& \ion{H}{I}	&	$0.0\pm0.4$	& $17.48\pm0.01$		& $18.3\pm0.2$ \\
        & \ion{C}{II}	&	$0.0$		& $<13.11$			& $10$\\
	& \ion{C}{III}	&	$0.0$		& $<12.19$			& $10$\\
	& \ion{N}{II}	&	$0.0$		& $<12.87$			& $10$ \\
	& \ion{N}{III}	&	$0.0$		& $<13.19$			& $10$ \\
	& \ion{O}{I}	&	$0.0$		& $<13.38$			& $10$ \\
	& \ion{O}{VI}	&	$0.0$		& $<13.19$			& $30$ \\
	& \ion{Mg}{I}	&	$0.0$		& $<11.18$			& $10$ \\
	& \ion{Mg}{II}	&	$0.0$		& $<11.71$			& $10$ \\
	& \ion{Si}{II}	&	$0.0$		& $<12.10$			& $10$ \\	
	& \ion{Si}{III}	&	$0.0$		& $<11.94$			& $10$ \\
	& \ion{Fe}{II}	&	$0.0$		& $<12.31$			& $10$ \\ 	
\hline
\end{tabular}}
\end{center}
\end{table}
\end{subtables}

\subsection{SDSS\, J0910$+$1014 at $d=140$ kpc}

\begin{subfigures}
\begin{figure*}
\includegraphics[scale=1.08]{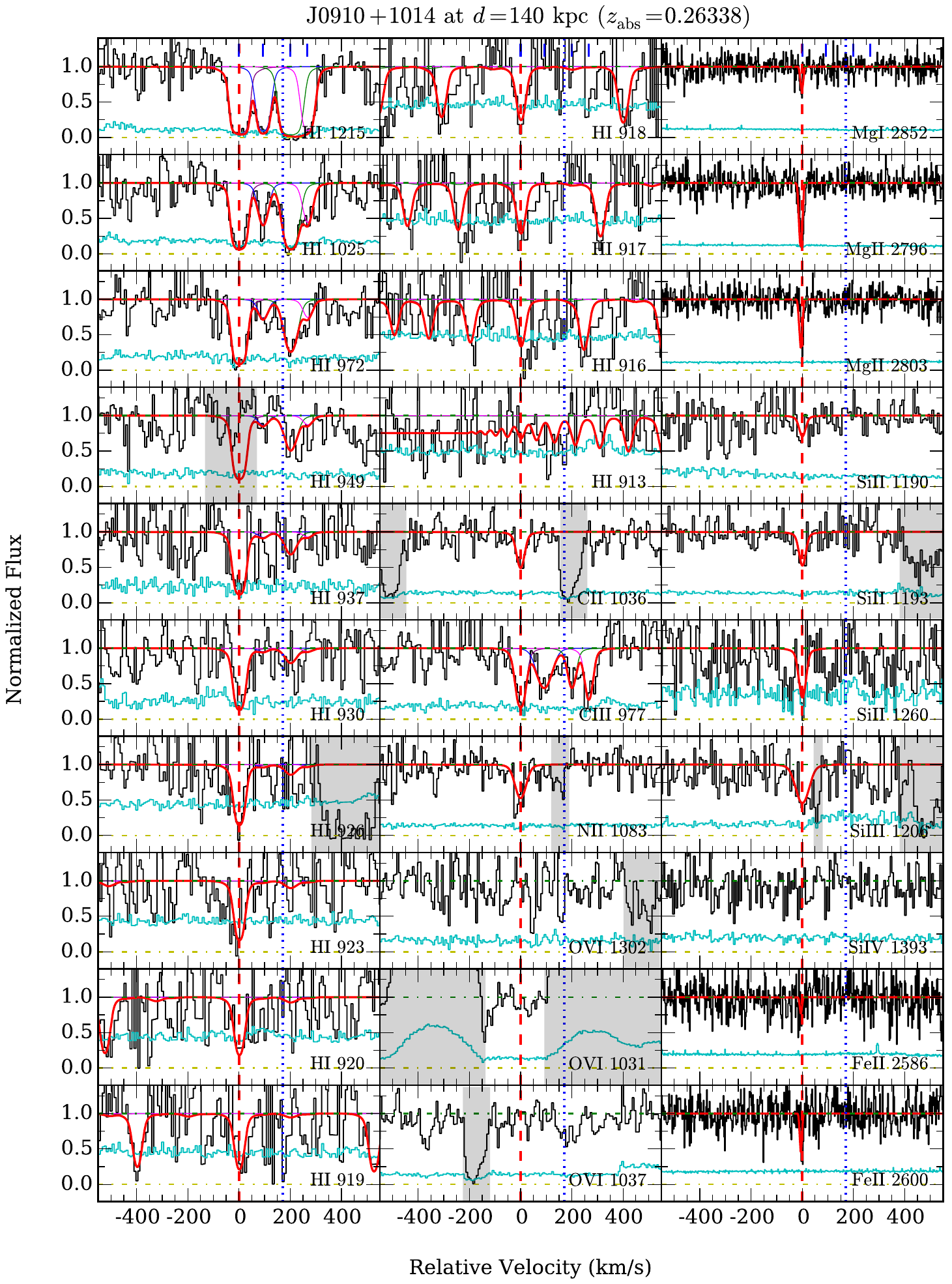}
\vspace{-0.75em}
\caption{Similar to Figure A1a, but for SDSS\,J0910$+$1014 at $d=140$ kpc from the LRG. The observed lack of \ion{H}{I} $\lambda949$ absorption is due to  contamination from the geocoronal  \ion{N}{I} $\lambda1199$ emission line.}
\label{figure:ions}
\end{figure*}

\begin{figure}
\centering
\hspace{-0.8em}
\includegraphics[scale=0.68]{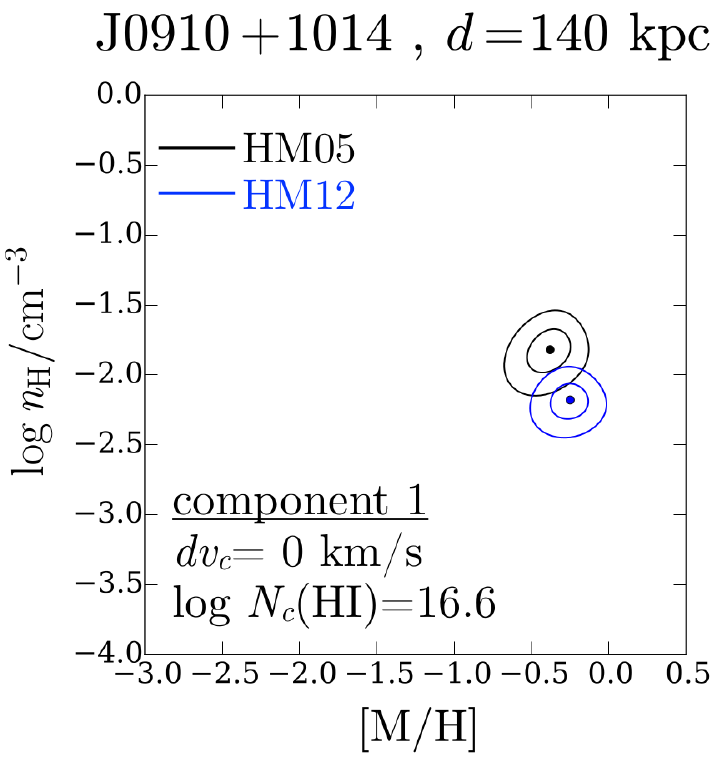}
\vspace{-1em}
\caption{Probability distribution contours of gas metallicity and density for the absorbing gas detected along SDSS\,J0910$+$1014, at $d=140$ kpc from the LRG. Contour levels are the same as in Figure A1b. Not shown here are the probability distributions for components 2 to 4, which have lower \ion{H}{I} column densities of log\, $N_c\mathrm{(\ion{H}{I})}\lesssim15$ and show only \ion{C}{III} absorption, resulting in weak constraints on the gas metallicity and density (see Table 1A2b).}
\label{figure:ions}
\end{figure}
\end{subfigures}

This LRG occurs at $z_\mathrm{LRG}=0.2641$, and it has a pLLS with a total $N\mathrm{(\ion{H}{I})}$ of log\,$N\mathrm{(\ion{H}{I})/\cmjj}=16.65^{+0.34}_{-0.22}$. In addition to \ion{H}{I}, we also detect corresponding ionic absorption of \ion{C}{II}, \ion{C}{III}, \ion{N}{II}, \ion{Mg}{I}, \ion{Mg}{II}, \ion{Si}{II}, \ion{Si}{III}, and \ion{Fe}{II} (Figure A12a).

The HI absorption profile consists of four individual components (Figure A12a and Table A12a). The strongest \ion{H}{I} absorption is in component 1 at $z_\mathrm{abs}=0.26338$ or $170$ \kms\ blueward of the LRG redshift. This component contributes to most ($98\%$) of the absorber's total $N\mathrm{(\ion{H}{I})}$. In addition, component 1 also exhibits strong absorption of low- and intermediate-ionization metal species (Figure A12a), suggesting the gas is chemically enriched. In contrast, \ion{C}{III} is the only metal species detected in the other three components, components 2 to 4. 

The observed velocity spread of the absorber is $\Delta v\approx270$ \kms\ from the bluest to reddest component. As shown in Figure A12a, our Voigt profile analysis demonstrates that both low- and intermediate-ionization metal absorption profiles are well-matched to the component structure of \ion{H}{I}. By comparing the measured linewidths for \ion{H}{I} and  \ion{Mg}{II}, we find that the bulk of the absorbing gas (component 1) is cool, $T\sim 10^4$ K, with a modest amount of non-thermal line broadening, $b_\mathrm{nt}\approx6$\,\kms. For components 2 to 4, the observed \ion{H}{I} linewidths of $b_c\mathrm{(\ion{H}{I})}\lesssim25$ \kms\ place an upper limit on the gas temperature of $T\lesssim4\times10^4$ K.

As shown in Figure A12b and Table A12b, our ionization analysis finds that the bulk of the absorbing gas (component 1) has a density of log\,$n_\mathrm{H}/ \cmjjj=-1.8\pm0.2$ under HM05 and log\,$n_\mathrm{H}/ \cmjjj=-2.2\pm0.1$ under HM12. To reproduce the observed ionic column densities, the required gas metallicity is sub-solar, between $\mathrm{[M/H]}=-0.4\pm0.1$ unwed HM05 UVB and $\mathrm{[M/H]}=-0.3\pm0.1$ under HM12 UVB. 

For the weaker components 2, 3, and 4, our analysis determines that the gas is low-density, with an upper limit of log\,$n_\mathrm{H}/ \cmjjj\lesssim-2.5$ under HM05 and log\,$n_\mathrm{H}/ \cmjjj\lesssim-2.7$ under HM12, constrained by the detection of \ion{C}{III} and non-detections of other metal species. Because \ion{C}{III} the only metal detection, the inferred metallicity of the gas is subject to large uncertainties, ranging from between $\mathrm{[M/H]}=-1.7^{+0.7}_{-0.3}$ (HM05) and $\mathrm{[M/H]}=-1.0^{+0.6}_{-0.3}$ (HM12) for component 3, to between $\mathrm{[M/H]}=-0.5\pm0.5$ (HM05) and $\mathrm{[M/H]}=0.2\pm0.5$ (HM12) for components 2 and 4. 

This absorber was also studied in the COS-Halos survey. Using the  Haardt \& Madau (2001) UVB (which is similar of the HM05 UVB at energies $\lesssim2$ Ryd), Werk \etal\ (2014) inferred a mean metallicity of $\mathrm{[M/H]}=-0.7\pm0.5$ and density of between log\,$n_\mathrm{H}/ \cmjjj=-4.5$ and  log\,$n_\mathrm{H}/ \cmjjj=-3.0$ under the Haardt \& Madau (2001) UVB. In an updated analysis using the HM12 UVB, Prochaska \etal\ (2017) found a mean metallicity of $\mathrm{[M/H]}=-0.17^{+0.13}_{-0.08}$ and density of  log\,$n_\mathrm{H}/ \cmjjj=-2.6\pm0.2$. The Prochaska \etal\ (2017) values are consistent with what we find in our analysis using using the single-clump model, after accounting for the difference in the adopted redshift of the HM12 UVB spectrum used in both studies.

\begin{subtables}
\begin{table}
\begin{center}
\caption{Absorption properties along QSO sightline SDSS\,J0910$+$1014 at $d=140$ kpc from the LRG}
\hspace{-2.5em}
\vspace{-0.5em}
\label{tab:Imaging}
\resizebox{3.5in}{!}{
\begin{tabular}{clrrr}\hline
Component	&	Species		&\multicolumn{1}{c}{$dv_c$} 		& \multicolumn{1}{c}{log\,$N_c$}	&\multicolumn{1}{c}{$b_c$}		\\	
 			&				&\multicolumn{1}{c}{(km\,s$^{-1}$)}	&   		   					& \multicolumn{1}{c}{(km\,s$^{-1}$)}  \\ \hline \hline

all	& \ion{H}{I}	&	$...$					& $16.65^{+0.34}_{-0.22}$& $...$ \\
	& \ion{C}{II}	&	$...$					& $13.97\pm0.12$		& $...$	\\	
	& \ion{C}{III}	&	$...$					& $>14.07$			& $...$\\
	& \ion{N}{II}	&	$...$					& $14.08\pm0.12$		& $...$ \\
	& \ion{N}{V}	&	$...$					& $<13.33$			& $...$ \\
	& \ion{O}{I}	&	$...$					& $<14.20$			& $...$ \\
	& \ion{O}{VI}	&	$...$					& $<13.53	$			& $...$	\\
	& \ion{Mg}{I}	&	$...$					& $11.59\pm0.10$		& $...$	\\
	& \ion{Mg}{II}	&	$...$					& $12.84\pm0.04$		& $...$\\
	& \ion{Si}{II}	&	$...$					& $13.22^{+0.11}_{-0.08}$	& $...$	\\	
	& \ion{Si}{III}	&	$...$					& $12.98\pm0.09$		& $...$	\\	
	& \ion{Si}{IV}	&	$...$					& $<12.98$			& $...$	\\	
	& \ion{Fe}{II}	&	$...$					& $12.75\pm0.13$		& $...$	\\ \hline	
	
1	& \ion{H}{I}	&	$0.0^{+0.5}_{-0.4}$		& $16.64^{+0.35}_{-0.23}$& $14.8^{+0.9}_{-1.4}$ \\
	& \ion{C}{II}	&	$0.0$				& $13.97\pm0.12$		& $15.6^{+6.6}_{-6.8}$	\\	
	& \ion{C}{III}	&	$0.0$				& $>13.52$			& $<25.7$ 	\\
	& \ion{N}{II}	&	$0.0$				& $14.08\pm0.12$		& $23.8\pm9.7$	\\
	& \ion{O}{I}	&	$0.0$				& $<13.71$			& 10	\\
	& \ion{Mg}{I}	&	$-1.9\pm1.1$			& $11.59\pm0.10$		& $6.0\pm2.2$	\\
	& \ion{Mg}{II}	&	$0.0\pm0.4$			& $12.84\pm0.04$		& $6.5\pm0.7$ \\
	& \ion{Si}{II}	&	$0.0$				& $13.22^{+0.11}_{-0.08}$	& $12.7^{+8.5}_{-4.1}$	\\	
	& \ion{Si}{III}	&	$0.0$				& $12.98\pm0.09$		& $28.8^{+12.0}_{-6.3}$	\\	
	& \ion{Si}{IV}	&	$0.0$				& $<12.50$			& 10	\\	
	& \ion{Fe}{II}	&	$0.0$				& $12.75\pm0.13$		& $4.2^{+3.0}_{-1.6}$	\\ \hline

2	& \ion{H}{I}	&	$+92.8^{+3.1}_{-2.8}$	& $14.34\pm0.15$		& $20.8\pm2.5$ \\
	& \ion{C}{II}	&	$+92.8$				& $<13.36$			& 10 \\		
	& \ion{C}{III}	&	$+92.8$				& $13.47^{+0.18}_{-0.16}$& $32.4\pm15.5$ \\
	& \ion{N}{II}	&	$+92.8$				& $<13.38$			& 10	\\
	& \ion{O}{I}	&	$+92.8$				& $<13.78$			& 10	\\
	& \ion{Mg}{I}	&	$+92.8$				& $<11.17$			& 10	\\
	& \ion{Mg}{II}	&	$+92.8$				& $<11.67$			& 10 \\	
	& \ion{Si}{II}	&	$+92.8$				& $<12.44	$			& 10 \\	
	& \ion{Si}{III}	&	$+92.8$				& $<12.19$			& 10 \\	
	& \ion{Si}{IV}	&	$+92.8$				& $<12.41$			& 10 \\	
	& \ion{Fe}{II}	&	$+92.8$				& $<12.36$			& 10	\\ \hline

3	& \ion{H}{I}	&	$+200.1^{+5.4}_{-4.6}$	& $15.08^{+0.14}_{-0.13}$& $26.3^{+3.8}_{-2.7}$ \\
	& \ion{C}{II}	&	$+200.1$				& $<13.09$			& 10 \\		
	& \ion{C}{III}	&	$+200.1$				& $13.28^{+0.22}_{-0.14}$& $16.2^{+16.9}_{-6.7}$ \\
	& \ion{N}{II}	&	$+200.1$				& $<13.40$			& 10	\\
	& \ion{O}{I}	&	$+200.1$				& $<13.66$			& 10	\\
	& \ion{Mg}{I}	&	$+200.1$				& $<11.14$			& 10	\\
	& \ion{Mg}{II}	&	$+200.1$				& $<11.65$			& 10 \\	
	& \ion{Si}{II}	&	$+200.1$				& $<12.44	$			& 10 \\	
	& \ion{Si}{III}	&	$+200.1$				& $<12.17$			& 10 \\	
	& \ion{Si}{IV}	&	$+200.1$				& $<12.47$			& 10 \\	
	& \ion{Fe}{II}	&	$+200.1$				& $<12.36$			& 10	\\ \hline

4	& \ion{H}{I}	&	$+266.5^{+7.5}_{-9.3}$	& $14.29^{+0.21}_{-0.27}$& $19.9^{+5.6}_{-3.6}$ \\
	& \ion{C}{II}	&	$+266.5$				& $<13.35$			& 10 \\		
	& \ion{C}{III}	&	$+266.5$				& $13.56^{+0.25}_{-0.19}$& $17.1^{+8.8}_{-3.9}$ \\
	& \ion{N}{II}	&	$+266.5$				& $<13.43$			& 10	\\
	& \ion{O}{I}	&	$+266.5$				& $<13.63$			& 10	\\
	& \ion{Mg}{I}	&	$+266.5$				& $<11.16$			& 10	\\
	& \ion{Mg}{II}	&	$+266.5$				& $<11.67$			& 10 \\	
	& \ion{Si}{II}	&	$+266.5$				& $<12.41	$			& 10 \\	
	& \ion{Si}{III}	&	$+266.5$				& $<12.14$			& 10 \\	
	& \ion{Si}{IV}	&	$+266.5$				& $<12.47$			& 10 \\	
	& \ion{Fe}{II}	&	$+266.5$				& $<12.39$			& 10	\\ \hline

\end{tabular}}
\end{center}
\end{table}

\begin{table}
\begin{center}
\caption{Ionization modeling results for the absorber along SDSS\,J0910$+$1014 at $d=140$ kpc from the LRG}
\hspace{-2.5em}
\label{tab:Imaging}
\resizebox{3.5in}{!}{
\begin{tabular}{@{\extracolsep{3pt}}ccrrrr@{}}\hline
Component	&$N_\mathrm{metal}$& \multicolumn{2}{c}{$\mathrm{[M/H]}$} 	& \multicolumn{2}{c}{$\mathrm{log\,}n_\mathrm{H}/\cmjjj$}		\\
\cline{3-4} \cline {5-6}
	& &\multicolumn{1}{c}{HM05}&	\multicolumn{1}{c}{HM12}	&\multicolumn{1}{c}{HM05} 	& 	\multicolumn{1}{c}{HM12}			\\	\hline \hline

SC	&8& $-0.39^{+0.12}_{-0.14}$	& $-0.25^{+0.11}_{-0.12}$	& $-1.88^{+0.10}_{-0.18}$	& $-2.26^{+0.10}_{-0.14}$ \\ \hline
1	&8& $-0.39^{+0.11}_{-0.15}$	& $-0.26^{+0.11}_{-0.13}$ & $-1.84^{+0.12}_{-0.16}$	& $-2.20^{+0.10}_{-0.12}$ \\
2	&1& $-0.68^{+0.49}_{-0.30}$	& $0.10^{+0.48}_{-0.28}$	& $<-2.92$			& $<-3.14$	\\	
3	&1& $-1.73^{+0.68}_{-0.27}$	& $-0.99^{+0.63}_{-0.26}$	& $<-2.50$			& $<-2.70$	\\	
4	&1& $-0.51^{+0.52}_{-0.45}$	& $0.23^{+0.42}_{-0.46}$	& $<-2.90$			& $<-3.10$	\\

\hline
\end{tabular}}
\end{center}
\end{table}
\end{subtables}

\subsection{SDSS\, J1413$+$0920 at $d=149$ kpc}

The LRG occurs at $z_\mathrm{LRG}=0.3584$. As shown in Figure A13a, no \ion{H}{I} absorption is detected within the adopted search window of $\pm500$ \kms\ from the LRG redshift. We are able to place a sensitive 2-$\sigma$ column density upper limit of log\,$N\mathrm{(\ion{H}{I})/\cmjj}<12.5$, calculated for an \ion{H}{I} line with $b\mathrm{(\ion{H}{I})}=15$ \kms\ that is centered at the LRG redshift (Table A13a).

\begin{subfigures}
\begin{figure*}
\includegraphics[scale=1.08]{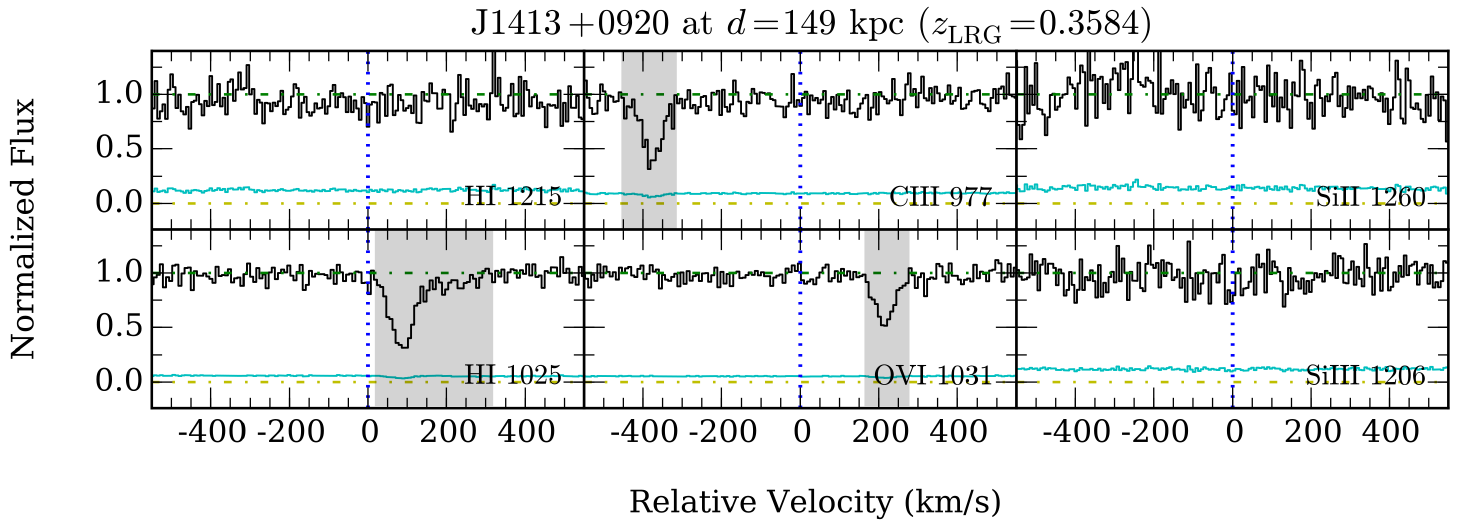}
\vspace{-0.75em}
\caption{Similar to Figure A1a, but for SDSS\,J1413$+$0920 at $d=149$ kpc from the LRG.}
\label{figure:ions}
\end{figure*}
\end{subfigures}

\begin{subtables}
\begin{table}
\begin{center}
\caption{Constraints on absorption properties along QSO sightline SDSS\,J1413$+$0920 at $d=149$ kpc from the LRG}
\hspace{-2.5em}
\vspace{-0.5em}
\label{tab:Imaging}
\resizebox{3.3in}{!}{
\begin{tabular}{clcrc}\hline
Component	&	Species		&\multicolumn{1}{c}{$dv_c$} 		& \multicolumn{1}{c}{log\,$N_c$}	&\multicolumn{1}{c}{$b_c$}		\\	
 			&				&\multicolumn{1}{c}{(km\,s$^{-1}$)}	&   		   					& \multicolumn{1}{c}{(km\,s$^{-1}$)}  \\ \hline \hline

...	& \ion{H}{I}	&	$0.0$					& $<12.49$			& $15$ \\
	& \ion{C}{II}	&	$0.0$					& $<12.68$			& $10$	\\	
	& \ion{C}{III}	&	$0.0$					& $<12.15$			& $10$\\
	& \ion{N}{II}	&	$0.0$					& $<12.87$			& $10$ \\
	& \ion{N}{III}	&	$0.0$					& $<13.02$			& $10$ \\
	& \ion{N}{V}	&	$0.0$					& $<13.16$			& $30$ \\
	& \ion{O}{I}	&	$0.0$					& $<13.37$			& $10$ \\
	& \ion{O}{VI}	&	$0.0$					& $<12.92$			& $30$ \\
	& \ion{Si}{II}	&	$0.0$					& $<12.09$			& $10$	\\	
	& \ion{Si}{III}	&	$0.0$					& $<11.74$			& $10$	\\
	& \ion{Fe}{III}	&	$0.0$					& $<12.98$			& $10$	\\ 	
\hline
\end{tabular}}
\end{center}
\end{table}
\end{subtables}

\subsection{SDSS\, J1553$+$3548 at $d=156$ kpc}

\begin{subfigures}
\begin{figure*}
\includegraphics[scale=1.08]{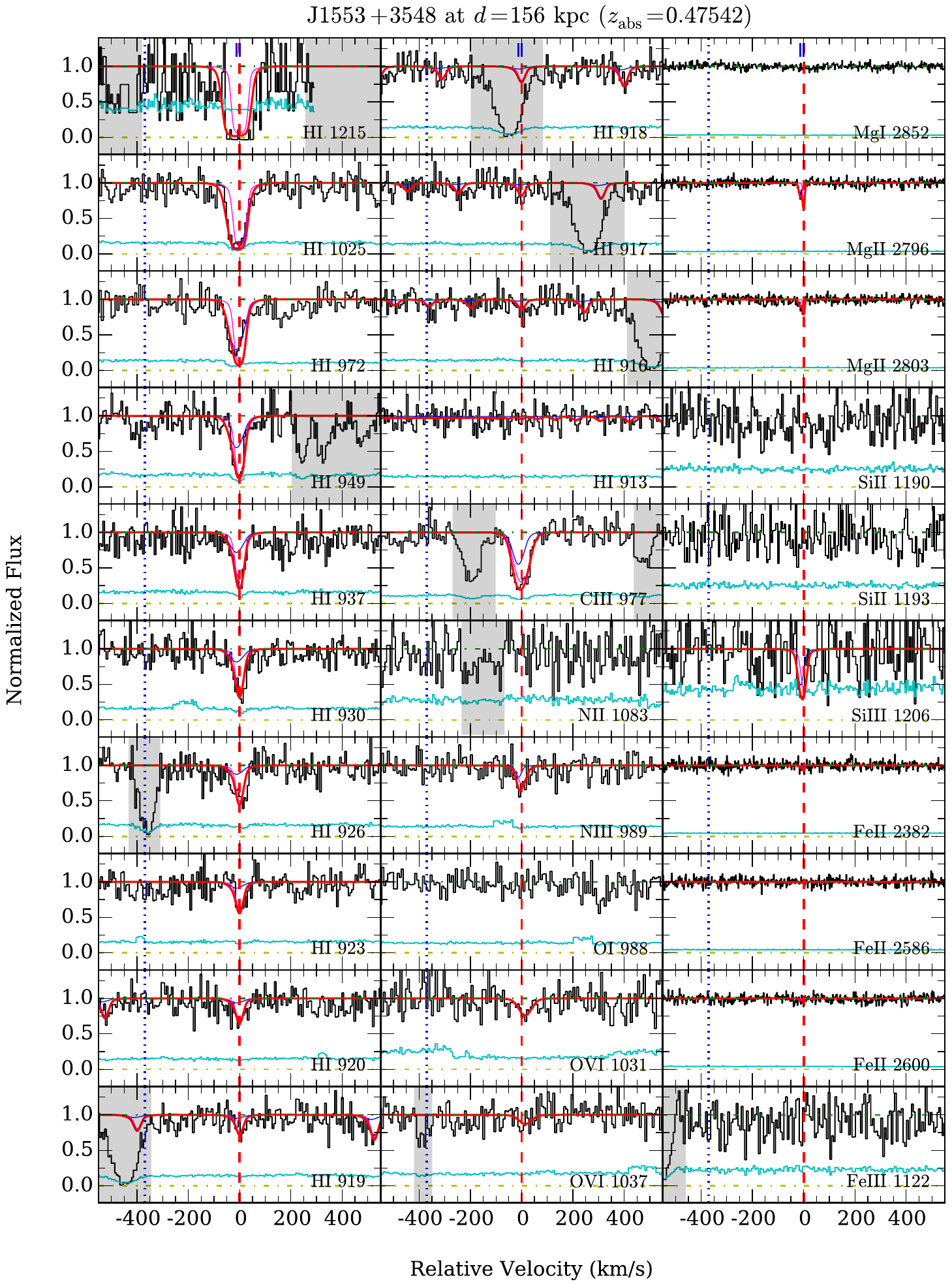}
\vspace{-0.75em}
\caption{Similar to Figure A1a, but for SDSS\,J1553$+$3548 at $d=156$ kpc from the LRG. Note that the 
The \ion{H}{I} $\lambda972$ line is likely contaminated by \ion{O}{III} $\lambda832$ and \ion{O}{II} $\lambda834$ emission lines from the background QSO.}
\label{figure:ions}
\end{figure*}

\begin{figure}
\includegraphics[scale=0.61]{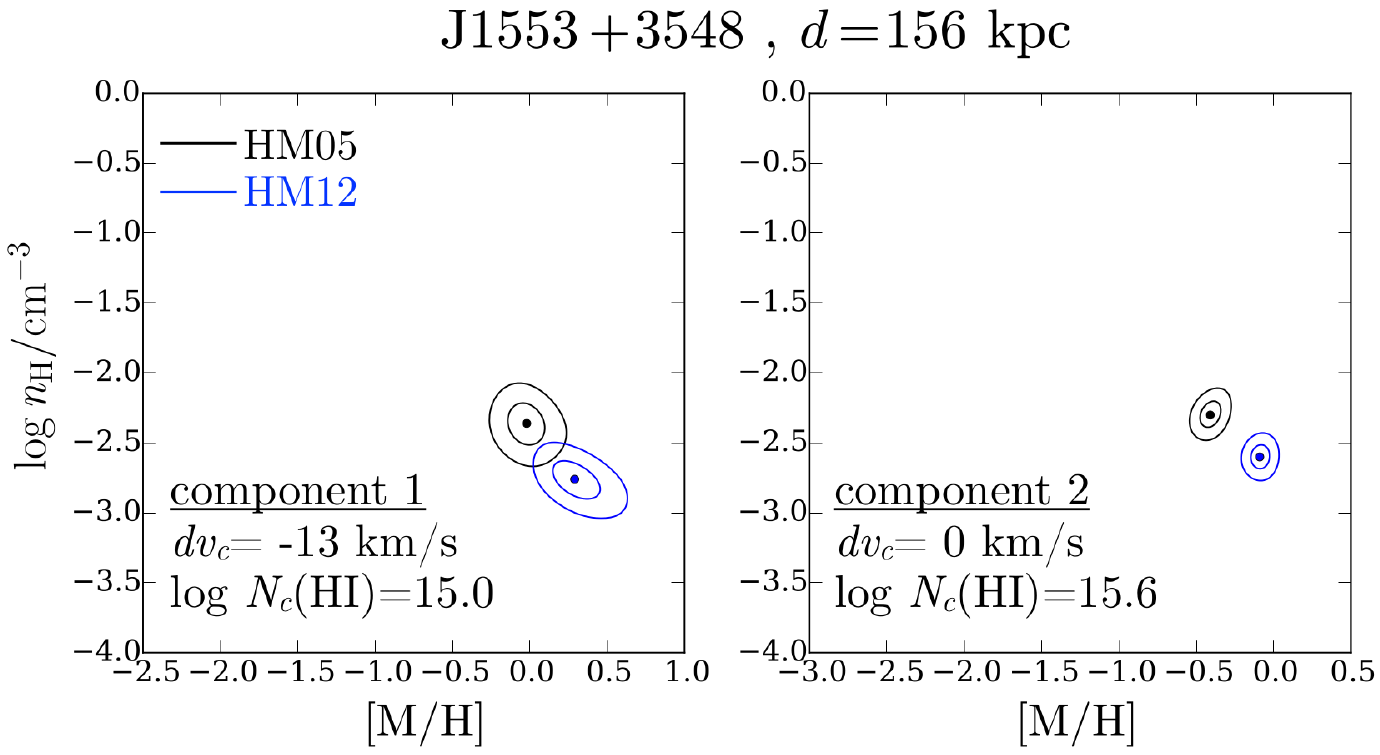}
\vspace{-1.5em}
\caption{Probability distribution contours of gas metallicity and density for individual absorption components identified along SDSS\,J1553$+$3548, at $d=156$ kpc from the LRG. Contour levels are the same as in Figure A1b.}
\label{figure:ions}
\end{figure}
\end{subfigures}

This LRG is at $z_\mathrm{LRG}=0.4736$. A \ion{H}{I} Lyman absorption series is present near the LRG redshift, with a total $N\mathrm{(\ion{H}{I})}$ of log\,$N\mathrm{(\ion{H}{I})/\cmjj}=15.69\pm0.04$. In addition to \ion{H}{I}, we detect corresponding absorption of metal species \ion{C}{III}, \ion{N}{III}, \ion{O}{VI}, \ion{Mg}{II}, \ion{Si}{III}, and \ion{Fe}{II} (Figure A14a).

As presented in Figure A14a and Table A14a, our Voigt profile analysis shows that the \ion{Mg}{II} absorption profile consists of two individual components which are separated by only 12 \kms\ in line-of-sight velocity. The stronger of these two components, component 2, occurs at $z_\mathrm{abs}=0.47542$ or $+370$ \kms\ from the LRG redshift. To accurately measure the corresponding $N\mathrm{(\ion{H}{I})}$ of each component given the close velocity separation between the two components, we perform a Voigt profile analysis on the \ion{H}{I} Lyman series absorption by tying the velocity structure of \ion{H}{I} absorption to that of \ion{Mg}{II}. The best-fit $N\mathrm{(\ion{H}{I})}$ for the two components are log\,$N_c\mathrm{(\ion{H}{I})/\cmjj}=15.04^{+0.08}_{-0.18}$ for component 1, and log\,$N_c\mathrm{(\ion{H}{I})/\cmjj}=15.57\pm0.07$ for component 2. 

The observed \ion{H}{I} and \ion{Mg}{II} Doppler linewidths for component 2 indicate that the absorbing gas is relatively cool, $T\approx1\times10^4$ K, with negligible non-thermal broadening. For the weaker component 1, the $b$ values imply a higher temperature,  $T\approx5\times10^4$ K, with a modest non-thermal line broadening of $b_\mathrm{nt}\sim5$\,\kms.

Our ionization analysis finds little variation ($<0.1$ dex) in gas densities across the two components, as shown in Figure A14b and Table A14b. For both components, the observed ionic column densities can be reproduced by models with log\,$n_\mathrm{H}/ \cmjjj\approx-2.3$ under HM05 and log\,$n_\mathrm{H}/ \cmjjj\approx-2.7$ under HM12. Furthermore, the gas appears to have been significantly enriched by heavy metals. For component 2, we find a sub-solar metallicity of $\mathrm{[M/H]}=-0.4\pm0.1$ under HM05 and $\mathrm{[M/H]}=-0.1\pm0.1$ under HM12. An even higher chemical abundance is inferred for component 1, where our \textsc{Cloudy} models constrain the metallicity to $\mathrm{[M/H]}=0.0\pm0.1$ under HM05 or $\mathrm{[M/H]}=+0.3\pm0.2$ under HM12. In addition, the strength of \ion{N}{III} absorption suggests that the gas is likely to be nitrogen-rich. The inferred $\mathrm{[N/\alpha]}$ ratios for the two components are $\mathrm{[N/\alpha]}= 0.3\pm0.3$ and $0.5\pm0.2$ for components 1 and 2, respectively.

A  modest absorption of \ion{O}{VI} gas is seen in this system at $dv_c=+10$ \kms. The \ion{O}{VI} absorption consists of a single component with log\,$N\mathrm{(\ion{O}{VI})/\cmjj}=13.69^{+0.13}_{-0.18}$ and a $b$ parameter of 29 \kms. No low- or intermediate-ionization metal  \ion{H}{I}  or component is found to match the \ion{O}{VI} absorption in velocity space (Figure A14a), with the \ion{O}{VI} doublet situated $10$ \kms\ away from the nearest low-ionization component.

\begin{subtables}
\begin{table}
\begin{center}
\caption{Absorption properties along QSO sightline SDSS\, J1553$+$3548 at $d=156$ kpc from  the LRG}
\hspace{-2.5em}
\vspace{-0.5em}
\label{tab:Imaging}
\resizebox{3.5in}{!}{
\begin{tabular}{clrrr}\hline
Component	&	Species		&\multicolumn{1}{c}{$dv_c$} 		& \multicolumn{1}{c}{log\,$N_c$}	&\multicolumn{1}{c}{$b_c$}		\\	
 			&				&\multicolumn{1}{c}{(km\,s$^{-1}$)}	&   		   					& \multicolumn{1}{c}{(km\,s$^{-1}$)}  \\ \hline \hline

all	& \ion{H}{I}	&	$...$					& $15.69\pm0.04$		& $...$ \\
	& \ion{C}{III}	&	$...$					& $13.78^{+0.08}_{-0.05}$& $...$	\\
	& \ion{N}{II}	&	$...$					& $<13.68$			& $...$ \\
	& \ion{N}{III}	&	$...$					& $13.87^{+0.12}_{-0.09}$& $...$	\\
	& \ion{O}{I}	&	$...$					& $<13.78$			& $...$ \\
	& \ion{O}{VI}	&	$...$					& $13.69^{+0.13}_{-0.18}$& $...$	\\
	& \ion{Mg}{I}	&	$...$					& $<10.80$			& $...$	\\
	& \ion{Mg}{II}	&	$...$					& $12.22^{+0.03}_{-0.02}$& $...$ \\
	& \ion{Si}{II}	&	$...$					& $<12.85$			& $...$	\\	
	& \ion{Si}{III}	&	$...$					& $13.00^{+0.24}_{-0.29}$& $...$\\	
	& \ion{Fe}{II}	&	$...$					& $11.81^{+0.08}_{-0.17}$& $...$	\\
	& \ion{Fe}{III}	&	$...$					& $<13.69$			& $...$	\\ \hline	
	
1	& \ion{H}{I}	&	$-12.3$				& $15.04^{+0.08}_{-0.17}$& $30.5^{+5.1}_{-3.2}$ \\
	& \ion{C}{III}	&	$-12.3$				& $13.23^{+0.22}_{-0.27}$& $25.1^{+8.6}_{-8.2}$	\\
	& \ion{N}{II}	&	$-12.3$				& $<13.48$			& 10	\\
	& \ion{N}{III}	&	$-12.3$				& $13.34^{+0.27}_{-0.37}$& $10.0\pm1.9$	\\
	& \ion{O}{I}	&	$-12.3$				& $<13.57$			& 10	\\
	& \ion{Mg}{I}	&	$-12.3$				& $<10.60$			& 10	\\
	& \ion{Mg}{II}	&	$-12.3\pm1.9$			& $11.90\pm0.05$		&  $8.0^{+1.8}_{-1.0}$	\\
	& \ion{Si}{II}	&	$-12.3$				& $<12.63$			& 10	\\	
	& \ion{Si}{III}	&	$-12.3$				& $12.67^{+0.29}_{-0.42}$& $12.0^{+4.3}_{-3.9}$ \\	
	& \ion{Fe}{II}	&	$-12.8^{+2.2}_{-1.4}$	& $11.47^{+0.13}_{-0.31}$& $7.1^{+5.1}_{-5.5}$ \\
	& \ion{Fe}{III}	&	$-12.3$				& $<13.51$			& 10	\\ \hline	

2	& \ion{H}{I}	&	$0.0^{+3.2}_{-2.0}$		& $15.57^{+0.07}_{-0.07}$& $14.2^{+3.1}_{-1.4}$ \\
	& \ion{C}{III}	&	$0.0$				& $13.63^{+0.11}_{-0.10}$& $31.9^{+3.7}_{-4.7}$ \\
	& \ion{N}{II}	&	$0.0$				& $<13.50$			& 10	\\
	& \ion{N}{III}	&	$0.0$				& $13.72^{+0.14}_{-0.19}$& $19.3^{+7.1}_{-5.7}$	\\
	& \ion{O}{I}	&	$0.0$				& $<13.62$			& 10	\\
	& \ion{Mg}{I}	&	$0.0$				& $<10.60$			& 10	\\
	& \ion{Mg}{II}	&	$0.0\pm0.6$			& $11.94\pm0.04$		& $3.0\pm0.7$ \\	
	& \ion{Si}{II}	&	$0.0$				& $<12.64	$			& 10 \\	
	& \ion{Si}{III}	&	$0.0$				& $12.73^{+0.21}_{-0.48}$& $11.1^{+6.4}_{-2.4}$ \\	
	& \ion{Fe}{II}	&	$0.3^{+1.5}_{-1.6}$		& $11.55^{+0.13}_{-0.24}$	& $4.5^{+4.7}_{-2.9}$	\\
	& \ion{Fe}{III}	&	$0.0$				& $<13.50$			& 10	\\ \hline
	
high-1	& \ion{O}{VI}	&	$+9.5\pm8.1$		& $13.69^{+0.13}_{-0.18}$& $29.2^{+28.5}_{-11.8}$ \\
\hline
\end{tabular}}
\end{center}
\end{table}

\begin{table}
\begin{center}
\caption{Ionization modeling results for the absorber along SDSS\, J1553$+$3548 at $d=156$ kpc from the LRG}
\hspace{-2.5em}
\label{tab:Imaging}
\resizebox{3.5in}{!}{
\begin{tabular}{@{\extracolsep{3pt}}ccrrrr@{}}\hline
Component	&$N_\mathrm{metal}$& \multicolumn{2}{c}{$\mathrm{[M/H]}$} 	& \multicolumn{2}{c}{$\mathrm{log\,}n_\mathrm{H}/\cmjjj$}		\\
\cline{3-4} \cline {5-6}
	& &\multicolumn{1}{c}{HM05}&	\multicolumn{1}{c}{HM12}	&\multicolumn{1}{c}{HM05} 	& 	\multicolumn{1}{c}{HM12}			\\	\hline \hline

SC	&5& $-0.28\pm0.04$			& $+0.05^{+0.03}_{-0.04}$		& $-2.28^{+0.04}_{-0.06}$		& $-2.60^{+0.04}_{-0.06}$ \\ \hline
1	&5& $-0.03^{+0.13}_{-0.11}$	& $+0.30^{+0.17}_{-0.14}$		& $-2.36^{+0.12}_{-0.16}$		& $-2.78\pm0.12$ \\
2	&5& $-0.41\pm0.06$			& $-0.08^{+0.05}_{-0.07}$			& $-2.30\pm0.08$			& $-2.60^{+0.06}_{-0.08}$	\\

\hline
\end{tabular}}
\end{center}
\end{table}
\end{subtables}

\subsection{SDSS\, J1259$+$4130 at $d=159$ kpc}

The LRG occurs at $z_\mathrm{LRG}=0.2790$. As shown in Figure A15a, no \ion{H}{I} absorption is detected within the adopted search window of $\pm500$ \kms\ from the LRG redshift. We are able to place a sensitive 2-$\sigma$ column density upper limit of log\,$N\mathrm{(\ion{H}{I})/\cmjj}<12.5$, calculated for an \ion{H}{I} line with $b\mathrm{(\ion{H}{I})}=15$ \kms\ that is centered at the LRG redshift (Table A15a).

\begin{subfigures}
\begin{figure*}
\includegraphics[scale=1.08]{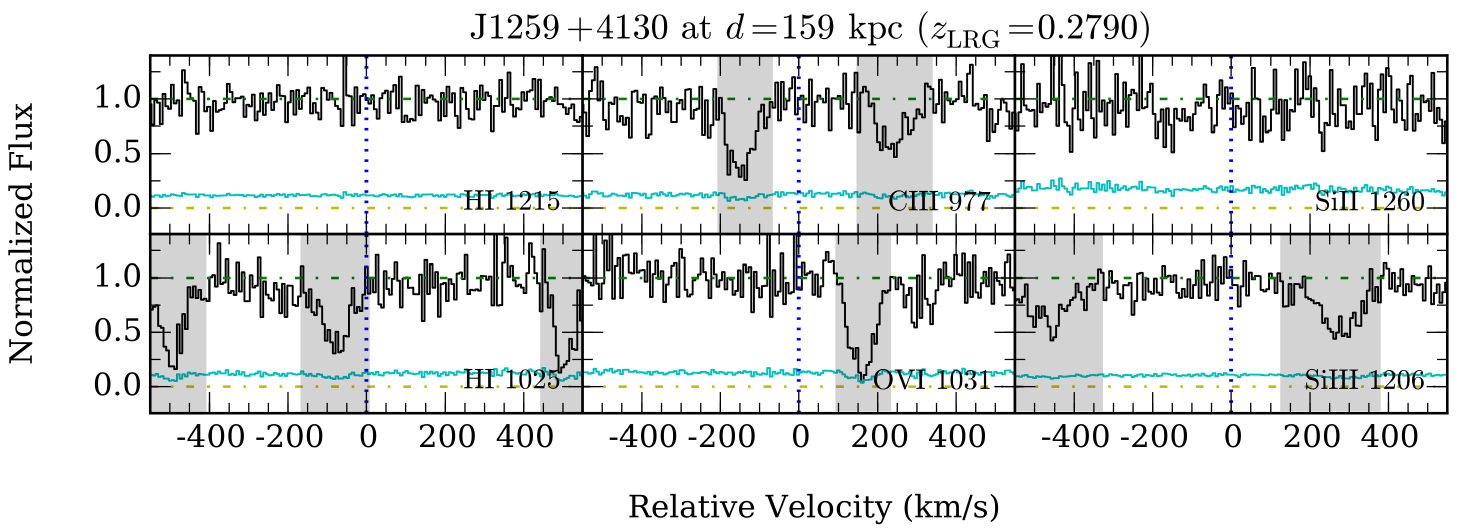}
\vspace{-0.75em}
\caption{Similar to Figure A1a, but for SDSS\,J1259$+$4130 at $d=159$ kpc from the LRG.}
\label{figure:ions}
\end{figure*}
\end{subfigures}

\begin{subtables}
\begin{table}
\begin{center}
\caption{Constraints on absorption properties along QSO sightline SDSS\,J1259$+$4130 at $d=159$ kpc from  the LRG}
\hspace{-2.5em}
\vspace{-0.5em}
\label{tab:Imaging}
\resizebox{3.3in}{!}{
\begin{tabular}{clcrc}\hline
Component	&	Species		&\multicolumn{1}{c}{$dv_c$} 		& \multicolumn{1}{c}{log\,$N_c$}	&\multicolumn{1}{c}{$b_c$}		\\	
 			&				&\multicolumn{1}{c}{(km\,s$^{-1}$)}	&   		   					& \multicolumn{1}{c}{(km\,s$^{-1}$)}  \\ \hline \hline

...	& \ion{H}{I}	&	$0.0$					& $<12.51$			& $15$ \\
	& \ion{C}{II}	&	$0.0$					& $<13.03$			& $10$	\\	
	& \ion{C}{III}	&	$0.0$					& $<12.33$			& $10$\\
	& \ion{N}{III}	&	$0.0$					& $<13.06$			& $10$ \\
	& \ion{N}{V}	&	$0.0$					& $<13.31$			& $30$ \\
	& \ion{O}{VI}	&	$0.0$					& $<13.27$			& $30$ \\
	& \ion{Si}{II}	&	$0.0$					& $<12.20$			& $10$	\\	
	& \ion{Si}{III}	&	$0.0$					& $<11.83$			& $10$	\\
	& \ion{Fe}{III}	&	$0.0$					& $<13.08$			& $10$	\\ 	
\hline
\end{tabular}}
\end{center}
\end{table}
\end{subtables}

\subsection{SDSS\, J1244$+$1721 at $d=160$ kpc}

\begin{subfigures}
\begin{figure*}
\includegraphics[scale=1.08]{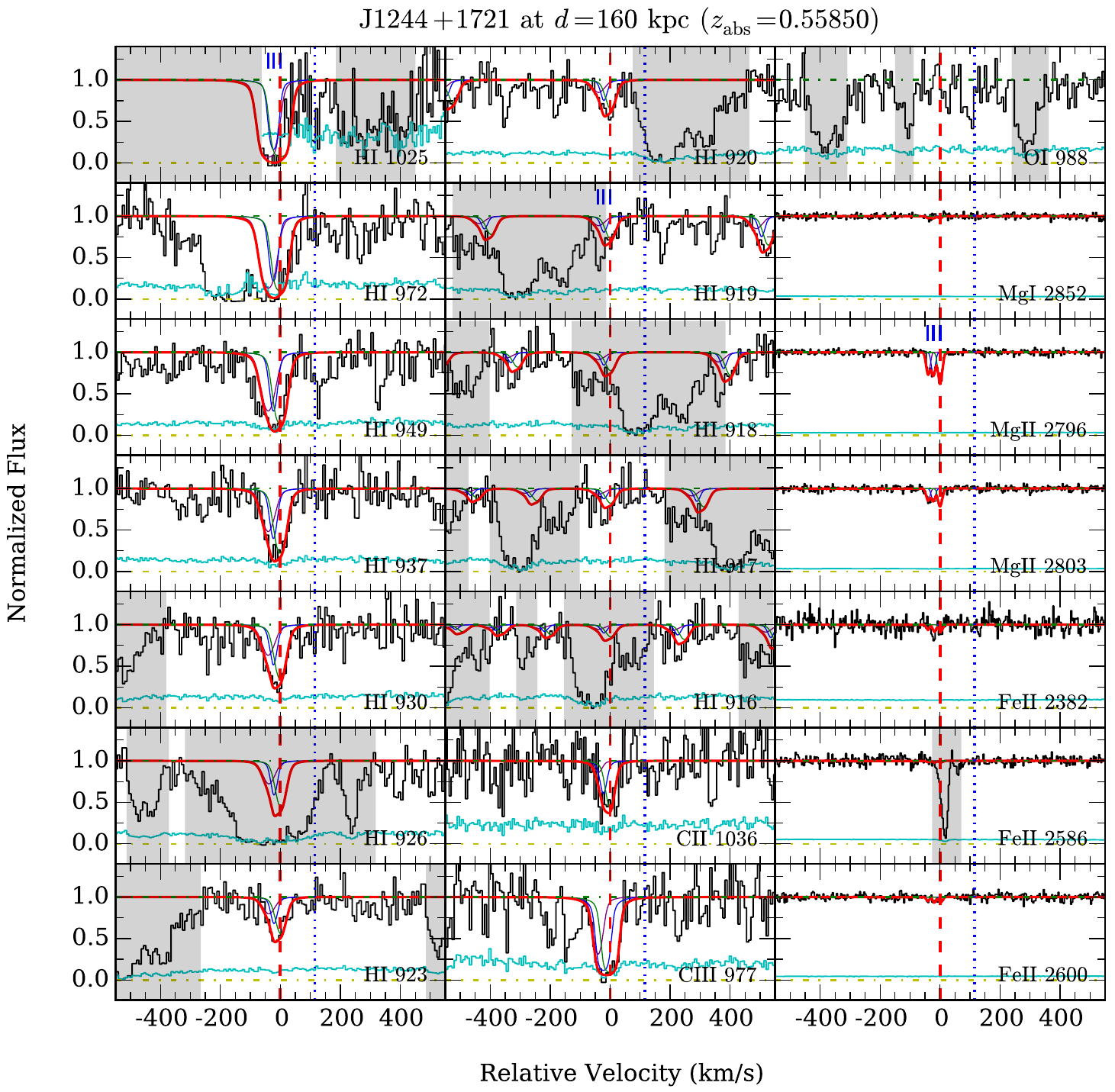}
\vspace{-0.75em}
\caption{Similar to Figure A1a, but for SDSS\,J1244$+$1721 at $d=160$ kpc from the LRG.}
\label{figure:ions}
\end{figure*}

\begin{figure}
\hspace{-0.8em}
\includegraphics[scale=0.62]{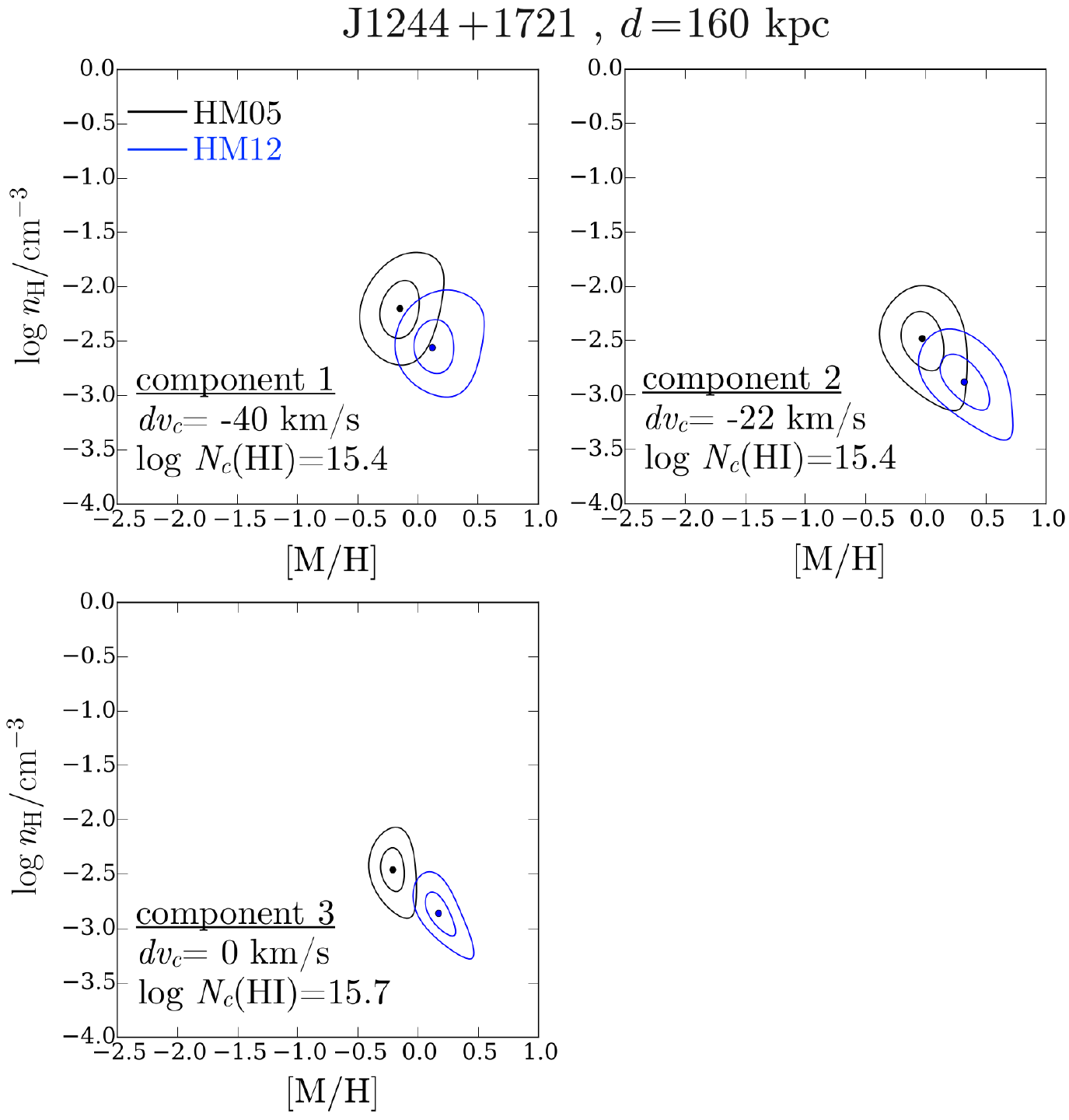}
\vspace{-1.5em}
\caption{Probability distribution contours of gas metallicity and density for individual absorption components identified along SDSS\,J1244$+$1721, at $d=160$ kpc from the LRG. Contour levels are the same as in Figure A1b.}
\label{figure:ions}
\end{figure}
\end{subfigures}

This LRG is located at $z_\mathrm{LRG}=0.5591$. We detect absorption from the \ion{H}{I} Lyman series near the LRG redshift, with a total $N\mathrm{(\ion{H}{I})}$ of log\,$N\mathrm{(\ion{H}{I})/\cmjj}=15.99\pm0.04$. In addition to \ion{H}{I}, the following metal species are present in absorption: \ion{C}{II}, \ion{C}{III}, \ion{Mg}{II}, and \ion{Fe}{II} (see Figure A16a). Due to the relatively high redshift of the LRG, all transitions above rest-frame $\lambda \approx 1140$\,\AA\ fall outside the wavelength coverage of the FUV COS spectrum, including Ly$\alpha$, \ion{Si}{II} $\lambda1260$ and \ion{Si}{II} $\lambda1206$. We further note that while there is possible \ion{N}{III} $\lambda 989$ absorption in this system, this transition is often blended/contaminated by the adjacent \ion{Si}{II} $\lambda989$ line. Because no other \ion{Si}{II} transition is covered by the data, we are unable to assess how much the \ion{N}{III} absorption is contaminated by \ion{Si}{II}. For that reason, we chose to exclude this possible \ion{N}{III} detection from our subsequent analysis of the absorber. 

Our Voigt profile analysis identifies three components in the absorber, which can be seen clearly in the \ion{Mg}{II} absorption profile (Figure A16a and Table A16a). The observed velocity spread of the absorber is $\Delta v\approx40$ \kms\ from the bluest to reddest component. The strongest \ion{H}{I} absorption occurs in component 3 at $z_\mathrm{abs}=0.55850$, or $-115$ \kms\ from the LRG. This component comprises half of the total $N\mathrm{(\ion{H}{I})}$ of the absorber, with the rest distributed equally between the other two components. As shown in Figure A16a, the component structure of \ion{H}{I} is in very good agreement with those of the metal ions (e.g., \ion{Mg}{II} and \ion{Fe}{II}). The ratio of the Doppler linewidths of \ion{H}{I} and \ion{Mg}{II} in each component is consistent with the expectation for a cool gas with $T\sim (1-3)\times10^4$ K and a modest amount of non-thermal line broadening, $b_\mathrm{nt}\approx6$\,\kms.

Our ionization analysis indicates a modest variation ($\sim0.3$ dex) in gas densities across the three components (see Figure A16b and Table A16b). For all components, the observed absorption profile can be reproduced by the models over a gas density range of from log\,$n_\mathrm{H}/ \cmjjj\approx-2.5$ to log\,$n_\mathrm{H}/ \cmjjj\approx-2.2$ under the HM05 UVB, and log\,$n_\mathrm{H}/ \cmjjj\approx-2.9$ to log\,$n_\mathrm{H}/ \cmjjj\approx-2.5$ under the HM12 UVB. The observed metal column densities in the absorber are consistent with a high degree of chemical enrichment. The inferred metallicities range from $\mathrm{[M/H]}=-0.2\pm0.1$ (component 3) to $\mathrm{[M/H]}=0.0\pm0.2$ (component 2) under HM05, and $\mathrm{[M/H]}=0.1\pm0.2$ (component 1) to $\mathrm{[M/H]}=0.3\pm0.2$ (component 2) under HM12. Finally, we note that the observed column densities of carbon ions imply that the absorber is carbon-rich, with estimated $\mathrm{[C/\alpha]}$ ratios of $\mathrm{[C/\alpha]}= 0.0-0.5$.

\begin{subtables}
\begin{table}
\begin{center}
\caption{Absorption properties along QSO sightline SDSS\, J1244$+$1721 at $d=160$ kpc from  the LRG}
\hspace{-2.5em}
\vspace{-0.5em}
\label{tab:Imaging}
\resizebox{3.5in}{!}{
\begin{tabular}{clrrr}\hline
Component	&	Species		&\multicolumn{1}{c}{$dv_c$} 		& \multicolumn{1}{c}{log\,$N_c$}	&\multicolumn{1}{c}{$b_c$}		\\	
 			&				&\multicolumn{1}{c}{(km\,s$^{-1}$)}	&   		   					& \multicolumn{1}{c}{(km\,s$^{-1}$)}  \\ \hline \hline

all	& \ion{H}{I}	&	$...$					& $15.99\pm0.04$		& $...$ \\
	& \ion{C}{II}	&	$...$					& $14.26^{+0.15}_{-0.10}$& $...$	\\	
	& \ion{C}{III}	&	$...$					& $>14.42$			& $...$	\\
	& \ion{O}{I}	&	$...$					& $<13.83$			& $...$ \\
	& \ion{O}{VI}	&	$...$					& $<13.74$			& $...$ \\
	& \ion{Mg}{I}	&	$...$					& $<10.88$			& $...$	\\
	& \ion{Mg}{II}	&	$...$					& $12.61\pm0.02$		& $...$ \\
	& \ion{Fe}{II}	&	$...$					& $12.26^{+0.05}_{-0.20}$& $...$	\\ \hline
	
1	& \ion{H}{I}	&	$-39.9$				& $15.35^{+0.07}_{-0.21}$& $23.8^{+10.5}_{-5.5}$ \\
	& \ion{C}{II}	&	$-39.9$				& $<13.25$			& 10 \\
	& \ion{C}{III}	&	$-39.9$				& $13.50^{+0.20}_{-0.36}$& $15.0^{+9.3}_{-2.8}$	\\
	& \ion{O}{I}	&	$-39.9$				& $<13.66$			& 10	\\
	& \ion{Mg}{I}	&	$-39.9$				& $<10.60$			& 10	\\
	& \ion{Mg}{II}	&	$-39.9\pm1.4$			& $12.04\pm0.07$		&  $6.2^{+4.4}_{-2.2}$	\\
	& \ion{Fe}{II}	&	$-40.7\pm2.9$			& $11.67^{+0.14}_{-0.46}$& $3.5^{7.3}_{-1.8}$ \\ \hline

2	& \ion{H}{I}	&	$-22.4$				& $15.38^{+0.23}_{-0.22}$& $11.6^{+5.5}_{-3.0}$ \\
	& \ion{C}{II}	&	$-22.4$				& $13.73^{+0.36}_{-0.27}$& $10.0^{+10.2}_{-2.5}$ \\		
	& \ion{C}{III}	&	$-22.4$				& $>13.62$			& $<25.6$ \\
	& \ion{O}{I}	&	$-22.4$				& $<13.65$			& 10	\\
	& \ion{Mg}{I}	&	$-22.4$				& $<10.60$			& 10	\\
	& \ion{Mg}{II}	&	$-22.4\pm1.1$			& $12.06\pm0.05$		& $6.1^{+2.0}_{-1.2}$ \\	
	& \ion{Fe}{II}	&	$-22.4$				& $11.81^{+0.12}_{-0.44}$	& $4.0^{+7.2}_{-1.9}$	\\ \hline

3	& \ion{H}{I}	&	$0.0$				& $15.71^{+0.06}_{-0.11}$& $19.0^{+4.3}_{-2.2}$ \\
	& \ion{C}{II}	&	$0.0$				& $14.05^{+0.17}_{-0.19}$& $24.3^{+7.5}_{-6.5}$ \\		
	& \ion{C}{III}	&	$0.0$				& $>13.88$			& $<24.2$ \\
	& \ion{O}{I}	&	$0.0$				& $<13.70$			& 10	\\
	& \ion{Mg}{I}	&	$0.0$				& $<10.60$			& 10	\\
	& \ion{Mg}{II}	&	$0.0\pm0.4$			& $12.27\pm0.03$		& $7.0\pm0.7$ \\	
	& \ion{Fe}{II}	&	$1.5\pm2.2$			& $11.86^{+0.11}_{-0.32}$	& $5.8^{+5.1}_{-3.3}$	\\
	
\hline
\end{tabular}}
\end{center}
\end{table}

\begin{table}
\begin{center}
\caption{Ionization modeling results for the absorber along SDSS\, J1244$+$1721 at $d=160$ kpc from the LRG}
\hspace{-2.5em}
\label{tab:Imaging}
\resizebox{3.5in}{!}{
\begin{tabular}{@{\extracolsep{3pt}}ccrrrr@{}}\hline
Component	&$N_\mathrm{metal}$& \multicolumn{2}{c}{$\mathrm{[M/H]}$} 	& \multicolumn{2}{c}{$\mathrm{log\,}n_\mathrm{H}/\cmjjj$}		\\
\cline{3-4} \cline {5-6}
	& &\multicolumn{1}{c}{HM05}&	\multicolumn{1}{c}{HM12}	&\multicolumn{1}{c}{HM05} 	& 	\multicolumn{1}{c}{HM12}			\\	\hline \hline

SC	&4& $-0.14^{+0.03}_{-0.04}$	& $+0.21^{+0.05}_{-0.04}$		& $-2.40\pm0.12$			& $-2.84^{+0.10}_{-0.12}$ \\ \hline
1	&3& $-0.14^{+0.19}_{-0.16}$	& $+0.15^{+0.24}_{-0.15}$		& $-2.20\pm0.26$			& $-2.54^{+0.26}_{-0.22}$ \\
2	&4& $0.00^{+0.19}_{-0.18}$	& $+0.33^{+0.22}_{-0.18}$		& $-2.50^{+0.22}_{-0.32}$		& $-2.86^{+0.22}_{-0.26}$	\\	
3	&4& $-0.20\pm0.10$			& $+0.17^{+0.14}_{-0.09}$		& $-2.46^{+0.16}_{-0.22}$		& $-2.86^{+0.16}_{-0.20}$	\\

\hline
\end{tabular}}
\end{center}
\end{table}
\end{subtables}

\label{lastpage}

\end{document}